# Wide Binary Stars in the Galactic Field

## A Statistical Approach

**Inauguraldissertation**

zur
Erlangung der Würde eines Doktors der Philosophie
vorgelegt der
Philosophisch-Naturwissenschaftlichen Fakultät
der Universität Basel
von

Marco Longhitano

aus Reinach, BL

Genehmigt von der Philosophisch-Naturwissenschaftlichen Fakultät

auf Antrag von Prof. Dr. Bruno Binggeli und Dr. Jean-Louis Halbwachs

Basel, den 21. September 2010

Prof. Dr. Martin Spiess
Dekan

*Per Nicole,*

*che mi fatto vedere*
*la bellezza delle scienze umane.*

# Contents

# Zusammenfassung

Diese Arbeit untersucht die statistischen Eigenschaften von weiten Doppelsternsystemen (WB) im Galaktischen Feld. Mit Separationen von über 200 AU und ihrer folglich geringen Bindungsenergien, reagieren WB empfindliche auf gravitative Störungen. Dies macht sie zu einem interessanten Werkzeug, um Rückschlüsse auf die Natur der dunklen Materie (DM) in unserer Galaxie zu ziehen. Für unsere Studie wählen wir einen knapp 675 Quadratgrad grossen Himmelsausschnitt in Richtung des nördlichen Galaktischen Pols. Dieser enthählt etwa 670 000 Hauptreihensterne mit scheinbaren Helligkeiten zwischen 15 und 20.5 mag und Spektralklassen später als G5. Die Daten stammen vom Sloan Digital Sky Survey. Wir konstruieren die Zweipunkt Korrelationsfunktion (2PCF) für Winkelseparationen zwischen 2 und 30 Bogensekunden. Das resultierende Signal wird mit Hilfe der Wasserman-Weinberg Technik modelliert. Wir zeigen, dass die Verteilung der grossen Halbachsen konsistent ist mit Öpiks Gesetz und leiten ab, dass etwa 10% aller sonnennaher Sterne Mitglied eines WBs sind. Die 2PCF-Methode ist allerdings stark eingeschränkt durch das von optischen Paaren verursachte statistische Rauschen; besonders bei den weitesten Systemen, die die vielversprechensten Rückschlüsse auf die DM zuliessen. Um das Rauschen zu reduzieren und die Empfindlichkeit unserer Analyse bei grösseren Separationen zu steigern, rechnen wir Distanzinformationen von photometrischen Parallaxen mit ein. Wir führen ein neuartiges Gewichtungsverfahren ein, das auf der Bindungswahrscheinlichkeit eines gegebenen Paares beruht. Damit leiten wir die Verteilung der Farben und der Massenverhältnisse ab, wobei wir Auswahleffekte sorgfältig berücksichtigen. Statistisch wurden etwa 4 000 WBs mit Massen zwischen 0.2 und 0.85 Sonnenmassen in unserer Analyse mit einbezogen. Wir stellen fest, dass deren Farbverteilung mit jener der Feldeinzelsterne übereinstimmt. Es scheint jedoch, dass Paare mit einer Massendifferenz von über 0.5 Sonnenmassen verglichen mit einer Zufallspaarung von Feldsternen systematisch unterrepräsentiert sind. Wir haben eine "Rankliste" von WB-Kandidaten zusammengestellt, die sich für Folgestudien als nützlich erweisen könnte. Aufgrund fehlender Daten über die relativen Geschwindigkeiten der Paare, konnte die Einschränkung betreffend der weitesten Systeme nicht vollständig überwunden werden. Ein weiterer Nachteil unseres Ansatzes ist die Notwendigkeit eines komplizierten Models, um Auswahleffekten Rechnung zu tragen. Andererseits gelang es uns die Einschränkungen von Studien, die sich auf die Eigenbewegungen stützen, zu umgehen. Das neuartige Verfahren, das in der vorliegenden Arbeit vorgestellt wird, kann daher als zu den Eigenbewegungsstudien komplementär angesehen werden und stellt eine gangbare Herangehensweise zur Erforschung weiter Doppelsterne dar.

# Abstract

This thesis focuses on the statistical properties of wide binary (WB) star systems in the Galactic field. With projected separations larger than 200 AU and, consequently, having very low binding energies, WB are sensitive probes of the Galactic gravitational potential making them an interesting tool to constrain the dark matter (DM) content in the Milky Way Galaxy. For the present study we select a homogeneous sample covering about 675 square degrees in the direction of the Northern Galactic Pole. It contains nearly 670 000 main sequence stars with apparent magnitudes between 15 and 20.5 mag and spectral classes later than G5. The data were taken from the Sloan Digital Sky Survey. We construct the two-point correlation function (2PCF) for angular separations between 2 and 30 arcseconds. The resulting clustering signal is modeled by means of the Wasserman-Weinberg technique. We show that the distribution of semi-major axis is consistent with the canonical Öpik law and infer that about 10% of all stars in the solar neighbourhood belong to a WB system. The 2PCF method is, however, seriously limited by the noise from optical pairs, especially for the widest systems, which would provide the most stringent constraints of the DM's nature. To reduce the noise from optical pairs and to increase the sensitivity of the analysis at larger separations, we include distance information from photometric parallaxes. Introducing a novel weighting procedure based on the binding probability of a double star, we infer the distribution of colours and mass ratios, which were carefully corrected for observational selection effects. About 4 000 WBs were taken into account statistically, whose components have masses between 0.2 and 0.85 solar masses. We find that the WB colour distribution is in accord with the colour distribution of single field stars. However, pairs with a mass difference exceeding 0.5 solar masses seem to be systematically underrepresented as compared to a random pairing of field stars. Our results are broadly in agreement with prior studies but a direct comparison is often difficult or even impossible. We compiled a 'ranked list' of WB candidates that will prove to be useful for follow-up studies. Due to lack of information about the relative velocities of the pairs, the limitation concerning the widest pairs could not be entirely overcome. A further drawback of our approach is clearly the need for a sophisticated modeling to allow for selection effects. The method, however, successfully circumvents the limitations of studies based on proper motions. The novel procedure presented in this thesis can therefore be regarded as complementary to common proper motion studies, and constitutes a viable approach to study the statistical properties of WBs in the Galactic field.

# Vorwort

In den vergangenen vier Jahren hatte ich das Glück ein Ideal der wissenschaftlichen Arbeit zu leben, wie es heute, so scheint mir, immer seltener anzutreffen ist. Die humanistische Tradition der Stadt Basel und ihre damit verbundene Geschichte des freien Denkens ist an der ältesten Universität der Schweiz noch immer spürbar. Frei von wirtschaftlichen und politischen Interessen, fern von Karriereplanung und – dank des Schweizerischen Nationalfonds – ohne grössere finanzielle Sorgen hatte ich das Privileg mich voll und ganz auf die Forschung konzentrieren zu können. Was mich zum Doktoratsstudium in Astronomie bewegte, kann ich nicht mehr genau sagen. Vielleicht war es die Faszination, die von diesem Fachgebiet ausgeht und mein Interesse für Naturphänomene aller Art, welches mich seit Kindestagen begleitet; vielleicht die tiefe Sympathie, die ich seit den Anfängen meines Studiums für meinen Doktorvater Herrn Prof. Dr. Bruno Binggeli hege; oder aber vielleicht das Erhabene, das vom gestirnten Himmel und den dahinter liegenden scheinbar unendlichen Weiten des Weltalls ausgeht und mich mit tiefer Ehrfurcht erfüllt.

Dass ich hier die universitäre Unabhängikeit hervorgehoben habe, soll nicht ein Desinteresse oder gar eine ablehnende Haltung meinerseits gegenüber weltlichen Belangen andeuten. Eine langfristig stabile wirtschaftliche und politische Lage ist sicherlich eine Voraussetzung um Grundlagenforschung betreiben zu können und eine vielseitige universitären Bildung zu ermöglichen. Gleichzeitig glaube ich, dass es der grösstmöglichen akademischen Freiheit bedarf, um das volle wissenschaftliche Potential nutzbar zu machen. Die bildungspolitische Schwerpunktsetzung birgt die Gefahr der Verarmung der Hochschullandschaft. Die damit verbundene Reduktion der fachlichen Diversität geht auf Kosten von Fachgebieten, die von Natur aus weniger darauf bedacht sind in näherer Zukunft vermarktbare Produkte hervorzubringen und somit einen absehbaren finanziellen Profit abzuwerfen. Es tut mir sehr Leid, dass das ehemalige Astronomische Institut der Universität Basel dem Sog dieser Reformen nicht standhalten konnte und darin unter ging.

Von dieser tristen Episode abgesehen, konnte ich mir keine besseren Bedingungen für mein Doktoratsstudium wünschen, als jene die mir in den letzten vier Jahren zuteil wurden. Massgeblich verantwortlich dafür war Prof. Binggeli. Er ermunterte mich von Anfang an einen breiten Zugang zur Astronomie zu wählen und öffnete mir die Augen für die zahlreichen Verflechtungen der verschiedenen wissenschaftlichen Zweige. Zu jeder Zeit konnte ich auf Prof. Binggelis Erfahrung und Wissen zurückgreifen; nie verspürte ich von seiner Seite einen Druck schneller mit der Arbeit voranzukommen oder baldmöglichst zu veröffentlichen; immer stand die Qualität der Arbeit und die wissenschaftliche Integrität im Vordergrund. Ich bedanke mich an dieser Stelle ganz herzlich bei Prof. Binggeli für die viele Zeit, die er mir widmete, die unzähligen, spannenden Gespräche, die oft weit über die Astronomie hinaus gingen, und seine Wärme und Herzlichkeit, die er mir immer entgegenbrachte. Vielen Dank auch an den externen Begutachter, Dr. Jean-Louis Halbwachs vom Observatoire de Strasbourg, der gleich Interesse an meiner Arbeit zeigte und sich dazu bereit erklärte das Korreferat zu übernehmen.

In den letzten vier Jahren habe ich nicht nur Wissenschaft betrieben, sondern – was vielleicht noch wichtiger war – etwas über die Wissenschaft gelernt. Die meiner Ansicht nach wichtigste Lektion scheint zunächst eine banale Erkenntnis zu sein: Wissenschaft wird von Menschen gemacht. Bevor ich das Doktoratsstudium begann, hatte ich eine romantische, beinahe puristische Vorstellung von der Wissenschaft als vom Menschlichen völlig losgelöste, idealistische Unternehmung zur Auffindung einer verborgenen, absoluten Wahrheit über die Natur. Der wissenschaftliche Betrieb spiegelt jedoch die menschliche Natur wider – mit all ihren Stärken und Schwächen, ihren Höhen und Tiefen. Ob den Naturgesetzen selbst eine vom Beobachter unabhängige Existenz zugeschrieben werden kann, möchte ich mir nicht anmassen zu urteilen. Doch der gesamte Prozess zur Auffindung von Naturgesetzen sowie deren Formulierung – Wissenschaft eben – ist von Menschen gemacht. Diese Erkenntnis hat meine Meinung über die Wissenschaft keineswegs geschmälert. Vielleicht ist sie die viel gesuchte Wendeltreppe, die aus dem Elfenbeinturm führt.

Viele Menschen haben bei der Entstehung der vorliegenden Arbeit durch anregende Diskussionen und Ratschläge, durch konstruktive Kritik und aufmunternden Worte, oder einfach durch entspannte Plaudereien über Gott und die Welt mitgewirkt. Vielen herzlichen Dank an Karin Ammon, Andreas Aste, Selçuk Bilir, Christoph Bruder, Roland Buser, Heinz Breitenstein, Daniel Cerrito, Isabelle Cherchneff, Stefano Chesi, Didier Curty, François Erkadoo, Jan Fischer, Tobias Fischer, Urs Frischkecht, Peter M. Garnavich, Alfred Gautschy, Kuno Glanzmann, Katharina Glatt, Beat Glatz, Eva K. Grebel, Bernd Heimann, Kai Hencken, Helmut Jerjen, Katrin Jordi, Jürg Jourdan, Astrid Kalt, Barbara Kammermann, Roger Käppeli, Salih Karaali, Stefan Kautsch, Andrea Kayser, Ralf Klessen, Andreas Koch, Bernd Krusche, Thijs Kouwenhoven, Matthias Liebendörfer, Thorsten Lisker, Wolfgang Löffler, Phani Peddibhotla, Albino Perego, Damien Quinn, Thomas Rauscher, Peter Reimann, Beat Röthlisberger, Niranjan Sambhus, Simon Scheidegger, Ingo Sick, Michael Steinacher, Roland Steiner, Gustav A. Tammann, Karl-Friedrich Thielemann, Dirk Trautmann, Mircea Trief, Cyrill von Arx, Kevin van Hoogdalem, Pieter Westera, Stuart Whitehouse, Alex Willand, Christian Winteler, Tobias Zesiger und Tobias Zingg.

Ganz herzlich bedanke ich mich bei Nicole Peduzzi, die immer bei mir war und auch in schwierigen Momenten es stets verstand mir Kraft zu geben. Ihre zahllosen Ratschläge und Anregungen sind an jeder Stelle dieser Arbeit zu finden und ihr sorgfältiges Lektorat der gesamten Arbeit ist von unschätzbarem Wert. Diese Dissertation ist ihr gewidmet. An dieser Stelle sei auch Alida, Dante und Stefano Peduzzi gedankt, deren Herzlichkeit und Gastfreundschaft ich in den vergangenen vier Jahren so oft in Anspruch nehmen durfte. Ein nicht unwesentlicher Teil der vorliegenden Arbeit wurde bei ihnen in Cama (GR) geschrieben. Auch meiner Familie gebührt grösster Dank. Besonders möchte ich meinem

Onkel, Norbert Klenner, und seiner Familie danken, die mich zum Physikstudium inspiriert und motiviert haben. Nicht zuletzt bin ich meinen Eltern, Lucie und Giuseppe Longhitano, die immer an mich glaubten, zu grösstem Dank verpflichtet. Sie ermöglichten mir das Studium und unterstützten mich stets auf jede erdenkliche Weise. Ich möchte hier auch meiner Grossmutter, Anna Klenner, herzlich danken, die immer zu mir hielt. Sie durfte den Abschluss dieser Arbeit leider nicht mehr erleben. Schliesslich sei hier all meinen lieben Freunden und Verwandten gedankt, die mir auf so viele Arten halfen und mir immer zur Seite standen.

Die vorliegende Arbeit stützt sich auf Daten des Sloan Digital Sky Surveys (SDSS), dessen Finanzierung von der Alfred P. Sloan Foundation, den teilnehmenden Institutionen, der NASA, der National Science Foundation, dem US Energieministerium, dem Japanischen Ministerium für Bildung, Kultur, Sport, Wissenschaft und Technologie (Monbukagakusho) und der Max-Planck-Gesellschaft. Die Internetadresse des SDSS lautet `http://www.sdss.org/`.

Der SDSS wird vom Astrophysical Research Consortium (ARC) und den teilnehmenden Institutionen verwaltet. Die teilnemenden Institutionen sind die University of Chicago, das Fermilab, das Institute for Advanced Study, die Japan Participation Group, die Johns Hopkins University, das Los Alamos National Laboratory, das Max-Planck-Institut für Astronomie (MPIA), das Max-Planck-Institut für Astrophysik (MPA), die New Mexico State University, die University of Pittsburgh, die Princeton University, das United States Naval Observatory und die University of Washington.

Während meiner Forschungsarbeit machte ich umfassenden Gebrauch vom Astrophysics Data System (ADS) der NASA, von der Suchmaschine Google und von Wikipedia. Weiterhin nahm ich die Dienste der Universitätsbibliothek und des Universitätsrechenzentrum Basel in Anspruch, die den Zugriff auf eine Vielzahl von elektronischen Fachzeitschriften gewähren und eine moderne und einwandfrei funktionierende IT-Infrastruktur zur Verfügung stellen. Ein Teil der während dieser Arbeit generierten Daten wurde beim Centre de Données astronomiques de Strasbourg (CDS) veröffentlicht. Diese Arbeit wurde unterstützt durch den Schweizerischen Nationalfond, der Schweizerischen Gesellschaft für Astrophysik und Astronomie und der Jubiläumstiftung der Basellandschaftlichen Kantonalbank.

Marco Longhitano
*Basel, September 2010*

# Preface

During the past four years I was lucky enough to live an ideal of scientific work that nowadays, I think, is becoming rare. The humanistic tradition of the city of Basel is still noticeable at the oldest university of Switzerland. Free from economic and political interests, far from career planning and – thanks to the Swiss National Science Foundation – without major financial worries, I had the privilege to be able to fully concentrate on research. I cannot say precisely what has motivated me to start a doctorate in astronomy. Maybe it was the fascination for this field and my interest in all kinds of natural phenomena that had been accompanying me since childhood. Maybe it was the deep sympathy towards my doctoral advisor Prof. Dr. Bruno Binggeli, or perhaps also the sublime coming from the starry sky and the seemingly endless expanse of the Universe that strikes me with deep reverence.

That I have emphasised here the universities' independency should not be interpreted as a lack of interest or even a negative attitude towards worldly matters. A long-term stable economic and political situation is certainly a prerequisite to make fundamental research possible and to enable a variegated education at the universities. At the same time I think that, to make the full scientific potential available, it is necessary that the highest possible academic freedom is guaranteed. The setting of educational priorities holds the danger of the impoverishment of higher education and is connected to a reduction of professional diversity, penalising those disciplines that by nature are less determined to produce marketable products in the short run, and to yield a foreseeable financial profit. I am sorry that the former Astronomical Institute of the University of Basel could not withstand these reforms and was forced to close its doors.

Apart from this sad episode, I could not have wished better conditions for my doctoral studies, and this is mainly to ascribe to Prof. Binggeli. From the beginning he encouraged me to choose a broad approach to astronomy and indicated the multitude of interdependencies among the various scientific branches. At any time I could resort to Prof. Binggelis experience and knowledge; he never put me under pressure to progress more quickly or to publish extensively. It was the research quality and the scientific integrity that stood to the fore. I would like to thank Prof. Binggeli for the time he devoted to me, for the countless fascinating conversations that often went far beyond astronomy, and for his warmth and cordiality. Thanks also to the external assessor, Dr Jean-Louis Halbwachs from the Observatoire de Strasbourg, who immediately showed interest in my work and agreed to be the second examiner.

Over the last four years I have pursued not only science, but – perhaps more importantly – I have learned something *about* science. My principal lesson in this regard was an apparently trivial insight: that science is made by human beings. Before I began the doctorate, I had a romantic, almost puristic idea of science as something completely disconnected from the human, idealistic enterprise aimed at discovering a hidden, absolute truth about nature. Scientific operations, however, do reflect human nature with all its strengths and weaknesses, its ups and downs. Whether the laws of nature themselves exist independently from the observer, I would not dare to judge. Yet the whole process of discovery of natural laws and their formulation is made by human beings. This insight has not diminished my opinion of science at all. Perhaps this even constitutes the much sought spiral staircase leading out of the ivory tower.

Many people have participated in the development of this work with stimulating discussions and advice, constructive criticism and encouraging words, or simply with relaxing chats about everything under the Sun. Many thanks to Karin Ammon, Andreas Aste, Selçuk Bilir, Christoph Bruder, Roland Buser, Heinz Breitenstein, Daniel Cerrito, Isabelle Cherchneff, Stefano Chesi, Didier Curty, François Erkadoo, Jan Fischer, Tobias Fischer, Urs Frischkecht, Peter M. Garnavich, Alfred Gautschy, Kuno Glanzmann, Katharina Glatt, Beat Glatz, Eva K. Grebel, Bernd Heimann, Kai Hencken, Helmut Jerjen, Katrin Jordi, Jürg Jourdan, Astrid Kalt, Barbara Kammermann, Roger Käppeli, Salih Karaali, Stefan Kautsch, Andrea Kayser, Ralf Klessen, Andreas Koch, Bernd Krusche, Thijs Kouwenhoven, Matthias Liebendörfer, Thorsten Lisker, Wolfgang Löffler, Phani Peddibhotla, Albino Perego, Damien Quinn, Thomas Rauscher, Peter Reimann, Beat Röthlisberger, Niranjan Sambhus, Simon Scheidegger, Ingo Sick, Michael Steinacher, Roland Steiner, Gustav A. Tammann, Karl-Friedrich Thielemann, Dirk Trautmann, Mircea Trief, Cyrill von Arx, Kevin van Hoogdalem, Pieter Westera, Stuart Whitehouse, Alex Willand, Christian Winteler, Tobias Zesiger und Tobias Zingg.

I sincerely thank Nicole Peduzzi, who was always with me and supported me in difficult moments. Her countless tips and suggestions can be found at any point in this work and the careful proofreading of the entire manuscript is of inestimable value. I dedicated this work to her. Many thanks also to Alida, Dante and Stefano Peduzzi, for their warmth and hospitality. A significant part of this work has been written at their house in Cama (GR). My family deserves the greatest thanks. I would particularly like to thank my uncle, Norbert Klenner, and his family, who inspired and motivated me to study physics. I am deeply indebted to my parents, Lucie and Giuseppe Longhitano, who always believed in me. They enabled me to study and have always supported me in every possible way. I would also like to thank my grandmother, Anna Klenner, who always stood by me. Unfortunately, she could not live to see the completion of this work. Finally, I thank all my dear friends and relatives, who helped me in so many ways and were always on my side.

Funding for the Sloan Digital Sky Survey (SDSS) has been provided by the Alfred P. Sloan Foundation, the Participating Institutions, the National Aeronautics and Space Administration, the National Science Foundation, the U.S. Department of Energy, the Japanese Monbukagakusho, and the Max Planck Society. The SDSS Web site is http://www.sdss.org/.

The SDSS is managed by the Astrophysical Research Consortium (ARC) for the Participating Institutions. The Participating Institutions are The University of Chicago, Fermilab, the Institute for Advanced Study, the Japan Participation Group, The Johns Hopkins University, Los Alamos National Laboratory, the Max-Planck-Institute for Astronomy (MPIA), the Max-Planck-Institute for Astrophysics (MPA), New Mexico State University, University of Pittsburgh, Princeton University, the United States Naval Observatory, and the University of Washington.

This research has made use of NASA's Astrophysics Data System Bibliographic Services, the Google search engine and Wikipedia. Furthermore, I utilised the services of the library and the Computing Center of the Basel University, who grant access to a variety of electronic journals and provide a modern and fully functioning IT infrastructure. Some of the data generated during this work were published at the Centre de Données astronomiques de Strasbourg (CDS). This work was supported by the Swiss National Science Foundation, the Swiss Society for Astrophysics and Astronomy and the Jubiläumsstiftung of the Basellandschaftliche Kantonalbank.

Marco Longhitano
*Basel, September 2010*

# Chapter 1

# Introduction and motivation

## 1.1 Historical sketch of double stars

"It's one of the beautiful things in the sky and I don't believe that in our pursuit one could desire better", remarked Benedetto Castelli (1578–1643) in a letter[1] sent to his friend and former teacher Galileo Galilei (1564–1642) on January 7, 1617. Castelli did not explain their "pursuit" more in detail in that letter – for good reasons: Only one year before, early in 1616, the Pope Paul V. declared that the Copernican doctrine was contrary to the Bible. Galilei received the papal order not to "hold or defend" the idea that the Earth moves and the Sun stands still at the centre. But in fact, the "pursuit" of Castelli and Galilei was to find observational evidence in favour of the Copernican system. They believed that such evidence may be provided by "one of the beautiful things in the sky": the double star Mizar.

Mizar ($\zeta$ *Ursae Majoris*) belongs to the Great Bear constellation, also called the Great Dipper. It is the middle star of the three located in the bear's tail. From ancient time it is known that Mizar has a dimmer companion, Alcor, which is separated from Mizar by about 12 arcminutes, resolvable to the naked eye under good observing conditions. Mizar itself consists of two stars, Mizar A and Mizar B, separated by about 14.4 arcseconds, and resolvable only with the aid of a telescope.

Even if Galilei's responses to Castelli have not been preserved, it may be concluded from his observational notes that the double nature of Mizar was the cause for Castelli's enthusiasm. Indeed, just one week after Castelli had written to him Galilei pointed his telescope at Mizar. Galilei wrote a detailed record[2] of his observations, where he derived

[1] Castelli (Pisa) to Galilei (Florence), 7 Jan. 1617, in *Le Opere di Galileo Galilei*, Edizione Nazionale, ed. by Antonio Favaro (20 vols., Florence, 1890–1909; hereafter Galilei, *Opere*), XII, 301 (Letter 1241): "Desiderarei che V. S. Ecc.$^{ma}$, concedendoglielo la sanità, una sera desse un'occhiatina a quella stella di mezo delle tre che sono nella coda dell'Orsa maggiore, perchè è una delle belle cose che sia in cielo, e non credo che per il nostro servizio si possa desiderar meglio in quelle parti."

[2] Galilei, *Opere*, III, Part II, 877. The record is not dated, but the ecliptical longitude of Earth given by Galilei corresponds to January 15 and there are good reasons to believe that the year was 1617 (Fedele 1949; Siebert 2005).

an angular separation between Mizar A and B of 15 arcseconds – an excellent result for that time.

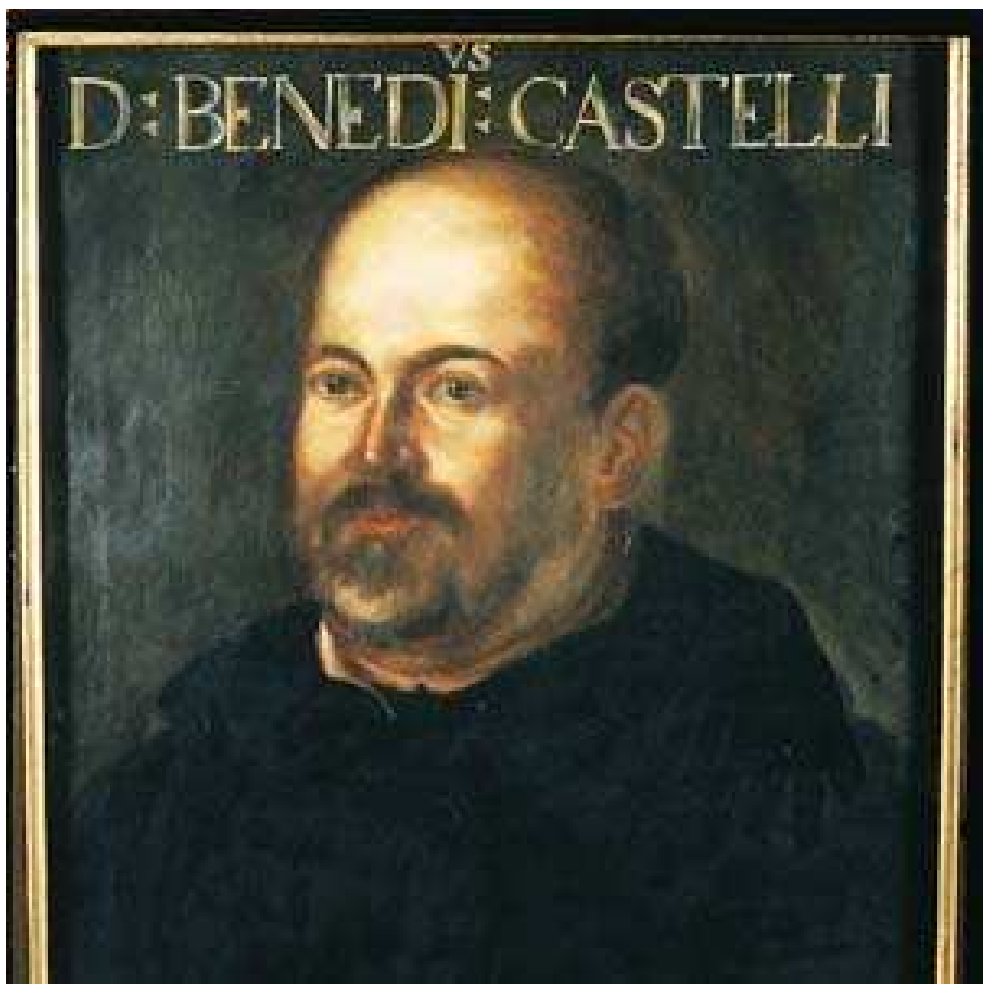

Figure 1.1: Benedetto Castelli (1578–1643) studied natural sciences in Padua and later became an abbot at the Benedictine monastery in Monte Cassino. He was most likely the first to resolve Mizar ($\zeta$ UMa) with a telescope.

It is nowadays believed that Castelli, and shortly after him Galilei following Castelli's suggestion, were the first to split Mizar with a telescope (Ondra 2004; Siebert 2006) – the first double star ever resolved by telescope. Some older double star literature (e.g Aitken 1964) mistakenly attributes this achievement to Giambattista Riccioli (1598–1671), the Jesuit astronomer and geographer of Bologna, who briefly mentioned the double appearance of Mizar in his *Almagestum novum* printed in 1651[3]. This misconception was already noted by Fedele (1949) but he did not receive much attention, since his article[4] was published in Italian in the little-known journal *Coelum* of the Bologna Observatory.

But how can a double star like Mizar provide any evidence in favour of the Copernican world picture? In a letter[5] to Galilei, Ludovico Ramponi (c. 1577–?) describes a method to measure the annual parallax using double stars: if the Earth is orbiting around the Sun, then two stars with very different distances from Earth should change their relative positions. This parallactic effect is much easier to measure when the two stars appear close together in the sky as observed from Earth, forming a so-called optical pair. Double stars of unequal brightness seemed to be especially suitable, since such pairs were suggesting very different distances of the two components from Earth.

The detection of the annual parallax may be regarded as the 'experimentum crucis' of the cosmological controversy of the 17$^{\text{th}}$ century and the failure to observe it had been one of the most important objections against the motion of Earth (Siebert 2005). Galilei and Castelli tried to proof that the Earth is moving by measuring the relative parallactic displacement of close optical pairs. Galilei explained this method later in the *Dialogo*, through his alter ego, Salviati, the defender of the Copernican system[6].

---

[3]Riccioli, *Almagestum novum* (2 vols., Bologna, 1651), I, 422a (Lib. 6, cap. 9): "...adeo ut stella unica videatur illa, quae media est in cauda Ursae maioris, cum tamen sint duae, ut Telescopium prodidit...."

[4]A reproduction of Fedele's article in Italian can be found on Leoš Ondra's homepage: `http://www.leosondra.cz/en/mizar/fedele/`

[5]Ramponi (Bologna) to Galilei (Florence), 23 July, 1611, in Galilei, *Opere*, XI, 159–62 (Letter 561).

[6]Galileo Galilei (1632), *Dialogo sopra i due Massimi Sistemi del Mondo Tolemaico e Copernicano*, in Galilei, *Opere*, VII, 409.

Even if correct in principle, the reasoning was flawed for two reasons: first, most close double stars, such as Mizar A and B, are *not* optical pairs, but true binary systems[7], bound to each other by their mutual gravity. Their different apparent magnitudes – Mizar A has a magnitude of 2.3 and Mizar B of 4 – misled Castelli and Galilei, who assumed that all stars have more or less the same brightness, and thought that the two components of Mizar have very different distances from Earth. But in fact, they are at the same distance from us and, therefore, there is no relative change in the positions of Mizar A and Mizar B due to the motion of the Earth around the Sun. Second, unaware of the phenomenon of diffraction of light, Galilei thought that the observed angular radius of a star corresponds to its real radius[8]. Assuming furthermore that Mizar A is as big as the Sun ("si stella ponatur tam magna ut ⊙"[9]), Galilei heavily underestimated the distance of Mizar A from the Earth (he got only 300 AU). Consequently, he overestimated the expected parallactic displacement. With the resolution of the telescopes used at that time, it was impossible to observe any parallax - even that of the closest stars[10].

Castelli and Galilei took it for granted that they were looking at purely perspective phenomena, stars that accidentally come to lie close together in the sky. Real binary and multiple stellar systems seemed to be contradictory to their conception of the Copernican doctrine. Like Giordano Bruno (1548–1600), Galilei and Castelli believed that an infinite number of worlds similar to our exists. According to them, every star in the sky was a sun like our own, all being equal in size and supposed to stand still at the centre of other possible planetary systems. In their view, only planets, but no stars could revolve around stars (Siebert 2005, 2006).

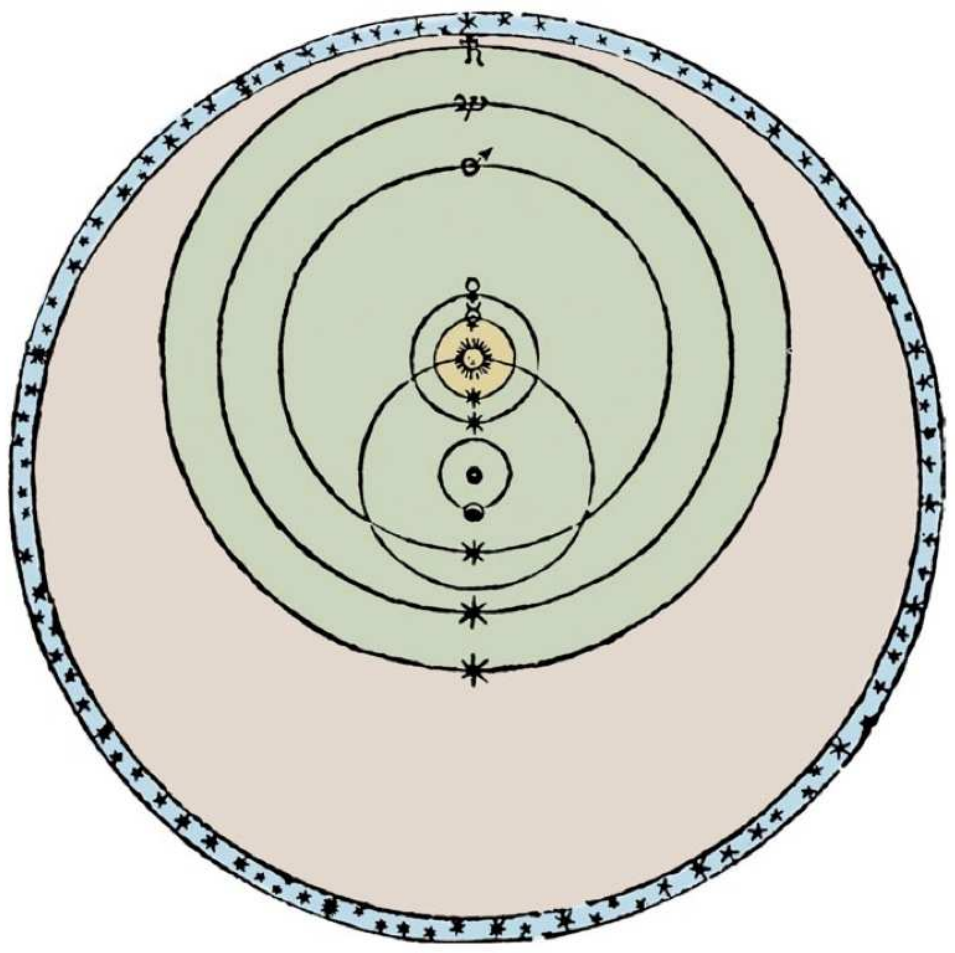

Figure 1.2: The Tychonic system created by Tycho Brahe (1546–1601). The Moon and the Sun revolve around the Earth, which stands still at the centre. The other planets revolve around the Sun. Image from Siebert (2006).

In addition, abandoning the traditional idea of fixed stars in favour of physically associated stellar systems would undermine Galilei's method to proof the Earth's motion around the Sun, which was accepted from both sides of the cosmological debate. An eventually observed change in the relative positions of two stars must then not be necessarily interpreted as a proof of the Earth's motion but could also be attributed to a real orbiting of the one star around the other.

[7]Nowadays we know from spectroscopic observations that both Mizar A and B consist of two components. Mizar is, thus, a quadruple system.

[8]The Airy disk formed by diffraction through a telescope similar to the one used by Galilei has a diameter of a few arcseconds (Graney and Sipes 2009).

[9]Galilei, *op. cit.* (footnote 2).

[10]The first detection of a stellar parallax succeeded more than two centuries later in 1838 with the observation of 61 *Cygni* by the German astronomer and mathematician Friedrich Wilhelm Bessel.

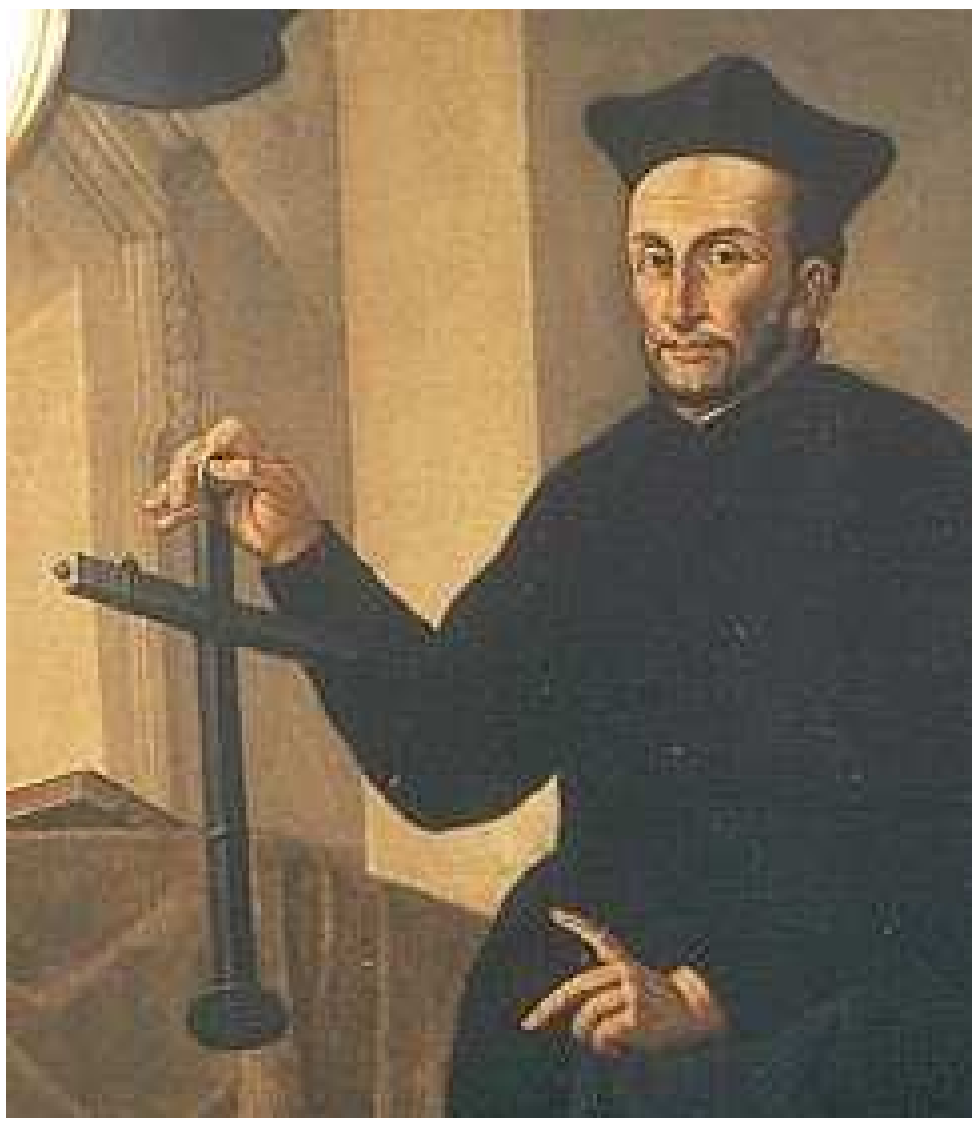

Figure 1.3: Johann Baptist Cysat (1587–1657) holding a Jacob's staff. Born in Lucerne, he joined the Jesuits in Ingolstadt where he later became a theology student. He was the first to write about the perception of double stars as physically related systems.

The argument that, if real double stars orbiting each other exists, the detection of a small variation in the relative positions of fixed stars might not be the definite proof in favour of the heliocentric world system was well-known to Galilei but he dismissed it as quibbling. Interestingly, the opponents of Copernicus were more open-minded to the idea of physically related stellar systems, since in their world picture the Sun was not the only organising principle. As noted by Siebert (2005, 2006), supporters of the geo-heliocentric system by Tycho Brahe (1546–1601), had less problems imagining stars building systems, in which the components are orbiting each other. In Brahe's system the Moon and Sun are orbiting the Earth, whereas all other planets revolve around the Sun (Fig. 1.2). It is mathematically and empirically – regarding the observed (retrograde) motion of the planets and the phases of Venus discovered by Galilei – equivalent to the Copernican system.

One of its supporters was the Swiss Jesuit mathematician and astronomer Johann Baptist Cysat (1587–1657). In his most important work, the *Mathemata astronomica*[11] printed in 1619, he gives detailed descriptions of one of the earliest telescopic comet observations. Cysat believed to have seen the fourth comet that appeared in 1618 as a composed celestial body consisting of many small 'stars'. To illustrate this bizarre aspect, Cysat compares it to the open star cluster Praesepe – also called the Beehive Cluster – in the constellation Cancer. The five brightest stars of the Praesepe cluster resemble, according to Cysat, to the cometary nucleus he observed. Furthermore, Cysat also compares his comet observations to globular cluster, multiple stellar systems as well as the systems of Jupiter and Saturn with their moons. These comparisons suggest that Cysat, contrary to Castelli and Galilei, interpreted multiple stellar system like Praesepe as stars of different size belonging together in space – a radically new picture of stars that will be generally accepted only 150 year later.

Galilei was right that the Earth is moving around the Sun but he was wrong in dismissing the idea of double stars as physical systems – an idea suggested by the loosers of the great cosmological debate. Declining the Copernican picture they were considered in retrospect as opponents of scientific progress. At least in the context of double stars this thinking appears to be wrong (Siebert 2006).

The term 'double star' ('stella dublex') in its modern meaning (a 'star' that appears single to the naked eye but double in a telescope) was used for the first time by the Italian astronomer and architect Giovan Battista Hodierna (1597–1660) – another supporter of the Tychonic system. Born in Ragusa, Sicily, Hodierna worked in Palma di Montechiaro near Agrigento as court scientist of the Duke's family di Lampedusa[12]. In his *De Systemate*[13] he compiled a catalogue of more than 40 objects, including at least 19 nebulous objects such as stellar clusters and galaxies, where he anticipated a host of observations that should only appear in the famous catalogues of Charles Messier (1730–1817) and John Louis Emil Dreyer (1852–1926). Hodierna gives, for example, the earliest known record of the Orion Nebula – a discovery that was attributed a long time to Christiaan Huygens (1629–1695), who saw the Orion Nebula in 1656 and published his observations in 1659, five years after Hodierna.

[11] Johann Baptist Cysat, *Mathematica astronomica de loco, motu, magnitudine et causis cometae qui sub finem anni 1618 et initium anni 1619 in coelo fulsit*, Ingolstadt, 1619.

In his book from 1654, Hodierna dedicated for the first time a whole section[14] to the double star phenomenon. He also gives the first list of double stars of 13 pairs only resolvable with a telescope. Like Cysat, Hodierna considered it thoroughly possible that stars do not only appear close in the sky, but that they could also be close in space, since they might well be of different intrinsic size and luminosity.

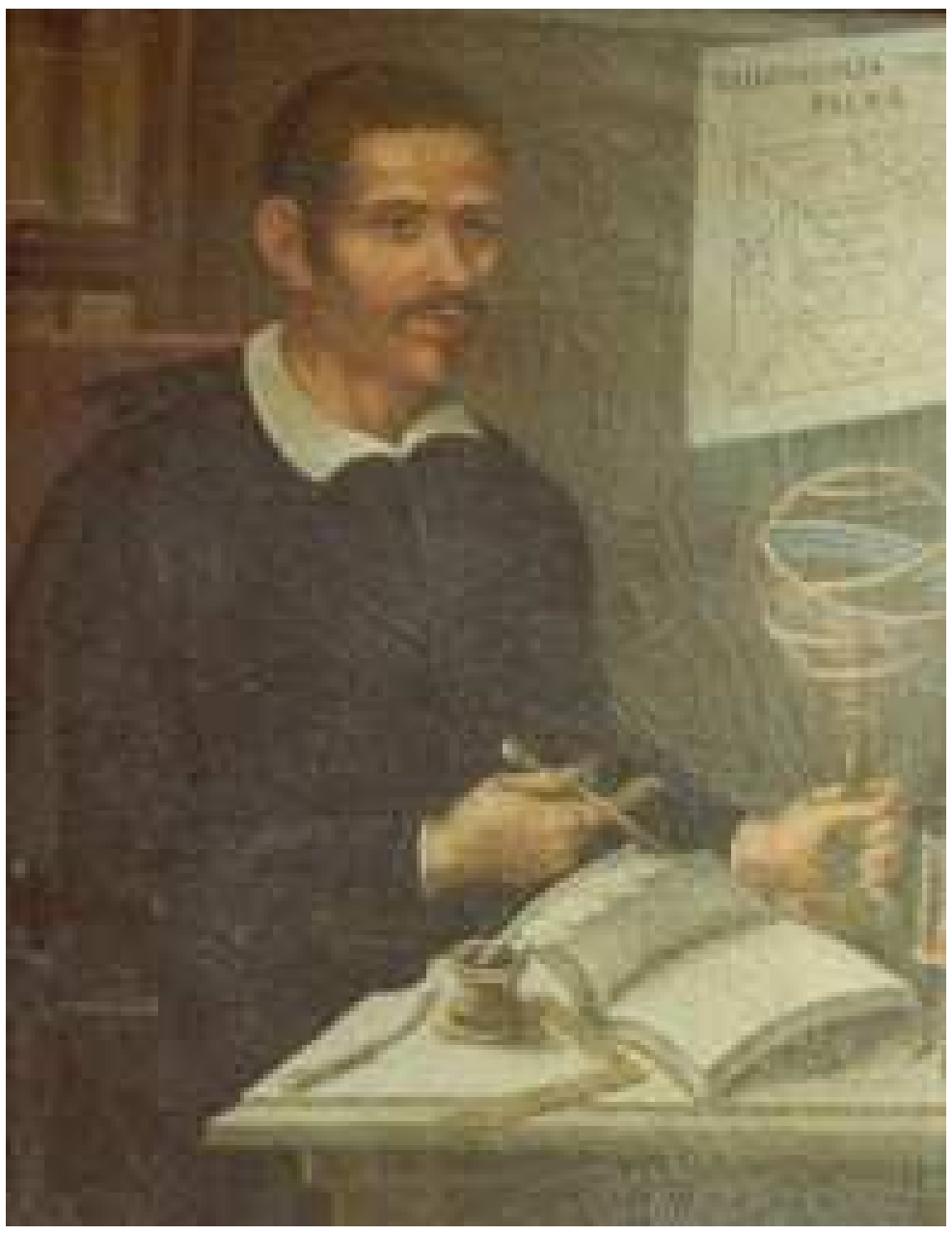

Figure 1.4: Giovan Battista Hodierna (1597–1660), born in Ragusa, was an Italian astronomer at the court of the Duke of Montechiaro. He published the first list of 13 double stars.

Most astronomers of the 18$^{th}$ and 19$^{th}$ centuries, however, had no doubt on the optical nature of the known double stars, which were consequently regarded as mere curiosities. The first well-founded argument in favour of bound stellar pairs was due to Reverend John Michell (1724–1793). Michell was an English natural philosopher and geologist and "one of the most brilliant and original scientists of his time" (Soter and deGrasse Tyson 2001). Unfortunately, no portrait of him exists, but a contemporary diarist describes him as "a little short Man, of a black Complexion, and fat"[15].

Michell's work spanned a wide range of subjects. He demonstrated that the magnetic force decreases with the square of the distance (Michell 1750). After the catastrophic earthquake in Lisbon in 1755, he developed a theory of earthquakes as wave motions in the interior of the Earth (Michell 1759) and since then is regarded as the father of seismology. Michell invented the torsion balance[16] used by Henry Cavendish (1731–1810) in the famous experiment where the average density of the Earth was inferred for the first time (Cavendish 1798). More recently, he has also become famous for anticipating the concept of black holes – celestial bodies so dense that their escape velocity exceeds the speed of light – in a letter to Cavedish, where he also pointed out that the existence of such non-luminous objects might be inferred "if any other luminous bodies should happen to revolve about them" (Michell 1784, p. 50).

[12]Famous through Giuseppe Tomasi di Lampedusa (1896–1957), author of the *Gattopardo*.

[13]Giovan Battista Hodierna, *De Systemate Orbis Cometici, Deque Admirandis Coeli Characteribus*, Palermo, 1654.

[14]The section is entitled: *de Stellis Contiguis Duplicibus, seu Geminis, deque Mondani Systematis Copernicaeorum implicantia, ratiocinandum venit.* Hodierna, *De Systemate*, p. 29.

[15]William Cole, MSS XXXIII, 156, British Library

The argument Michell put forward in 1767 in favour of binary stars was of a statistical nature (Michell 1767). He argued that the probability that a stellar clusters like the Pleiades were due to a chance alignment was very small[17], and that there were far more double star in the sky as would be expected if the stars were distributed randomly in the sky. Michell concludes (Michell 1767, p. 249)

> that the stars are really collected together in clusters in some places, where they form a kind of systems, whilst in others there are either few or none of them, to whatever cause this may be owing, whether to their mutual gravitation, or some other law or appointment of the Creator. And the natural conclusion from hence is, that it is highly probable in particular, and next to a certainty in general, that such double stars, &c. as appear to consist of two or more stars placed very near together, do really consist of stars placed near together, and under the influence of some general law, whenever the probability is very great, that there would not have been any such stars so near together, if all those, that are not less bright than themselves, had been scattered at random through the whole heavens.

Michell has the credit to have applied as first the new theory of probability to astronomy. His work, however, did not receive much attention, and nearly 40 years passed until the scientific community became convinced of the existence of binary stars systems.

The systematic search and observation of double stars started with Christian Mayer (1719–1783). In 1771 he initiated the construction of the Mannheim observatory and started to observe double stars systematically in 1776. In the following year, Mayer presented his results to the Electoral Academy of Sciences in Mannheim. He suggested a new method to study the at that time ill-understood phenomenon of the proper motion of the stars[18] by measuring a change in the relative positions of the components in close double stars. Mayer called the fainter components of the double stars 'Fixsterntrabanten' ('satellites of fixed stars'), which suggests that he considered at least some of the double stars as gravitationally bound systems. The then director of the Vienna Observatory, Maximilian Hell (1720–1792), based, in part, on a misunderstanding[19], openly criticised Mayer's view of double stars and the use of the term 'Fixsterntrabant'. Mayer, in turn, saw himself obliged to write in 1778 a "Thorough vindication of the new observations of satellites of fixed stars" (Mayer 1778), wherein he speculates upon the possibility of physically related stellar systems[20].

[16] The torsion balance was invented independently by Charles-Augustin de Coulomb (1736–1806).

[17] For the five brightest stars in the Pleiades Michell calculated a probability of about 1 in 496 000 to find such a group as a chance alignment among 1 500 stars anywhere in the sky.

[18] Stellar proper motions were discovered more than half a century before by Edmond Halley (1656–1742).

[19] An outline in German of the controversy between Hell and Mayer is given by J. S. Schlimmer, *Christian Mayer und die Fixsterntrabanten*, 2006, accessible at `http://www.epsilon-lyrae.de/`.

From his book it emerges that for Mayer the idea of stars with varying size was natural[21] and he was already aware of the right connection between apparent luminosity, diameter and distance of a star – an important step towards modern stellar astronomy.

Mayer also tried to proof by means of their proper motion that the double stars he observed are real binary stars and compares his observed positions with various older observations. Even though Mayer believed that he succeeded, a more accurate investigation shows that the uncertainties in Mayer's data were too large to draw any firm conclusion on the binarity of 'his' double stars[22] and, hence, he confused several optical pairs with physical ones.

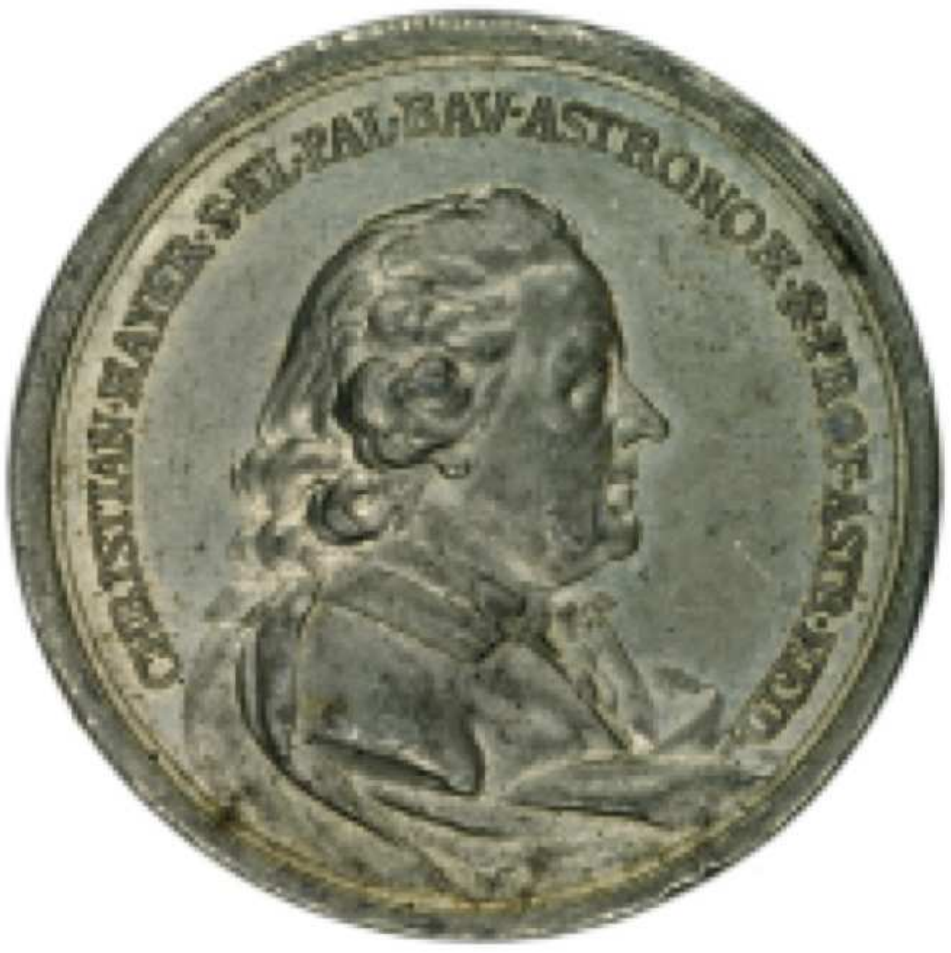

Figure 1.5: Christian Mayer (1719–1783) was born in Modřice, Moravia. After studying theology in Mainz he became a Jesuit. In 1763 he was appointed Court Astronomer at Mannheim and was the first to observe double stars systematically. Image taken from the city panel of the Mannheim observatory available at `www.mannheim.de`. © Stadt Mannheim.

A year later, Mayer published a further book (Mayer 1779), in which he continued his work on double stars and gave a list of 72 pairs – the first double star catalogue. This book brought Friedrich Wilhelm Herschel's (1738–1822) attention to the double stars. In 1781 Mayer's catalogue was published in the *Berliner Astronomisches Jahrbuch für 1784* under the caption *Verzeichnis aller bisher entdeckten Doppeltsterne*[23], where 8 additional already known pairs were added. Thus, Mayer's catalogue lists 80 pairs together with their angular separations and position angles.

Mayer's work on double stars may be regarded as the beginning of double stars astronomy. However, his work has been completely dwarfed by the work of Herschel published only a few years later. Nevertheless, Mayer gave the actual impetus to the systematic study of double stars and has proved himself as one of the most important and farsighted scientist of his time.

In 1779 Herschel began a systematic search for double stars assisted by his sister Caroline. Also Herschel assumed that double stars are merely optical phenomena. Like Galilei and Castelli long before him, Herschel aimed at measuring the stellar parallax. At that time little doubt was casted on the correctness of the heliocentric system[24] but the success of the 'experimentum crucis' – the detection of the parallax – was still owing.

[20] "endlich kann in einem Doppeltsterne der grösere so wohl als der kleinere eine an sich selbst leuchtende und bewegliche Sonne seyn, die in ihrem eigenem Systeme um einen allgemeinen Ruhepunkt angezogen werden." (Mayer 1778, p. 112).

[21] "gehöhret derselben [the size of the stars] Verschiedenheit nicht zur Schönheit unsers Weltbaues?" (Mayer 1778, p. 237f.).

[22] Schlimmer, *op. cit.* (footnote 19).

[23] A reproduction of Mayer's catalogue can be found at `http://www.epsilon-lyrae.de/`.

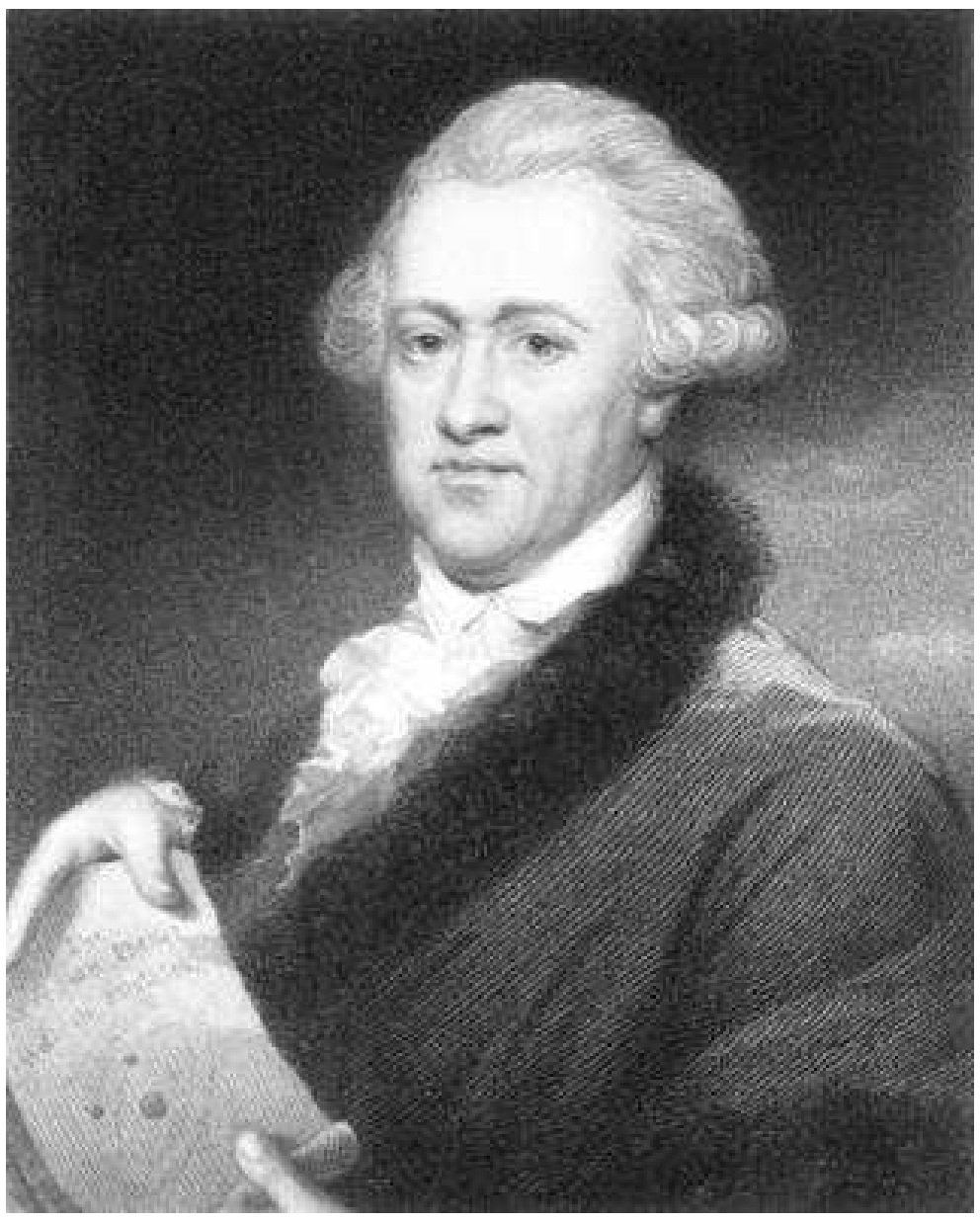

Figure 1.6: Friedrich Wilhelm Herschel (1738–1822) was a German-British astronomer and composer. Born in Hannover, he emigrated to Britain in 1757. Herschel has for the first time convincingly demonstrated the existence of bound stellar pairs by observing their orbital motion, and has introduced the term 'binary system'.

Herschel too recognised the advantages of using the double star method to measure the annual parallax (Herschel 1782b, p. 97). His telescopes were more powerful than any previous one and he soon discovered much more double stars than he had anticipated. In 1782 Herschel published his first *Catalogue of double stars* with 269 pairs (Herschel 1782a). A second catalogue containing 434 additional double stars appeared in 1785 (Herschel 1785). Unlike Mayer, however, Herschel was more cautious in choosing his terms for the stellar pairs he observed. He regarded the time as not ripe to speculate "about small stars revolving round large ones" and avoided using expressions "such as Comes, Companion, or Satellite" (Herschel 1782a, p. 161).

Applying his earlier statistical argument to Herschel's first catalogue, Michell concluded that most of the double stars listed by Herschel were physical systems (Michell 1784, p. 36):

> The very great number of stars that have been discovered to be double, triple, &c. particularly by Mr. Herschel, if we apply the doctrine of chances, ..., cannot leave a doubt with any one, who is properly aware of the force of those arguments, that by far the greatest part, if not all of them, are systems of stars so near to each other, as probably to be liable to be affected sensibly by their mutual gravitation; ....

It was, however, only in 1802 that Herschel expressed similar views, giving a minute distinction between optical and real pairs and introducing the term 'binary system' (Herschel 1802, p. 480f.).

The actual demonstration that some double stars are true binary systems, is given by Herschel in the following year. In the fundamental paper, entitled *Account of the Changes that have happened, during the last Twenty-five Years, in the relative Situations of Double-*

[24]The Copernican world system already received confirmation by the mathematical investigations of Johannes Kepler (1571–1630) and Isaac Newton (1643–1727) and by the observations of James Bradley (1693–1762) who, in 1725, discovered and correctly interpreted the phenomenon of stellar aberration.

*stars; with an Investigation of the Cause to which they are owing*, Herschel gives (Herschel 1803, p. 340)

> an account of a series of observations on double stars, comprehending a period of about 25 years, which, if I am not mistaken, will go to prove, that many of them are not merely double in appearance, but must be allowed to be real binary combinations of two stars, intimately held together by the bond of mutual attraction.

He was not mistaken. The first double star on his "account" is $\alpha$ *Geminorum* (Castor). By a detailed analysis he shows that orbital motion is the most probable explanation of the change in the position of the components. He repeats the analysis for five further systems ($\gamma$ *Leonis*, $\varepsilon$ *Bootis*, $\zeta$ *Herculis*, $\delta$ *Serpentis*, and $\gamma$ *Virginis*) and concludes that the only reasonable conclusion is that all these double stars are binary systems.

The next important advancement in double star astronomy was due to Friedrich Georg Wilhelm Struve (1793–1864) using the celebrated Fraunhofer refractor[25]. Equipped with an excellent driving clock, this telescope was far superior to any previously constructed. With this instrument Struve conducted from 1824 to 1837 more than 10 000 micrometric measurements with unprecedented accuracy of nearly 3 000 double stars. Many of his results, contained in his principal work *Mensurae Micrometricae* (Struve 1837), are still in harmony with modern investigations. For example, Struve argues, on the basis of the theory of probability, that practically all the pairs with separation smaller than 4″ and the great majority of those with separations under 12″ are real double stars, whereas the probability that an optical pair is included increases with angular separation, especially for the fainter pairs in his catalogue (both components fainter than 8.5 mag).

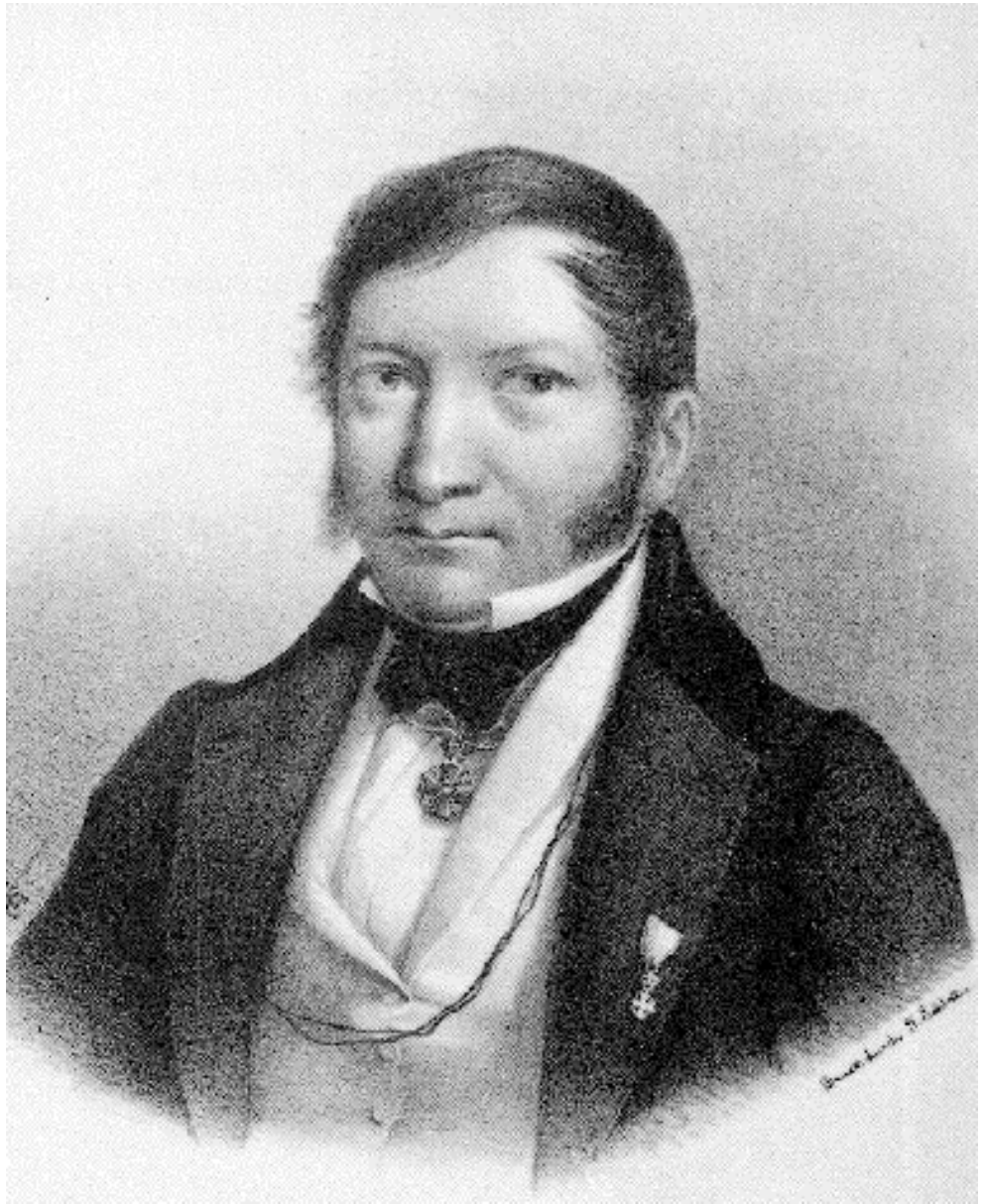

Figure 1.7: Friedrich Georg Wilhelm Struve (1793–1864) was born in Altona, a borough of Hamburg and studied at the University of Tartu. Using the largest refractors of his time, he observed a large number of binary systems with unprecedented accuracy.

In the following decades many astronomers continued the work of Herschel and Struve on binary star systems. They are far too numerous to name them all here and I refer to Aitken (1964), Heintz (1971) and Zinnecker (2001) for more detailed treatises of the double star history, especially regarding the modern period, when so important technologies as photography and spectrography were developed and applied to astronomy. It is a remarkable historical coincidence that the first binary system ever resolved by telescope, Mizar, in 1617 by Castelli and Galiliei, was also the first double star to be observed photographically by George Philips Bond (1825–1865) in 1857, and its primary component, Mizar A, was the first spectroscopic binary to be discovered by Edward Charles Pickering (1846–1919) in 1889.

[25]A detailed description of the Fraunhofer refractor is given by Struve himself (Struve 1825).

## 1.2 Definition and classification of wide binaries

The distinction between close and wide binary systems is not well defined. From which separation on a binary is now called 'wide', differs from author to author and what aspects of the binary systems are studied. One required property of wide binary systems is that the formation and evolution of the two components occur largely independent from each other. This is usually fullfilled for separations of more than about 50 AU – just beyond Plutos orbit.

A dynamical boundary between close and wide binaries can be drawn by means of *Heggie's law*, named after Douglas Heggie. Heggie investigated the formation, evolution, and destruction of binary stars resulting from gravitational encounters with single stars (Heggie 1975). In the simplest case he considered a homogeneous stellar system consisting of single and binary stars. He assumed that all single stars have mass $m$ and a Maxwellian velocity distribution with velocity dispersion $\sigma$.

Heggie found that two classes of binary stars can be distinguished: *hard* binaries, where the absolute value of the internal energy of the binary is larger than $m\sigma^2$ and *soft* binaries, where the average kinetic energy of a perturber exceeds the binding energy of the binary. The behaviour of hard and soft binaries during encounters are quite different. While soft binaries on average gain energy from encounters with field stars and therefore have their orbit widened to become even less bound, the opposit is true for hard binaries. Hence, Heggie's law can be formulated as follows: *Hard binaries get harder and soft binaries get softer*. It can be interpreted in terms of energy equipartition (see also Binney and Tremaine 2008).

Assuming that all stars involved in an encounter have solar masses and taking typical values for the velocity dispersion in the solar neighbourhood ($\sim$ 20 km s$^{-1}$), the watershed (binding) energy $-m\sigma^2$ corresponds to a semi-major axis of the binary of a few astronomical units. One could now identify the hard binaries with close binaries and soft with wide binaries. But this is not common practice and we too follow a different approach in the present study.

Our working definition is to call a binary system 'wide' when its semi-major axis larger than 200 AU. As we show in §2.4.1, separations smaller than 200 AU lie outside of our observational window, i.e. we are not sensitive to smaller separations. Binaries with such a large separation are surely soft in the sense of Heggie.

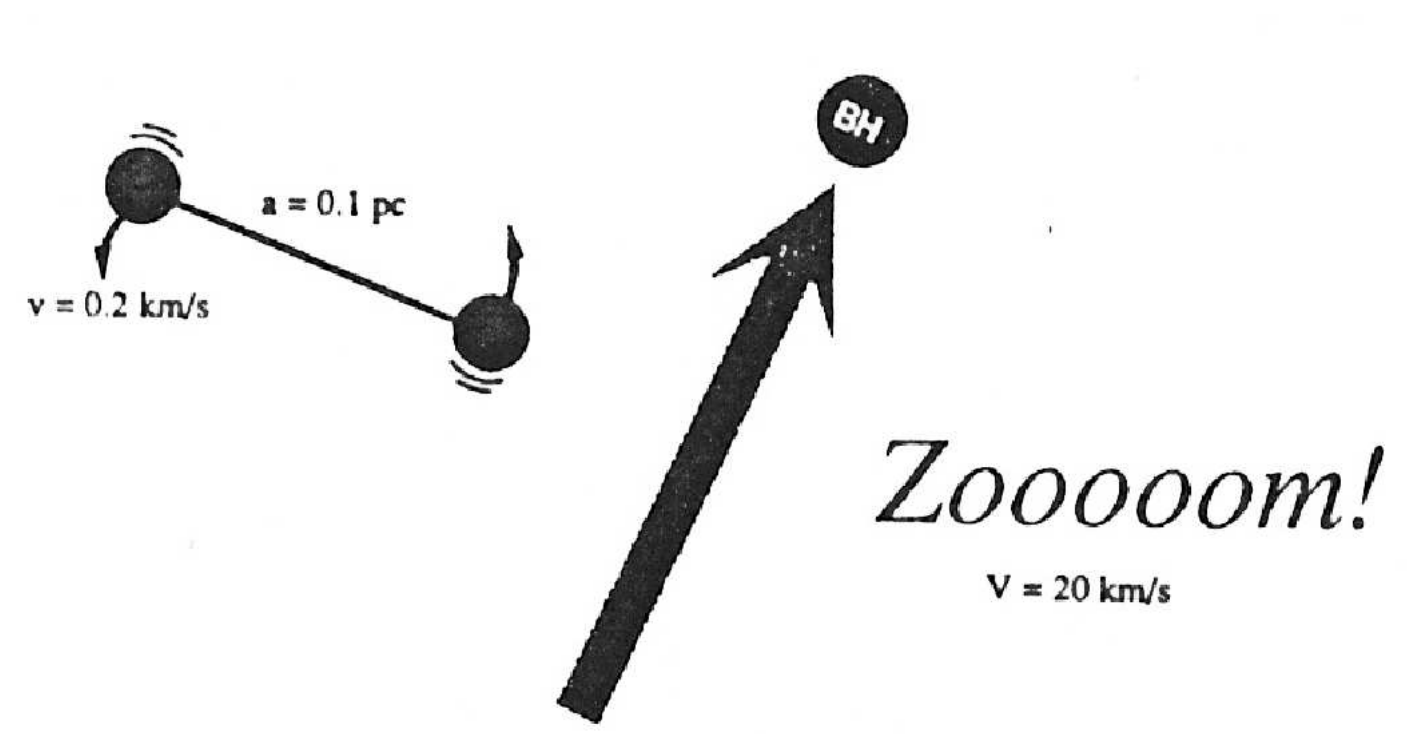

Figure 1.8: Depiction of an encounter between a binary system and a perturber (here a black hole) adapted from Weinberg (1990). Fiducial parameters are indicated. © Kluwer Academic Publishers.

# 1.3 Why study wide binary stars?

## 1.3.1 Constraints on MACHOs

Possibly wide binaries would have remained just a curiosity if it would not have been realised that they may shed light to one of the most pressing mysteries of modern astronomy: the nature of dark matter. In 1985 John Bahcall, Piet Hut and Scott Tremaine published a seminal paper, where they used data from the widest binaries to constrain the mass of individual unseen disk objects to be less than 2 solar masses (Bahcall et al. 1985a).

The widest then known binary stars in the Galactic field have separations of about 0.1 pc (Bahcall and Soneira 1981). These are huge orbits with periods of millions of years. Such wide binaries are only weakly bound and are easily disrupted by encounters with other stars, molecular clouds or even massive non-luminous objects like black holes (Fig. 1.8). Thus, wide pairs constitute a sensitive probe of the Galactic gravitational potential and the distribution of their semi-major axis might contain fossil information about the dynamical history if the Galaxy.

Bahcall et al. attribute the absence of binaries with separations larger than 0.1 pc to gravitational encounters that have disrupted them (see also Retterer and King 1982). Furthermore, they showed that if the mass of these unseen “disk thing”, as they call them, would exceed 2 solar masses, also binaries with a semi-major axis smaller than 0.1 pc must have been disrupted.

The study of Bahcall et al. has been subsequently critisised, mainly because of the sparse data they rely on. In the following years Ira Wasserman and Martin Weinberg performed a more sophisticated analysis (for a review see Weinberg 1990). In their opinion the data available at that time is not conclusive and no firm constraints on the mass of non-luminous objects can be drawn unless much larger wide binary samples become available.

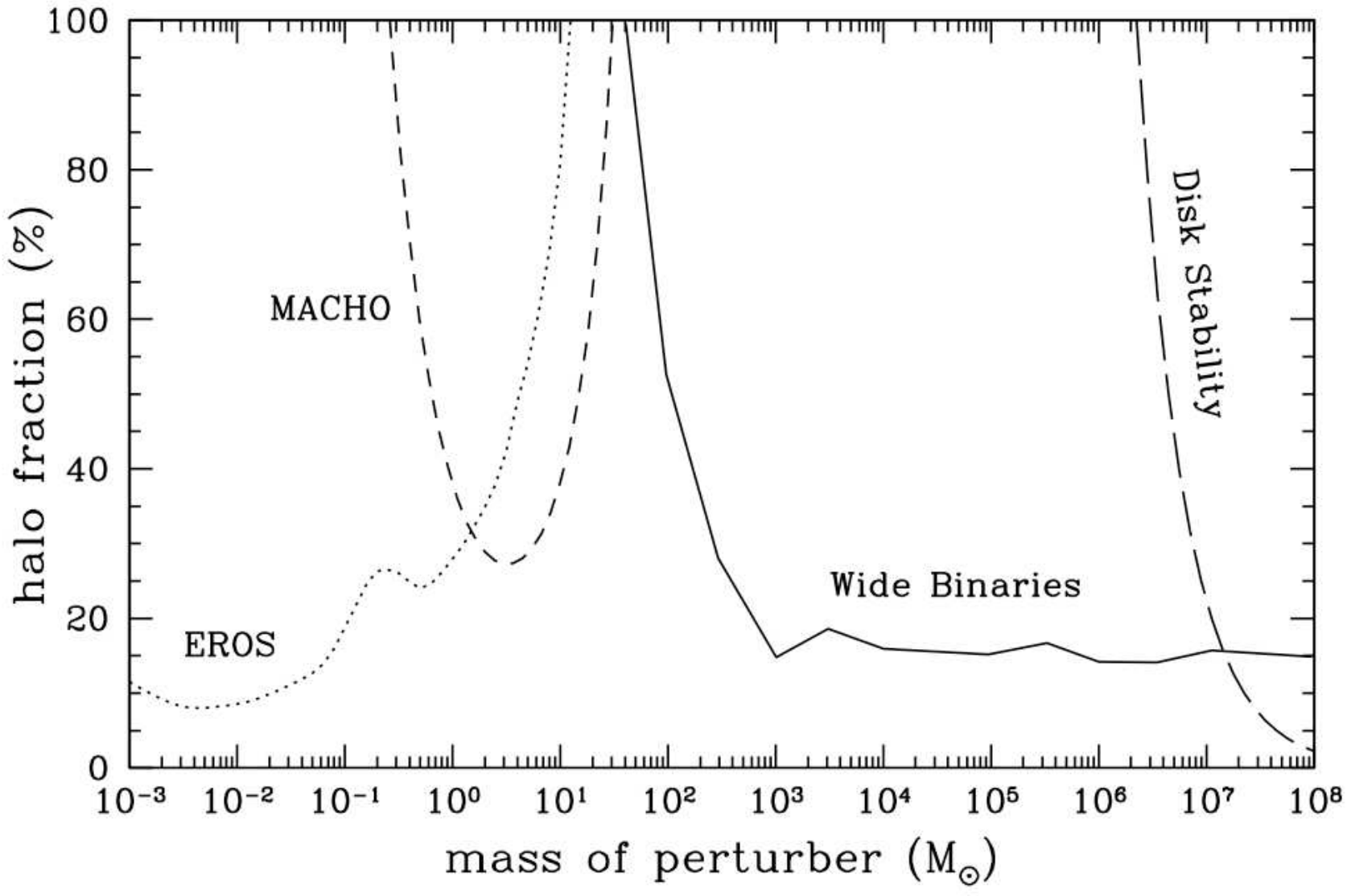

Figure 1.9: Exclusion contour plot at 95 % confidence level reproduced from Yoo et al. (2004). Perturber masses excluded by wide binaries are indicated by the solid line. The dotted and dashed lines are from the EROS (Afonso et al. 2003) and MACHO (Alcock et al. 2001) microlensing surveys. The long-dashed line from a disk stability criterion (Lacey and Ostriker 1985). © The American Astronomical Society.

In recent years some interest has returned to this topic. With new, precise proper motion data of a large number of stars, it became also possible to construct larger samples of wide binaries that belong to the halo (Chanamé and Gould 2004). Studying the halo wide binary population has several advatages. First, the halo is dominated by dark matter to much larger degree than the disk, enhancing the chance to constrain the mass of hypothetical MAssive Compact Halo Objects, so called 'MACHOs'. Second, the kinmatics of the halo population is not influenced by molecular clouds, whose density distribution is only poorly known. And third, no star and binary formation has to be taken into account, which complicates the analysis considerably.

In 2004 Yoo, Chanamé, and Gould announced "The end of the MACHO era". In thier paper they showed that the sample of halo wide binaries excludes MACHOs with masses greater than about 40 solar masses. MACHOs with smaller masses have already been excluded by microlensing surveys. So it seem that, together with microlensing experiments, wide binaries leave only a small windows for haloes composed entirely of baryonic MACHOs (Fig. 1.9).

The constraints of MACHOs masses from halo wide binaries stands, however, not on a firm basis as was recently pointed out by Quinn et al. (2009). The constraints rely heavily on the genuineness of the four widest binaries of Yoo et al. sample. Using radial velocity measurements, Quinn et al. showed that one of these four pairs is most likely not real. Omitting this pair, substantially relaxes the constraints on MACHO masses. This sensitivity clearly means that again larger samples of wide binaries are needed to put severe constraints on MACHOs.

### 1.3.2 A probe for dark matter in dwarf spheroidal galaxies

Nowadays it is widely believed that the vast majority of dark matter is of non-baryonic nature. An often quoted candidate are WIMPs – Weakly Interacting Massive Particles. There are also alternative theories, such as MOND (MOdified Newtonian Dynamics, Milgrom 1983; Milgrom and Bekenstein 1987) that proposes a modification of Newtons law of gravitation at very low accelerations. It turns out to be difficult to distinguish observationally between dark matter theories and MOND.

An interesting possibility to do this involving wide binary stars was recently put forward by Hernandez and Lee (2008). According to their calculations, very wide binaries, with separations larger than 0.1 pc, should be absent in low velocity dispersion, high-density dark matter haloes as inferred for the local dwarf spheroidal (dSph) galaxies. There, wide binary stars should have evolved into tighter binaries because of the dynamical friction caused by dark matter particles.

Of course, there would be no such orbital thightening in a purely MONDian Universe, where no dark matter exists. Therefore, Hernandez and Lee conclude that if "plentiful wide binaries were to be found in local dSph galaxies, the dark matter scenario would be very seriously challenged."

In a similar vein, Peñarrubia et al. (2010) examined by analytical and $N$-body methods the survival of wide binaries during repeated encounters with dark substructures in dSphs as expected from the present cosmological paradigm ($\Lambda$CDM). According to their calculations, a truncation in the semi-major axis distribution around 0.1 pc should be present in most local dSphs beyond which the distribution falls of as $a^{-2.1}$.

The ACS camera of the *Hubble Space Telescope* (HST) might be able to test these predictions for the nearest dSphs. In particular, Peñarrubia et al. estimate that in *Coma*, *Ursa Minor*, *Bootes I*, *Ursa Major II*, *Sculptor*, and *Draco* several deep ACS exposures are needed to place significant limits on the wide binary fractions[26]. In view of upcoming surveys (*Pan-STARRS*, *LSST*, *Gaia*,...) the study of wide binary stars in nearby dShps may pose a stringent test through which the $\Lambda$CDM model soon has to pass.

### 1.3.3 Clues to star formation

The observed number and properties of binary and multiple stellar system have traditionally been used to constrain star formation theory. The additional parameters provided by binary systems, such as angular momentum, eccentricity, and mass ratio, allow in principle to place stronger constraints on the nature of star formation process as single stars alone (Larson 2001). In the light that probably most stars from as binary and higher-order multiple systems (Goodwin 2010, and references therein), the study of the statistical properties of binary systems receives an even greater importance.

[26] This is in line with the results of an (unpublished) feasibility study performed by Marc Horat in a project work at the University of Basel.

Current theories on star formation (e.g. Bate 2009) can not explain the population of the widest binary stars. For example, Parker et al. (2009) examined the dynamical destruction of binary systems in stellar clusters of different densities. They found that most binaries with a separation larger than 1 000 AU disrupted after a few crossing times – even in low-density cluster. Since most stars (75% to 90%) form in stellar clusters (Lada and Lada 2003), the origin of the wide binary population in the Galactic field (e.g. Duquennoy and Mayor 1991) is a mystery.

Very recently a formation mechanism has been suggested independently by Moeckel and Bate (2010) and by Kouwenhoven et al. (2010). The basic idea is that wide binaries form in the expanding halo of young clusters, where the stars are formed. During the dissolution phase of open clusters, sometimes two stars leave the cluster in almost the same direction with almost the same velocity. These two stars can become bound after they have left the cluster and form a wide binary star. $N$-body simulations of this dissolution process have shown that a considerable number of wide stellar pairs can be formed in this way despite that the original star-forming molecular cloud is too small to produce such pair by direct fragmentation. Thus, the scenarios suggested by Moeckel and Bate and by Kouwenhoven et al. potentially solve the mystery of the wide binary field population.

The study of the statistical properties of wide binaries may shed some light on their formation process and the environment in which they were born. One has to bear in mind, however, that it is probably not appropriate to directly compare the outcome of current star formation simulation to the stellar field population (Goodwin 2010). The field population must be regarded as a mixture of many star forming regions with different initial conditions. Furthermore, the simulations were stopped right after the dissolution of the cluster when the binary population was released into the field. Thus, the dynamical processing the binaries experiences in the field is not taken into account in the simulations. This 'gap' between simulations and observations must be bridged in order to place meaningful constraints on star formation theory.

## 1.4 How study wide binary stars?

### 1.4.1 Common proper motion

The orbital periods of wide binaries range from a few 1 000 to millions of years – far too long to be observed directly. Wide binaries are thus generally identified by means of statistical techniques. Because of their low orbital velocities, wide binaries are expected to have very similar proper motions. Indeed, a common proper motion (CPM) of two stars is excellent indication for the genuineness of that pair, especially if it has a small angular separation (e.g. Lépine and Bongiorno 2007).

Most previous studies made use of proper motion information to distinguish genuine wide pairs from optical ones. One of the pioneers of using proper motions to study wide binary systems was Willem Jacob Luyten (1899–1994). He discovered more than 6 200 CPM pairs with proper motions $\mu \gtrsim 0.1''\mathrm{yr}^{-1}$ in the course of over 50 years. They are listed in the Luyten Double-Star (LDS) Catalogue that was completed in 1987 and made available online through the CDS[27] (Luyten 1997). The only IAU colloquium so far focusing especially on *Wide components in double and multiple stars* (Dommanget et al. 1988), held in Brussels, Belgium, in 1987, was then also dedicated to Luyten.

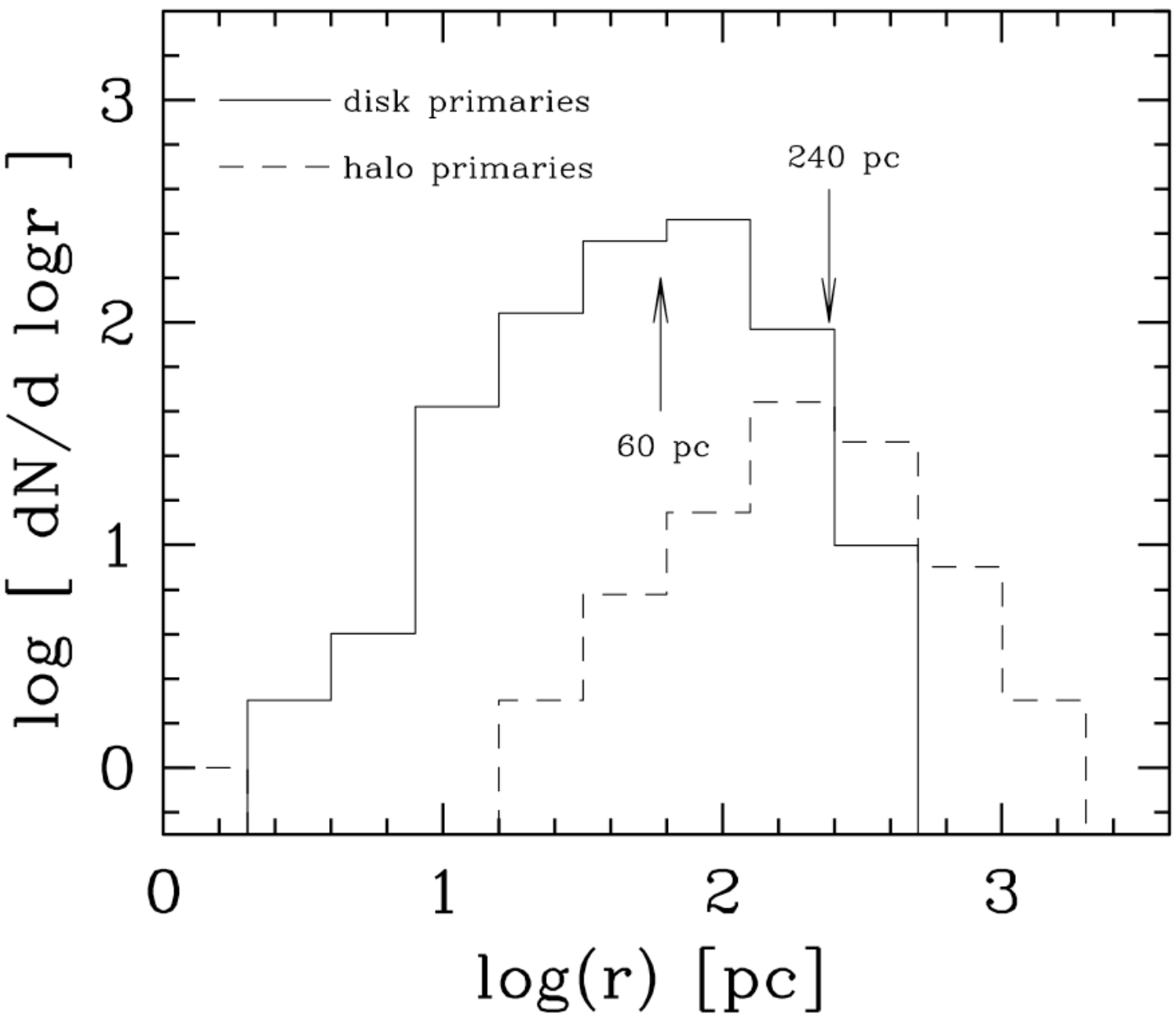

Figure 1.10: Distance distribution of disk and halo primaries of the CMP pairs identified by Chanamé and Gould (2004). Average distances for disk and halo sample are indicated. Halo stars can be detected at larger distances because they have a larger velocity dispersion than that of disk stars. Figure adapted from Chanamé and Gould (2004). © The American Astronomical Society.

More recently, Chanamé and Gould (2004) analysed the revised New Luyten Two-Tenths (rNLTT) Catalogue assembled by (Salim and Gould 2003), which includes the fastest stars originally identified by Luyten ($\mu \gtrsim 0.2''\mathrm{yr}^{-1}$) and has nearly 60 000 entries. They found 1 247 CPM pairs classified into two groups: those wide binaries belonging to the local disk and those belonging the local halo. It appears that both population of wide binaries have similar distributions of semi-major axis, luminosity, and mass ratio. Chanamé and Gould therefore conclude that disk and halo wide binaries have probably formed under similar conditions.

The CPM method is the most successful approach in identifying *individual* genuine wide pairs. To be reliably identified, the components of a CPM pair must have relatively high proper motions and, therefore, tend to be relatively nearby (Fig. 1.10). The proximity has several advantages: Separations as small as 200 AU are resolved in most imaging surveys and intrinsically faint companions, such as M dwarfs, can be identified. Furthermore, many high proper motion stars have precise parallax measurements providing a further stringent test of the genuineness of the pair.

[27] *Centre de Données astronomiques de Strasbourg* accessible at `http://cds.u-strasbg.fr/`. The CDS-ViZier catalogue number of the LDS is I/130.

On the other hand, the restriction to nearby stars limits the size of current samples of wide binary candidates. Larger wide binary sample are needed, especially regarding the widest pairs and their implications on dark matter. In the present study, we decided, therefore, not to use proper motion information but to take a different approach. We do *not* attempt to identify individual wide binary stars in first place, but we look for a statistical signal stemming from real pairs by exploiting position measurements only.

### 1.4.2 Two-point correlation function

The two-point correlation function (2PCF) is a straightforward and well-established clustering measure and is widely used to study the distribution of galaxies. It is the principal tool for studying the large-scale structure of the Universe (e.g. Peebles 1980; Saslaw 2000).

The 2PCF compares the observed distribution of the positions (of the stars or galaxies) with the distribution expected if the objects would have been placed by chance in the sky. *The 2PCF measure the excess probability of observing two objects with a certain separation with respect to a random distribution.* The measured excess is then explained by the presence of pairs bound by gravity. This is just like the argument put forward already in 1767 by Michell.

A mathematical definition of the 2PCF can be given as follows (see also §2.3): let $\Omega$ be an arbitrarely shaped area containing $N$ stars as illustrated in Fig. 1.11. If the stars were distributed randomly over $\Omega$ then the number of pairs having an angular separation between $\theta$ and $\theta + \mathrm{d}\theta$, $P$, equals the total number of distinct pairs in $\Omega$ times the fraction of the area that lies within an annulum of radius $\theta$ and width $\mathrm{d}\theta$

$$P(\theta)\mathrm{d}\theta = \frac{N(N-1)}{2}\frac{\mathrm{d}\Omega(\theta)}{\Omega}\,. \tag{1.1}$$

For small angular separations we can ignore the spherical nature of the sky. If we furthermore neglect edge corrections due to the finite sample size (see §2.3.2), the area of the annulum $\mathrm{d}\Omega$ is simply $2\pi\theta\mathrm{d}\theta$.

The number of *observed* pairs, $F$, can then be expressed by an analogous formula

$$F(\theta)\mathrm{d}\theta = \frac{N(N-1)}{2}(1 + w(\theta))\frac{\mathrm{d}\Omega(\theta)}{\Omega}\,, \tag{1.2}$$

where we have introduced $w$, the 2PCF. This equation can be regarded as the defining formula of the 2PCF. Combining the Eqs. 1.1 and 1.2 we get the simplest estimator of the 2PCF

$$w(\theta) = \frac{F(\theta)}{P(\theta)} - 1\,. \tag{1.3}$$

For a random distribution we have, of course, $w = 0$ at all separations A positive $w$ implies clustering, attributed to the mutual attraction by gravity, a negative $w$ would be an indication that the stars are avoiding each other.

The first application of the 2PCF to study the clustering properties of stars, as far as known to the author, is by Shanks et al. (1980). They constructed the angular 2PCF for a sample of 20 000 stars and found evidence the bluer and the redder stars in their sample form two distinct populations.

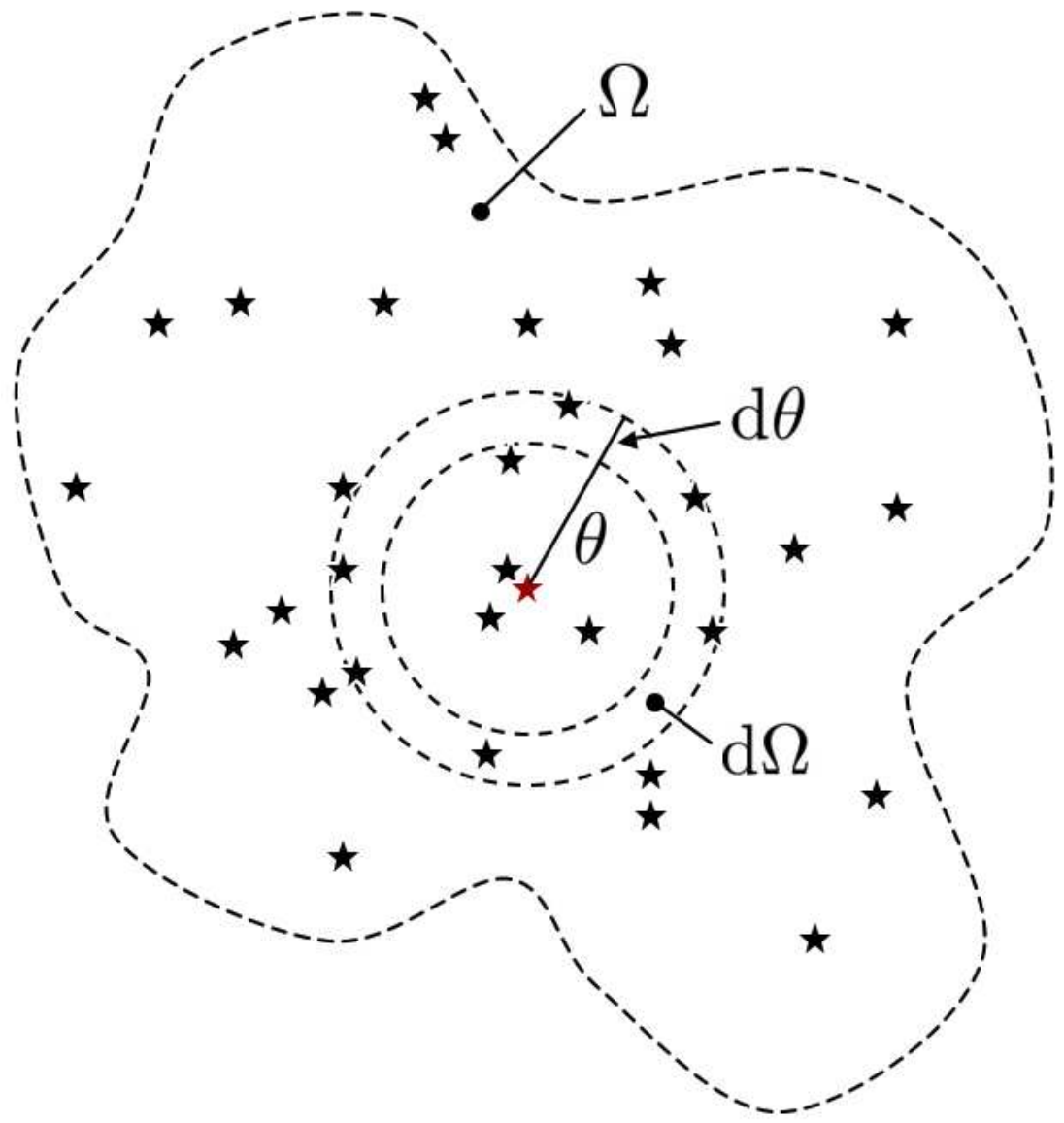

Figure 1.11: Illustration of the geometry for constructing the 2PCF (see text).

The study of wide binary stars using the 2PCF was pioneered by Bahcall and Soneira (1981). They investigated the clustering properties of a sample of nearly 3 000 stars brighter than $V = 16$ mag from the Weistrop (1972) catalogue covering 13.5 square degrees in the direction of the NGP. Beside the 2PCF, Bahcall and Soneira also constructed nearest-neighbour distributions – a related clustering measure – and demonstrated "that the stars which seem close together in the sky are in fact close together in space". They found that a significant fraction ($\sim 15\%$) of the stars are members of binary or triple system with a typical separation of about 0.1 pc. Using distance estimates by photometric parallaxes, they compiled a list of 19 candidate pairs brighter than $V = 12$ mag, from which six have been shown to be physical binaries using accurate radial velocity measurements and spectroscopic parallaxes (Latham et al. 1984).

Garnavich (1988, 1991) used for the first time the technique developed by Wasserman and Weinberg (1987) (WW-technique) to model the observed angular 2PCF. The WW-technique makes some simple assumption on the statistical properties of the WB population, such as the semi-major axis distribution (single power law), the distribution of eccentricities ('thermal': $f(e) = 2e$), the orientations of the orbital planes (random), etc., and projects them onto the sky using the selection criteria of the given star catalogue (magnitude limits, resolution limits, etc.) and the density distribution and luminosity function of the stars in the Milky Way Galaxy (see also §2.4).

The modeling allowed Garnavich to infer the local number density of the WBs $n_{\rm WB}$ as well as the power-law index of the semi-major axis distribution $\lambda$ from a large sample of stars brighter than $V = 14$ mag. For the NGP sample covering nearly 240 square degrees, he found $\lambda = 0.7 \pm 0.2$ and an unphysically large density of $\sim 0.06$ pc$^{-3}$, which would require that every star in the solar neighbourhood is a member of a WB.

This overdensity was already noted by Wasserman and Weinberg. They pointed out that the six pairs confirmed by Latham et al. imply a uncomfortably large WB density of $n_{\rm WB} \sim 0.3$ pc$^{-3}$, which would require a single star density 4 times the observed value of $n_* \sim 0.15$ pc$^{-3}$. So, it seemed that there were too many WBs in the Weistrop sample (see also Weinberg 1988). Most likely this overdensity is due to a large statistical fluctuation, i.e. the limited samples studied by Wasserman and Weinberg and by Garnavich are not representative regarding the WB frequency around the NGP as a whole (Saarinen and Gilmore 1989). The conjecture that the overdensity is due to an undetected star cluster could not be confirmed by Saarinen and Gilmore, even though there is observational evidence that some WBs are the remnants of disrupted clusters (Loden 1988). In the light of the formation scenarios suggested by Moeckel and Bate (2010) and by Kouwenhoven et al. (2010) a possible connection between WBs and open clusters is certainly expected and worth to be further investigated.

Gould et al. (1995) applied the 2PCF to the HST Snapshot Survey and obtained 13 candidate pairs. They found that the distribution of angular separations is well fitted by a single power law with index $1.2 \pm 0.4$. Furthermore, their study suggests that the WBs have bluer colours than would be expected if both components were drawn randomly from the field star population.

In their appendix Gould et al. give an excellent review of the work on WBs done so far. This review seem to mark the end of the use of the 2PCF to study WBs; no follow-up study is known to the author. In the last 20 years an enormous progress in deep photometric sky surveys took place and a systematic search for wide binary candidates should be much more rewarding now.

The next two Chapters describe in detail the author's PhD project focusing on the statistical properties of WBs in the Galactic field. The study is based on the data from the Sloan Digital Sky Survey (SDSS, York et al. 2000), from which the 2PCF is constructed (Chap. 2). With the aid of a novel weighting procedure based on the binding probability of each pair, bias-corrected colour and mass-ratio distributions of the WBs are inferred (Chap. 3). Finally, a 'ranking list' of WB candidates is compiled.

# Chapter 2

# The stellar correlation function from SDSS

## A statistical search for wide binary stars

**Abstract.** We study the statistical properties of the wide binary population in the Galaxy field with projected separations larger than 200 AU by constructing the stellar *angular* two-point correlation function (2PCF) from a homogeneous sample of nearly 670 000 main sequence stars. The selected stars lie within a rectangular region around the Northern Galactic Pole and have apparent $r$-band magnitudes between 15 and 20.5 mag and spectral classes later than G5 ($g - r > 0.5$ mag). The data were taken from the Sixth Data Release of the Sloan Digital Sky Survey. We model the 2PCF by means of the Wasserman-Weinberg technique including several assumptions on the distribution of the binaries' orbital parameters, luminosity function, and density distribution in the Galaxy. In particular, we assume that the semi-major axis distribution is described by a single powerlaw. The free model parameters – the local wide binary number density $n_{\rm WB}$ and the power-law index $\lambda$ of the semi-major axis distribution – are inferred simultaneously by least-square fitting. We find the separation distribution to follow Öpik's law ($\lambda = 1$) up to the Galactic tidal limit, without any break and a local density of 5 wide binaries per 1 000 pc$^3$ with both components having spectral type later than G5. This implies that about 10% of all stars in the solar neighbourhood are members of such a late-type wide binary system. With a relative statistical ($2\sigma$) error of about 10%, our findings are in general agreement with previous studies of wide binaries. The data suggest that about 800 very wide pairs with projected separations larger than 0.1 pc exist in our sample, whereas none are found beyond 0.8 pc. Modern large-scale surveys make the 2PCF method a viable tool for studying wide binary stars and a true complement to common proper motion studies. The method is, however, seriously limited by the noise from optical pairs and the (over)simplifying assumptions made to model the selection effects and to interpret the measured clustering signal.

## 2.1 Introduction

Binary stars have traditionally been central for astronomy, especially *close* binary systems (with typical orbital periods of days to years), because they are genuine laboratories of stellar evolution and its exotic remnants, making them a cornerstone for determining masses, distances, and many other fundamental astrophysical parameters. On the other hand, *wide* binary systems (orbital periods of many thousands to millions of years) are of particular interest, too.

Because they are only weakly bound by gravitation, wide binaries are prone to tidal disruption by passing massive objects, such as massive stars, molecular clouds, MACHOs (massive compact halo objects), or dark matter (DM) substructure. The shape of their separation distribution, in particular towards the most extreme, widest separations, should therefore allow constraints on the mass and frequency of the disruptive perturbers (Retterer and King 1982), as well as to estimate the age of a population (Poveda and Allen 2004). Wide binary-based MACHO constraints have been placed (but subsequently criticised) by several authors (Bahcall et al. 1985a; Weinberg et al. 1987; Weinberg 1990; Wasserman and Weinberg 1991; Yoo et al. 2004; Quinn et al. 2009).

In a different context, Hernandez and Lee (2008) propose using the tightening of wide binaries in dwarf spheroidal galaxies through dynamical friction as a test for DM. Moreover, binaries with separations over 0.1 pc, which are known to exist in the Galaxy (e.g. Lépine and Bongiorno 2007, also this paper), are in the 'weak-accelaration' regime where, in principle, one could test for possible deviations from Newtonian gravity, such as MOND (Milgrom and Bekenstein 1987; Close et al. 1990).

Wide binaries are not only probes of dynamical evolution and Galactic structure, but they also provide important clues to star formation (Larson 2001). The exact outcome of the binary population in a star-formation event is an exceedingly complex and still unsolved problem (Goodwin et al. 2007, and references therein). The formation of extremely wide binaries is particularly difficult to understand (Allen et al. 2007; Parker et al. 2009). A good knowledge of the wide binary frequency and separation distribution is primary for a whole host of problems in astrophysics (e.g. Chanamé 2007).

Unfortunately, the long orbital periods of wide binary systems make their identification very difficult in the first place. There are basically two different methods for detecting a wide binary system: (1) by the number excess of neighbours around a given star with respect to a random distribution, or (2) by the common proper motion (CPM) of two well-separated stars. Although there is no way to observe the orbital motion of a wide binary pair, the CPM is nevertheless a reliable indicator of a physical relationship between two individual stars (e.g. Lépine and Bongiorno 2007). In the present study we use the *angular two-point correlation function* (2PCF) to measure the excess of neighbours compared to a random distribution. This method has the advantage that larger samples of more distant stars can be used, as only the stars' positions are required. It only allows, however, statistical statements on the pairing and is limited to relatively small angular separations, because the noise due to randomly associated pairs increases rapidly (linearly) with angular separation.

To date, our knowledge on wide binaries essentially had rested on studies using the CPM method. There are two excellent recent CPM studies of wide binaries by Chanamé and Gould (2004) and Lépine and Bongiorno (2007). Based on updated proper motion (Luyten and Hipparcos) catalogues of stars, 917 and 521 wide binary systems, respectively, have been identified over the entire Northern sky and with a typical median distance of about 100 pc. Lépine and Bongiorno found the separation distribution of the pairs to follow Öpik's (1924) law, i.e. frequency being proportional to the inverse of separation, out to separations of around 3 500 AU. Beyond this characteristic scale, however, the separation distribution seems to be falling by a steeper power law, without an obvious cut-off. Lépine and Bongiorno find wide binaries out to separations of almost 100 000 AU, or 0.4 pc! They quote a number of 9.5% for nearby ($D < 100$ pc) Hipparcos stars belonging to a wide binary system with a separation greater than 1 000 AU, again demonstrating the ubiquity of the (wide) binary phenomenon. Chanamé and Gould, in addition, achieve a distinction between wide binaries belonging to the Galactic disc and those belonging to the Galactic halo. No significant difference in the separation distribution has been found, suggesting that the disc and halo binaries were formed under similar conditions, despite the very different metallicities and ages.

The angular 2PCF (also called “covariance function”) is one of the most useful tools for studying the clustering properties of *galaxies*, and it has been widely used to probe the large-scale structure of the Universe (e.g. Peebles 1980, for a more recent study of the angular clustering of galaxies see Scranton et al. 2002 and references therein.). The 2PCF can also be, and has been, used to measure the clustering properties of stars. While on large scales the stars are clearly randomly distributed in the Galaxy, as expected in a collisionless system, correlations up to the 10 pc scale (or even beyond) can be found in moving groups and halo streamers (Doinidis and Beers 1989), star-forming regions (e.g. Gomez and Lada 1998), and open star clusters (López-Corredoira et al. 1998). On very small scales (sub-pc, observationally: sub-arcmin), there is a strong signal due to (visual) binary stars.

The 2PCF method of probing wide binary stars in the Galactic field was pioneered by Bahcall and Soneira (1981), who studied the distribution of stars down to a limiting magnitude V = 16 in a 10 square degree field at the NGP, and found very significant clustering at a separation of 0.1 pc. Of the 19 binary candidates, 6 turned out to be real (Latham et al. 1984). The theoretical implications of these observations for the frequency and separation distribution of the binaries, and a general method for modelling them, was worked out by Wasserman and Weinberg (1987). To date, there are only few follow-up studies of the stellar 2PCF (Saarinen and Gilmore 1989; Garnavich 1988, 1991; Gould et al. 1995). Given this surprising lack of further work on the stellar 2PCF to study wide binaries, and in view of the enormous progress in deep photometric sky surveys in the past 20 years that should render the 2PCF method – as a true complement to the CPM method – much more rewarding now, we have started a project to use the huge stellar database of the Sloan Digital Sky Survey (SDSS York et al. 2000, `www.sdss.org`) for a detailed stellar correlation analysis to very faint magnitudes.

An independent study of wide binaries in the SDSS database by Sesar et al. (2008) takes a different approach, based on the "Milky Way Tomography" of Jurić et al. (2008), where approximate distances are ascribed to all stars by adopting a photometric parallax relation. Candidate binaries are selected by the requirement that the difference between two potential components in apparent magnitude is within a certain error equal to the difference in absolute magnitude.

Using distance information has the fundamental advantage of filtering out the disturbing noise from chance projections and of avoiding the need for a complex, Wasserman and Weinberg-type modelling to calculate integrated, sky-projected quantities. Although we plan to include distance information in future work, we show here that the *angular* correlation analysis is in principle still a viable approach. Our main results agree with those of the CPM studies. We also show, however, what the limitations of the method are.

The paper is organised as follows. In Sect. 2.2 we describe the SDSS data input and define the cleansed sample used for the analysis. Section 2.3 is devoted to the angular correlation function apparatus, followed by an extensive description of the modelling of the correlation function drawing on a modified Wasserman and Weinberg technique in Sect. 2.4. Our results for the total sample and for a number of subsamples are presented in Sect. 2.5 and are critically discussed and compared to previous work in Sect. 2.6. Concluding remarks are given in Sect. 2.7.

## 2.2 Data

For our analysis we selected a homogeneous sample of stars from the Sixth SDSS Data Release (DR6; Adelman-McCarthy et al. 2008). We took a rectangular area around the Northern Galactic Pole (NGP) covering approximately $\Omega_{\rm tot} \simeq 675$ square degrees (dec: $22° - 42°$, RA: $165° - 205°$; see Fig. 2.1). It contains 966 656 *primary*[1] point-like objects – including quasars, asteroids, and possibly some misidentified galaxies – selected from the *Star* view[2], having an apparent PSF magnitude[3] in the $r$-band between 15 and 20.5 mag. Following the SDSS recommendations[4], we required the stars to have *clean photometry*[5] in the $g$, $r$, and $i$ band. Although the effect of interstellar dust on the measurements is waek in the direction of the NGP, we corrected the data using the Schlegel et al. (1998) maps, which can be easily done, as the `extinction` at the position of each object is stored in the SDSS database.

The magnitude limits were chosen to be within the saturation limit of the SDSS CCD cameras ($r \sim 14$ mag; Gunn et al. 1998)[6] and the limit where the star-galaxy separation becomes unreliable ($r \sim 21.5$ mag; Lupton et al. 2001)[7]. Additionally, we avoid close stars being overcorrected by the adopted extinction correction, since most stars with $r > 15$ mag are behind the entire dust column (Jurić et al. 2008). On the other hand, with the somewhat conservative faint limit of $r = 20.5$ we make sure that only very few stars have a magnitude in the $g$ or $i$ band beyond the 95% completeness limit of 22.2 mag or 21.3 mag, respectively (Adelman-McCarthy et al. 2007, Table 1)[8].

[1]Due to overlaps in the imaging, many objects are observed more than once. The best of those observations is called the *primary object*, all the others are called *secondary objects*.

[2]See `cas.sdss.org/dr6/en/help/browser/browser.asp`

[3]See `sdss.org/dr6/algorithms/photometry.html#mag_psf`

[4]See `sdss.org/dr6/products/catalogs/flags.html`

[5]See `cas.sdss.org/dr6/en/help/docs/realquery.asp#flags`

[6]See also `astro.princeton.edu/PBOOK/camera/camera.htm`

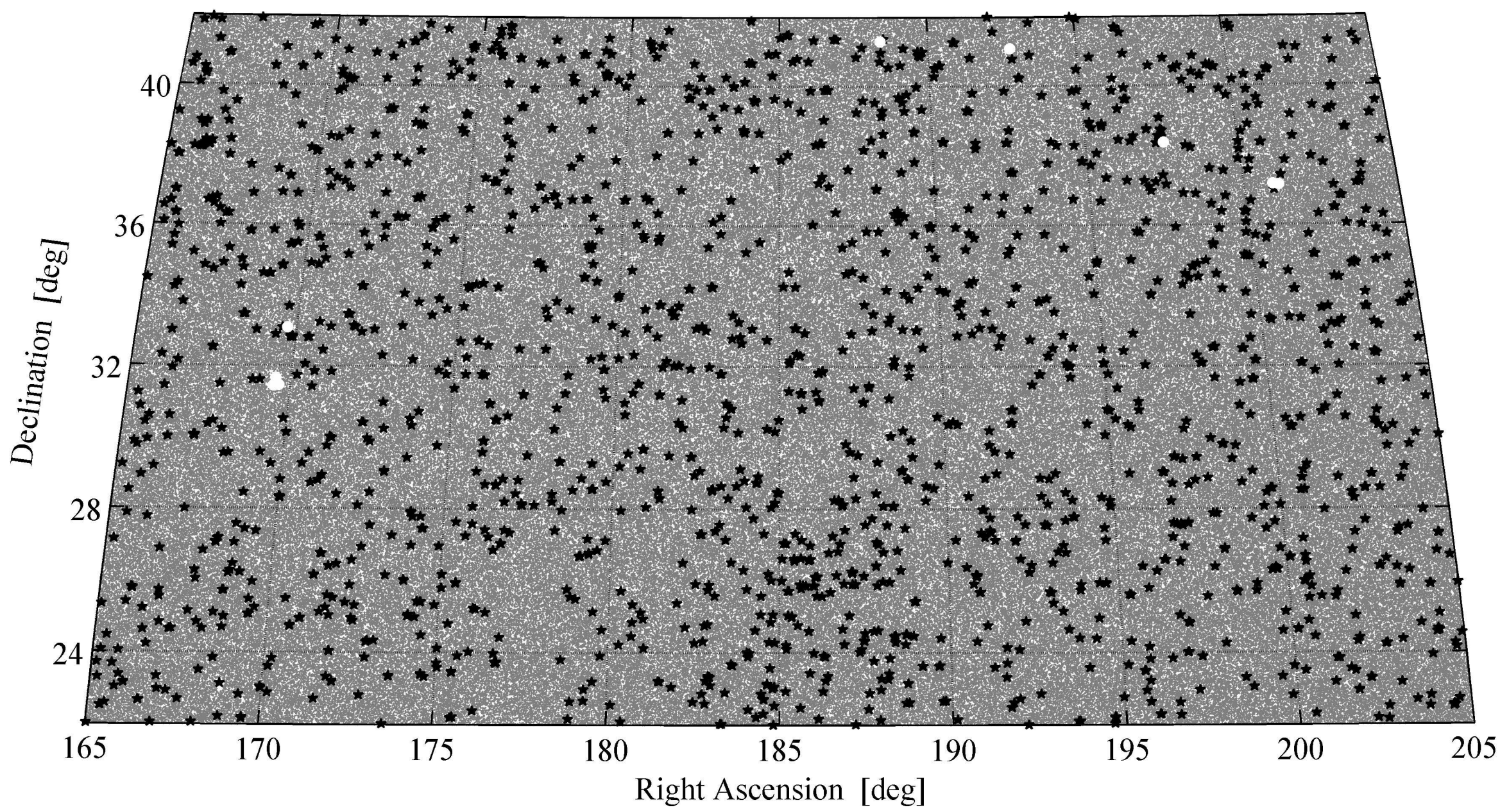

Figure 2.1: Distribution of stars (grey points filling the background) in our total sample. Black asterisks show the positions of bright star masks, white circles those of hole masks. The size of the symbols are not to scale.

The average seeing of the SDSS imaging data (median PSF width) is $1.4''$ in the $r$-band (Adelman-McCarthy et al. 2008)[9]. To be on the safe side, we took the minimum angular separation to be $\theta_{\rm min} = 2''$.

### 2.2.1 Contaminations

Matching our sample with the *QsoBest* table[10] resulted in the exclusion of 10 041 quasar candidates. Most of them (8 157) have $g - r \lesssim 0.5$ mag and are scattered in a colour-colour diagram around the otherwise narrow stellar locus as shown in Fig. 2.2. Even after removing the quasar candidates, the remaining objects in the blue part of our sample show a suspicious scatter, which is probably caused by further quasars and misidentified galaxies.

[7]See also `sdss.org/dr6/products/general/stargalsep.html`
[8]See also `sdss.org/dr6/`
[9]See also `http://www.sdss.org/dr6/` and the DR5 paper (Adelman-McCarthy et al. 2007)
[10]See `cas.sdss.org/dr6/en/help/docs/algorithm.asp?key=qsocat`

We therefore decided to exclude all objects with $g-r < 0.5$ mag, removing a further 286 227 objects from our sample. Furthermore, we reject all moving objects (asteroids) by cutting on the `DEBLENDED_AS_MOVING` flag. This leaves us with 670 388 objects classified as "stars" in the sample.

The large majority of the stars observed by the SDSS are main sequence (MS) stars. Finlator et al. (2000) estimate the fraction of stars that are not on the MS to be $\sim 1.0\%$, most of them giants and subgiants ($\sim 90\%$), but also horizontal branch stars ($\sim 10\%$). The number of white dwarfs observed by the SDSS is negligible compared to the number of MS stars (e.g. Harris et al. 2006). Thus, it seems well-justified to assume that *all* the stars in our sample are on the MS.

The cut discussed above at $g - r = 0.5$ mag implies that our sample contains only stars from spectral type later than about G5 (Finlator et al. 2000). In addition, this cut is appropriate for our purposes for the following three reasons:

*i)* Because they are very young, the bluest MS stars are mostly members of loose associations, so their clustering properties still represent the peculiarities of their birth places. Being an interesting subject to study (e.g. Kobulnicky and Fryer 2007), excluding them assures that our clustering signal is predominantly due to wide binaries in the field that have lost their memory of their birth places.

*ii)* As the MS becomes more sparsely populated towards the blue end, and a significant fraction could be made up by metal-poor halo giants, the assumption that all stars in our sample are on the MS might not be valid for the bluest stars.

*iii)* For magnitudes $M_V \gtrsim 4.5$ mag, the shape of the halo luminosity function agrees well with that of the disc luminosity function (e.g. Bahcall et al. 1985b, their Fig. 2). Thus, the cut on $g - r = 0.5$ mag, which corresponds to a cut at $M_V \simeq 5.6$ mag, allows us to use the disc luminosity function for the halo component.

### 2.2.2 Survey holes and bright stars

The sample chosen contains some regions where, for various reasons, no object could be observed. These regions are masked[11] as *hole* in the SDSS database and can therefore be easily identified. Holes in the sample affect our analysis in two ways: first, they diminish the total area of the sample. Second, they introduce an edge effect, as stars near a hole show a lack of neighbouring stars. The former effect can be easily corrected for in an approximate way as the SDSS provides the radius $\theta_{\rm M}^{(i)}$ of its bounding circle for every mask $i$. The residual area of the sample is then

$$\Omega \simeq \Omega_{\rm tot} - \sum_{i=1}^{N_{\rm M}} \pi \left(\theta_{\rm M}^{(i)}\right)^2 \tag{2.1}$$

where $N_{\rm M}$ denotes the number of hole masks. We discuss the correction for edge effects in Sect. 2.3.2 in the context of the correlation function.

[11] See `sdss.org/dr6/algorithms/masks.html`

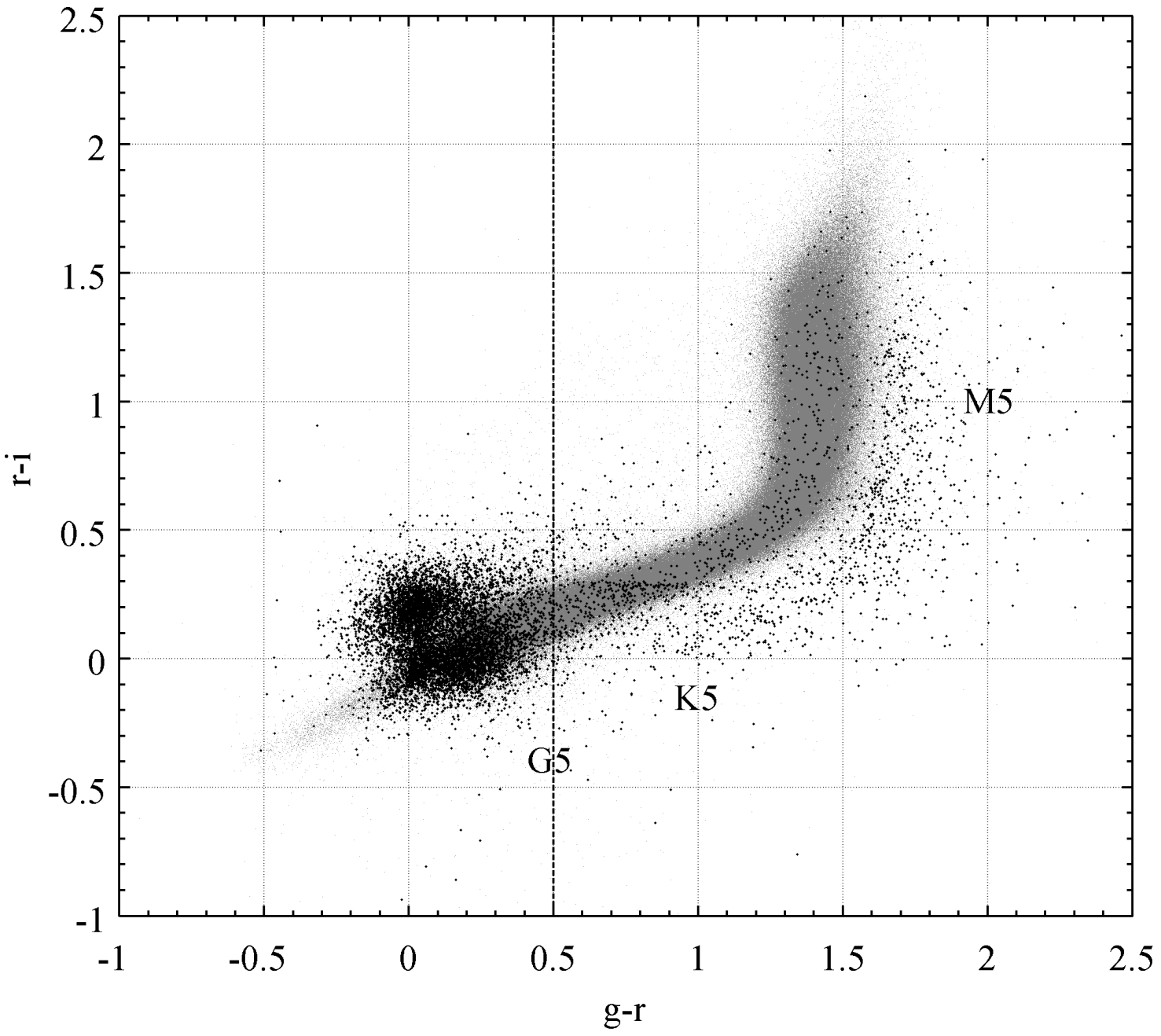

Figure 2.2: Colour-colour diagram of all point-like objects within our coordinate's and apparent magnitude's limits (grey points). Black points are SDSS quasar candidates. The vertical dashed line indicates the cut at $g - r = 0.5$ mag (see text). Approximate spectral classes are indicated.

Restricting ourselves to the masks defined in the $r$-band, we find $N_{\rm SH} = 9$ regions masked as "hole" in our sample (survey holes: SH) with an average radius of $\bar{\theta}_{\rm SH} \simeq 7.76$ arcmin. These survey holes diminish the total area of the sample by approximately 0.46 square degrees, constituting a marginal correction that could be safely ignored.

Very bright (saturated) stars cause similar problems: a very bright object may appear like a hole in our sample, when the underlying fainter stars blended with this object cannot be revealed. Within our sample there are $N_{\rm BS} = 1\,790$ *bright star* (BS) masks in the $r$-band. We exclude all objects inside such a mask, removing 545 stars from our sample. The bounding circles of bright star masks have an average radius of $\bar{\theta}_{\rm BS} \simeq 1.95$ arcmin, resulting in a further diminishing of the total area of the sample of circa 5.96 square degrees. By correcting the sample's total area for both hole and bright star masks, we get a residual area $\Omega \simeq 675 - 0.46 - 5.96\ \mathrm{deg}^2 = 668.58\ \mathrm{deg}^2$.

### 2.2.3 Final sample

Our final sample contains $N_{\rm obs} = 669\,843$ MS stars with *clean photometry*, a spectral type later than about G5 ($g - r \geq 0.5$ mag), and an apparent magnitude in the range $15 \leq r \leq 20.5$ mag. The stars are distributed over a solid angle $\Omega \simeq 668.58$ square degrees (after correcting for bright star masks and minor survey holes), corresponding to a mean surface density of $n = N_{\rm obs}/\Omega \simeq 1\,000$ stars per square degree. Throughout this study, we use the terms “total sample” and “final sample” synonymously.

## 2.3 Stellar correlation function

### 2.3.1 Estimation of the correlation function

The angular two-point *auto*-correlation function (2PCF) $w(\theta)$ is defined via the joint probability $\mathrm{d}\mathcal{P}$ of finding objects in both the solid angles $\mathrm{d}\Omega_1$ and $\mathrm{d}\Omega_2$

$$\mathrm{d}\mathcal{P} = n^2\,(1 + w(\theta))\,\mathrm{d}\Omega_1 \mathrm{d}\Omega_2\,, \tag{2.2}$$

where $n$ is the mean density of objects in the sky and $\theta$ is the separation between the two areas (e.g. Peebles 1980, §45). The number of distinct stellar pairs, $F(\theta)\mathrm{d}\theta$, with an angular separation between $\theta$ and $\theta + \mathrm{d}\theta$, *observed* in a region of angular size $\Omega$, can then be written as

$$F(\theta)\mathrm{d}\theta = \frac{N_{\rm obs}(N_{\rm obs}-1)}{2}\,(1 + w(\theta))\,\frac{\mathrm{d}\Omega(\theta)}{\Omega} \tag{2.3}$$

where $N_{\rm obs}$ is the number of stars observed in $\Omega$ and $\mathrm{d}\Omega(\theta)$ is approximately equal to the solid angle of a ring with (middle) radius $\theta$ and width $\mathrm{d}\theta$

$$\mathrm{d}\Omega(\theta) \simeq 2\pi\theta\mathrm{d}\theta\,. \tag{2.4}$$

Solving for $w$ in Eq. 2.3 yields a simple estimator for the 2PCF:

$$\hat{w}(\theta) = \frac{F(\theta)}{P(\theta)} - 1\,, \tag{2.5}$$

where $P(\theta)$ is the number of pairs expected from a random sample with $N_{\rm obs}$ points ($w = 0$ in Eq. 2.3):

$$P(\theta)\mathrm{d}\theta \simeq \frac{\pi N_{\rm obs}(N_{\rm obs}-1)}{\Omega}\theta\mathrm{d}\theta\,. \tag{2.6}$$

The 2PCF estimate $\hat{w}(\theta)$ is a measure for the excess of observed pairs separated by an angle $\theta$ with respect to a randomly distributed sample.

Another statistical measure equivalent to the 2PCF, which we use for visualising the data, is the *cumulative difference distribution* (CDD) $\hat{\gamma}(\theta)$, giving the cumulative number of pairs in excess of a random distribution:

$$\hat{\gamma}(\theta) \equiv \int_{\theta_{\rm min}}^{\theta} (F(\theta') - P(\theta'))\,\mathrm{d}\theta' = \int_{\theta_{\rm min}}^{\theta} \hat{w}(\theta')P(\theta')\mathrm{d}\theta'\,, \tag{2.7}$$

with $\theta_{\rm min} = 2''$ (see above). The CDD is completely equivalent to the 2PCF, but is – as for all cumulative distributions – more sensitive to statistical trends caused over a wider range of angular separations (i.e. over several bins), at the expense of a strong correlation of its values at different separations. The use of the simple estimator in Eq. 2.5 is appropriate as long as boundary effects due to the finite sample size are negligible (Bernstein 1994). We discuss boundary effect in the next section. The number of pairs observed $F(\theta)$ is determined very efficiently *inside* the SDSS database using the precalculated *Neighbors* table which contains all the objects within $\theta_{\max} = 30''$ of any given object.

### 2.3.2 Boundary effects

Stars close to the sample's boundary have a somewhat truncated ring $\mathrm{d}\Omega(\theta)$ and consequently show a lack of neighbours. However, as in the present study the probed angular scale is small compared to the sample's size, it turns out that these edge effects have a negligible impact on our results: we estimate the relative error introduced in $\mathrm{d}\Omega(\theta)$ by omitting the edge correction to be less than $\sim 0.04\%$ for any given $\theta \leq 30''$.

The stars near the boundary of a hole or a bright star mask (in the following simply "hole") show a lack of neighbours, too. Therefore, only a fraction $F_{\mathrm{H}}^{\mathrm{tot}}(\theta)$ of all pairs separated by an angular distance $\theta$ has been observed, and we need to correct the observed number of pairs $F(\theta)$. The "true" number of pairs, corrected for edge effects, then reads as

$$F_{\mathrm{corr}}(\theta) = \frac{F(\theta)}{F_{\mathrm{H}}^{\mathrm{tot}}(\theta)}\,. \tag{2.8}$$

In calculating the fraction $F_{\mathrm{H}}^{\mathrm{tot}}$ we proceeded in the same way as López-Corredoira et al. (1998). The calculation is outlined in Appendix 2.A. We find that the relative error in $F$ for $\theta = 30''$ amounts to

$$\Delta_F \equiv \frac{F_{\mathrm{corr}} - F}{F_{\mathrm{corr}}} = 1 - F_{\mathrm{H}}^{\mathrm{tot}} \simeq 0.7\%\,. \tag{2.9}$$

For smaller angular separations, the effect is even less.

Taking into consideration these edge effects, we rewrite the 2PCF estimate

$$\hat{w}_{\mathrm{corr}}(\theta) = \frac{F_{\mathrm{corr}}(\theta)}{P(\theta)} - 1 \tag{2.10}$$

and the CDD

$$\hat{\gamma}_{\mathrm{corr}}(\theta) = \int_{\theta_{\min}}^{\theta} \hat{w}_{\mathrm{corr}}(\theta')P(\theta')\mathrm{d}\theta'\,. \tag{2.11}$$

### 2.3.3 Uncertainty of the correlation function estimate

As the values of $\hat{w}_{\mathrm{corr}}$ at different separations are *not* independent, Poisson errors may underestimate the uncertainty in $\hat{w}_{\mathrm{corr}}$, especially when the angular scales under consideration are large (Hamilton 1993; Bernstein 1994). However, as we see in Sect. 2.5, the clustering of wide binary stars occurs on small angular scales ($\theta \leq 15''$) compared to the size of our sample (20 deg $\times$ 40 deg). We therefore expect that using Poisson errors only underestimates the true errors by a small amount in our case. Thus, we adopt Poissonian errors on $F(\theta)$:

$$\delta F(\theta) = \sqrt{F(\theta)} \quad \text{and} \quad \delta F_{\rm corr}(\theta) = \sqrt{\frac{F_{\rm corr}(\theta)}{F_{\rm H}^{\rm tot}(\theta)}}\,. \tag{2.12}$$

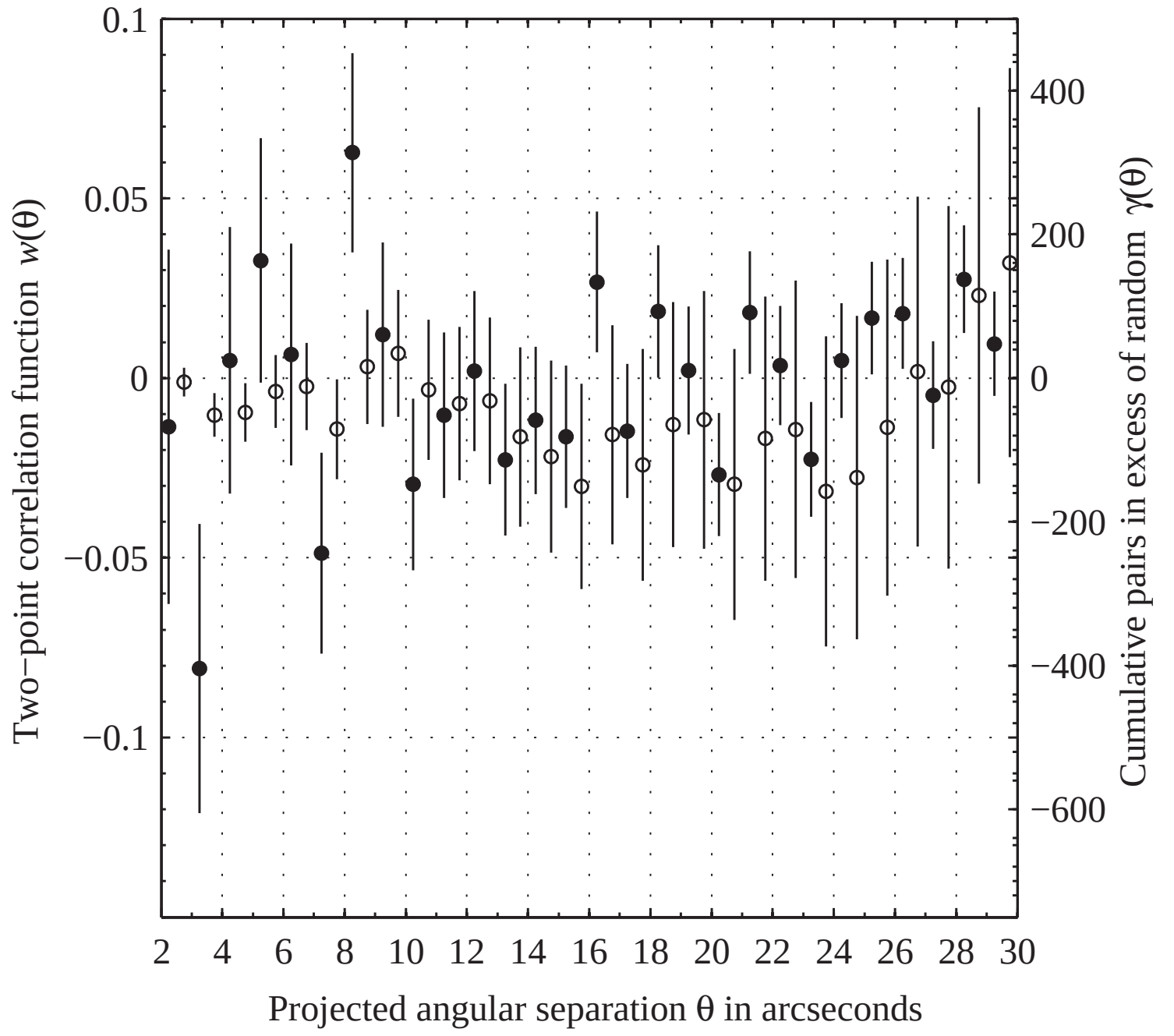

Figure 2.3: 2PCF as inferred from a random sample (solid circles, left ordinate) and the corresponding CDD (open circles, right ordinate). Poisson errors are indicated as vertical lines.

Using Eq. 2.10 and Gauss' error propagation formula, we may write the uncertainty of the CF estimate $\hat{w}_{\rm corr}$ as

$$\delta\hat{w}_{\rm corr} = \frac{\delta F_{\rm corr}}{P} = \sqrt{\frac{\hat{w}_{\rm corr}+1}{F_{\rm H}^{\rm tot}P}}\,, \tag{2.13}$$

where we omitted the dependencies on $\theta$ for the sake of brevity.

The uncertainty in the CDD $\hat{\gamma}$ is easily obtained in the same way:

$$\delta\hat{\gamma}_{\rm corr}(\theta) = \sqrt{\int_{\theta_{\rm min}}^{\theta} \delta F_{\rm corr}^2(\theta')\mathrm{d}\theta'}\,. \tag{2.14}$$

### 2.3.4 Testing the procedure for a random sample

To test the validity of the procedure to estimate the 2PCF described in the previous sections, we generated a random sample having the same number of "stars" and distributed over the same area as the stars in our final sample. In addition, we made certain that the random sample contains the same number of holes with appropriate radii.

For the analysis of the random sample we wrote a dedicated computer program that lays a grid with a mesh size of 1 arcmin over the sample before counting the pairs. This grid acts as a distance filter and avoids calculating the distances between all possible pairs: only pairs with both their components within a cell or with them in two adjacent cells, respectively, were taken into account.

If our procedure is correct we would expect a zero signal in both the 2PCF and the CDD when analysing a sample of randomly distributed stars. Figure 2.3 shows the result of the analysis of our random sample. There is indeed no clustering signal evident, neither in the 2PCF nor in the CDD. The results are consistent throughout with a zero signal out to the separation limit of 30"(compare also with Fig. 2.5). This shows that our procedure for estimating the 2PCF is reliable.

## 2.4 The model

Our approach to modelling the angular 2PCF is based on a technique developed by Wasserman and Weinberg (1987), hereafter WW87. As we describe in more detail in the following sections, this technique makes some simple assumptions on the basic statistical properties of wide binaries, and projects these theoretical distributions on the observational plane using the selection criteria of a given (binary) star catalogue and the geometry of the Milky Way galaxy (see also Weinberg 1988).

Their long periods make it virtually impossible to distinguish wide binary stars from mere chance projections (optical pairs) by their orbital motion. Therefore, we do not attempt to identify individual wide binaries, but we look for a *statistical* signal stemming from physical wide pairs in the sample, solely exploiting precise stellar position measurements provided by the SDSS.

The original Wasserman and Weinberg (WW) technique calculates the projected separation distribution to compare it with a sample of wide binaries of known distance and angular separation. In the present study we are only dealing nwith angular separations. The calculation of wide binary counts as a function of angular separation alone requires an appropiate modification of the WW-technique (see also Garnavich 1991), which we discuss in Sect. 2.4.3.

### 2.4.1 Wasserman-Weinberg technique

WW87 developed “a versatile technique for comparing wide binary observations with theoretical semimajor axis distributions”. According to WW87 we may write the number of observed binaries $\psi(s)\mathrm{d}s$ with projected physical separations between $s$ and $s+\mathrm{d}s$ in a given catalogue of stars as

$$\psi(s)\,\mathrm{d}s = n_{\mathrm{WB}}Q(s)V(s)\,\mathrm{d}s\,, \tag{2.15}$$

where $n_{\mathrm{WB}}$ is the total number density of wide binaries in the solar neighbourhood[12], $Q(s)$ is the reduced distribution of projected separations, and $V(s)$ the "effective volume".

The total number density $n_{\mathrm{WB}}$ is one of the two free parameters in the model that will be inferred by fitting the model to the observational data. The reduced separation distribution $Q(s)$ contains the physical properties from the wide binaries (semi-major axis distribution, distribution of eccentricities, orientation of the orbital planes) projected onto the observational plane, whereas the effective volume $V(s)$ takes into account the characteristics of the sample under consideration (covered area, range of angular separation examined, magnitude limits), as well as the stellar density distribution in the Galaxy and the luminosity function.

As we show in Sect. 2.4.3, this neat formal splitting in physical properties and selection effects will not be possible anymore after the required modification on the WW-technique mentioned above. At this point, we discuss the two parts – $Q(s)$ and $V(s)$ – more in detail.

### Reduced separation distribution

The reduced separation distribution $Q(s)$ is given by the projection of the reduced (present-day) semi-major axis distribution $q(a)$ against the sky. Gravitational perturbations due to stars, giant molecular clouds, and (hypothetical) DM particles cause the semi-major axes of (disc) wide binaries to evolve. Little is known about the initial semi-major axis distribution, and usually a single power-law is assumed, because of its simplicity (Weinberg 1988) but also because of theoretical considerations (Valtonen 1997; Poveda et al. 2007, and references therein). The evolution of the semi-major axis distribution of disc wide binaries has been modelled by Weinberg, Shapiro, and Wasserman (1987) in terms of the Fokker-Plack equation. Their numerical simulations suggest that the semi-major axis distribution evolves in a self-similar way for reasonable choices of the initial power-law index and the wide binary birth rate function. We decided, therefore, to model the present-epoch semi-major axis distribution of wide binary stars $q(a)$ by a single powerlaw

$$q(a) \propto \left(\frac{a}{\mathrm{pc}}\right)^{-\lambda} \ \mathrm{pc}^{-1} \tag{2.16}$$

with the power-law index $\lambda$, the second free parameter of the model. The range in $a$ where Eq. 2.16 is a valid description of the observed semi-major axis distribution at the same time provides the *definition* of what we consider as a "wide binary star". We specify this range by an upper and a lower limit, $a_{\mathrm{min}}$ and $a_{\mathrm{max}}$, respectively. There is observational evidence that the distribution of the semi-major axis of binary stars has a break at 0.001 pc (Abt 1983). Following WW87 and Weinberg (1988), we thus take the lower limit, providing the division between the close and wide binary populations, to be $a_{\mathrm{min}} = 0.001$ pc ($\approx 200$ AU). On the other hand, the Galactic tides provide a natural maximum semi-major axis $a_{\mathrm{T}}$ beyond which no bound orbits can exist. Details on the calculation of $a_{\mathrm{T}}$ are given in Appendix 2.B. We find it to be of the order of 1 pc.

[12]We use the term "solar neighbourhood" for local quantities.

We normalise $q(a)$ – hence the name "*reduced* semi-major axis distribution" – such that

$$\int_{a_{\min}}^{a_{\mathrm{T}}} q(a)\,\mathrm{d}a = 1 \tag{2.17}$$

giving

$$q(a) = c_\lambda \left(\frac{a}{\mathrm{pc}}\right)^{-\lambda} \mathrm{pc}^{-1} \tag{2.18}$$

where

$$c_\lambda = \begin{cases} (1-\lambda)\left[a_{\mathrm{T}}^{1-\lambda} - a_{\min}^{1-\lambda}\right]^{-1}, & \text{for } \lambda \neq 1 \\ \left[\ln a_{\mathrm{T}} - \ln a_{\min}\right]^{-1}, & \text{for } \lambda = 1 \end{cases} \tag{2.19}$$

with $a_{\min}$ and $a_{\mathrm{T}}$ in pc.

In a more general treatment, $q(a)$ might also depend on the luminosity classes of the binary star members, as well as on their magnitudes. Like WW87, we restrict our analysis to models where $q(a)$ only depends on the semi-major axis $a$.

It is somewhat disputed whether the semi-major axis distribution has a break at larger $a$ attributed to the disruptive effects of the environment on the widest binary stars. In their extensive work Wasserman and Weinberg (1991) (see also Weinberg 1990) conclud that "although the data *suggest* a break in the physical distribution of wide binary separations, they do not *require* a break with overwhelming statistical significance", whereas the more recent study by Lépine and Bongiorno (2007) shows evidence for a break at $s \sim 3\,500$ AU (statistically, we have $\langle s\rangle \simeq a$; see Eq. 2.22). In this context, the question arises whether the assumption that the data is described by a single powerlaw (Eq. 2.16) up to the tidal limit $a_{\mathrm{T}}$ can be rejected with confidence. We address this question in Sect. 2.5.

Assuming that "the binaries' orbital planes are randomly oriented and that their eccentricities $e$ are distributed uniformly in $e^2$" (WW87) we can project $q(a)$ against the sky by (Weinberg 1988)

$$Q(s) = \int_{s/2}^{\infty} \frac{\mathrm{d}a}{a} q(a)\mathcal{F}(s/a)\,, \tag{2.20}$$

where $\mathcal{F}(x)$ takes the eccentricity and angular averaging into account for the pairs, which may be written as (WW87, Weinberg and Wasserman 1988)

$$\mathcal{F}(x) = \frac{4x}{\pi} \int_0^{\sqrt{2-x}} \mathrm{d}u \sqrt{\frac{(u^2+x)(2-x-u^2)}{u^2+2x}}\,. \tag{2.21}$$

With the above assumptions on the orientations of the orbital planes and the distribution of eccentricities it can be shown that the *average* observed projected separation $\langle s\rangle$ for a given semi-major axis $a$ is basically equal to $a$ (e.g. Yoo et al. 2004):

$$\langle s\rangle = \frac{5\pi}{16} a \simeq 0.98a\,. \tag{2.22}$$

From the normalisations of $q(a)$ it also follows that $Q(s)$ is normalised to unity in the range $[\langle s\rangle_{\min}, \langle s\rangle_{\max}]$.

Substituting $\eta = \sqrt{s/2a}$ in Eq. 2.20, as well as $\xi = u/\sqrt{2}$ in Eq. 2.21 leads to a convenient form for numerical integration

$$Q(s) = \left(\frac{s}{\mathrm{pc}}\right)^{-\lambda} C_\lambda(s)\ \mathrm{pc}^{-1}, \tag{2.23}$$

where

$$C_\lambda(s) = 2^{\lambda+5}\,\frac{c_\lambda}{\pi} \int\limits_{\min\left(1,\,\sqrt{s/2a_{\mathrm{T}}}\right)}^{\min\left(1,\,\sqrt{s/2a_{\min}}\right)} \mathrm{d}\eta\,\eta^{2\lambda+1} \int\limits_0^{\sqrt{1-\eta^2}} \mathrm{d}\xi \sqrt{\frac{(\xi^2+\eta^2)(1-\eta^2-\xi^2)}{\xi^2+2\eta^2}}\,. \tag{2.24}$$

The integrals in Eq. 2.24 are evaluated using Gauss quadrature (Press et al. 1992). For $s$ far from $a_{\min}$ or $a_{\mathrm{T}}$ the distribution of the projected separations $Q(s)$ is approximately powerlaw

$$Q(s) \propto \left(\frac{s}{\mathrm{pc}}\right)^{-\lambda} \mathrm{pc}^{-1}. \tag{2.25}$$

**Effective volume**

As we deal with a magnitude-limited sample, we are plagued by selection effects that must be properly taken into account. Following WW87, we do this by means of the *effective volume*. It allows for the solid angle $\Omega$ covered by the sample, the angular separation range $[\theta_{\min}, \theta_{\max}]$ we are examining, and the apparent magnitude limits $m_{\min}$ and $m_{\max}$. Furthermore, the effective volume takes the stellar density distribution $\rho$ into account, as well as the normalised (single star) luminosity function $\widetilde{\Phi}$ (both described more in detail in Sect. 2.4.2).

"Making the plausible assumptions that the intrinsic luminosities of the two stars in a wide binary are independent and distributed in the same way as the luminosities of field stars" (WW87), and assuming further that the stellar density distribution $\rho$ is independent of the stars' absolute luminosities, we may write the effective volume as

$$V(s) = \Omega \int\limits_{s/\theta_{\max}}^{s/\theta_{\min}} \mathrm{d}D D^2\,\tilde{\rho}(D) \iint\limits_{M_{\min}(D)}^{M_{\max}(D)} \mathrm{d}M_1 \mathrm{d}M_2\,\widetilde{\Phi}(M_1)\widetilde{\Phi}(M_2)\,, \tag{2.26}$$

where we have the standard relations

$$M_{\max}(D) = m_{\max} - 5\log_{10}(D/10\ \mathrm{pc}) \tag{2.27}$$

$$M_{\min}(D) = m_{\min} - 5\log_{10}(D/10\ \mathrm{pc})\,, \tag{2.28}$$

and $\tilde{\rho}(D)$ is normalised such that $\tilde{\rho}(0) = 1$.

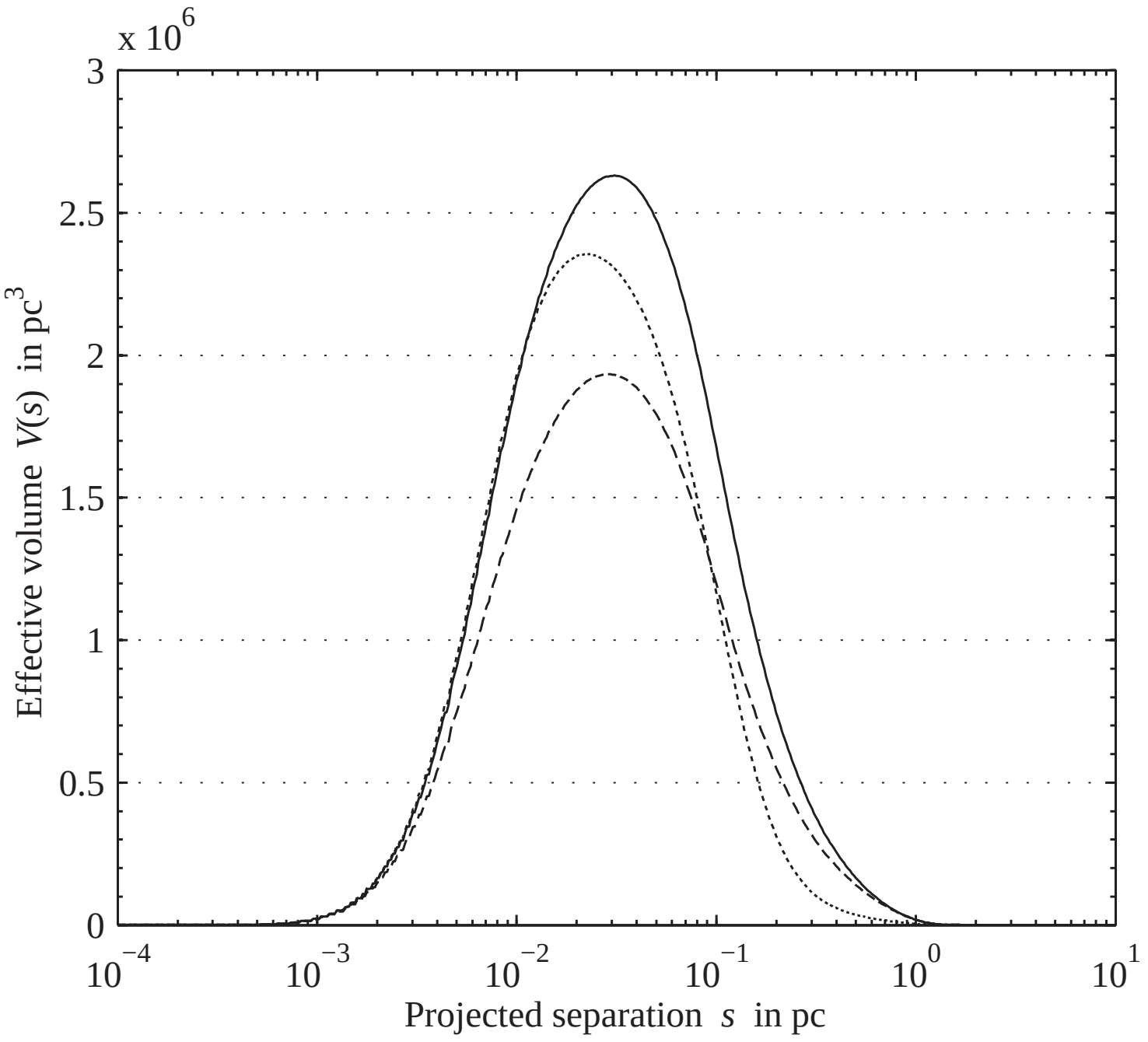

Figure 2.4: Effective volumes calculated using the parameters that correspond to our final sample (Table 2.2) for the three Galactic structure parameter sets as discussed in §2.4.2. Long dashed line: set 1; solid line: set 2; short dashed line: set 3.

In Fig. 2.4 the effective volumes for the three Galactic structure parameter sets (described in the next section) are plotted using the parameters that correspond to our final sample (see Table 2.2). From the effective volume $V(s)$, it is evident that our study is most sensitive to projected separations in 0.01 pc $\lesssim s \lesssim$ 0.1 pc, whereas our study is absolutely insensitive when $s \lesssim$ 0.001 pc or $s \gtrsim$ 1 pc. Referring to Eq. 2.22, we see that the shape of the effective volume also indicates that we are insensitive to semi-major axis $a \lesssim 0.001$ pc and for $a \gtrsim$ 1 pc, which nicely fits the range in $a$ where we assume that the single power-law model holds. But it also shows that our analysis is not too sensitive to very wide binary stars with semi-major axes larger than 0.1 pc.

### 2.4.2 Galactic model

#### Stellar density distribution

Following earlier work (especially Jurić et al. 2008; Chen et al. 2001, and references therein), we modelled the stellar density distribution of the Milky Way Galaxy by including two exponential disc components – a thin and a thick disc – and an elliptical halo component whose density profile obeys a powerlaw. The contribution of the Galactic bulge is negligible in the direction of the NGP and it is therefore ignored in the following.

The three components are added to yield the total density distribution

$$\rho(D) = \rho(0)\tilde{\rho}(D) \tag{2.29}$$

with

$$\tilde{\rho}(D) = \frac{\tilde{\rho}^{\text{thin}}(D) + n^{\text{thick}}\tilde{\rho}^{\text{thick}}(D) + n^{\text{halo}}\tilde{\rho}^{\text{halo}}(D)}{1 + n^{\text{thick}} + n^{\text{halo}}}\,, \tag{2.30}$$

which is normalised such that $\tilde{\rho}(0) = 1$. The thick disc and the halo component are normalised with respect to the thin disc via the normalisation constants $n^{\text{thick}}$ and $n^{\text{halo}}$, respectively. The overall normalisation $\rho(0)$ is calibrated to produce the observed star counts.

Both the thin and the thick disc populations obey a double exponential density law of the form (e.g. Bahcall and Soneira 1980)

$$\exp\left(-\frac{r - r_0}{h_r}\right)\exp\left(-\frac{|z|}{h_z}\right) \tag{2.31}$$

where $h_r$ and $h_z$ are the scale length and the scale height of the disc, $z$ is the object's height above the Galactic midplane, and $r$ is its Galactocentric distance in the Galactic plane. We take the Sun's distance from the centre of the Galaxy in the Galactic plane to be $r_0 = 8$ kpc. Jurić et al. (2008) find that the Sun is located $z_0 \simeq 25$ pc above the Galactic midplane (Chen et al. (2001) give $z_0 \simeq 27$ pc). It is straightforward to derive the following useful relations

$$z = z_0 + D\sin b \tag{2.32}$$
$$r^2 = r_0^2 + d^2 - 2r_0 d\cos\ell \tag{2.33}$$
$$d = D\cos b\,, \tag{2.34}$$

where $d$ is its distance from the Sun in the Galactic plane and $\ell$ and $b$ are its Galactic longitude and latitude, respectively. The stellar density is fairly constant within the sample, so we neglect the direction-dependent density variations. For the sake of simplicity, we adopt the coordinates of our sample's centre: $\ell = 175.6°$ and $b = 81.6°$. (Dividing our sample into subsamples and summing over them taking their centres hardly influences our results.)

In line with Jurić et al. (2008) and Chen et al. (2001) we assume the scale heights to be independent of absolute magnitude $M$. Using capitals for the thick disc's and lower case letters for the thin disc's scale height and length, we may write the normalised number density distribution of the stars in the Galactic thin disc as

$$\tilde{\rho}^{\text{thin}}(D) = \exp\left(-\frac{r - r_0}{h_r} - \frac{|z|}{h_z}\right)\cdot\exp\left(\frac{z_0}{h_z}\right) \tag{2.35}$$

and that of the thick disc as

$$\tilde{\rho}^{\text{thick}}(D) = \exp\left(-\frac{r - r_0}{H_r} - \frac{|z|}{H_z}\right)\cdot\exp\left(\frac{z_0}{H_z}\right)\,, \tag{2.36}$$

Table 2.1: Galactic structure parameters

| structure parameter | set 1 | set 2 | set 3 |
|---|---|---|---|
| $z_0$ [pc] | 25 | 25 | 27 |
| $h_z$ [pc] | 245 | 300 | 330 |
| $h_r$ [pc] | 2 159 | 2 600 | 2 250 |
| $n^{\text{thick}}$ | 0.13 | 0.12 | 0.0975 |
| $H_z$ [pc] | 743 | 900 | 665 |
| $H_r$ [pc] | 3 261 | 3 600 | 3 500 |
| $n^{\text{halo}}$ | 0.0051 | 0.0051 | 0.00125 |
| $\kappa$ | 0.64 | 0.64 | 0.55 |
| $k$ | 2.77 | 2.77 | 2.5 |

where the rightmost factors are for normalisation purposes and ensure that $\tilde{\rho}^{\text{thin}}(0) = \tilde{\rho}^{\text{thick}}(0) = 1$.

The density distribution of the stellar halo population is modelled by a powerlaw with index $k$. The observational data prefer a somewhat oblate halo giving an ellipsoid, flattened in the same sense as the Galactic disc, with axes $a = b$ and $c = \kappa a$, where $\kappa$ controls the ellipticity of the halo (cf. Jurić et al. 2008)

$$\tilde{\rho}^{\text{halo}}(D) = \left[ \frac{r^2 + \left(\frac{z}{\kappa}\right)^2}{r_0^2 + \left(\frac{z_0}{\kappa}\right)^2} \right]^{-\frac{k}{2}} , \tag{2.37}$$

which, of course, also satisfies $\tilde{\rho}^{\text{halo}}(0) = 1$.

We compare three different sets of structure parameters:

*Set 1:* The *measured* values from Jurić et al. (2008) (see their Table 10). These values are best-fit parameters as *directly* measured from the apparent number density distribution maps using their "bright" photometric parallax relation (their Eq. 2). They are not corrected for bias caused by, e.g., unresolved stellar multiplicity, hence the term "apparent".

*Set 2:* The *bias-corrected* values from Jurić et al. (2008) (see again their Table 10). These values were corrected for unrecognised stellar multiplicity, Malmquist bias, and systematic distance determination errors by means of Monte Carlo-generated mock catalogues (Jurić et al. 2008, §4.3.). Following Reid and Gizis (1997), the fraction of "stars" in the local stellar population that in fact are unresolved binaries is taken to be 35%. However, the halo component was not included in the mock catalogues, and its structure parameters were therefore not corrected, but the measured values are used instead. This set of parameters will be referred to as our standard set.

*Set 3:* As a third independent set of Galactic structure parameters, we refer to the somewhat earlier work of Chen et al. (2001). It is also based on observations obtained with the SDSS, but Chen et al. (2001) infer the density distribution of the stars by inverting the fundamental equation of stellar statistics (e.g. Karttunen et al. 1996).

These three sets of Galactic structure parameters are summarised in Table 2.1. Jurić et al. (2008) quote unrecognised multiplicity as one of the dominant sources of error in the distance determination of the stars. (Only the uncertainties in absolute calibration of the photometric parallax relation might be even more important, but little can be done to increase its accuracy at the moment; see, however, Sesar et al. (2008).) It is therefore not surprising that the scale heights and lengths of the bias-corrected parameter set are larger than those of the measured parameters as the misidentification of a binary star as a single star results in an underestimation of its distance. Regarding the values derived by Chen et al. (2001), we find the largest difference in the normalisation of the halo component that is about four times smaller than that given by Jurić et al. (2008), whereas the other parameters are not too dissimilar.

**Stellar luminosity function**

For our study we refer to the luminosity function (LF) inferred by Jahreiß and Wielen (1997) using Hipparcos parallaxes, which gives reliable values for a wide magnitude range ($-1 \leq M_V \leq 19$). (The faint end of the LF is somewhat uncertain and Jahreiß and Wielen (1997) give only a lower limit in the range $20 \leq M_V \leq 23$. We take that lower limit at the faint end to be the true value of the LF in that magnitude range.)

We need to transform the Jahreiß and Wielen (1997) LF from visual ($V$-band) into $r$-band magnitudes, i.e. from $\Phi(M_V)$ to $\Phi(M_r)$. We perform this transformation in an unsophisticated way by combining the photometric parallax relation from Laird et al. (1988), which was calibrated using the Hyades and is linear in $(B-V)$ (their Eq. 1a)

$$M_V = 5.64(B-V) + 1.11 \text{ mag} \tag{2.38}$$

with a transformation equation that is obtained by subtracting Eq. 1 from 3 in Bilir et al. (2005)

$$V - r = 0.491(B-V) - 0.144 \text{ mag}\,. \tag{2.39}$$

The linearity of the photometric parallax relation removes any inversion problem, since it assures – in a mathematical sense – that a colour index $(B-V)$ exists for every $M_V$. For a given distance we then have

$$M_r = 0.913 M_V + 0.0476 \text{ mag} \tag{2.40}$$

and the transformed LF is simply

$$\Phi(M_r) = \Phi(M_V)\frac{\mathrm{d}M_V}{\mathrm{d}M_r}\,. \tag{2.41}$$

Using linear interpolation (the brightest bin at $M_r = -1$ mag is an *extra*polation), we finally have rebinned $\Phi(M_r)$ such that the bins are centred on integer values of absolute magnitude $M_r$ and have a width of $\Delta M_r = 1$ mag.

We normalise the LF by integrating over the absolute magnitude range corresponding to our total sample

$$\widetilde{\Phi}(M_r) = \frac{\Phi(M_r)}{\int_{5.2}^{\infty} \Phi(\mu)\mathrm{d}\mu} = \frac{\Phi(M_r)}{n'_*} \tag{2.42}$$

with $n'_* \simeq 0.094$ pc$^{-3}$ being the total number density of stars in the solar neighbourhood having $g-r > 0.5$ mag. Assuming that all stars are on the MS, we find, using the ("bright") photometric parallax relation from Jurić et al. (2008), that the cut at $g - r = 0.5$ mag corresponds to $M_r \simeq 5.2$ mag. (We first transformed it from $g - r$ into $r - i$ using Eq. 4 from Jurić et al.)

### 2.4.3 Modification of the Wasserman-Weinberg technique

To derive a theoretical angular 2PCF we need to calculate the expected number of wide binaries as a function of *angular* separation. To this end, the above-described WW-technique requires the modification that we address now.

Let $\varphi(\theta)\mathrm{d}\theta$ be the number of wide binaries observed with an angular separation between $\theta$ and $\theta+\mathrm{d}\theta$. For a given distance $D$, we have $\varphi(\theta)\mathrm{d}\theta = \psi(s)\mathrm{d}s$ and $s = D\theta$. (Note that our unit for $s$ and $D$ is pc, whereas $\theta$ is in rad.) The latter expression can be used to write the reduced separation distribution $Q(s)$ as $Q(D\theta)$. This introduces an explicit dependency on $D$, so we need to incorporate $Q(D\theta)$ into the integration over $D$ in the formula for the effective volume (Eq. 2.26). This incorporation is the reason it is no longer possible to separate the reduced separation distribution from the effective volume in a formal way like in Eq. 2.15.

Furthermore, we need to modify the limits in the integration over $D$ in Eq. 2.15. Recalling that for a given semi-major axis $a$, the average observed projected separation is $\langle s \rangle = 0.98a$ (Eq. 2.22), it appears to be appropriate to let the integration limits run from $\langle s \rangle_{\min}/\theta$ to $\langle s \rangle_{\max}/\theta$ (Garnavich 1991). We take $\langle s \rangle_{\min} = 0.98a_{\min} = 9.8 \cdot 10^{-4}$ pc and $\langle s \rangle_{\max} = 0.98a_{\mathrm{T}}$ (for the calculation of the tidal limit $a_{\mathrm{T}}$ see Appendix 2.B).

Putting it all together, we may write the number of observed wide binaries as a function of angular separation as

$$\varphi(\theta) = n_{\mathrm{WB}}\Omega \int_{\langle s \rangle_{\min}/\theta}^{\langle s \rangle_{\max}/\theta} \mathrm{d}D D^3 \tilde{\rho}(D) Q(D\theta) \iint_{M_{\min}(D)}^{M_{\max}(D)} \mathrm{d}M_1 \mathrm{d}M_2 \widetilde{\Phi}(M_1)\widetilde{\Phi}(M_2)\,, \tag{2.43}$$

where we have included an additional factor $D$ into the integration over $D$, because $\mathrm{d}s = D\mathrm{d}\theta$ for a given $D$.

The model-2PCF is now determined by adding the number of physical pairs, calculated by Eq. 2.43, to the number of pairs expected from a random sample given by Eq. 2.6

$$w_{\mathrm{mod}}(\theta) = \frac{\varphi(\theta) + P(\theta)}{P(\theta)} - 1 = \frac{\Omega\, \varphi(\theta)}{\pi N_{\mathrm{obs}}(N_{\mathrm{obs}} - 1)\theta}\,. \tag{2.44}$$

The two free parameters – $n_{\rm WB}$ and $\lambda$ – are determined in a least-square sense by fitting the measured 2PCF $\hat{w}_{\rm corr}(\theta)$ to the model-2PCF $w_{\rm mod}(\theta)$ as described more in detail in the next section.

### 2.4.4 Fitting procedure

We determine the two free parameters of the model, $n_{\rm WB}$ and $\lambda$, by means of a Levenberg-Marquardt nonlinear least-square algorithm (Lourakis 2004)[13], which minimises the value of the $\chi^2$ defined as

$$\chi^2 = \sum_{i=1}^{N} \left( \frac{\hat{w}_{\rm corr}(\theta_i) - w_{\rm mod}(\theta_i)}{\delta \hat{w}_{\rm corr}(\theta_i)} \right)^2 . \tag{2.45}$$

The data is binned in steps of $\Delta\theta \equiv \theta_{i+1} - \theta_i = 1''$, where $\theta_1 = \theta_{\rm min} = 2''$ and $\theta_N = \theta_{\rm max} = 30''$ as imposed by the resolution limit of the SDSS and maximal distance in the *Neighbors* table, respectively. As a result, we have $N = 28$ here. Given the approximative character of our study, the use of the this standard definition of the $\chi^2$ is appropriate, even though the values of the 2PCF at different angular separations are not strictly independent of each other.

Following Press et al. (1992), we use the incomplete gamma function $Q(\chi^2|\nu)$ with $\nu = N - 2 = 26$ degrees of freedom as a quantitative measure of the goodness-of-fit. (N-2 because the model has two free parameters: $n_{\rm WB}$ and $\lambda$.) Values of $Q$ near unity indicate that the model adequately represents the data.

### 2.4.5 Confidence intervals

To estimate the uncertainties of our best-fit values, we use Monte Carlo confidence intervals (MCCRs) (e.g. Press et al. 1992, §15.6). Assuming Poissonian errors, we generate 10 000 *synthetic* data sets by drawing the number of unique pairs "observed" in the $k$-th synthetic sample $F^{(k)}_{\rm syn}(\theta)$ from a Poisson distribution with mean $F(\theta)$, where $F(\theta)$ is the observed number of pairs, *not* corrected for edge effects due to survey holes. The "true" number of pairs in a synthetic sample is then given by dividing $F^{(k)}_{\rm syn}$ by $F^{\rm tot}_{\rm H}$ (see §2.3.2).

For each synthetic sample, we determine best-fit values for the model's free parameters, $n^{(k)}_{\rm WB}$ and $\lambda^{(k)}$, in a least-square sense as described above. A $p$%-MCCR is defined by the line of constant $\chi^2$ which encloses $p$% of the best-fit values in the $n_{\rm WB}$ versus $\lambda$ plane (see Fig. 2.6). The confidence intervals of $n_{\rm WB}$ and $\lambda$ are then given by the orthogonal projection of the MCCR onto the corresponding axis.

We include in our error estimate only the uncertainties stemming from the pair counts in $F(\theta)$. Neither the uncertainties in the Galactic structure parameters nor those in the LF are taken into account.

[13]We use *levmar-2.2*.

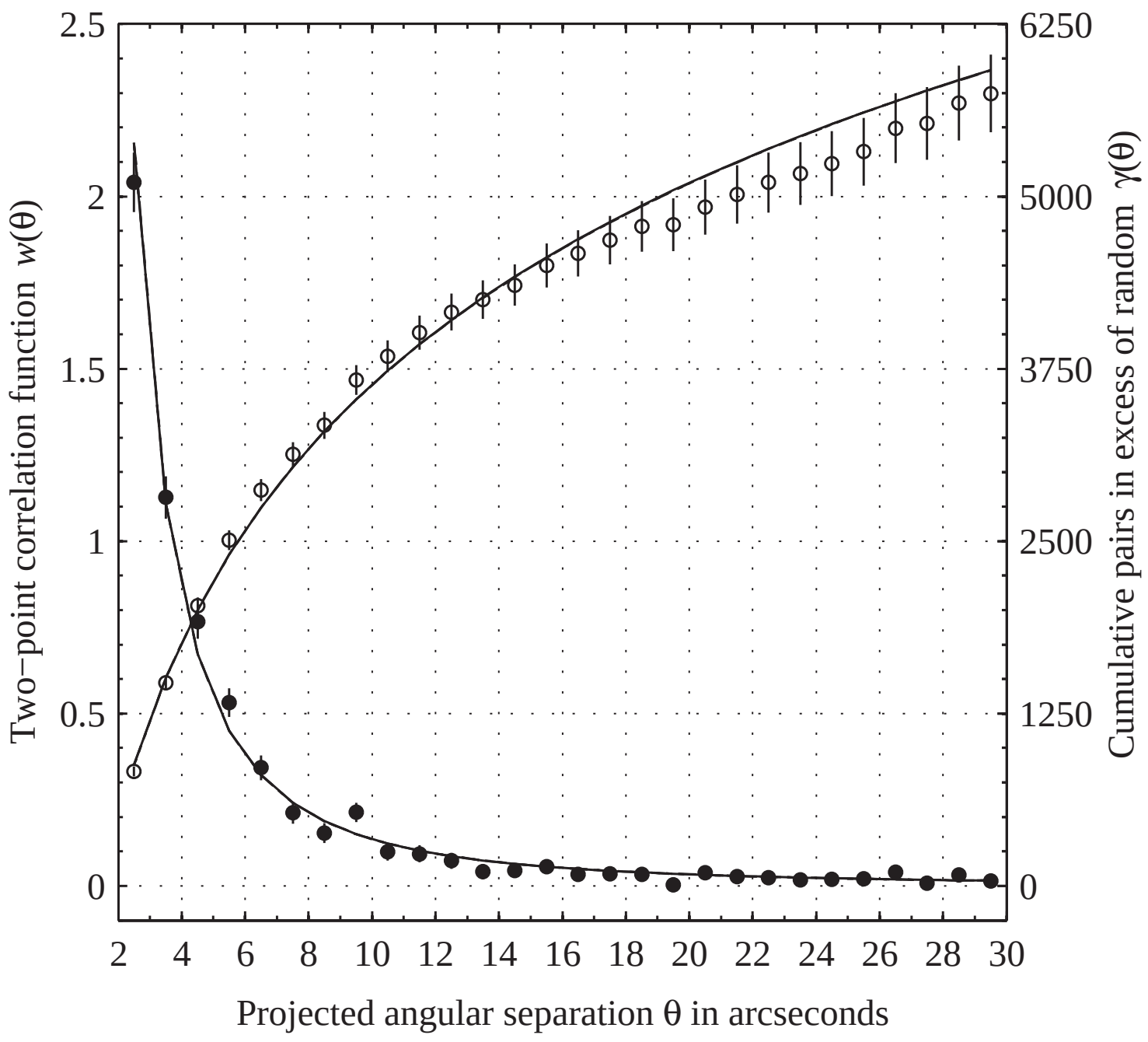

Figure 2.5: 2PCF as inferred from the total sample (solid circles, left ordinate) and the corresponding CDD (open circles, right ordinate). Poisson errors are indicated as vertical lines. Model curves for the three Galactic structure parameter sets are plotted, too, but as the differences between them are marginal they lie one upon the other, giving a single solid line.

## 2.5 Results

### 2.5.1 Analysis of the total sample

The results of the analysis for the total sample are shown in Fig. 2.5 and Table 2.3. Figure 2.5 shows the 2PCF estimate $\hat{w}_{\mathrm{corr}}(\theta)$ and the CDD $\hat{\gamma}(\theta)$. The statistical uncertainties calculated according to Eq. 2.13 are shown as vertical lines. A strong clustering signal out to at least $\theta = 10''$ is evident, whereas the CDD suggest that there are pairs in excess of a random distribution up to maximum angular separation examined, that is, up to 30". The outlier at $\theta = 9''$ is probably a random fluctuation. We also plot in Fig. 2.5 best-fitting models using the three Galactic structure parameter sets described in Sect. 2.4.2.

The best-fit values of the two free parameters, $n_{\mathrm{WB}}$ and $\lambda$, are tabulated in Table 2.3 for the three Galactic structure parameter sets. The power-law index $\lambda$ appears to be quite independent of the set we choose. The wide binary density $n_{\mathrm{WB}}$, on the other hand, shows some variation. The difference between the sets 1 and 2 reflects, for the most part, the difference in the overall normalisation $\rho(0)$ of the density distribution, whereas in set 3 the unequal halo normalisation with respect to the other two sets also contributes to the difference in $n_{\mathrm{WB}}$.

Table 2.2: Parameters of the final sample and of the subsamples

| (Sub)sample | $\Omega$ [deg$^2$] | $\ell$ [deg] | $b$ [deg] | $N_{\mathrm{obs}}$ | $D_{\mathrm{eff}}$[a] [pc] | $a_{\mathrm{T}}$[a] [pc] | $f_{\mathrm{Halo}}$[a] % |
|---|---|---|---|---|---|---|---|
| total | 668.58 | 175.6 | 81.6 | 669 843 | 1 555 | 1.07 | 30.0 |
| $r < 20.0$ mag | 668.58 | 175.6 | 81.6 | 535 595 | 1 380 | 1.06 | 25.2 |
| $r < 19.5$ mag | 668.58 | 175.6 | 81.6 | 425 674 | 1 235 | 1.06 | 20.6 |
| left | 334.57 | 191.2 | 73.7 | 326 333 | 1 495 | 1.08 | 28.3 |
| right | 334.01 | 102.5 | 84.8 | 343 510 | 1 625 | 1.05 | 31.0 |
| A | 78.94 | 180.1 | 68.3 | 77 008 | 1 460 | 1.09 | 26.8 |
| B | 78.10 | 167.5 | 75.3 | 76 011 | 1 510 | 1.08 | 28.6 |
| C | 78.99 | 136.0 | 79.8 | 76 686 | 1 565 | 1.06 | 29.9 |
| D | 78.81 | 93.3 | 78.4 | 82 780 | 1 635 | 1.05 | 30.6 |
| E | 88.23 | 208.0 | 69.7 | 87 751 | 1 485 | 1.08 | 27.4 |
| F | 88.30 | 210.6 | 78.6 | 85 563 | 1 545 | 1.07 | 29.5 |
| G | 87.95 | 215.2 | 87.5 | 88 440 | 1 615 | 1.05 | 31.0 |
| H | 88.27 | 33.4 | 83.6 | 95 604 | 1 700 | 1.04 | 31.8 |

[a] Calculated using the Galactic structure parameter set 2.

Table 2.3: Best-fit values: total sample

| set of structure parameters | $n_{\mathrm{WB}}$ [pc$^{-3}$] | $\lambda$ | $Q$ | $\chi^2/\nu$ | $f_{\mathrm{WB}}$ % |
|---|---|---|---|---|---|
| set 1 | 0.0052 | 1.00 | 0.29 | 1.14 | 10.2 |
| set 2 | 0.0069 | 1.01 | 0.30 | 1.13 | 13.6 |
| set 3 | 0.0061 | 1.01 | 0.29 | 1.14 | 12.1 |

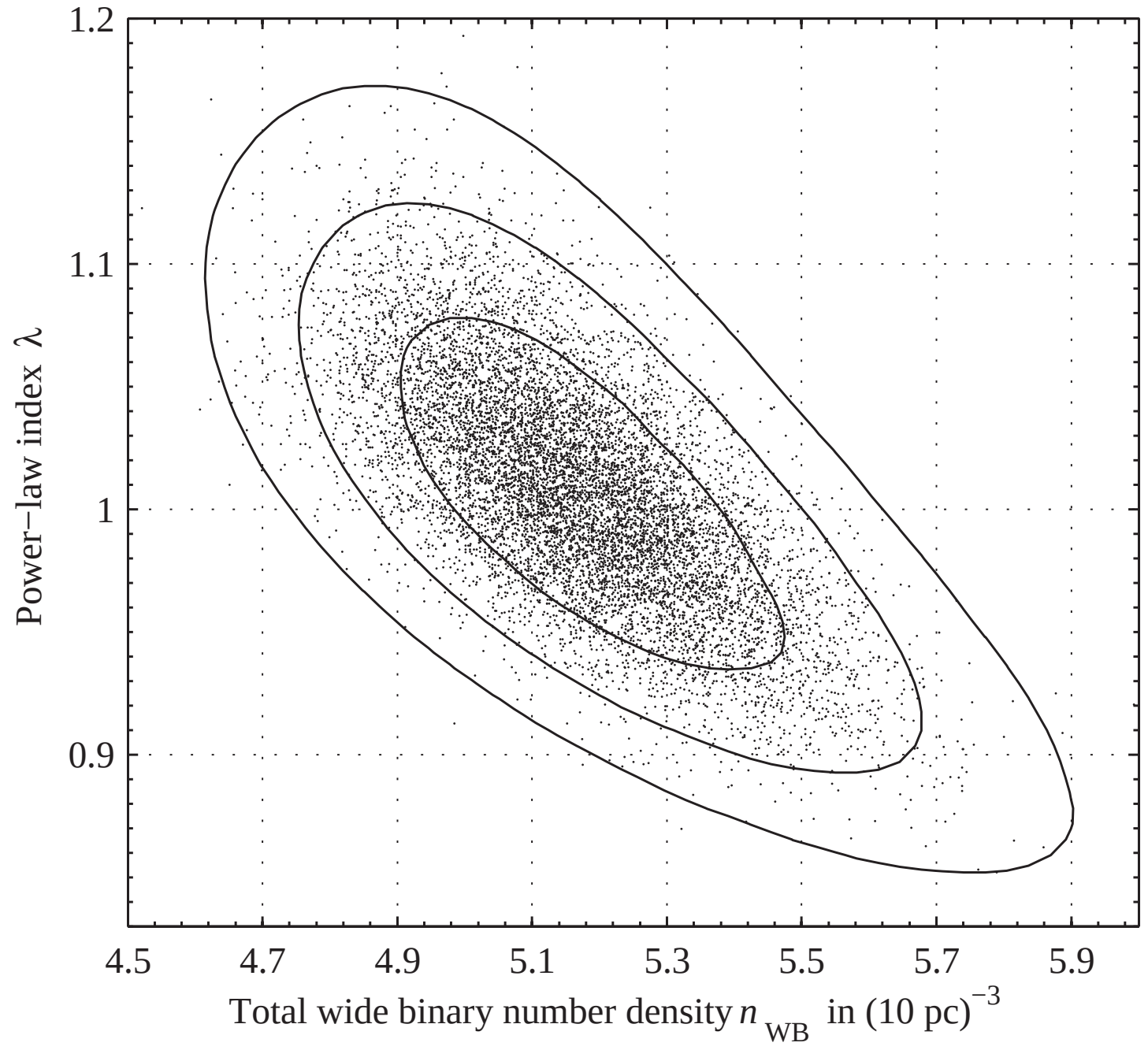

Figure 2.6: Distribution of the best-fit values for the total sample using the structure parameter set 2 obtained by the Monte Carlo procedure described in the text. The solid contours (MCCRs) are lines of constant $\chi^2$ and enclose 68.3%, 95.4%, and 99.7% of the best-fit values.

We also list the goodness-of-fit $Q$ and the corresponding reduced chi-square values ($\chi^2$ divided by the degrees of freedom $\nu$) in Table 2.3. All three sets of Galactic structure parameters give equally good fits, whereas the reduced chi-square values, which are only slightly more than unity, indicate that we have not severely underestimated the uncertainty in the 2PCF.

In Fig. 2.6 we show, representative of the other structure parameter sets, the distribution of the best-fit values from the synthetic samples using set 2, our standard set. In the same figure the 68.3% ($1\sigma$), 95.4% ($2\sigma$), and 99.7% ($3\sigma$) MCCRs are also plotted. Quoting 95.4% confidence intervals throughout, we find for our final sample using the Galactic structure parameter set 2

$$n_{\rm WB} = 0.0052^{+0.0006}_{-0.0005}\ {\rm pc}^{-3} \qquad \text{and} \qquad \lambda = 1.00^{+0.13}_{-0.12}\,. \tag{2.46}$$

The power-law index $\lambda$ of semi-major axis distribution is consistent with Öpik's law ($\lambda = 1$) up to the Galactic tidal limit, whereas the number density $n_{\rm WB}$ corresponds to a wide binary fraction with respect to *all* stars (*no* colour-cut) in the solar neighbourhood of

$$f_{\rm WB} \equiv \frac{2n_{\rm WB}}{n_*} = 10.2^{+1.2}_{-1.0}\,\%, \tag{2.47}$$

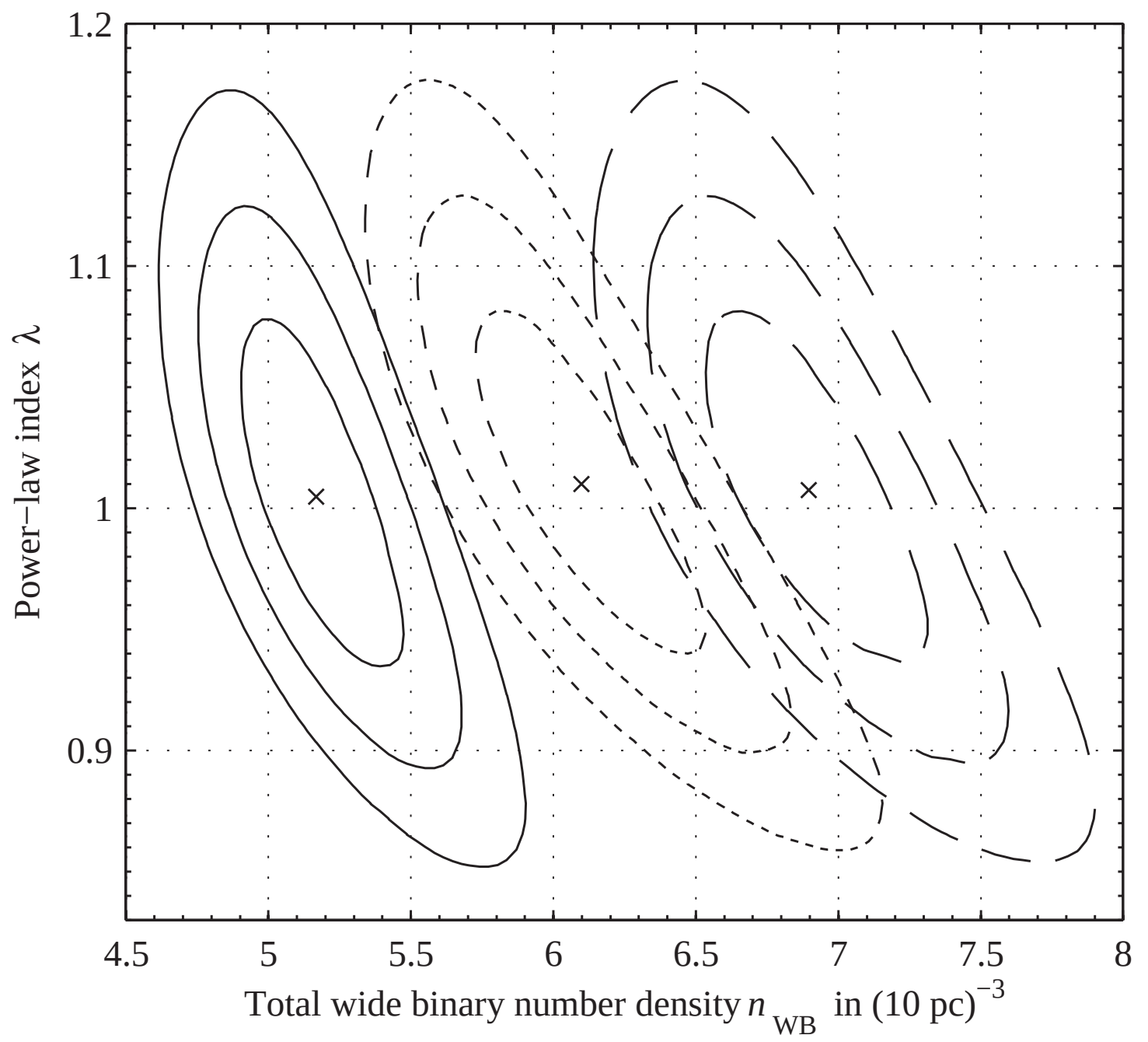

Figure 2.7: MCCRs for the three Galactic structure parameter sets corresponding to the total sample. Long dashed line: set 1; solid line: set 2; short dashed line: set 3. The crosses indicate the best-fit values inferred from the true *observed* number of pairs, $F_{\rm corr}(\theta)$.

where $n_* \simeq 0.10$ pc$^{-3}$ is the *total* local stellar number density, i.e. the integral over the *whole* Jahreiß and Wielen LF. (The local wide binary density $n_{\rm WB}$ corresponds only to wide binaries with both components having $g - r > 0.5$ mag.)

We show the confidence regions corresponding to the three sets of Galactic structure parameters in Fig. 2.7. The confidence intervals for $\lambda$ agree for all the sets, which demonstrates that we can determine $\lambda$ and its confidence intervals reliably – provided that no systematic error has crept into our analysis. The wide binary density $n_{\rm WB}$, on the other hand, is more sensitive to the exact values of the Galactic structure parameters than $\lambda$ is, so the values we have derived for the wide binary density $n_{\rm WB}$ and the fraction $f_{\rm WB}$ should be viewed with caution. We are, however, confident that the true value of the wide binary density $n_{\rm WB}$ is within a factor of 2 of our derived value.

Since the structure parameter set 2 is – as far as known to the authors – the only one systematically corrected for unresolved multiplicity, we give that set more weight. From now on, we use the bias-corrected values from Jurić et al. (2008), i.e. our standard set 2, exclusively.

Having determined the two free parameter in our model, we can calculate the number of wide binaries in our total sample as a function of the projected separation $s$. We show the distribution in Fig. 2.8 whose shape is largely dominated by the effective volume, i.e. by selection effects. In particular, this distribution implies that we have observed $830^{+250}_{-215}$ very wide binaries with a projected separation larger than 0.1 pc in our total sample, while none are expected to be found beyond 0.8 pc, given our selection criteria. However, that extremely wide binaries with projected separations of more than 1 pc can exist in the Galactic halo has recently been confirmed by Quinn et al. (2009).

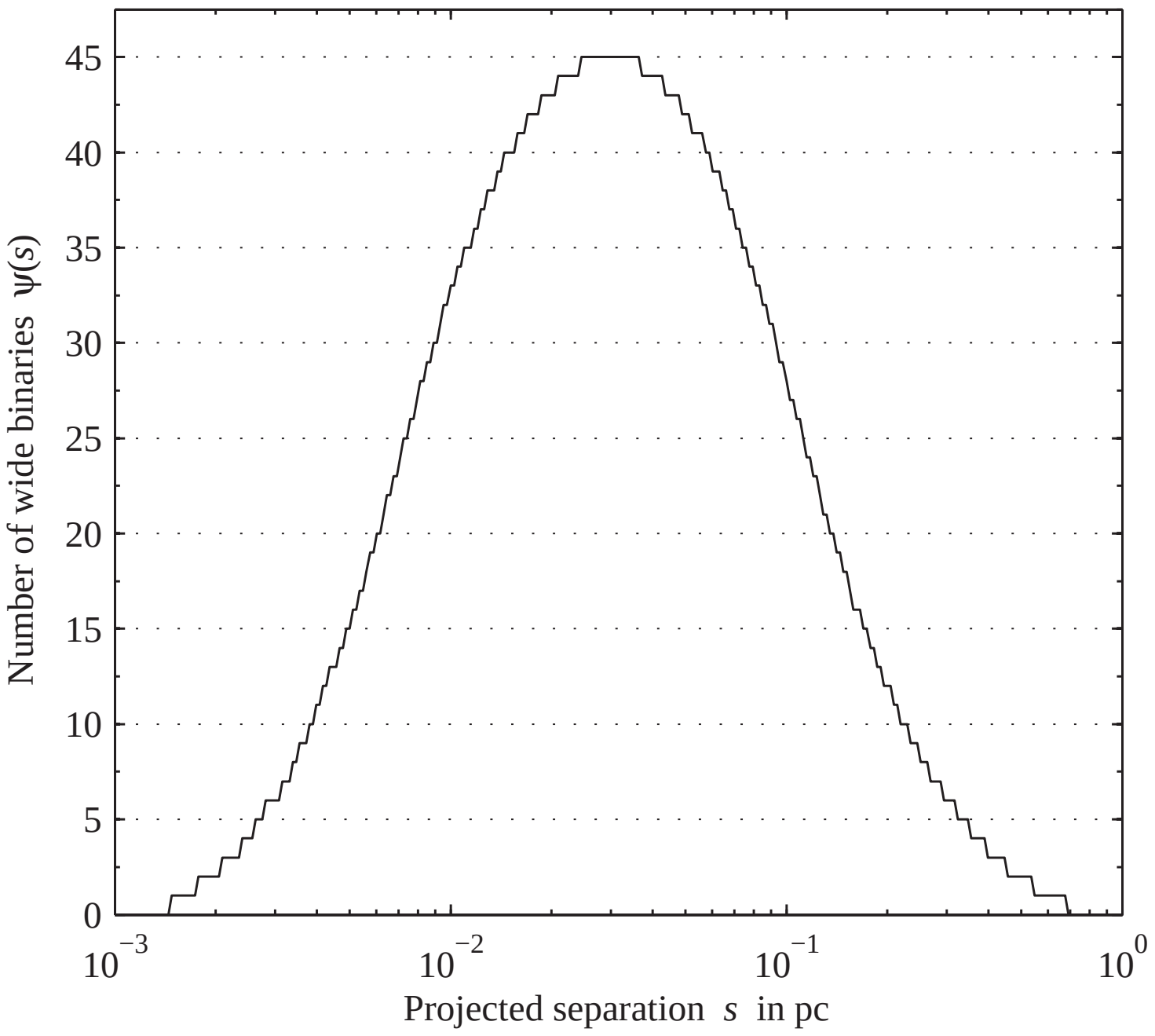

Figure 2.8: Distribution of wide binaries as a function of projected separation $s$ that is expected to be observed in the total sample, given the model assumptions and the selection criteria we have adopted (see text).

Table 2.4: Best-fit values: $r < 20.0$ mag and $r < 19.5$ mag

| (sub)sample | $n_{\rm WB}$ [pc$^{-3}$] | $\lambda$ | $Q$ | $\chi^2/\nu$ | $f_{\rm WB}$ % |
|---|---|---|---|---|---|
| $r < 20.0$ mag | 0.0057 | 1.15 | 0.18 | 1.25 | 11.3 |
| $r < 19.5$ mag | 0.0064 | 1.22 | 0.34 | 1.09 | 12.7 |

### 2.5.2 Differentiation in terms of apparent magnitude

To test the consistency of the standard model, we used two subsamples having brighter upper apparent magnitude limits than our total sample and analysed them in the same way as the total sample itself. We set the upper apparent magnitude limit to $r = 20.0$ mag and $r = 19.5$ mag – all other selection criteria (including the lower apparent magnitude

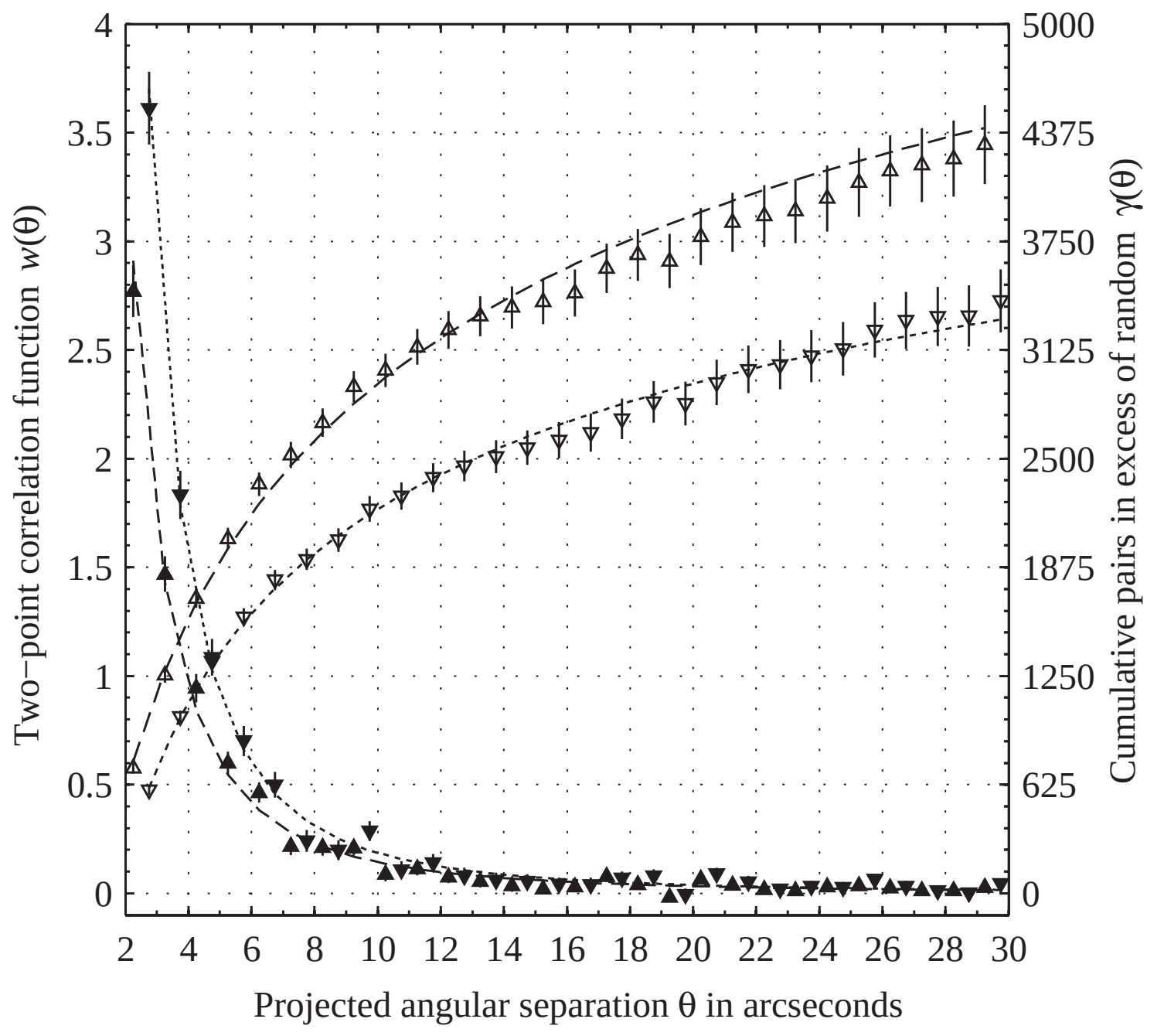

Figure 2.9: 2PCF and CDD for the $r < 20.0$ mag and $r < 19.5$ mag subsamples together with the corresponding model curves. Up-pointing triangles (long-dashed lines) are for the $r < 20.0$ mag subsample, whereas down-pointing triangles (short-dashed lines) are for the $r < 19.5$ mag subsample. The symbols were shifted apart by $0.25''$ for better visibility.

limits) remain unchanged. The main parameters of these subsamples are listed in Table 2.2.

In Fig. 2.9 we show the 2PCF and the CCD as in Fig. 2.5, together with the corresponding model curves for standard set 2 (see caption). The best-fit values are listed in Table 2.4. The MCCRs of the $r < 20.0$ mag and $r < 19.5$ mag subsamples are shown in Fig. 2.10.

If the model were self-consistent, we would expect that the best-fit values agree with each other within their uncertainties. Figure 2.10 indicates that this is not the case: It appears that the best-fit values are systematically shifted to higher densities and larger power-law indices when using a brighter upper magnitude limit. As we discuss in Sect. 2.6.2, this inconsistency is most likely an artefact of the (oversimplifying) model assumptions and does not entirely undermine our results.

## 2.5.3 Differentiation in terms of direction

In a previous study, Saarinen and Gilmore (1989) found that the binaries appear to be highly clumped in the NGP. However, it has not become entirely clear whether this patchiness of the wide binary distribution in the sky is a real physical characteristic of the wide binary population or if it is due to statistical fluctuations. In principle, we can check whether the wide binary density varies with position in the sky by dividing our sample into subareas.

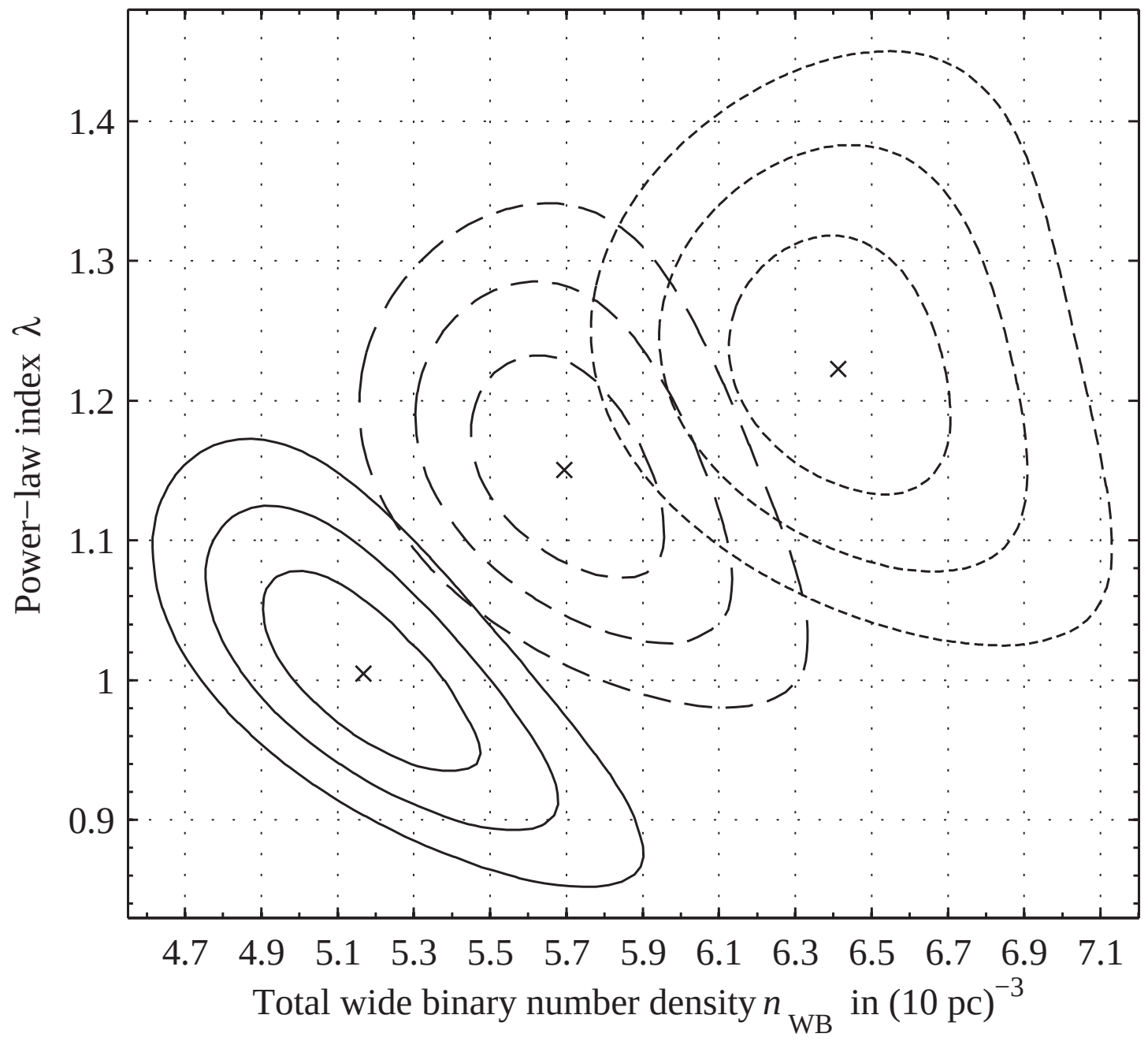

Figure 2.10: MCCRs for the $r < 20.0$ mag (long dashed contours) and $r < 19.5$ mag (short-dashed contours) subsamples. For comparison, the MCCR for our final sample is also shown (solid-contours). The crosses have the same meaning as in Fig. 2.7.

Table 2.5: Best-fit values: left and right

| subsample | $n_{\mathrm{WB}}$ [pc$^{-3}$] | $\lambda$ | $Q$ | $\chi^2/\nu$ | $f_{\mathrm{WB}}$ % |
|---|---|---|---|---|---|
| left | 0.0055 | 1.00 | 0.55 | 0.94 | 10.8 |
| right | 0.0046 | 1.03 | 0.82 | 0.74 | 9.1 |

To begin with, we divide our sample in two halves by cutting it along the $\alpha = 185°$ meridian. In the following, we refer to the subarea with $\alpha < 185°$ as the "left" subsample and the other half as the "right" subsample. The subsamples' main parameters are listed in Table 2.2.

In Fig. 2.11 we show the 2PCFs inferred from the left and the right sample, as well as the corresponding CDDs. It appears that there are a few more pairs in excess of random in the left half. In the same figure also the corresponding model curves are plotted. The best-fit values are listed in Table 2.5.

In Fig. 2.12 we show the MCCRs of the left and the right subsamples. The left subsample indeed shows a higher wide binary density than the right one. The difference is significant at the $3\sigma$ level. The power-law indices, on the other hand, do agree in both subsamples. Regarding the total sample, the right half differs significantly (at $3\sigma$), whereas the left half is more consistent with it.

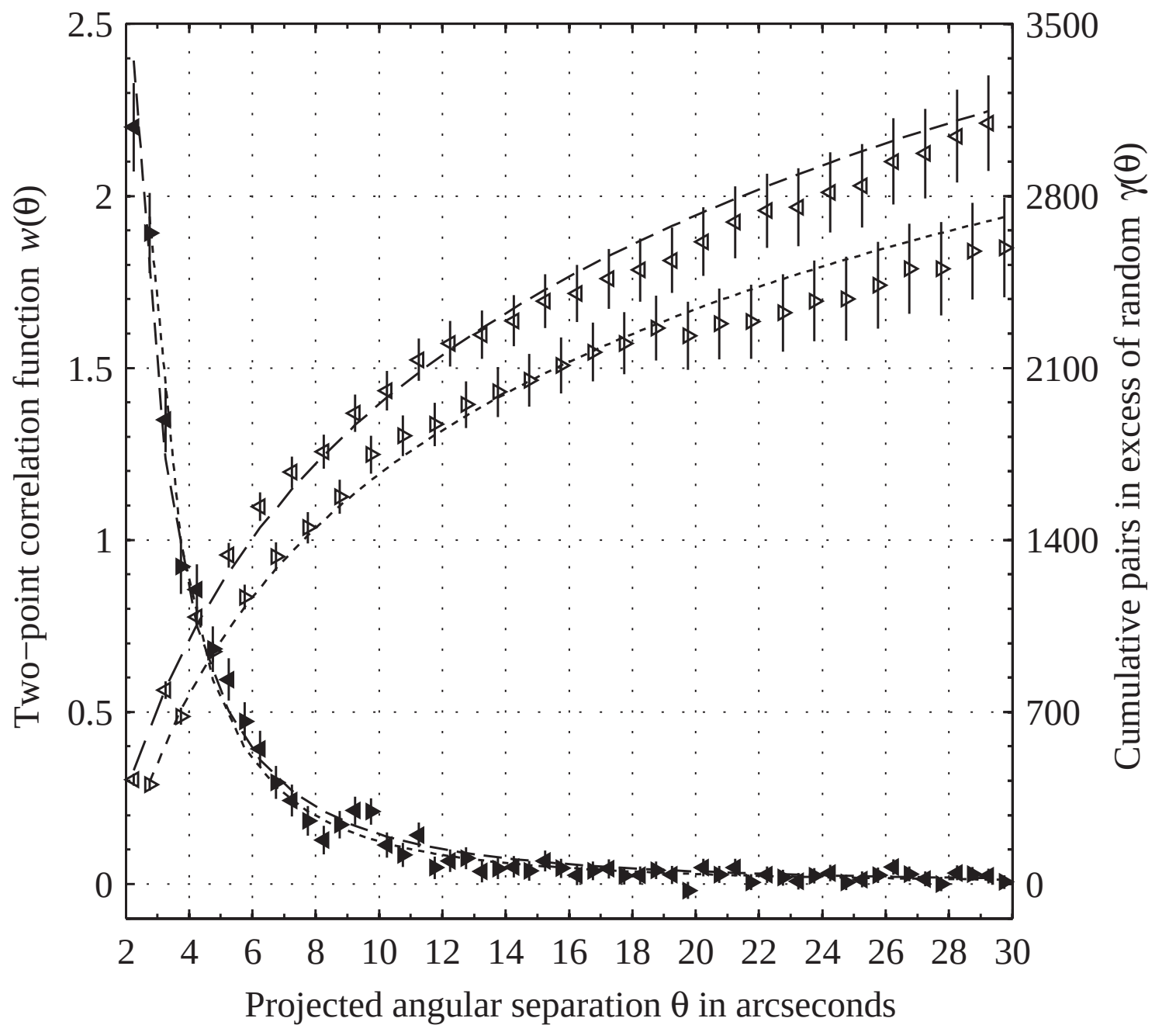

Figure 2.11: 2PCF and CDD for the left and the right subsamples, together with the corresponding model curves. Left-pointing triangles (long-dashed lines) are for the left subsample, whereas right-pointing triangles (short-dashed lines) are for the right subsample. The symbols were shifted apart by $0.25''$ for better visibility.

The difference in the wide binary density between the left and the right halves is probably a real feature, as any inadequacies of our model (e.g. inaccuracies of the stellar density distribution we use) should affect both halves in almost the same manner. To examine this apparent positional dependency of the wide binary density in more detail, we divide our sample further into eight subsamples, each covering $10^\circ \times 10^\circ$. They are labelled from A ("upper left") to H ("lower right") as suggested in Fig. 2.13, where the abscissa can thought of representing the right ascension from 165° ("left") to 205° ("right") and similarly the ordinate represents the declination from 22° ("bottom") to 42° ("top"). The subsample's main parameters are again summarised in Table 2.2.

Figure 2.13 shows the 2PCFs estimate and the CDDs with the best-fit model curves. The corresponding best-fit parameters are listed in Table 2.6. Some subtle differences are apparent between different subsamples. While the CDD in A, B, C, and G are reproduced well by the model, the CDD in D, F, and H appears to be too flat. In those subsamples it seems that almost no physical pairs are present at angular separations over $15''$, in agreement with the findings of Sesar et al. (2008) (see Sect. 2.6.1).

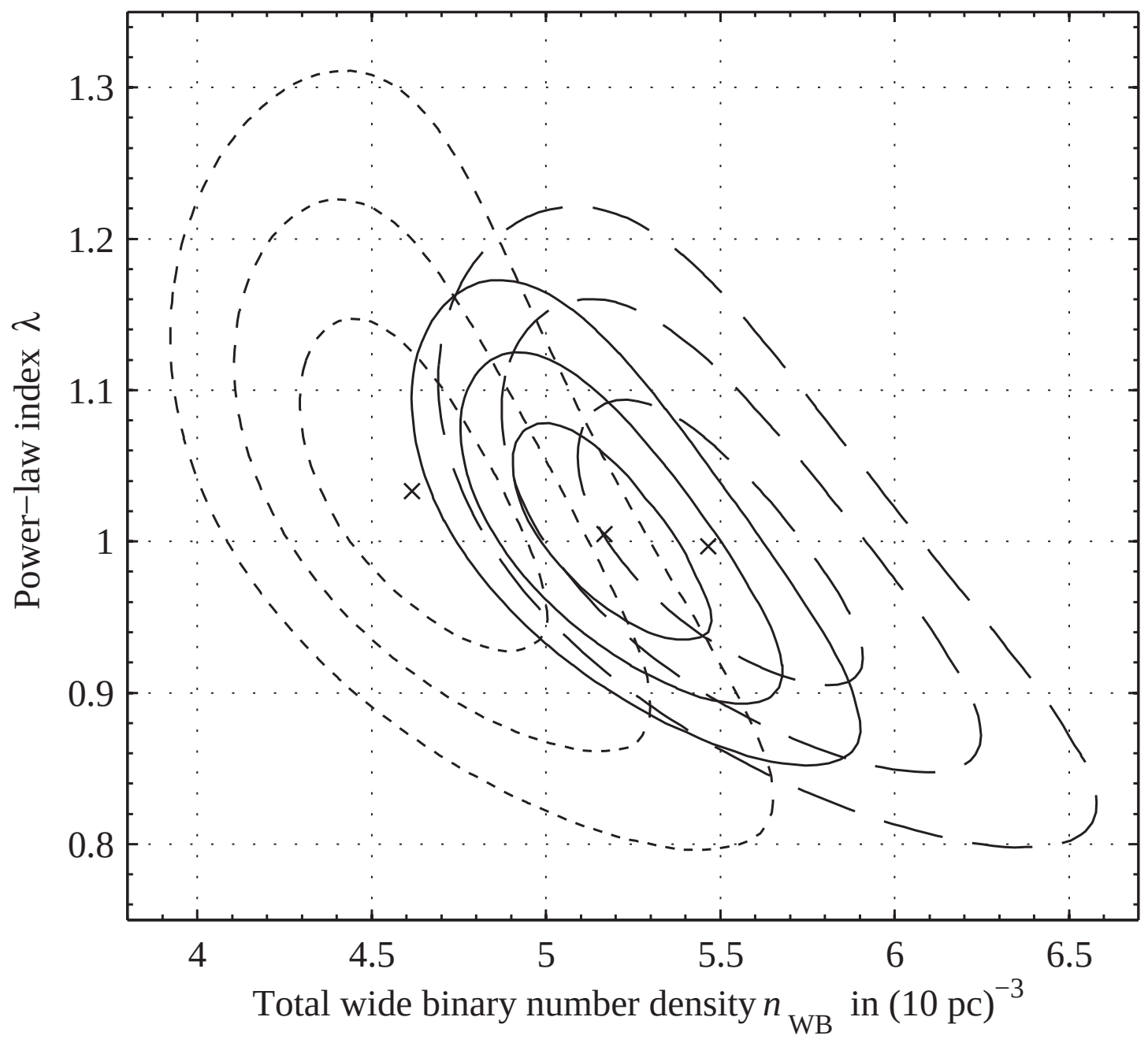

Figure 2.12: MCCRs for the left (long-dashed contours) and right (short-dashed contours) subsamples. For comparison, also the MCCR for our final sample is shown (solid contours). The crosses have the same meaning as in Fig. 2.7.

The subsample E appears to be an outlier, because it contains almost twice as many pairs in excess of random as the seven other subsamples. The listed best-fit values confirm the peculiar character of subsample E having the highest wide binary density of all eight subsamples, together with the lowest power-law index. As far as the authors can judge, there is no obvious special feature (e.g. open star cluster) in subsample E that could cause this anomaly.

The $2\sigma$ MCCR are shown in Fig. 2.14. Except for the outlier E, all the other subsamples are quite consistent with each other. Also, no obvious trend, e.g. with Galactic latitude $b$, is apparent. To what extent is the subsample E responsible for the difference between the right and the left subsamples?

To answer this question, we repeated the analysis of the total and left (sub)sample excluding subsample E. The results are listed in Table 2.7. The wide binary density drops significantly to a value more consistent with the right subsample. Thus, we conclude that the difference between the right and the left subsamples is largely caused by the outlier E. Apart from subsample E, the wide binary densities in different directions appear to be consistent with each other. The reason for the high density in subsample E remains unclear. However, a statistical fluctuation cannot be ruled out to a level better than $2\sigma$.

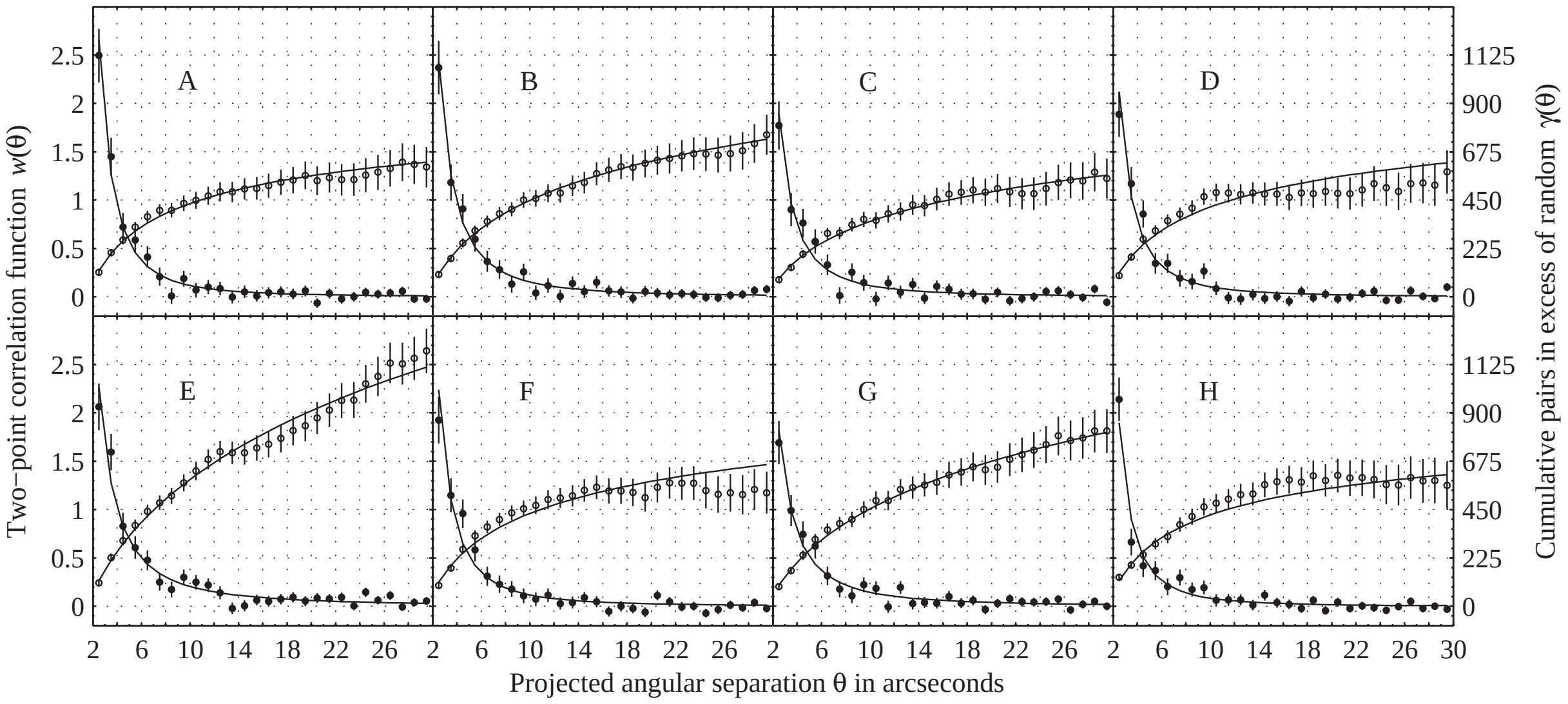

Figure 2.13: 2PCF (solid circles) and CDD (open circles) for the subsamples A to H, together with the corresponding model curves. The subfigures are arranged as the corresponding subareas would be seen on the sky.

Table 2.6: Best-fit values: A to H

| subsample | $n_{\mathrm{WB}}$ [pc$^{-3}$] | $\lambda$ | $Q$ | $\chi^2/\nu$ | $f_{\mathrm{WB}}$ % |
|---|---|---|---|---|---|
| A | 0.0050 | 1.27 | 0.92 | 0.64 | 9.9 |
| B | 0.0054 | 1.00 | 0.96 | 0.57 | 10.7 |
| C | 0.0041 | 1.02 | 0.36 | 1.07 | 8.2 |
| D | 0.0046 | 1.19 | 0.29 | 1.13 | 9.1 |
| E | 0.0077 | 0.73 | 0.24 | 1.18 | 15.2 |
| F | 0.0045 | 1.14 | 0.37 | 1.07 | 8.9 |
| G | 0.0054 | 0.83 | 0.67 | 0.86 | 10.6 |
| H | 0.0042 | 1.27 | 0.64 | 0.88 | 8.4 |

Table 2.7: Best-fit values: excluding subsample E

| subsample | $n_{\mathrm{WB}}$ [pc$^{-3}$] | $\lambda$ | $Q$ | $\chi^2/\nu$ | $f_{\mathrm{WB}}$ % |
|---|---|---|---|---|---|
| total\E | 0.0049 | 1.06 | 0.40 | 1.04 | 9.6 |
| left\E | 0.0049 | 1.12 | 0.76 | 0.79 | 9.8 |

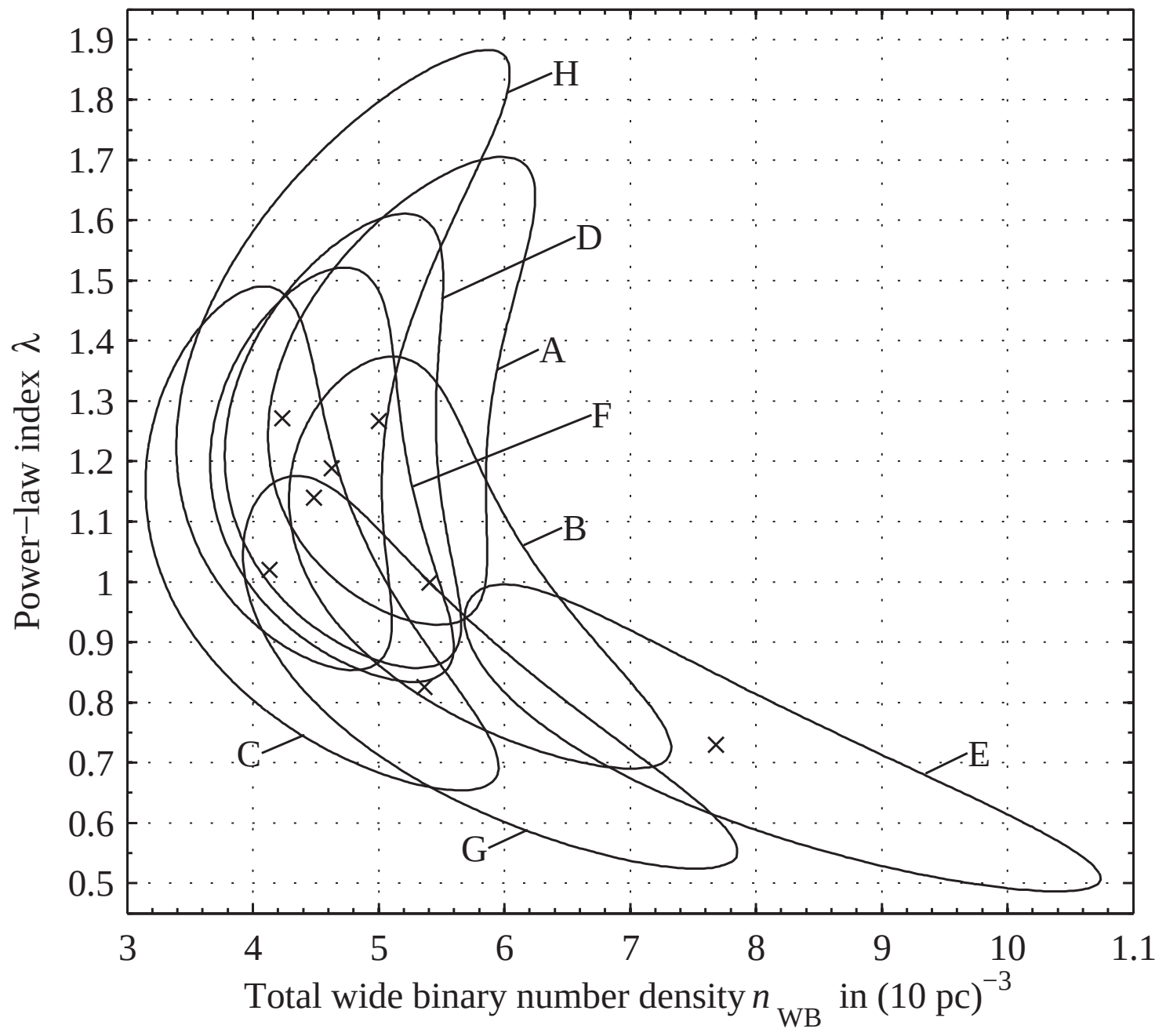

Figure 2.14: 95.4% MCCRs for the subsamples A to H. The crosses have the same meaning as in Fig. 2.7.

# 2.6 Discussion

## 2.6.1 External (in)consistencies

In this section we compare our results with those of previous studies. Our first point concerns the *observed* angular 2PCF. From the CDD of our total sample shown in Fig. 2.5 we have noted pairs in excess of random up to $\theta_{\max} = 30''$. This is contradictory to Sesar et al. (2008) who find that there are essentially no physically bound pairs with an angular separation $\theta > 15''$. (We note, however, that this is in accord with some of our *sub*samples, see Fig. 2.13.) We considered several possible reasons for this apparent discrepancy, such as differences in the selection criteria of the Sesar et al. sample with respect to our sample, underestimation of the total area of holes in our sample, or an overcorrection of edge effects due to those holes, but we convinced ourselves that none of them can account for this disagreement. A real physical difference in the wide binary population studied, e.g. caused by the fact that Sesar et al.'s sample is largely disc-dominated, while our final sample has a substantial halo contribution of $\sim 30\%$, can be excluded, as previous studies (Latham et al. 2002; Chanamé and Gould 2004) found that the disc and halo wide binary populations are reasonably consistent in their statistical characteristics.

Regarding the wide binary fraction, Sesar et al. also find a much smaller wide binary fraction of below 1% (decreasing with height above the Galactic plane), as compared to our $f_{\mathrm{WB}} \approx 10\%$. Even if all pairs with a semi-major axis larger than 3 000 AU (beyond the "break") were ignored, we would still find a wide binary fraction of 4.6%. On the other hand, Lépine and Bongiorno (2007) give a binary fraction of at least 9.5% for separations larger than 1 000 AU, which is in rough agreement with our results (setting $a_{\rm min} = 1\,000$ AU we find $f_{\rm WB} \approx 8.3\%$).

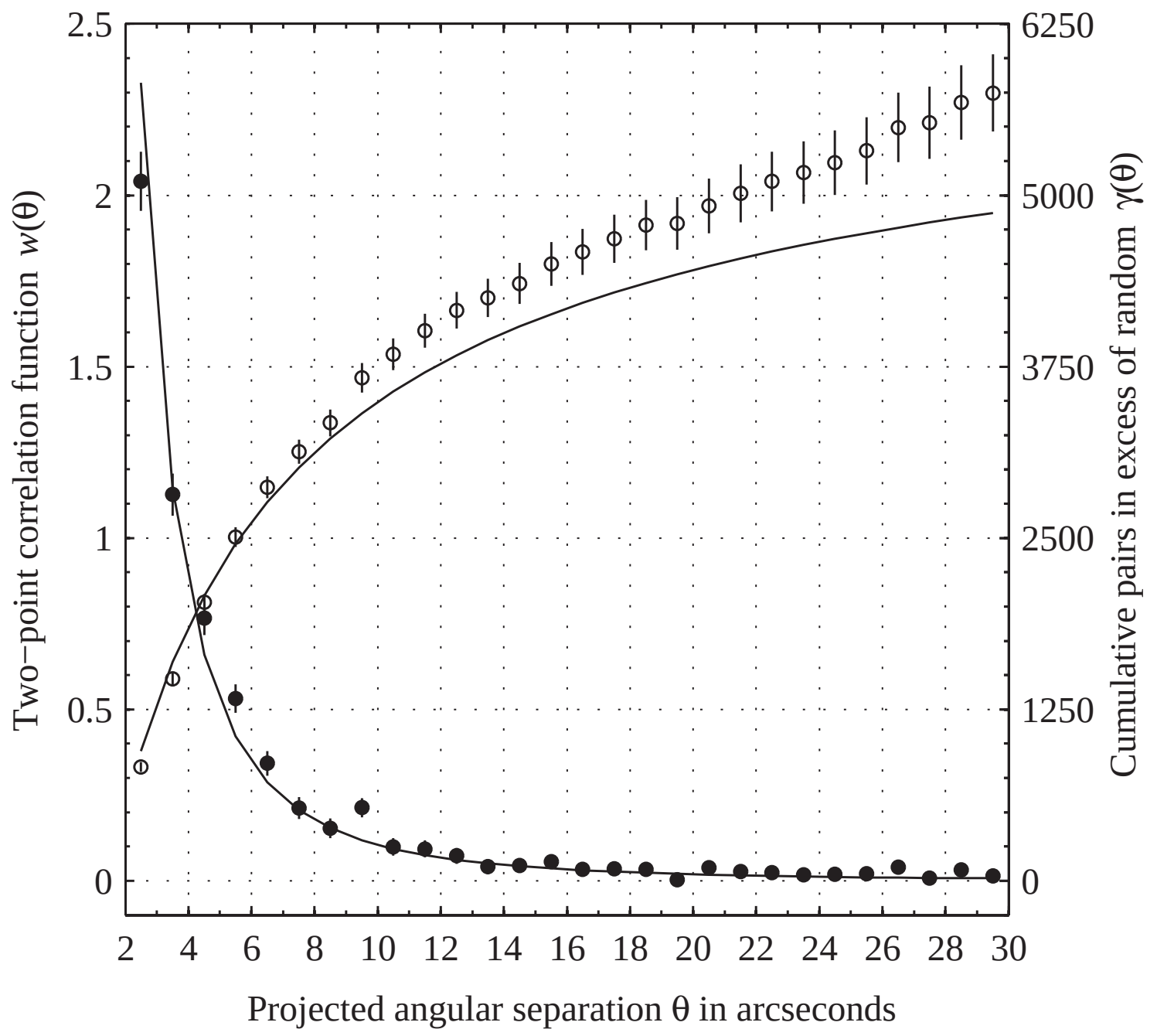

Figure 2.15: 2PCF and CDD as inferred from our total sample as in Fig. 2.5. The model curve is for the broken power-law model from Lépine and Bongiorno (2007) (see text).

Studying a much smaller region containing brighter stars as compared to our sample, WW87 and Garnavich (1988, 1991) found an unphysically large wide binary density in the direction of the NGP. Saarinen and Gilmore (1989) attribute this overdensity to a large statistical fluctuation. The region where this overdensity was found would be located in our subarea H. But as we probe fainter stars, no density enhancement is evident in that subarea. The slight overdensity we found in subarea E is different from that previously noted by WW87 and Garnavich.

As to the separation distribution, previous studies have found that the observational data are described by Öpik's law ($\lambda = 1$) up to a certain maximum separation ("break"): Lépine and Bongiorno (2007) find this break to be around 3 500 AU, beyond which a steeper slope, with $\lambda \approx 1.6$, should apply. Sesar et al. (2008) basically agree with Lépine and Bongiorno and find in addition that the maximum separation increases with height above the Galactic plane. Poveda and Allen (2004) divided their wide binary catalogue (Poveda et al. 1994) into two subsamples consisting of the oldest and youngest systems, respectively, and find that the "maximum major semiaxis for which Öpik's distribution holds is much larger for the youngest binaries ($a_m$ = 7862 AU) than for the oldest ($a_m$ = 2409 AU)".

Chanamé and Gould (2004) find the data to be described by a single powerlaw with an index of approximately 1.6 for their disc and halo samples. However, they note a "puzzling" flattening of the distribution of the disc binaries between 10"and 25". As already suggested by Sesar et al., this flat range might be the domain where Öpik's law is valid. In view of the substantial uncertainties inherent in studies by Lépine and Bongiorno and by Chanamé and Gould, the two studies appear to be broadly consistent. Similar to our study, Garnavich (1991) assumed that the semi-major axis distribution is described by a single powerlaw. He finds the power-law index to be $0.7 \pm 0.2$ for his NGP sample covering nearly 240 square degree. At intermediate Galactic latitudes he finds the slope to be steeper: $\lambda = 1.3 \pm 0.2$. His data seem to favour a (somewhat unrealistic[14]) "cutoff" around 0.1 pc, especially for the NGP sample.

The question of a break in the separation distribution has much been discussed in the context of DM constraints. Wasserman and Weinberg (1991), for example, show that the observational data suggest a break but they do not *require* one statistically. Looking at the CDD for our total sample in Fig. 2.5, we note that our best-fit model slightly overestimates the number of pairs in excess of random at larger angular separations, and a more curved model line would be preferred by the data. This could indeed be interpreted as a hint that the semi-major axis distribution is broken, because having a steeper power-law index from a certain semi-major axis on would result in a flatter CDD model curve. In principle, we could easily fit a broken powerlaw to our data as well. However, too many free parameters will only destabilise our modelling. Instead we check whether our data are also consistent with the specific broken power-law distribution found by Lépine and Bongiorno (2007). To this end we use a broken reduced semi-major axis distribution of the form

$$q(a) = q_1(a) + q_2(a) = c_{\lambda_1} a^{-\lambda_1} \Theta(a_\mathrm{c} - a) + c_{\lambda_2} a^{-\lambda_2} \Theta(a - a_\mathrm{c}) \tag{2.48}$$

with the Heaviside function $\Theta$ that introduces a break at semi-major axis $a_\mathrm{c}$. It is normalised, as in §2.4.1, according to

$$\int_{a_\mathrm{min}}^{a_\mathrm{T}} q(a)\,\mathrm{d}a = \int_{a_\mathrm{min}}^{a_\mathrm{c}} q_1(a)\,\mathrm{d}a + \int_{a_\mathrm{c}}^{a_\mathrm{T}} q_2(a)\,\mathrm{d}a = 1\,. \tag{2.49}$$

Using the values given by Lépine and Bongiorno (2007), $\lambda_1 = 1$, $\lambda_2 = 1.6$, and $a_\mathrm{c} = 0.02$ pc, we find for the normalisation constants

$$c_{\lambda_1} = \left[\ln a_\mathrm{c} - \ln a_\mathrm{min} + \frac{(a_\mathrm{T}/a_\mathrm{c})^{1-\lambda_2} - 1}{1-\lambda_2}\right]^{-1} \tag{2.50}$$

and

$$c_{\lambda_2} = c_{\lambda_1} a_\mathrm{c}^{\lambda_2 - 1}\,. \tag{2.51}$$

The distribution of the projected separations $Q(s)$ then splits into a sum, too:

$$Q(s) = \left(\frac{s}{\mathrm{pc}}\right)^{-\lambda_1} C_{\lambda_1}(s)\ \mathrm{pc}^{-1} + \left(\frac{s}{\mathrm{pc}}\right)^{-\lambda_2} C_{\lambda_2}(s)\ \mathrm{pc}^{-1}, \tag{2.52}$$

[14] "Physically, one doesn't expect to see a sharp cutoff." (Weinberg 1990)

where $C_{\lambda_i}(s)$ is defined as in Eq. 2.24 with the limits chosen appropriately when integrating over $\eta$.

We now keep the parameter given by Lépine and Bongiorno fixed so that our broken power-law model has only one free parameter: the wide binary density $n_{\rm WB}$. In Fig. 2.15 we show the best-fit model curves corresponding to $n_{\rm WB} = 0.0075$ pc$^{-3}$. The broken power-law model also gives a decent fit to the 2PCF inferred from our total sample. (One should not be misled by large discrepancy between the model and the CDD showing the *cumulative* difference between the 2PCF and the corresponding model curve.) With a goodness-of-fit of $Q \simeq 0.0047$ and a reduced chi-square of $\chi^2/\nu \simeq 1.85$, the broken power-law model provides a significantly worse bestfit to the data than our single power-law model. Nevertheless, as the errors in the 2PCF are slightly underestimated, the model is not put into question until $Q < 0.001$ (Press et al. 1992, their §15), and we conclude that the broken power-law distribution found by Lépine and Bongiorno (2007) can not be rejected with confidence either.

### 2.6.2 Internal (in)consistencies

Figure 2.10 shows that the best-fitting values for $\lambda$ and $n_{\rm WB}$ are inconsistent, within the adopted *random* errors, when the limiting magnitude is varied. Both parameters are systematically shifted by roughly 20% when the limiting magnitude is changed by 1 mag. Further inconsistencies were met when we tried to differentiate our total sample with respect to colour (not shown here). This very likely means that there are unaccounted for *systematic* errors in the modelling, resting on oversimplified assumptions.

A basic assumption on the properties of wide binaries was that both components are drawn at random from the same stellar LF. There is, however, some evidence that this is not quite correct: Gould et al. (1995) noticed that "binaries are bluer and more distant than one would expect if they were formed of random combinations of field stars". Also Lépine and Bongiorno (2007) find "that the luminosity function of the secondaries is significantly different from that of the single stars field population, showing a relative deficiency in low-luminosity ($8 < M_V < 14$) objects". Similarly, Sesar et al. (2008) report that blue stars ($g - i \lesssim 2.0$ mag, corresponding roughly to $g - r \lesssim 1.4$ mag) that are a member of a wide binary, have more blue companions than expected from the LF. (For red stars, however, they find that the components are drawn randomly from the LF.) Moreover, it has long been known (e.g. Bahcall and Soneira 1980, their Fig. 2) that stars of early spectral type have smaller scale heights than late type stars; therefore, the assumption that the stellar density distribution depends only on distance, and not on the luminosity of the stars, might be an oversimplification as well.

On the other hand, the internal inconsistency discussed is overemphasised in Fig. 2.10 because we only include the statistical errors stemming from the pair counts when determining the MCCRs. If we also included the errors in the Galactic structure parameters and those from the LF, the MCCRs would be considerably larger, and the inconsistency evident from Fig. 2.10 would not be as severe as it seems. That a variation in the Galactic structure parameter has a significant impact at least on $n_{\rm WB}$, is apparent from Fig. 2.7. The systematic drift towards larger $\lambda$ and $n_{\rm WB}$, when adopting a brighter upper apparent magnitude limit (cf. Fig. 2.10), still points to some inconsistencies in our model. We think, however, that these inconsistencies do not entirely undermine our major results, although the quoted uncertainties might be considerably larger.

## 2.7 Summary and conclusions

We have derived the angular 2PCF for nearly 670 000 SDSS stars brighter than $r = 20.5$ mag and redder than $g - r = 0.5$ mag in a region of approximately 670 square degrees around the NGP. Various corrections had to be made for quasar contamination, survey holes, and bright stars. There is an unambiguous correlation signal on small scales up to 30". We modelled this signal by a modified WW-technique, closely following a previous study of Garnavich (1988, 1991) that was based on a much smaller sample of stars. The modelling involved a number of parametrised distribution functions: the spatial density and LF of stars, as well as the separation distribution of binaries. The Galactic model used is based on the recent study by Jurić et al. (2008) and the stellar LF derived by Jahreiß and Wielen (1997). For the wide binary semi-major axis distribution we assumed a single powerlaw. Furthermore, essential assumptions are: (1) binary stars follow the same density distribution as single stars, (2) both components of a binary are randomly drawn from the single star luminosiy function. These assumptions allowed a significant simplification of the model. The Galactic structure parameters were fixed (we explore three different sets), while the local wide binary number density $n_{\rm WB}$ and the power-law index $\lambda$ of the semi-major axis distribution were left free; i.e., they have been determined by a least-squares fitting algorithm.

The best fit to the observed angular 2PCF of the total sample was obtained with $\lambda \approx 1.0$, which corresponds to the canonical Öpik law, and $n_{\rm WB} \approx 0.005$ pc$^{-3}$, meaning an overall local wide binary fraction of about 10.0% in the projected separation range of 0.001 pc (200 AU) to 1 pc (Galactic tidal limit). Previous studies (Poveda and Allen 2004; Lépine and Bongiorno 2007; Sesar et al. 2008) have also found the data to be consistent with Öpik's law, but only to a maximal separation that is considerably smaller than the Galactic tidal limit. Beyond that maximum separation, the distribution continues with a steeper decline. We have shown that our data are also consistent with the broken powerlaw found by Lépine and Bongiorno, although with a fit of lower quality (though involving only one free parameter, namely, $n_{\rm WB}$). Given this ambiguity, we are not able to put any strong constraints on the presence of a break in the wide binary separation distribution, which is regarded as one of the most interesting aspects of wide binaries. As to the wide binary fraction, we are in good accord with Lépine and Bongiorno (2007), whereas an apparent discrepancy with Sesar et al. (2008) remains.

A differentiation of the sample with respect to limiting apparent magnitude turned up a systematic dependence of the binary parameters on the sample depth, which is most probably an artefact caused by oversimplified model assumptions. This conjecture is strength-ened by the impossibility obtaining self-consistent results when differentiating with respect to colour. We conclude that one or more of the simplifying assumptions put into the model (e.g. that the luminosities of the binary components are independent of each other and draw randomly from the single-star LF or that binaries' density distribution follows exactly the single-star density distribution and that the density distribution is independent of the stars' luminosities) are not quite correct.

Differentiating the sample in terms of direction in the sky did yield some modest but non-significant variations. Only in one direction (subarea E) was an unexplained overdensity found.

While we have shown here that the stellar angular 2PCF, as a complement to common proper motion studies, basically works and remains a viable tool for the study of wide binaries, it has become clear that this method is severely limited by the need for – even more – complex modelling. To relax those simple model assumptions would unduly complicate the analysis much further and probably no longer yield unique solutions. Any more efficient progress will indeed have to involve distance information to discriminate against unwanted chance projections. We plan to include distance information in our future work. In spite of the limitations of the present, simple modelling of the angular 2PCF, we think that the general result for the total sample derived here, i.e. $\lambda \approx 1$ (Öpik law) and $f_{\rm WB} \approx 10\%$ among stars having a spectral type later than G5, to within an uncertainty of 10%-20%, is a robust result.

## 2.A Edge correction for holes

To quantify the edge effect from holes in our sample, we assume that the holes are circular and flat (Euclidian) and that effects due to intersecting holes are negligible. Let $\overline{A}_k(\theta)$ be the solid angle of the annulus of width $2\theta$ around a hole of radius $\bar{\theta}_k$

$$\overline{A}_k(\theta) = 4\pi\theta\left(\bar{\theta}_k + \theta\right)\,, \qquad \text{with} \quad k = \{\text{SH}, \text{BS}\}\,, \tag{2.53}$$

where $k$ stands for hole masks (survey holes: SH) or bright star masks (BS). Due to the holes in our sample we observe only a fraction $F_{\rm H}^{\rm tot}(\theta)$ of all stellar pairs in the residual solid angle $\Omega$ sparated by an angular distance $\theta$

$$F_{\rm H}^{\rm tot}(\theta) = 1 - \frac{1}{\Omega} \sum_{k=\{\text{SH},\text{BS}\}} N_k \overline{A}_k(\theta)\left(1 - F_{\rm H}(\theta;\bar{\theta}_k)\right)\,, \tag{2.54}$$

where $F_{\rm H}(\theta;\bar{\theta}_k)$ is the fraction of all pairs separated by $\theta$ we observe in $\overline{A}_k(\theta)$

$$F_{\rm H}(\theta;\bar{\theta}_k) = \frac{2\pi}{\overline{A}_k(\theta)} \int_{\bar{\theta}_k}^{\bar{\theta}_k+\theta} x f(x;\theta,\bar{\theta}_k)\mathrm{d}x\,. \tag{2.55}$$

Here, $f$ is the fraction of the area $\pi\theta^2$ of the disc with radius $\theta$ around a star that lies outside the hole (Fig. 2.16)

$$f(x;\theta,\bar{\theta}_k) = \frac{\Omega_{\rm out}(x;\theta,\bar{\theta}_k)}{\pi\theta^2} = 1 - \frac{\Omega_{\rm in}(x;\theta,\bar{\theta}_k)}{\pi\theta^2}\,, \tag{2.56}$$

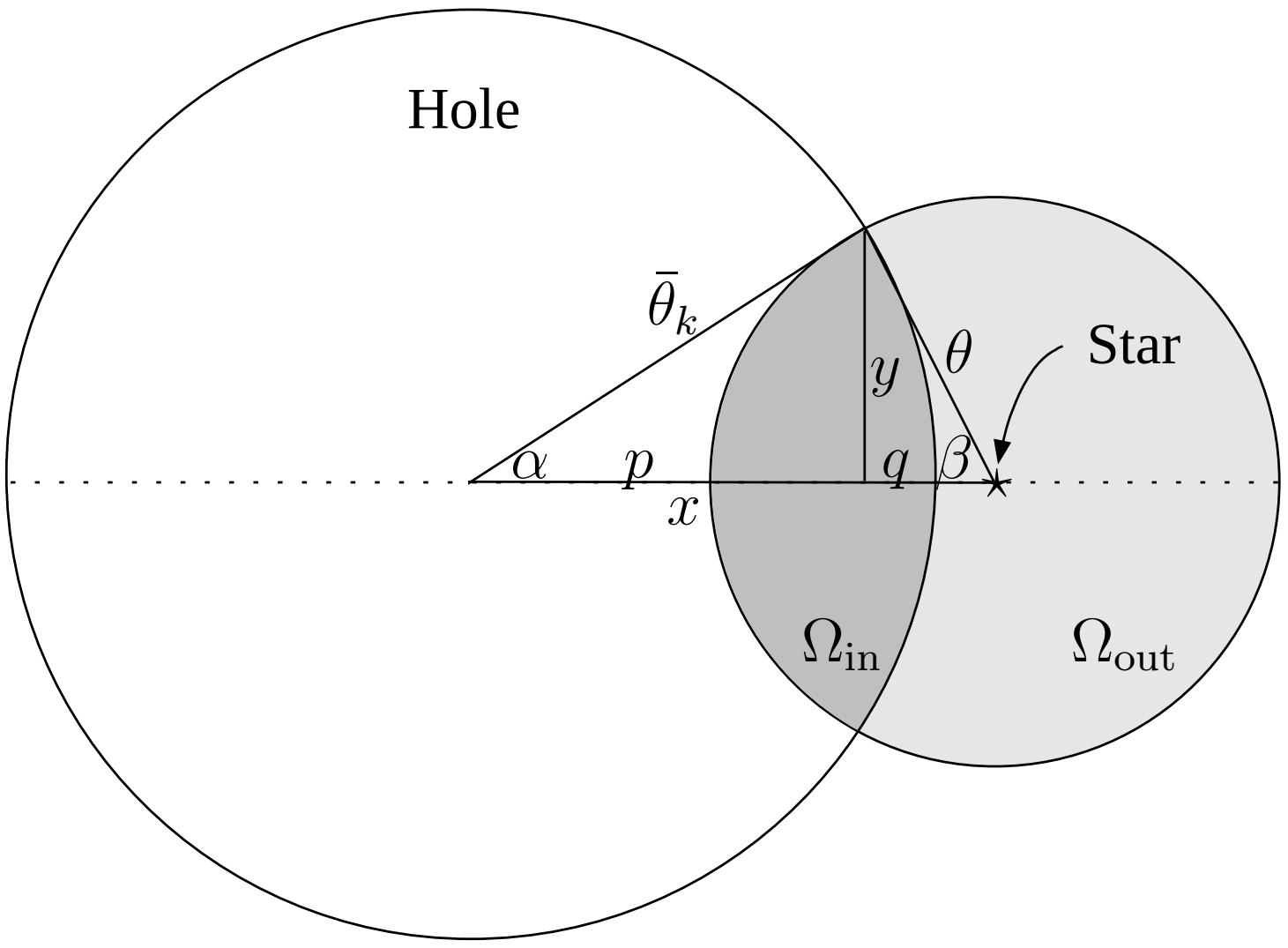

Figure 2.16: Illustration of the geometry used to correct for edge effects due to holes. The large circle represents the bounding circle of a hole or a bright star mask. A star lying close to the hole is indicated. The region within a distance $\theta$ of that star (highlighted in grey) is partially inside the hole (dark grey intersection).

whereas $\Omega_{\rm in} = \pi\theta^2 - \Omega_{\rm out}$ is the area of that disc within the hole, i.e. the intersection. With the nomenclature given in Fig. 2.16, we find

$$\Omega_{\rm out} = \bar{\theta}_k^2\alpha - yp + \theta^2\beta - yq\,, \qquad (2.57)$$

where

$$\alpha = \arcsin\left(\frac{y}{\bar{\theta}_k}\right) \;;\quad \beta = \arcsin\left(\frac{y}{\theta}\right) \qquad (2.58)$$

and

$$p = \sqrt{\bar{\theta}_k^2 - y^2}\,;\quad q = \sqrt{\theta^2 - y^2}\,. \qquad (2.59)$$

Solving $x = p + q$ for $y$ using *Maple* gives

$$y = \frac{1}{2x}\sqrt{2\theta_k^2x^2 - x^4 - \theta^4 + 2\theta^2x^2 + 2\theta^2\theta_k^2 - \theta_k^4}\,, \qquad (2.60)$$

where the integration in (2.55) was performed with *Maple*, too.

## 2.B Galactic tidal limit

A crude estimate of the Galactic tidal limit $a_{\rm T}$, i.e. of the maximum semi-major axis $a_{\rm max}$ of a binary star with total mass $\mathsf{M}$ orbiting in the Galactic tidal field is provided by the *Jacobi limit* $r_\mathrm{J}$ (sometimes also referred to as *zero-velocity surface* or *Roche sphere*) (e.g. Binney and Tremaine 2008, §8.3)

$$r_\mathrm{J} \simeq \left(\frac{\mathsf{M}}{3\mathcal{M}(<\widetilde{D})}\right)^{\frac{1}{3}} \widetilde{D}\,, \tag{2.61}$$

where $\widetilde{D}$ is the galactocentric distance of the binary system and $\mathcal{M}(<\widetilde{D})$ is the Galactic mass enclosed within radius $\widetilde{D}$. Then, $\mathcal{M}(<\widetilde{D})$ is given by

$$\mathcal{M}(<\widetilde{D}) = \frac{v_c^2}{G}\widetilde{D} \simeq 1.1\cdot 10^7 \left(\frac{\widetilde{D}}{\mathrm{pc}}\right)\mathcal{M}_\odot\,, \tag{2.62}$$

where we have adopted the canonical value of the circular speed $v_c = 220\ \mathrm{kms}^{-1}$.

For a rough estimate of this distance range, it is useful to define the *effective depth* of the sample as the most likely distance of an arbitrarly chosen star (Weinberg and Wasserman 1988). In a magnitude-limited sample the total number of stars expected to be observed in a (heliocentric) distance range between $D$ and $D + \mathrm{d}D$ is

$$\mathrm{d}N(D) = \Omega \mathrm{d}D D^2 \rho(D) \int_{M_\mathrm{min}(D)}^{M_\mathrm{max}(D)} \mathrm{d}M \Phi(M)\,, \tag{2.63}$$

where $\Omega$ denotes the solid angle covered by our sample, $\rho(D)$ the density distribution, and $\Phi(M)$ the stellar LF. The absolute magnitudes $M_\mathrm{min}(D)$ and $M_\mathrm{max}(D)$ are given by Eq. 3.38 and 3.39, respectively. The effective depth $D_\mathrm{eff}$ is then given by the median value of the distribution $\mathrm{d}N(D)$ (Weinberg and Wasserman 1988)

$$\frac{1}{2} = \frac{\int_0^{D_\mathrm{eff}} \mathrm{d}N(D)}{\int_0^\infty \mathrm{d}N(D)} = \frac{\int_{D_\mathrm{min}}^{D_\mathrm{eff}} \mathrm{d}N(D)}{\int_{D_\mathrm{min}}^{D_\mathrm{max}} \mathrm{d}N(D)}\,, \tag{2.64}$$

where $D_\mathrm{min}$ and $D_\mathrm{max}$ are used for the numerical integration and are chosen to bracket the theoretical distance range probed by our sample, which – neglecting interstellar dust extinction – is readily found from the apparent magnitude limits, $m_\mathrm{min}$ and $m_\mathrm{max}$, and the absolute magnitudes limits, $M_\mathrm{min}$ and $M_\mathrm{max}$, given by $\Phi(M)$. We take them to be

$$D_\mathrm{min} = 10^{(m_\mathrm{min} - M_\mathrm{max} + 5)/5} = 10^{(15-21.5+5)/5} \simeq 0.5\ \mathrm{pc} \tag{2.65}$$

and

$$D_\mathrm{max} = 10^{(m_\mathrm{max} - M_\mathrm{min} + 5)/5} = 10^{(20.5-(-0.5)+5)/5} \simeq 160\ \mathrm{kpc}\,. \tag{2.66}$$

In Table 2.2 we list the values of $D_\mathrm{eff}$ for our various (sub)samples calculated using the Galactic structure parameter set 2. With the notation of §2.4.2, the effective galactocentric distance $\widetilde{D}_\mathrm{eff}$ can be expressed as

$$\begin{aligned}\widetilde{D}_\mathrm{eff}^2 \;=\; & r_0^2 + D_\mathrm{eff}^2 \cos^2 b - 2 r_0 D_\mathrm{eff} \cos b \cos \ell \\ & + z_0^2 + 2 z_0 D_\mathrm{eff} \sin b + D_\mathrm{eff}^2 \sin^2 b\,.\end{aligned} \tag{2.67}$$

Its values are typically somewhat above 8 kpc.

The last parameter we need to specify in order to estimate the Galactic tidal limit $a_{\mathrm{T}}$, is the total mass of the binary star $\mathsf{M}$. Given the magnitude limits of our sample we may write the average mass of a star in the sample as

$$\langle \mathsf{m} \rangle = \frac{\int_{\mathsf{m}_{\min}}^{\mathsf{m}_{\max}} \mathsf{m}\xi(\mathsf{m})\mathrm{d}\mathsf{m}}{\int_{\mathsf{m}_{\min}}^{\mathsf{m}_{\max}} \xi(\mathsf{m})\mathrm{d}\mathsf{m}} = \frac{\int_{M_{\min}}^{M_{\max}} \mathsf{m}(M)\Phi(M)\mathrm{d}M}{\int_{M_{\min}}^{M_{\max}} \Phi(M)\mathrm{d}M}, \tag{2.68}$$

where $\xi(\mathsf{m})$ denotes the mass function and $\mathsf{m}(M)$ the mass-luminosity relation which we take from Kroupa et al. (1993). The transformation into $r$ magnitudes was performed as in §2.4.2.

Statistically, we expect the total mass of a binary star to be $\mathsf{M} = 2\langle \mathsf{m} \rangle$, and we finally estimate the Galactic tidal limit according to

$$\begin{aligned} a_{\mathrm{T}}(\widetilde{D}_{\mathrm{eff}}, \mathsf{M}) &\simeq \left( \frac{\mathsf{M}}{3\mathcal{M}(<\widetilde{D}_{\mathrm{eff}})} \right)^{\frac{1}{3}} \widetilde{D}_{\mathrm{eff}} \\ &\simeq 3.1 \cdot 10^{-3} \left( \frac{\mathsf{M}}{\mathcal{M}_{\odot}} \right)^{\frac{1}{3}} \left( \frac{\widetilde{D}_{\mathrm{eff}}}{\mathrm{pc}} \right)^{\frac{2}{3}} \mathrm{pc}\,. \end{aligned} \tag{2.69}$$

The values of $a_{\mathrm{T}}$ are around 1 pc and we list them for various (sub)samples in Table 2.2.

# Chapter 3

# Statistical properties of wide binary stars

## Distributions of colours and mass ratios

**Abstract.** We study the colour and mass ratio distribution of Galactic wide binary (WB) stars by combining the angular two-point correlation function for a large sample of SDSS stars with distance information from a photometric parallax method. A novel weighting procedure is applied based on the binding probability of a double star as inferred from its angular separation ($2'' \leq \theta \leq 30''$) and difference in parallax. About 4 000 WBs with separations larger than 200 AU are found statistically. Best fitting of the angular correlation function is achieved for a minimum binding energy that corresponds to an average maximum relative orbital velocity of 370 m/s. The resulting maximum separation, artificially introduced by the method adopted, is around 0.02 pc (4 000 AU). The colour range studied is restricted to $0.2 < r - i < 1.5$ mag, equivalent to a mass range of $0.2 \lesssim \mathcal{M} \lesssim 0.85\mathcal{M}_\odot$. The weighted, bias-corrected colour distribution of our WB candidates is in good accord with the colour distribution of single field stars, in line with previous findings. There is a significant lack of pairs with very different colours: pairs with a colour difference $\Delta(r - i) \gtrsim 1$ mag, corresponding to a mass difference $\Delta\mathcal{M} \gtrsim 0.5\mathcal{M}_\odot$, seem to be systematically underrepresented as compared to a random pairing of field stars. This preference for pairs with similar colour or mass is also reflected in the distribution of mass ratios $q$. For primary masses between 0.5 and 0.85 $\mathcal{M}_\odot$, our mass ratio distribution is peaking at $q > 0.8$, while it is nearly uniform in the range $0.4 < q < 0.8$. The secondary-mass distribution on the other hand is consistent with the field mass function. Previous observations tended to be broadly consistent with a more uniform $q$ and random pairing, but recent WB studies support our finding. Star-formation simulations show an inverse trend (low $q$ overabundance) for WBs. However, evolutionary effects could considerably change the original mass-ratio distribution.

## 3.1 Introduction

Wide binary stars (called WBs hereafter), i.e. binaries with separations typically larger than 100 AU, have only recently shifted into the focus of intense research. In the first place, these loosely bound stellar systems constitute sensitive probes for the gravitational potential of the Milky Way Galaxy. In particular, halo WBs are promising candidates to constrain the masses and densities of MACHOs (e.g. Yoo et al. 2004; Quinn et al. 2009), even though stringent constraints are still difficult to place, mainly because the samples of the widest binaries ($a \sim 0.1$ pc) are still too small. In the near future WBs should also permit to test the CDM paradigm, which predicts that stellar pairs with $a \gtrsim 0.1$ pc should be absent or strongly depleted in the dark haloes of dSph galaxies due to dynamical friction (Hernandez and Lee 2008; Peñarrubia et al. 2010).

The shape of the semi-major axis distribution of WBs, especially for the widest pairs, is therefore of particular interest. The canonical distribution is flat in $\log a$ ($f_a(a) \propto a^{-1}$; Öpik's 1924 law). Recent studies mostly agree that the semi-major axis distribution of the widest WB deviates from Öpik's law and fall off more steeply (e.g. Poveda and Allen 2004; Chanamé and Gould 2004; Lépine and Bongiorno 2007).

WBs are also highly relevant to the problem of star formation. The very presence of extremely wide (semi-major axis $a \gtrsim 10^4$ AU) binary stars in the Galactic field has been described as a 'mystery' (e.g. Parker et al. 2009). Most stars (75% to 90%) are born in stellar clusters (e.g. Lada and Lada 2003) where such loosely bound pairs cannot survive, beeing rapidly disrupted by dynamical encounters even in low-density clusters. Scenarios for the formation of extreme WBs have very recently been suggested by Moeckel and Bate (2010) and Kouwenhoven et al. (2010). According to their simulations, very wide binary stars can be formed in the expanding halo of a dissolving cluster when, loosely speaking, two stars leave the cluster in almost the same direction with almost the same velocity. Both studies make precise predictions on the expected binary properties, such as the semi-major axis distribution and the mass-ratio distribution.

In a previous paper (Longhitano and Binggeli 2010, Paper I hereafter) we constructed the angular two-point correlation function (2PCF) for a sample of about 670 000 stars selected from the Sloan Digital Sky Survey (SDSS, York et al. 2000) from which we derived a WB fraction of roughly 10% and a separation distribution that generally agrees with Öpik's law up to the tidal limit around 1 pc. Our data were, however, also consistent with a broken power-law distribution like that found by e.g. Lépine and Bongiorno. The large uncertainties involved with our purely statistical approach were, as with every angular correlation analysis, clearly caused by the noise from optical (non-physical) pairs.

Common proper motion studies of WBs are considered superior with this respect because they allow, with a high degree of certainty, the identification of individual pairs (Chanamé and Gould 2004; Lépine and Bongiorno 2007; Dhital et al. 2010). The disadvantage is the restriction to relatively nearby stars with large proper motions resulting in a limited sample size. The correlation method, on the other hand, can be significantly improved by including distance information.

The scope of the present study is to filter out optical pairs in the statistical analysis of Paper I by complementing it with a photometric parallax method. Most of the stars observed by the SDSS are main sequence (MS) stars (99%, Finlator et al. 2000) which have a fairly well defined colour-luminosity relation that can be used to estimate the luminosity and, hence, the distance of every star in our sample. We rely here on the photometric parallax relation (PPR) derived by Jurić et al. (2008) (J08 hereafter) who fit a fourth-order polynomial to recent PPRs from the literature. The major source of uncertainty in the PPR is the variance in stellar metallicity. The resulting uncertainty in distance is typically of the order of 100 pc – very large as compared to the typical separation of a WB. The PPR distance estimates derived in this way are therefore still too crude to allow a reliable distinction between optical and true pairs, individually. Thus the filtering procedure has to be of a statistical nature again.

We suggest and apply a statistical weighting procedure based on the binding probability of two stars, given their distance estimates from the PPR and their angular separation. Every stellar pair in our sample (with an angular separation between 2"and 30") gets a statistical weight assigned, which is large for true pairs but small for optical ones. Given the lack of information about the relative velocities of the pairs, we assign to every pair in our sample an *average* relative velocity. In this way a maximum separation is introduced, which is depriving us of constraining the separation distribution at very large separations. However, our procedure guarantees that the resulting colour distributions are dominated statistically by the real pairs in our sample, while optical pairs have a negligible influence. We therefore concentrate in this paper on the colour distributions of WB components, which are then translated into a mass-ratio distribution of WBs.

The mass-ratio distribution of binary stars is of primary importance for star formation theory. In their seminal multiplicity study of solar type stars, Duquennoy and Mayor (1991) found a mass-ratio distribution rising towards small mass ratios, whereas Fischer and Marcy (1992) found a more uniform distribution for M dwarfs (however, both studies are dominated by close binaries). For our WBs here we find a modest tendency for equal masses (high mass ratios), in good accord with the recently published study of Dhital et al. (2010), but in apparent disagreement with the simulations of Moeckel and Bate (2010).

The combination of a correlation analysis with a PPR was used by Sesar et al. (2008). By constructing a volume-complete sample from the SDSS using the PPR from J08, these authors derived a colour distribution of WB stars in a model-independent way. Our approach is different, as we rely on the Wasserman and Weinberg (1987) technique (see also Paper I) to correct the observed colour distribution for selection effects, i.e. we are not fully model-independent. On the other hand, Sesar et al. are limited to a relatively narrow distance interval (0.7 to 1 kpc) to avoid a bias against pairs with very different colours, while no such limitation is present in our study. We regard the two studies, whose results are in general agreement (see, however, Paper I), as complementary.

The paper is organised as follows. In Sect. 3.2 we describe the data and sample of WBs used. In Sect. 3.3 we explain the method of filtering out optical pairs by introducing weights based on the binding probability. In Sect. 3.4 we present the resulting colour and mass-ratio distributions, which then, in discussion Sect. 3.5, are compared with previous observational and theoretical work. Section 3.6 presents our concluding remarks.

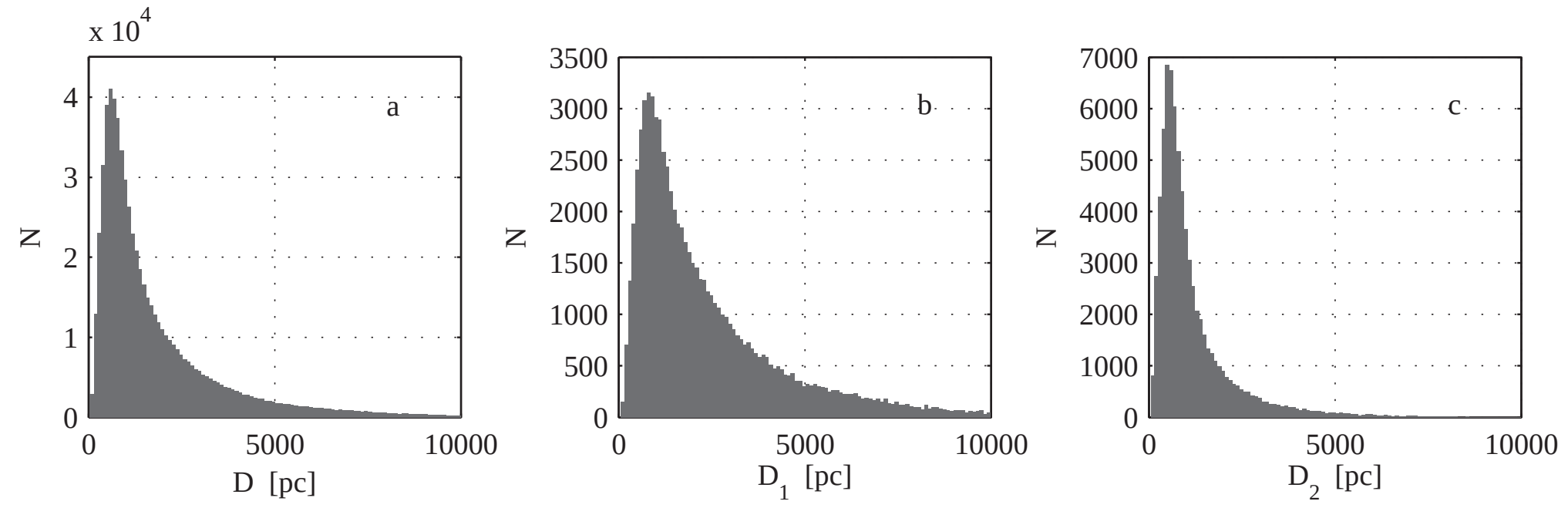

Figure 3.1: Distribution of distances: (a) 648 651 stars from our sample bluer than $r - i = 1.5$ mag; (b) 73 402 primary candidates; (c) 73 402 secondary candidates.

## 3.2 Data and sample

The present study is based on the total sample used in Paper I. The data are drawn from the Sixth SDSS Data Release (Adelman-McCarthy et al. 2008). To be consistent with the photometric parallax relation (PPR) of J08 that holds in the range $0.1 < ri < 1.5$ mag, we adopt a second colour cut removing all stars with $ri > 1.5$ mag, where we have defined $ri \equiv r - i$. The first cut was set at $g - r = 0.5$ mag, implying $ri \gtrsim 0.18$ mag, to avoid confusion with quasars (see Paper I). This leaves us with 648 651 main sequence stars having an apparent $r$-band magnitude between 15 and 20.5 mag and a spectral type later than G5. Using the PPR (Eq. 3.1) the colour cuts translate to the absolute magnitude range $5.2 \lesssim M_r \lesssim 12.0$ mag. The stars are distributed over the same solid angle as in Paper I, covering about 670 square degrees in the direction of the Northern Galactic Pole.

To each star in the sample is now assigned an absolute magnitude $M_r$ and a distance estimate $D$ using the 'bright' PPR from J08, which reads

$$M_r(ri) = 3.2 + 13.3ri - 11.5ri^2 + 5.4ri^3 - 0.7ri^4, \tag{3.1}$$

valid for $0.1 < ri < 1.5$ mag. The main source of uncertainty in the PPR stems from its variance in metallicity and is estimated to amount to $\hat{\sigma}_{M_r} \approx 0.3$ mag (J08). The distance estimate is then given by

$$D(r, M_r) = 10^{(r - M_r + 5)/5}\ \mathrm{pc}\,. \tag{3.2}$$

Following J08, we find a relative distance error $\sigma_D/D \approx 15\%$. A typical star in our sample lies in a distance of about 800 pc. Consequently, the uncertainty in distance is typically of the order of 100 pc. The distribution of distances is shown in Fig. 3.1a.

We tentatively consider every stellar pair having an angular separation $\theta_{ij}$ between $2''$ and $30''$ as a 'WB candidate'. There are 73 402 such pairs in our sample. We show the distance distribution of the primary candidates in Fig. 3.1b, and that of the secondary candidates in Fig. 3.1c. The primary candidate is the member of a WB candidate that has a brighter absolute magnitude assigned by Eq. 3.1. We see that, on average, the secondaries are observed at smaller distances than the primaries, simply because of their fainter intrinsic luminosity. Most of the secondaries have distances smaller than 2 000 pc. This also means that most of the primaries beyond 2 000 pc have no physical partner in our sample. We therefore exclude all pairs with distances larger than $D_{\rm max} = 2\,000$ pc.

This cut in distance leaves us with 456 089 stars and 37 610 WB candidates closer than 2 000 pc, hereafter called our *restricted* sample. Using the Galactic model from §4.2 in Paper I, we estimate that the halo contribution to our restricted sample is only about 7%. Thus most of the stars in our sample belong to the thin or the thick Galactic disk.

## 3.3 Method

Most of the 37 610 WB candidates are optical pairs, i.e. they are not gravitationally bound, and appear to be close to each other when viewed from Earth. In principle, one could distinguish, up to a certain degree, between optical and real pairs using accurate distances. (The identification of true pairs could be pushed further by exploiting proper motion and radial velocity data.) But with the rather crude distance estimates provided by the PPR, it is not possible to reliably distinguish individual optical from true stellar pairs, as the linear maximum separations of WBs is no more than ca. 1 pc – a hundred times smaller than the typical uncertainty in distance.

We therefore decided to adopt a probabilistic procedure: the idea is to establish a measure for the probability of a given visual pair to be bound based on the overlap of the (probabilistic) distance distributions of the two components. This measure allows us to assign a *weight* $w_{ij}$ to each pair consisting of the two stars $i$ and $j$. If these weights are chosen in a meaningful way, i.e. such that the angular correlation properties of the candidates are correctly reproduced, we can use the weights to transform the observed colour distribution of our WB candidates (being mostly optical pairs) into a probabilistic colour distribution of *true* WB stars. For the pair $(i, j)$ we then simply add $w_{ij}$ instead of 1 to the corresponding colour bin. Optical pairs should get very low weights (near 0), so that the resulting colour distribution is dominated by the true pairs. The difficulty lies in choosing the weights correctly. In the next Sections we describe the calculation of the weights in detail.

### 3.3.1 A measure for the binding probability

Following Jiang and Tremaine (2010), we call a binary bound if $E_{\rm J} < E_{\rm c}$, where $E_{\rm J}$ is the Jacobi constant and $E_{\rm c}$ its critical value given by Eq. 22 of Jiang and Tremaine. In a co-rotating coordinate system $(x, y, z)$ with origin in the Galactic midplane at the Sun's distance from the Galactic centre $R_{\rm g}$ and the $x$-axis pointing radially outward, the $y$-axis in the direction of the Galactic rotation and the $z$-axis towards the Southern Galactic Pole, the Jacobi constant for a WB in the Galactic field takes the following form (Eq. 21 in Jiang and Tremaine)

$$E_{\mathrm{J}} = \frac{1}{2}\left(\dot{x}^2 + \dot{y}^2 + \dot{z}^2\right) + \Phi_{\mathrm{eff}}(x, y, z)\,. \tag{3.3}$$

Here, $x, y$ and $z$ are *relative* coordinates, i.e. $x \equiv x_1 - x_2$, etc... The effective potential $\Phi_{\mathrm{eff}}$ includes the gravitational potential between the members of the binary system as well as terms representing the effect of the Galactic tides

$$\Phi_{\mathrm{eff}}(x, y, z) = 2\Omega_{\mathrm{g}} A_{\mathrm{g}} x^2 + \frac{1}{2}\nu_{\mathrm{g}} z^2 - \frac{G(\mathcal{M}_1 + \mathcal{M}_2)}{\sqrt{x^2 + y^2 + z^2}}\,, \tag{3.4}$$

where $\Omega_{\mathrm{g}}$ is the angular speed of the Galaxy, $A_{\mathrm{g}}$ the Oort constant and $\nu_{\mathrm{g}}$ the vertical frequency (see Jiang and Tremaine (2010) and §3.2 of Binney and Tremaine (2008) for more details). Here, the index 'g' means "at radius $R_{\mathrm{g}}$". Furthermore, $G$ is the gravitational constant and $\mathcal{M}_i$ is the mass of component $i$ of the binary system.

Strictly speaking, this expression for the Jacobi constant is only valid in the solar neighbourhood, not too far from Galactic midplane. The stars in our sample are typically about 800 pc above that midplane and Eq. 3.3 should be regarded only as a first approximation for the WB stars in our sample.

The small angular separation of the WB candidates and their positions around the NGP allows some further approximations. As $\theta_{ij} \ll 1$ we may approximate the physical separation $r_{ij}$ of the pair $(i, j)$ by

$$r_{ij} \approx \sqrt{(D_i - D_j)^2 + \theta_{ij} D_i D_j} \approx |D_i - D_j| \approx z\,. \tag{3.5}$$

Since most WB candidates are optical, $r_{ij}$ is typically much larger than the projected separation

$$s_{ij} \approx \theta_{ij} \min(D_i, D_j)\,, \tag{3.6}$$

which constitutes an upper limit to the relative coordinate $x$. Hence, $x \ll z$, allowing us to omit the first term in Eq. 3.4

$$\Phi_{\mathrm{eff}}(r_{ij}) \approx \frac{1}{2}\nu_{\mathrm{g}} r_{ij}^2 - \frac{G(\mathcal{M}_i + \mathcal{M}_j)}{r_{ij}}\,. \tag{3.7}$$

Taking the numerical value for the vertical frequency $\nu_g$ given in Table 1.2 of Binney and Tremaine (2008) we find

$$\Phi_{\mathrm{eff}}(r_{ij}) \approx 4.1 \cdot 10^3 \left(\frac{\mathrm{m}}{\mathrm{s}}\right)^2 \underbrace{\left(\frac{3}{5} r_{ij}^2 - \frac{\mathcal{M}_{ij}}{r_{ij}}\right)}_{\equiv \widehat{\Phi}_{\mathrm{eff}}(r_{ij})} \tag{3.8}$$

where $r_{ij}$ is in pc and the total mass of the binary system $\mathcal{M}_{ij} \equiv \mathcal{M}_i + \mathcal{M}_j$ in solar masses $\mathcal{M}_\odot$.

In the absence of any information on the relative velocities of the WB candidates in our sample, we replace the kinetic term in Eq. 3.3 by a *constant* average value. In this way we can use the effective potential $\Phi_{\rm eff}$ alone as a measure for the binding probability. More precisely, we take the probability that $\widehat{\Phi}_{\rm eff}$, defined in Eq. 3.8, is smaller than a certain limiting value $\Phi_{\rm lim}$ as a measure for the probability that a WB candidate is bound and set

$$w_{ij} \propto P(\widehat{\Phi}_{\rm eff} < \widehat{\Phi}_{\rm lim}) \equiv \int_{-\infty}^{\widehat{\Phi}_{\rm lim}} f_{\widehat{\Phi}_{\rm eff}}(\Phi)\,{\rm d}\Phi = \int_{0}^{r_{\rm lim}} f_{r_{ij}}(r)\,{\rm d}r\,. \tag{3.9}$$

Here, $r_{\rm lim} \equiv r_{ij}(\widehat{\Phi}_{\rm lim})$ and $f_X$ is the probability distribution function (PDF) of $X$. The weights $w_{ij}$ are normalised so that their sum (over all pairs $(i,j)$) adds up to unity

$$\sum_{(i,j)} w_{ij} = 1\,. \tag{3.10}$$

We solve $\widehat{\Phi}_{\rm eff} = \widehat{\Phi}_{\rm eff}(r_{ij})$ for $r_{ij}$ using *Maple* and find

$$\begin{aligned} r_{ij}(\Phi) \;&=\; \frac{20^{1/3}}{6}\left[\left(9\mathcal{M}_{ij} + \sqrt{81\mathcal{M}_{ij}^2 - 20\Phi^3}\right)^{2/3} + 20^{1/3}\Phi\right] \\ &\times\;\; \left[9\mathcal{M}_{ij} + \sqrt{81\mathcal{M}_{ij}^2 - 20\Phi^3}\right]^{-1/3}\,, \end{aligned} \tag{3.11}$$

where we have abbreviated $\widehat{\Phi}_{\rm eff}$ by $\Phi$. The imaginary parts that appear in Eq. 3.11 cancel, assuring $r_{ij}$ to be real for any $\Phi$.

For every WB candidate we infer an $r_{\rm lim} \equiv r_{ij}(\Phi_{\rm lim})$ through Eq. 3.11. We take for every pair the same $\Phi_{\rm lim}$, which is chosen to fit the observations (details of the fitting procedure for $\Phi_{\rm lim}$ are outlined in Sect. 3.3.3). However, $r_{\rm lim}$ varies from pair to pair as the total mass $\mathcal{M}_{ij}$ varies. The masses of the stars are assigned using the mass-luminosity relation (MLR) from Kroupa et al. (1993) (KTG93 hereafter) transformed into $r$-band magnitudes (see §4.2.2 in Paper I). We now discuss in detail how we determine the PDF $f_{r_{ij}}$.

## 3.3.2 Probability distribution of separations

We assume that the errors in absolute magnitude are normally distributed. This assumption is reasonable as long as the uncertainties in apparent magnitudes, $\sigma_r$ and $\sigma_i$ are small, i.e. $\lesssim 0.1$ mag, which is mostly the case. We denote with $\mathcal{N}(x;\mu,\sigma)$ the value of the normal distribution with mean $\mu$ and standard deviation $\sigma$ at position $x$. Then, the PDF of the absolute magnitude $M_r$ of a star, whose absolute magnitude was determined to be $\overline{M}_r = M_r(ri)$ is

$$f_{M_r}(M_r) \approx \mathcal{N}(M_r;\overline{M}_r,\sigma_{M_r})\,. \tag{3.12}$$

In calculating the standard deviation $\sigma_{M_r}$ we follow J08 and assume that $\sigma_{ri}^2 \approx 2\sigma_r^2$ and that the intrinsic scatter of the PPR $\hat{\sigma}_{M_r} = 0.3$ mag leading to

$$\sigma_{M_r}^2 \approx 2\left(\frac{\partial M_r}{\partial ri}\right)^2 \sigma_r^2 + \hat{\sigma}_{M_r}^2 \,. \tag{3.13}$$

Using Eq. 3.2 and neglecting the uncertainty in the apparent magnitude $r$, we can write the PDF of the distance $D$ of a star, whose distance was estimated to be $\overline{D} = D(r, \overline{M}_r)$, as

$$f_D(D) \approx f_{M_r}(M_r)\left|\frac{\partial M_r}{\partial D}\right| = \frac{5}{D \ln 10}\mathcal{N}(M_r(r, D); \overline{M}_r, \sigma_{M_r}), \tag{3.14}$$

with

$$M_r(r, D) = r - 5\log_{10}\frac{D}{10\mathrm{pc}}\,. \tag{3.15}$$

Since most of our WB candidates are just chance projections, we assume that the distance estimates of the two members of a WB candidate are to a good approximation independent from each other. Therefore, we may define the joint PDF of the distances $D_i$ and $D_j$ of the components of a pair $(i, j)$ as the product of the single PDFs

$$f_{(D_i,D_j)}(D_i, D_j) \approx f_{D_i}(D_i) \cdot f_{D_j}(D_j)\,. \tag{3.16}$$

The Jacobian determinant $|\mathbf{J}|$ relates the joint PDFs of the distances to that of $s_{ij}$ and $r_{ij}$

$$f_{(r_{ij},s_{ij})}(r_{ij}, s_{ij}) = |\mathbf{J}|^{-1} f_{(D_i,D_j)}(D_i, D_j)\,. \tag{3.17}$$

Here, $|\mathbf{J}|$ means the absolute value of the Jacobian determinant. We need to express the distances $D_i$ and $D_j$ as well as the Jacobian determinant $|\mathbf{J}|$ in Eq. 3.17 as functions of $r_{ij}$ and $s_{ij}$, respectively. In doing so, we have to distinguish the two cases $D_i < D_j$ and $D_i > D_j$ (for $D_i = D_j$ we have $r_{ij} = s_{ij}$ and therefore $|\mathbf{J}| = 0$). For the distances we find in the case $D_i < D_j$

$$D_i(r_{ij}, s_{ij}) \approx \frac{s_{ij}}{\theta_{ij}} \tag{3.18}$$

and

$$D_j(r_{ij}, s_{ij}) \approx s_{ij}\left(\frac{1}{\theta_{ij}} - \frac{\theta_{ij}}{2}\right) + \frac{r_{ij}}{\theta_{ij}}|\mathbf{J}|\,. \tag{3.19}$$

For the other case $D_i > D_j$ we simply need to interchange $D_i$ and $D_j$. For the Jacobian determinant the distinction results just in the opposite sign, which vanishes when taking the absolute value. We find for both cases

$$|\mathbf{J}| = \left|\frac{\partial r_{ij}}{\partial D_i}\frac{\partial s_{ij}}{\partial D_j} - \frac{\partial r_{ij}}{\partial D_j}\frac{\partial s_{ij}}{\partial D_i}\right| \approx \theta_{ij}\sqrt{1 - \left(\frac{s_{ij}}{r_{ij}}\right)^2}\,. \tag{3.20}$$

We split the joint PDFs of the distances, Eq. 3.16, into a sum to take into account the two cases

$$f_{(D_i,D_j)}(D_i, D_j) = f^{<}_{(D_i,D_j)}(D_i, D_j) + f^{>}_{(D_i,D_j)}(D_i, D_j)\,, \tag{3.21}$$

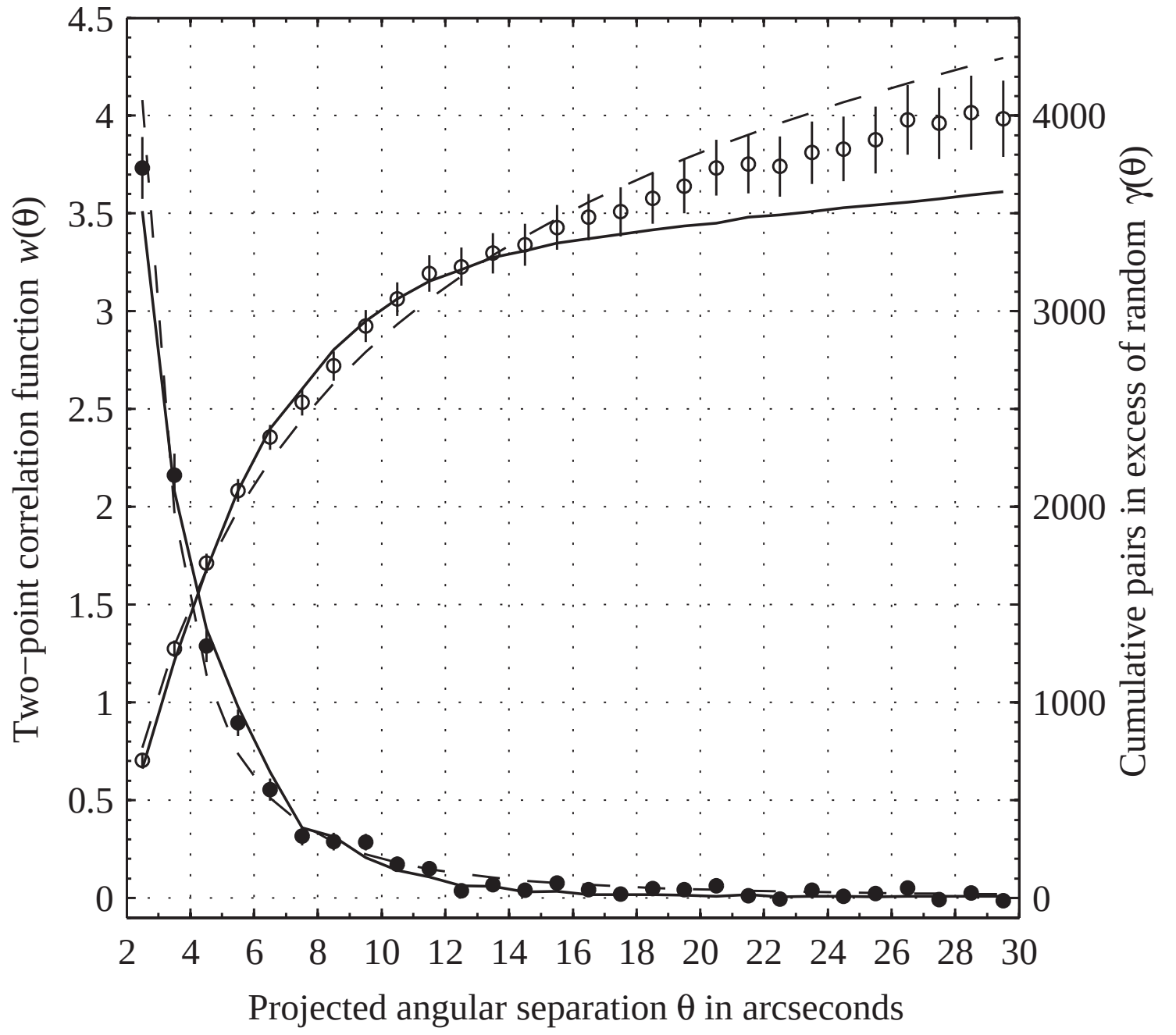

Figure 3.2: 2PCF as inferred from the restricted total sample with $D < 2\,000$ pc (solid circles, left ordinate) and the corresponding CDD (open circles, right ordinate). Poisson errors are indicated as vertical lines. Solid lines show the results from our weighting procedure (see text). Model curves calculated by the WW-technique are plotted as dashed lines.

where the supscript '$<$' stands for the case $D_i < D_j$ and, consequently, '$>$' for the case $D_i > D_j$. To infer the PDF of $r_{ij}$ alone, we need to integrate the joint PDF of $r_{ij}$ and $s_{ij}$ over all $s_{ij}$

$$f_{r_{ij}}(r_{ij}) = \int_0^{r_{ij}} f_{(r_{ij},s_{ij})}(r_{ij}, s_{ij})\,\mathrm{d}s_{ij}\,. \tag{3.22}$$

We only need to integrate up to $r_{ij}$ because $s_{ij} \leq r_{ij}$. This is the PDF we use to calculate the weights $w_{ij}$ in Eq. 3.9.

### 3.3.3 The choice of $\widehat{\Phi}_{\mathrm{lim}}$

The idea is to choose $\widehat{\Phi}_{\mathrm{lim}}$ in such a way that the *weighted* number of pairs derived from our WB candidates, i.e. the number of presumably *true* WB stars as a function of $\theta_{ij}$, reproduces the 2PCF as inferred from the restricted total sample. To this end, we first repeat the 2PCF analysis described in Paper I for the restricted sample to infer the corresponding best-fit values for the WB number density $n_{\mathrm{WB}}$ and the power-law index $\lambda$ of the semi-major axis distribution. We find about 1.2 WB stars per $1\,000$ pc$^3$ and $\lambda \approx 1.2$. This WB density implies that about 5.0% of all stars having $0.2 < ri < 1.5$ mag are a member of a WB star. Referred to *all* stars (no restriction in colour) the fraction amounts to 2.3%. The relative statistical ($2\sigma$) errors of these best-fit values are around 15%.

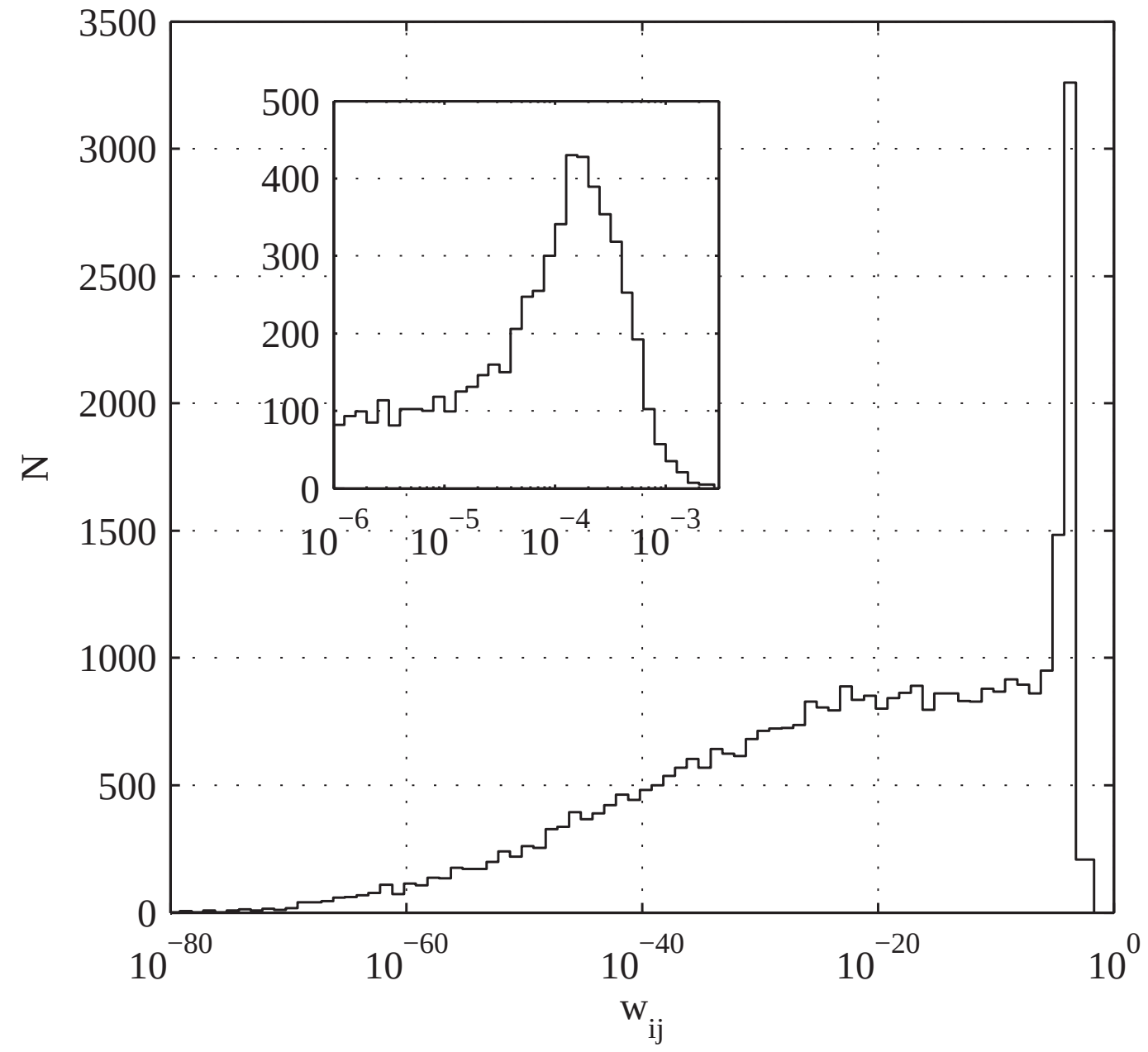

Figure 3.3: Distribution of weights, logarithmically binned. The insert shows an enlargement of the peak region.

The WB density is significantly smaller (by a factor 4) as compared to the density derived from the original total sample in Paper I. The reason for this is the additional colour cut at $ri = 1.5$ mag implying $M_r \lesssim 12$ mag (or, using Eq. 40 from Paper I, $M_V \lesssim 13$ mag). This roughly corresponds to the region where the Jahreiß and Wielen (1997) luminosity function (LF) has its maximum. Therefore, a large fraction of the LF – including the poorly known faint end – is excluded, resulting in the significant drop of the WB density. The restriction to distances smaller than 2 000 pc does not significantly affect the WB density. The power-law index, on the other hand, remains consistent with Öpik's (1924) law. In Fig. 3.2 we show the 2PCF $w(\theta_{ij})$ and the cumulative difference distribution $\gamma(\theta_{ij})$ (CDD; see Eq. 7 in Paper I) derived from the restricted sample together with the best-fit model curves as dashed lines (cf. Fig. 5 in Paper I).

Due to the normalisation of the weights (Eq. 3.10), the total weighted number of pairs, summed over $2'' \leq \theta_{ij} \leq 30''$ equals to 1. We therefore introduce a multiplicative factor $K$, which scales the total weighted number of pairs to fit the observed total excess of pairs with respect to a random distribution. Thus, we expect for the restricted sample $K \approx \gamma(30'') \approx 4\,000$.

Similar to the situation in Paper I (see there §4.4), we find ourselves with two free parameters, $K$ and $\widehat{\Phi}_{\rm lim}$ that we determine by fitting them to the observed 2PCF by means of a standard least square algorithm. In doing so, we find

$$\widehat{\Phi}_{\rm lim} \approx -32.2 \qquad \text{and} \qquad K \approx 3\,448 \tag{3.23}$$

as best-fit values of the two free parameters. Their relative $2\sigma$ errors are estimated to be about 20%. In Fig. 3.2 we also show the 2PCF and the CDD corresponding to these best-fit values as a solid line. Although inferred from a sample consisting of mostly optical pairs, the 2PCF and CDD constructed with our weighting procedure fit remarkably well the observed ones. At angular separations larger than $15''$ the observed 2PCF is slightly underestimated, resulting in the apparently large discrepancy in the CDD at larger $\theta_{ij}$. This discrepancy should, however, not be overemphasised, as the CDD shows the *cumulative* differences between the observed 2PCF and the 2PCF corresponding to the best-fit values in Eq. 3.23. The good agreement with the observation is also supported by the reduced chi-square[1] $\chi^2/\nu \approx 1.25$ and the goodness-of-fit $Q \approx 0.22$ measured by the incomplete gamma function (e.g. Press et al. 1992).

Using Eq. 3.8 we see that the best-fit value of $\widehat{\Phi}_{\rm lim}$ corresponds to $\Phi_{\rm lim} \approx -1.3 \cdot 10^5$ (m/s)$^2$. The difference of this value to the critical Jacobi constant $E_{\rm c}$ can be interpreted as the *average* (relative) kinetic energy per mass of a WB star. The associated average (RMS) relative velocity is

$$\bar{v}_{\rm rel} = \sqrt{2(E_{\rm c} - \Phi_{\rm lim})}\,. \tag{3.24}$$

This suggests that $\Phi_{\rm lim}$ has to be interpreted as an average limiting value as well: it is the largest value for the effective potential $\Phi_{\rm eff}$ where two stars with relative velocity $\bar{v}_{\rm rel}$ can remain bound to each other.

Using again the numerical values in Table 1.2 of Binney and Tremaine (2008) and Eq. 22 of Jiang and Tremaine (2010) we find for the critical Jacobi constant

$$\begin{aligned} E_{\rm c} &= -\frac{3}{2^{1/3}}\left(\Omega_{\rm g} A_{\rm g}\right)^{1/3}\left(G\mathcal{M}_{ij}\right)^{2/3} \\ &\approx -4.8\cdot 10^3\left(\frac{\rm m}{\rm s}\right)^2\left(\frac{\mathcal{M}_{ij}}{\mathcal{M}_\odot}\right)^{2/3} \end{aligned} \tag{3.25}$$

and for the average relative velocity

$$\bar{v}_{\rm rel} \approx 370\left(\frac{\rm m}{\rm s}\right)\left[1 - \frac{1}{27}\left(\frac{\mathcal{M}_{ij}}{\mathcal{M}_\odot}\right)^{2/3}\right]^{1/2}\,. \tag{3.26}$$

As all stars in our sample have a spectral type later than the Sun and are consequently also less massive, we expect the WB candidates' total masses $\mathcal{M}_{ij}$ to be below $2\mathcal{M}_\odot$. Indeed, the total masses $\mathcal{M}_{ij}$, as inferred from the KTG93 MLR, range from $0.4\mathcal{M}_\odot$ to $2\mathcal{M}_\odot$ with a median value of about $1\mathcal{M}_\odot$. In this mass range it is safe to ignore the weak mass dependency of the average relative velocity, giving $\bar{v}_{\rm rel} \approx 370$ m/s.

[1]As in Paper I, we have $\nu = 26$ degrees of freedom here.

Using Eq. 3.11 and the best-fit value of $\widehat{\Phi}_{\rm lim}$ (Eq. 3.23), this mass range translates in a limiting separation range of $0.01 \lesssim r_{\rm lim} \lesssim 0.06$ pc and a median value of approximately 0.03 pc. This limiting separation must be understood in a statistical sense, too. It does not mean that there are no WB stars with $r_{ij} > 0.06$ pc at all in the Galaxy. It just means that, to be bound, two stars with relative velocity $\bar{v}_{\rm rel}$ and a total mass of $1\mathcal{M}_\odot$ must have a separation smaller than 0.03 pc.

The weights $w_{ij}$ calculated in this way are shown as a histogram in Fig. 3.3. Most pairs get a negligible weight, $w_{ij} < 10^{-6}$, in line with the expectation that most pairs in our sample are optical. The distribution has a prominent peak around $w_{ij} \approx 10^{-4}$, which is enlarged in the insert. It is tempting to assign this peak to the presence of WB stars and the 'background' distribution rising up to ∼800 (or up to ∼100 with the binning used in the insert) to the presence of optical pairs. There are 4 751 pairs with $w_{ij} > 10^{-5}$ whose weights sum up to ∼0.996, i.e. these 4 751 pairs constitute about 99.6% of the total weight and, hence, dominate the distributions we infer in the next Section as well as in Sects. 3.4 and 3.4.4.

For the purpose of follow-up studies and cross identifications we list our top WB candidates sorted for their weights in Table 3.1. An extended table including the top 5 000 pairs is available at CDS.

### 3.3.4 Distribution of projected separations

As a further test of our weighting procedure, we compare the weighted *observed*[2] distribution of projected separation $s_{ij}$ with that expected from the Wasserman-Weinberg (WW) model (Wasserman and Weinberg 1987, see also §4 in Paper I). The *average* projected separation $\langle s_{ij}\rangle$ is related to the average separation $\langle r_{ij}\rangle$ by (e.g. Yoo et al. 2004)

$$\langle s_{ij}\rangle = \frac{\pi}{4}\langle r_{ij}\rangle\,. \tag{3.27}$$

This allows us to translate the range in $r_{\rm lim}$ (see the previous Section) into $0.008 \lesssim s_{\rm lim} \lesssim 0.05$ pc, which is the range where we expect the weighted distribution $s_{ij}$ to drop to zero.

Let us denote the weighted distribution of $s_{ij}$ by $\psi_{\rm obs}(s_{ij})$ and that predicted by the WW-model $\psi_{\rm WW}(s_{ij})$. In Fig. 3.4 we show $\psi_{\rm obs}$ and $\psi_{\rm WW}$ as solid and dashed line, respectively. If our weighting procedure is correct, $\psi_{\rm obs}$ should agree with $\psi_{\rm WW}$ at least up to $s_{ij} \approx 0.008$ pc. We have therefore normalised $\psi_{\rm obs}$ so that the number of observed WB stars with $s_{ij} < 0.008$ pc agree with the expectation of the WW-model:

$$\int_0^{0.008\ \rm pc} \psi_{\rm obs}(s_{ij})\,{\rm d}s_{ij} \;=\; \int_0^{0.008\ \rm pc} \psi_{\rm WW}(s_{ij})\,{\rm d}s_{ij}\,. \tag{3.28}$$

[2]We use the term 'observed' for distributions not corrected for selection effects.

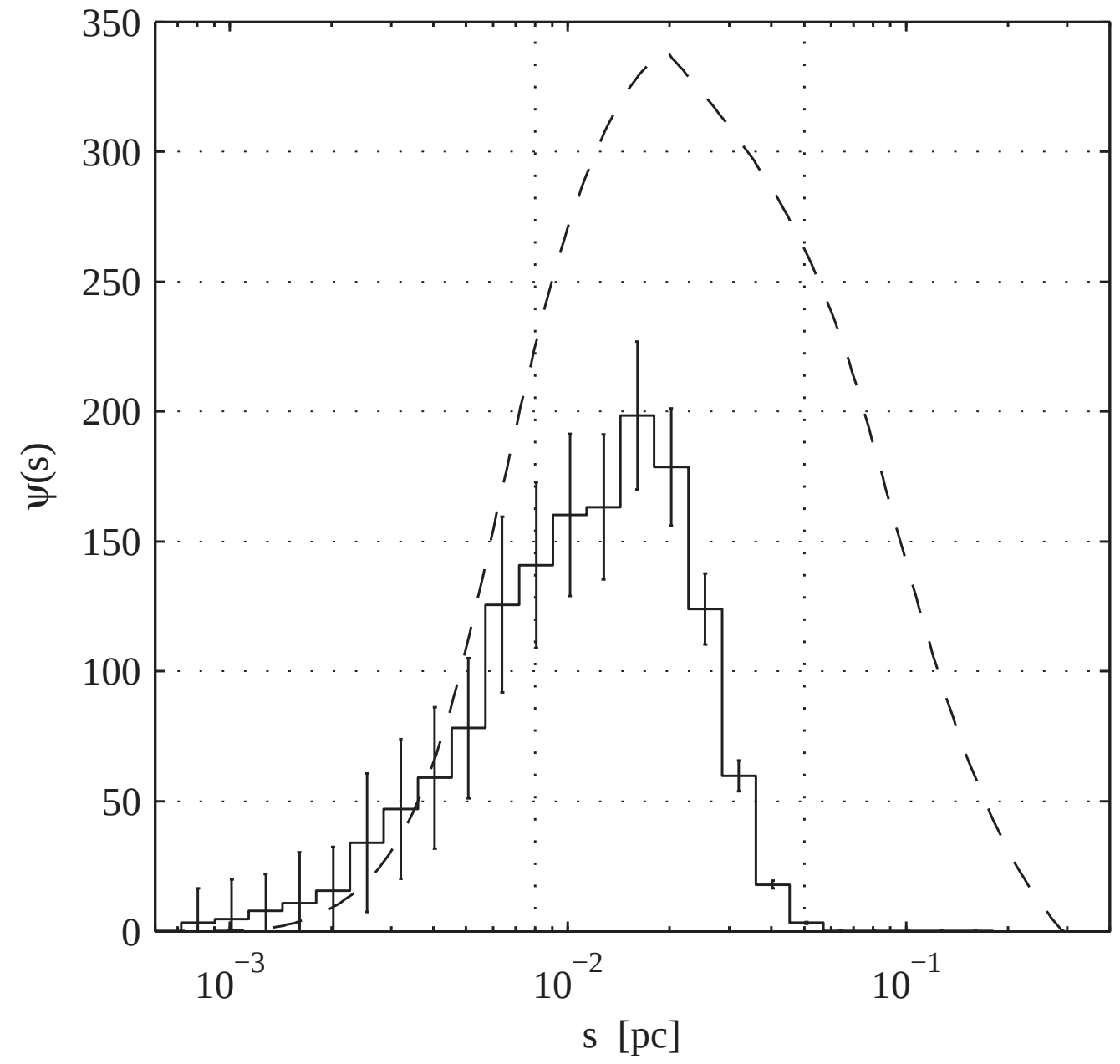

Figure 3.4: Observed distribution of projected separations $s$ inferred with our weighting procedure (solid line). Poisson errors are indicated as vertical lines. The dashed line shows the theoretical distribution expected from the WW-model. The $s_{\rm lim}$-range (see text) is indicated as dotted vertical lines.

From Fig. 3.4 we see that the overall agreement for $s_{ij} < 0.008$ is quite good given the approximative character of our study. While the (dashed) model curve lies within the Poisson errors (indicated by vertical lines), we note an overestimation at the smallest projected separations compensated by an underestimation in the range $0.005 \lesssim s_{ij} \lesssim 0.008$ pc. The subsequent deviation from the model prediction is expected, because we enter the $s_{\rm lim}$-range indicated by the two vertical dotted lines. The sharp drop of $\psi_{\rm WW}$ beyond $s_{ij} \approx 0.02$ pc is due to the cut in distances we have adopted at $D_{\rm max} = 2\,000$ pc. In the angular separation range we study, $2'' \leq \theta_{ij} \leq 30''$, the farthest pairs at $D = D_{\rm max}$ may have a projected separation $s_{ij}$ between 0.02 pc and 0.3 pc to be observed – the range where $\psi_{\rm WW}$ drops to zero.

All in all, the good agreement between $\psi_{\rm obs}$ and $\psi_{\rm WW}$ at $s_{ij} \lesssim 0.008$ pc shows that our weighting procedure is essentially correct, encouraging us to continue our statistical analysis of the properties of WBs. On the other hand, Fig. 3.4 also shows that we gain no information at all for pairs with projected separations beyond 0.06 pc. Not because such pairs do not exist, but because pairs with $s_{ij} \gtrsim 0.06$ pc must have relative velocities smaller than $\bar{v}_{\rm rel}$ to be bound. Consequently, these pairs get vanishing weights in our procedure.

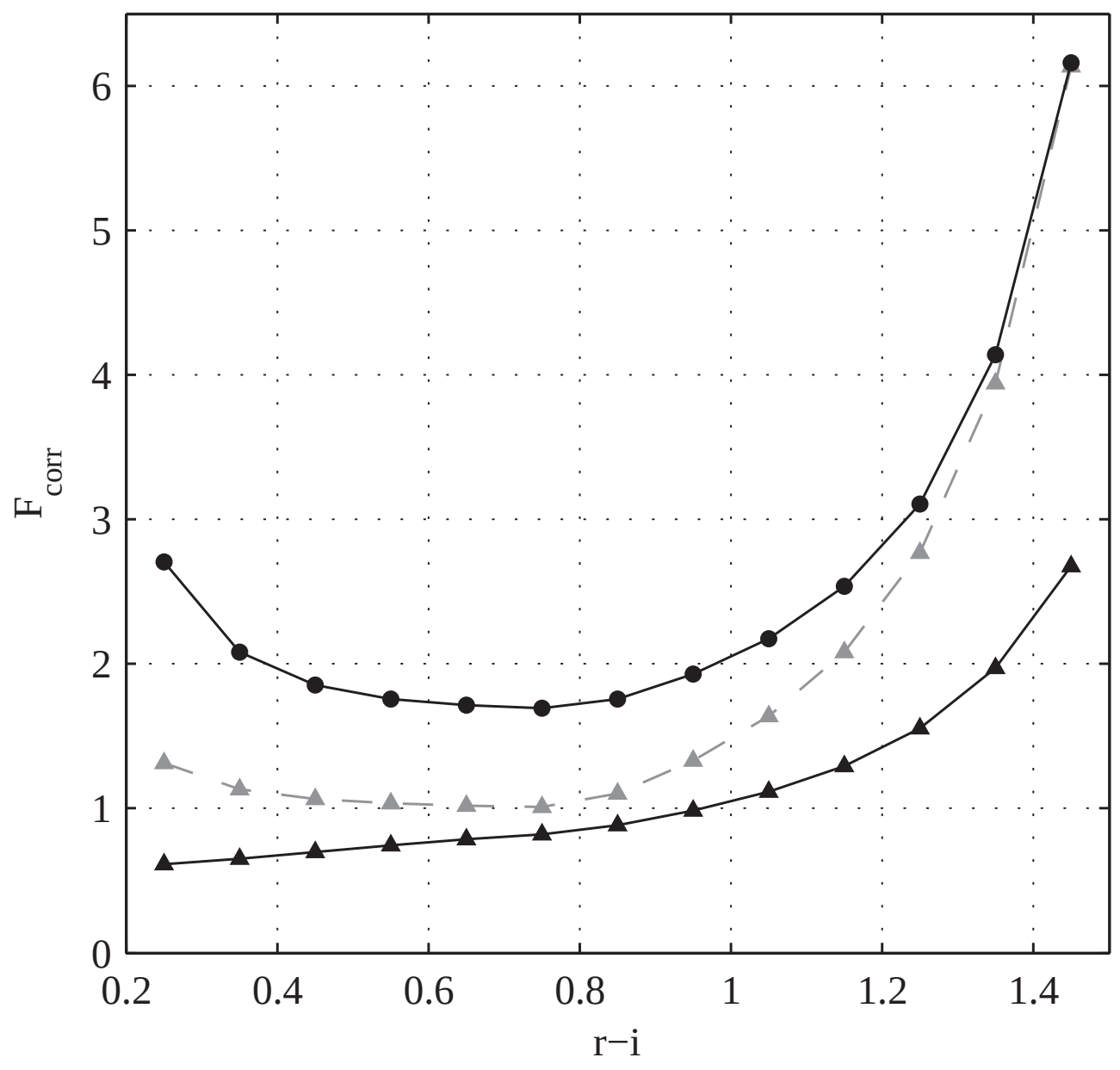

Figure 3.5: Correction functions for WB stars (circles), single field stars without "fairness factor" (grey triangles; see text), and single field stars with "fairness factor" (black triangles).

## 3.4 Results

All distributions of physical quantities we infer in the following, are based on the distribution of the colour index $ri$. The absolute magnitude is determined from the PPR $M_r = M_r(ri)$ (Eq. 3.1), which are then transformed into masses by the KTG93 MLR. Before drawing any conclusions from the colour distributions inferred by our weighting procedure, they have to be corrected for observational bias.

### 3.4.1 Corrections for selection effects

Our restriced sample is limited by apparent magnitudes and therefore subject to Malmquist bias (e.g. Binney and Merrifield 1998, §3.6.1): luminous stars can be observed out to larger distances than intrinsically faint stars. Consequently, bright stars, having a bluer colour index, are over-represented in magnitude-limited samples. Thus, the observed colour distributions differ from the true distributions so that it is necessary to apply a correction that takes into account this selection effect.

The basic idea of the correction we adopt is that the stars of every colour bin have to be normalised to the same (arbitrary) volume. Because our sample lies in the direction of the NGP with typical distances of the order of the scale height of the thick disk ($\sim$900 pc), we must also take into account the stellar density gradient towards the NGP. We use the same stellar density distribution as in Paper I (§4.2.1) based on the bias-corrected structure parameters from J08.

Let $N_{\rm obs}(ri)$ be the number of stars observed in the $ri$-bin that are a member of a WB star. As before, 'observed' stands for "weighted, but not corrected for selection effects". Then the true number of WB components in the $ri$-bin is

$$N_{\rm true}(ri) = N_{\rm obs}(ri) \cdot F_{\rm corr}(ri)\,, \tag{3.29}$$

with an appropriate correction function $F_{\rm corr}$. To infer $F_{\rm corr}$ we again rely on the WW-model. We calculate the ratio of the expected true number of stars in each bin $N_{\rm true}^{\rm WW}(ri)$ to the expected *observed* number of stars $N_{\rm obs}^{\rm WW}$:

$$F_{\rm corr}(ri) \equiv \frac{N_{\rm true}^{\rm WW}(ri)}{N_{\rm obs}^{\rm WW}(ri)}\,. \tag{3.30}$$

Recalling Eq. 43 of Paper I, we write the observed number of stars in a $ri$-bin being a member of a WB as

$$\begin{aligned} N_{\rm obs}^{\rm WW}(ri) &= n_{\rm WB}\Omega \int\limits_{\theta_{\rm min}}^{\theta_{\rm max}} {\rm d}\theta \int\limits_{D_1(\theta)}^{D_2(\theta)} {\rm d}D D^3 \tilde{\rho}(D) Q(D\theta) \\ &\quad \times \iint\limits_{M_1(D)}^{M_2(D)} {\rm d}M_A {\rm d}M_B \widetilde{\Phi}(M_A) \widetilde{\Phi}(M_B) \delta(M_A - M_r(ri)) \end{aligned} \tag{3.31}$$

with

$$\begin{aligned} D_1(\theta) &\equiv \max(\langle s\rangle_{\rm min}/\theta, D_{\rm min}(ri)) & (3.32)\\ D_2(\theta) &\equiv \min(\langle s\rangle_{\rm max}/\theta, D_{\rm max}(ri), D_{\rm max}) & (3.33) \end{aligned}$$

where, as in Paper I, $\langle s\rangle_{\rm min} = 0.98a_{\rm min} = 9.8 \cdot 10^{-4}$ pc and $\langle s\rangle_{\rm max} = 0.98a_{\rm T}$ (here, the tidal limit of the semi-major axis is $a_{\rm T} \approx 1.2$ pc). The distance range $[D_{\rm min}(ri), D_{\rm max}(ri)]$ in which a star of a given colour index $ri$ can be observed is defined by the range in apparent ($r$-band) magnitude $[m_{\rm min}, m_{\rm max}]$ we have chosen for our restricted sample and the PPR (Eq. 3.1)

$$\begin{aligned} D_{\rm min}(ri) &= 10^{(m_{\rm min}-M_r(ri)+5)/5}\ {\rm pc} & (3.34)\\ D_{\rm max}(ri) &= 10^{(m_{\rm max}-M_r(ri)+5)/5}\ {\rm pc}\,. & (3.35) \end{aligned}$$

Here, we have $m_{\rm min} = 15$ mag and $m_{\rm max} = 20.5$ mag. Finally, the integral over $D$ is constrained to distances smaller than $D_{\rm max} = 2\,000$ pc (see Sect. 3.2). To take into account the range in absolute magnitude $[M_{\rm min}, M_{\rm max}]$ corresponding to the adopted colour cuts (here, $M_{\rm min} \approx 5.2$ mag and $M_{\rm max} \approx 12$ mag; see again Sect. 3.2), we furthermore define

$$\begin{aligned} M_1(D) &\equiv \max(M_{\rm min}(D), M_{\rm min}) & (3.36)\\ M_2(D) &\equiv \min(M_{\rm max}(D), M_{\rm max}) & (3.37) \end{aligned}$$

where

$$M_{\max}(D) = m_{\max} - 5\log_{10}(D/10\ \mathrm{pc}) \tag{3.38}$$
$$M_{\min}(D) = m_{\min} - 5\log_{10}(D/10\ \mathrm{pc})\,. \tag{3.39}$$

The delta-function in Eq. 3.31 makes sure that one component of each WB star (here we arbitrarily choose component $A$) falls into the corresponding $ri$-bin. If $M_r(ri)$ is within the integration limits, we may replace the integral over $M_A$ by the value of the (normalised) LF at $M_r(ri)$. In doing so, we find for the number of stars being a component of a WB star and having a colour index within the $ri$-bin

$$N_{\mathrm{obs}}^{\mathrm{WW}}(ri) \propto \int\limits_{\theta_{\min}}^{\theta_{\max}} \mathrm{d}\theta \int\limits_{D_1(\theta)}^{D_2(\theta)} \mathrm{d}D D^3 \tilde{\rho}(D) Q(D\theta) \int\limits_{M_1(D)}^{M_2(D)} \mathrm{d}M_B \widetilde{\Phi}(M_B) \tag{3.40}$$

if $M_1(D) < M_r(ri) < M_2(D)$ and $N_{\mathrm{obs}}^{\mathrm{WW}}(ri) = 0$ else. The proportionality constant in Eq. 3.40 reads $n_{\mathrm{WB}}\Omega\widetilde{\Phi}(M_r(ri))$.

The *true* number of stars in a $ri$-bin expected from the WW-model can be regarded as the limiting case of Eq. 3.40 where $m_{\min} \to -\infty$ and $m_{\max} \to +\infty$. Proceeding in the same way as before, we find

$$N_{\mathrm{true}}^{\mathrm{WW}}(ri) \propto \int\limits_{\theta_{\min}}^{\theta_{\max}} \mathrm{d}\theta \int\limits_{\langle s\rangle_{\min}/\theta}^{\min(\langle s\rangle_{\max}/\theta,\, D_{\max})} \mathrm{d}D D^3 \tilde{\rho}(D) Q(D\theta) \tag{3.41}$$

if $M_{\min} < M_r(ri) < M_{\max}$ and $N_{\mathrm{true}}^{\mathrm{WW}}(ri) = 0$ else, where we have used the normalisation of the LF

$$\int\limits_{M_{\min}}^{M_{\max}} \widetilde{\Phi}(M)\,\mathrm{d}M = 1\,. \tag{3.42}$$

The proportionality constant in Eq. 3.41 is, of course, the same as in Eq. 3.40.

If either $N_{\mathrm{obs}}^{\mathrm{WW}}(ri)$ or $N_{\mathrm{true}}^{\mathrm{WW}}(ri)$ is zero, we set the correction function $F_{\mathrm{corr}}(ri) = 0$. Otherwise $F_{\mathrm{corr}}(ri)$ is the ratio as defined in Eq. 3.30. We see that $F_{\mathrm{corr}}$ is independent from the WB number density $n_{\mathrm{WB}}$, the solid angle $\Omega$ of our sample as well as the value of the LF at $M_r(ri)$. It does, however, depend on the power-law index $\lambda$ of the semi-major axis distribution through the reduced distribution of projected separations $Q(s)$ (see Eq. 23 in Paper I).

In Fig. 3.5 the correction function for binary stars $F_{\mathrm{corr}}$ is plotted as a solid line. By definition, $F_{\mathrm{corr}}(ri) \geq 1$ for all $ri$, because the observed number of stars in a given volume is always a lower limit to the true number of stars of that volume. The correction function $F_{\mathrm{corr}}$ has a minimum around $ri = 0.8$ mag and rises towards redder as well as bluer colours. The reason for the increase at bluer colour is the lower apparent magnitude limit at $r = 15$ mag excluding bright stars that are too close: According to the PPR (Eq. 3.1), stars at the blue end ($ri = 0.2$ mag) have an absolute magnitude of $M_r \approx 5.4$ mag. To have an apparent magnitude fainter than $r = 15$ mag, they have to be farther than 800 pc. The stars at the blue end are, however, over-represented compared to those at the red end ($ri = 1.5$ mag). The reddest stars are simply too faint to be observed at distances larger than about 500 pc given the upper limit in apparent magnitude at $r = 20.5$ mag.

We discuss at this point the correction for selection effects of field stars. Relying on von Seeliger's (1898) formula (see also Karttunen et al. 1996, §18.2), we write for the number of observed stars (not necessarily in a WB) in the $ri$-bin

$$\begin{aligned} n_{\rm obs}^{\rm WW}(ri) &= n'_*\Omega \int\limits_{D_{\rm min}(ri)}^{\min(D_{\rm max}(ri),D_{\rm max})} {\rm d}D D^2\tilde{\rho}(D) \\ &\times \int\limits_{M_1(D)}^{M_2(D)} {\rm d}M \widetilde{\Phi}(M)\delta(M - M_r(ri))\,, \end{aligned} \tag{3.43}$$

where $n'_*$ is the local stellar number density of the stars in our sample as inferred from the Jahreiß and Wielen LF $\Phi(M)$

$$n'_* \equiv \int\limits_{M_{\rm min}}^{M_{\rm max}} \Phi(M)\,{\rm d}M \approx 0.048 \text{ pc}^{-3}\,. \tag{3.44}$$

As before, we have

$$n_{\rm obs}^{\rm WW}(ri) = n'_*\Omega\widetilde{\Phi}(M_r(ri)) \int\limits_{D_{\rm min}(ri)}^{\min(D_{\rm max}(ri),D_{\rm max})} {\rm d}D D^2\tilde{\rho}(D) \tag{3.45}$$

if $M_1(D) < M_r(ri) < M_2(D)$ and $n_{\rm obs}^{\rm WW}(ri) = 0$ else. Consequently, the number of true stars in the $ri$-bin is then

$$n_{\rm true}^{\rm WW}(ri) = n'_*\Omega\widetilde{\Phi}(M_r(ri)) \int\limits_0^{D_{\rm max}} {\rm d}D D^2\tilde{\rho}(D) \tag{3.46}$$

if $M_{\rm min} < M_r(ri) < M_{\rm max}$ and $n_{\rm true}^{\rm WW}(ri) = 0$ else.

In analogy with the correction function for the WB stars, we define

$$f_{\rm corr}(ri) \equiv \frac{n_{\rm true}^{\rm WW}(ri)}{n_{\rm obs}^{\rm WW}(ri)} \tag{3.47}$$

if neither $n_{\rm obs}^{\rm WW}(ri)$ nor $n_{\rm true}^{\rm WW}(ri)$ equals zero. Otherwise, $f_{\rm corr}(ri) = 0$. We show the correction function of the Galactic field stars in Fig. 3.5 as a grey dash-dotted line. Of course, again by definition, $f_{\rm corr}(ri) \geq 1$ for all $ri$. The overall shape of $f_{\rm corr}$ is similar to that of the WB stars, $F_{\rm corr}$.

The comparison of the bias corrected colour distribution of all stars to that of the WB stars would not be 'fair', because not in every distance the whole projected separation range $[\langle s\rangle_{\rm min}, \langle s\rangle_{\rm max}]$ can be explored given the range in angular separation $[\theta_{\rm min}, \theta_{\rm max}]$. For example, in a distance of 1 kpc we can only explore the range 0.01 pc $\lesssim s \lesssim$ 0.15 pc. Thus, stars in a certain distance should be compared only to those WB stars that have a projected separation in the range corresponding to that distance. Using the reduced distribution of projected separations $Q(s)$, we may express the fraction of all WB stars with projected separations $s$ in $[\langle s\rangle_{\rm min}, \langle s\rangle_{\rm max}]$ that can be observed as

$$f_{\rm frac}(D) \equiv \int\limits_{\max(\theta_{\rm min}D,\,\langle s\rangle_{\rm min})}^{\min(\theta_{\rm max}D,\,\langle s\rangle_{\rm max})} Q(s)\,{\rm d}s \leq 1\,. \tag{3.48}$$

The colour index $ri$ corresponds to the distance range $[D_{\rm min}(ri), D_{\rm max}(ri)]$. So, we find for the fraction of observable WB stars with colour $ri$

$$f_{\rm fair}(ri) \equiv \left(\int\limits_{0}^{D_{\rm max}} f_{\rm frac}(D)\,{\rm d}D\right)^{-1} \cdot \int\limits_{D_{\rm min}(ri)}^{\min(D_{\rm max}(ri),\,D_{\rm max})} f_{\rm frac}(D)\,{\rm d}D \leq 1\,. \tag{3.49}$$

The corrected number of field stars in the $ri$-bin is then

$$n_{\rm true}(ri) = n_{\rm obs}(ri) \cdot f_{\rm corr}(ri) \cdot f_{\rm fair}(ri) \equiv n_{\rm obs}(ri) \cdot \widetilde{F}_{\rm corr}(ri)\,, \tag{3.50}$$

where $f_{\rm corr}$ corrects for Malmquist bias and $f_{\rm fair}$ makes sure that the comparison between the colour distribution of WB stars and field stars is 'fair' in the sense just described.

In Fig. 3.5 we show $\widetilde{F}_{\rm corr}$ – the correction function for the field stars – as (black) dash-dotted line. Due to the 'fairness factor', $f_{\rm fair}$, this correction function can be smaller than 1, as it is the case here for $ri \lesssim 1$ mag. The difference between $f_{\rm corr}$ and $\widetilde{F}_{\rm corr}$ is smallest around $ri \approx 0.8$ mag, while it increases towards both redder and bluer colours. The 'fairness correction' is important, especially at the red end.

### 3.4.2 Corrected colour distributions

We first construct the (corrected) joint probability density $p(ri_1, ri_2)$, which is the probability density that a WB has its primary component in the $ri_1$-bin and the secondary in the $ri_2$-bin. Assuming that the correction function $F_{\rm corr}$ for the primary component is independent from that of the secondary, we have

$$F_{\rm corr}(ri_1, ri_2) = F_{\rm corr}(ri_1) \cdot F_{\rm corr}(ri_2)\,. \tag{3.51}$$

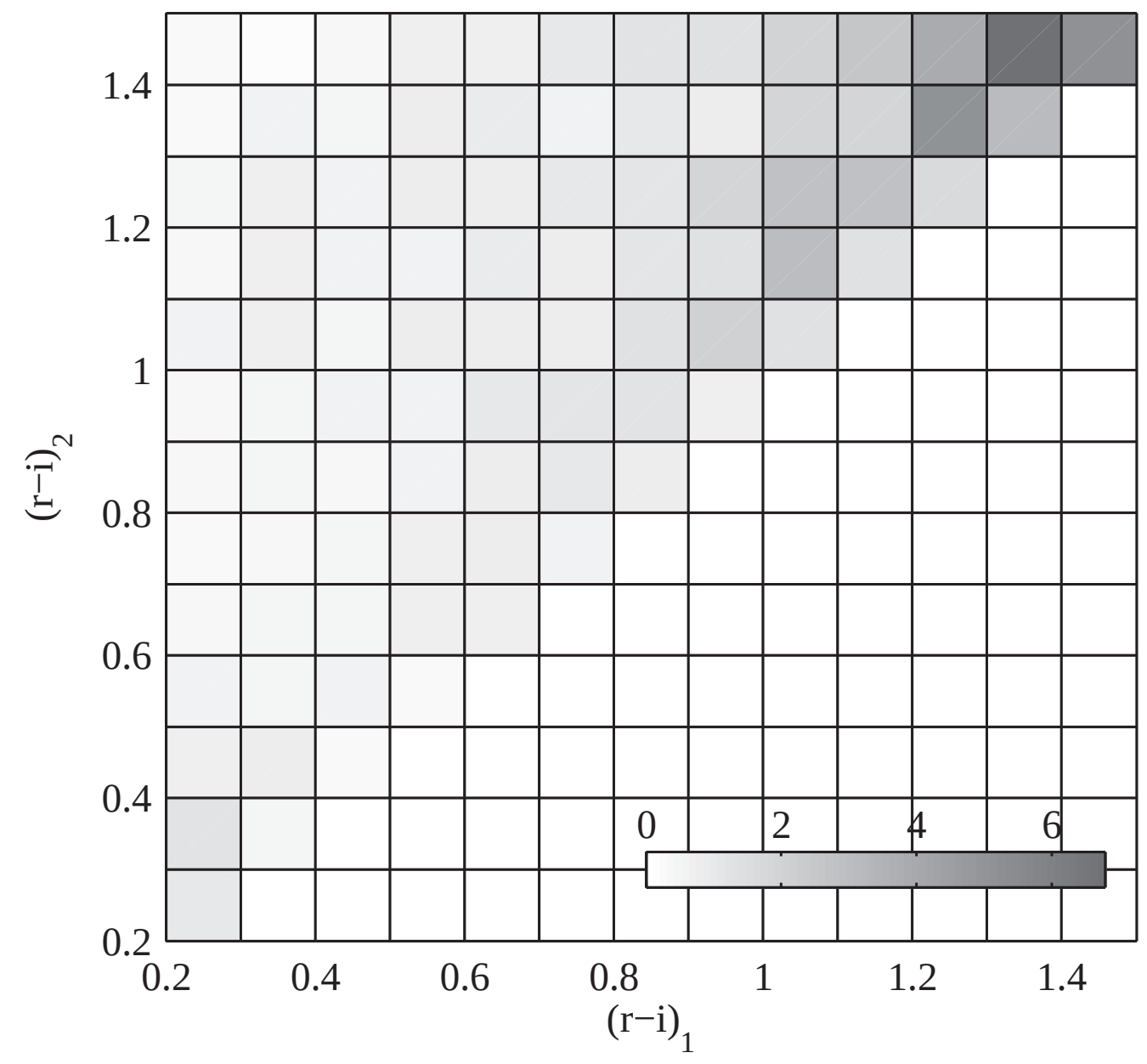

Figure 3.6: Corrected joint probability density of primary (index 1) and secondary (index 2) WB components. The relative error in each bin is of the order of 5%.

Let $N_{\rm obs}(ri_1, ri_2)$ be the weighted observed counts in a $(ri_1, ri_2)$-bin, i.e. the weighted number of observed pairs with a primary component whose colour falls in the $ri_1$-bin and a secondary component with colour in the $ri_2$-bin

$$N_{\rm obs}(ri_1, ri_2) \equiv \sum_{(ri_1,ri_2)\text{-bin}} w_{ij} \leq 1\,. \tag{3.52}$$

These are less than unity because of the normalisation of the weights (Eq. 3.10). The true weighted counts are then

$$N_{\rm true}(ri_1, ri_2) = N_{\rm obs}(ri_1, ri_2) \cdot F_{\rm corr}(ri_1, ri_2)\,, \tag{3.53}$$

which we expect to be proportional to the true number of WB stars in a $(ri_1, ri_2)$-bin. The bias corrected joint probability density that a WB star has a primary component with colour $ri_1$ and a secondary with colour $ri_2$ is then

$$p(ri_1, ri_2) = \frac{N_{\rm true}(ri_1, ri_2)}{(\Delta ri)^2 \sum\limits_{ri_1,\, ri_2} N_{\rm true}(ri_1, ri_2)}\,, \tag{3.54}$$

where $\Delta ri = 0.1$ mag is the width of a $ri$-bin and the sum goes over all combinations of $ri_1$ and $ri_2$. The resulting joint probability density map is shown in Fig. 3.6, where the $ri$ colour index of the primary component (index 1) is plotted on the abscissa and that of the secondary (index 2) on the ordinate.

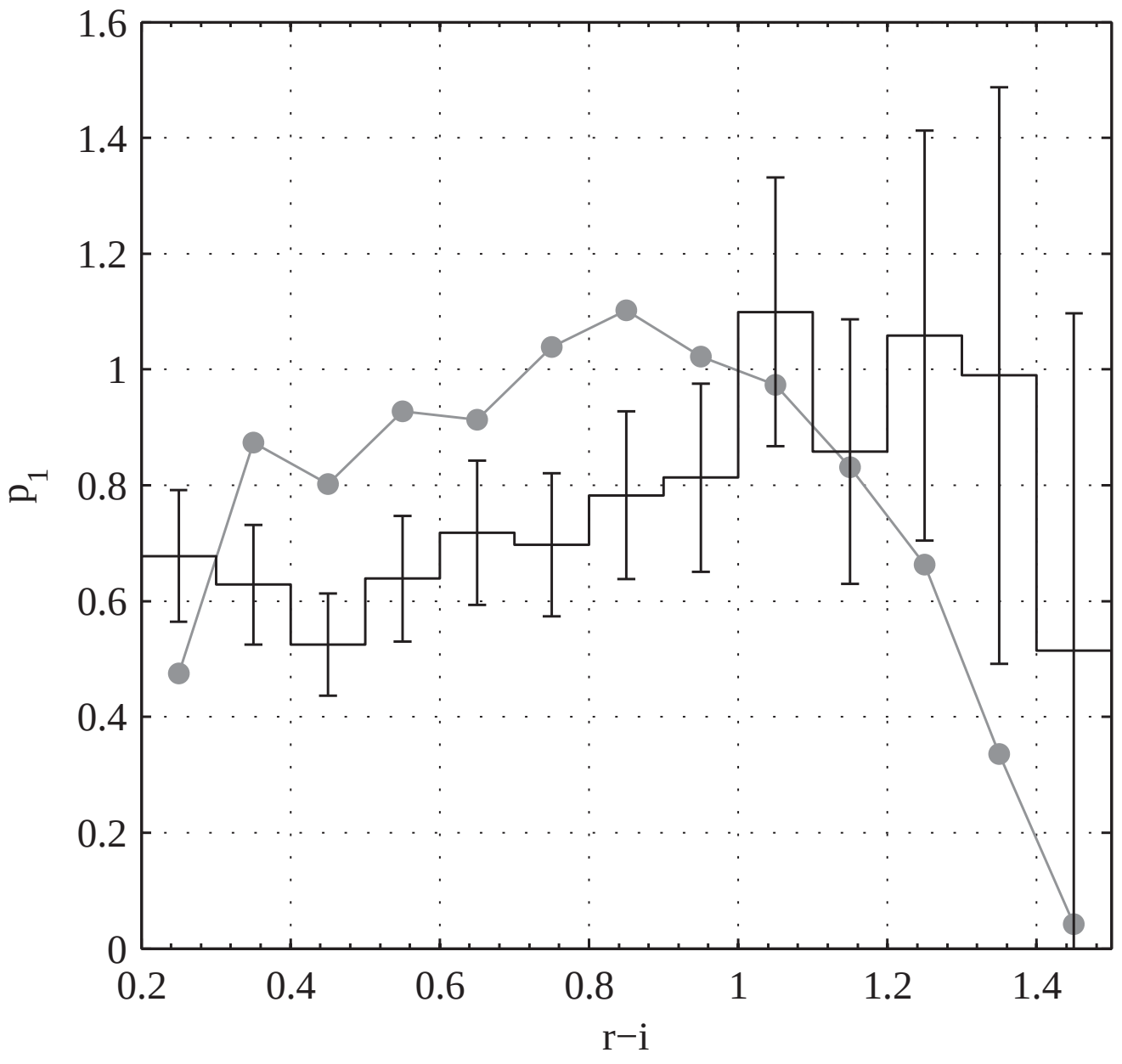

Figure 3.7: Corrected colour distribution (probability density) of *primary* WB components (histogram). The error bars reflect statistical uncertainties. The (corrected) distribution expected from random pairing of field stars is shown as grey circles.

To estimate the statistical errors, we rely on the *un*weighted number of pairs, $\widetilde{N}_{\rm obs}(ri_1, ri_2)$, giving the number of WB *candidates* in a $(ri_1, ri_2)$-bin. Throughout this work, we quote statistical (Poissonian) errors only. Uncertainties stemming from the PPR, the Galactic model, from observations or any other source are not included. Our error estimates must therefore be regarded as a lower limit to the true uncertainties. The statistical error of a bin with $\widetilde{N}_{\rm obs}(ri_1, ri_2)$ counts is

$$\delta\widetilde{N}_{\rm obs}(ri_1, ri_2) = \sqrt{\widetilde{N}_{\rm obs}(ri_1, ri_2)}\,. \tag{3.55}$$

Then, the statistical errors of the weighted observed counts are

$$\delta N_{\rm obs} = \delta\widetilde{N}_{\rm obs} \cdot \frac{N_{\rm obs}}{\widetilde{N}_{\rm obs}} = N_{\rm obs} \cdot \left(\widetilde{N}_{\rm obs}\right)^{-1/2} \tag{3.56}$$

and, consequently, the errors of the true weighted counts

$$\delta N_{\rm true} = \delta N_{\rm obs} \cdot F_{\rm corr} = N_{\rm true} \cdot \left(\widetilde{N}_{\rm obs}\right)^{-1/2}\,. \tag{3.57}$$

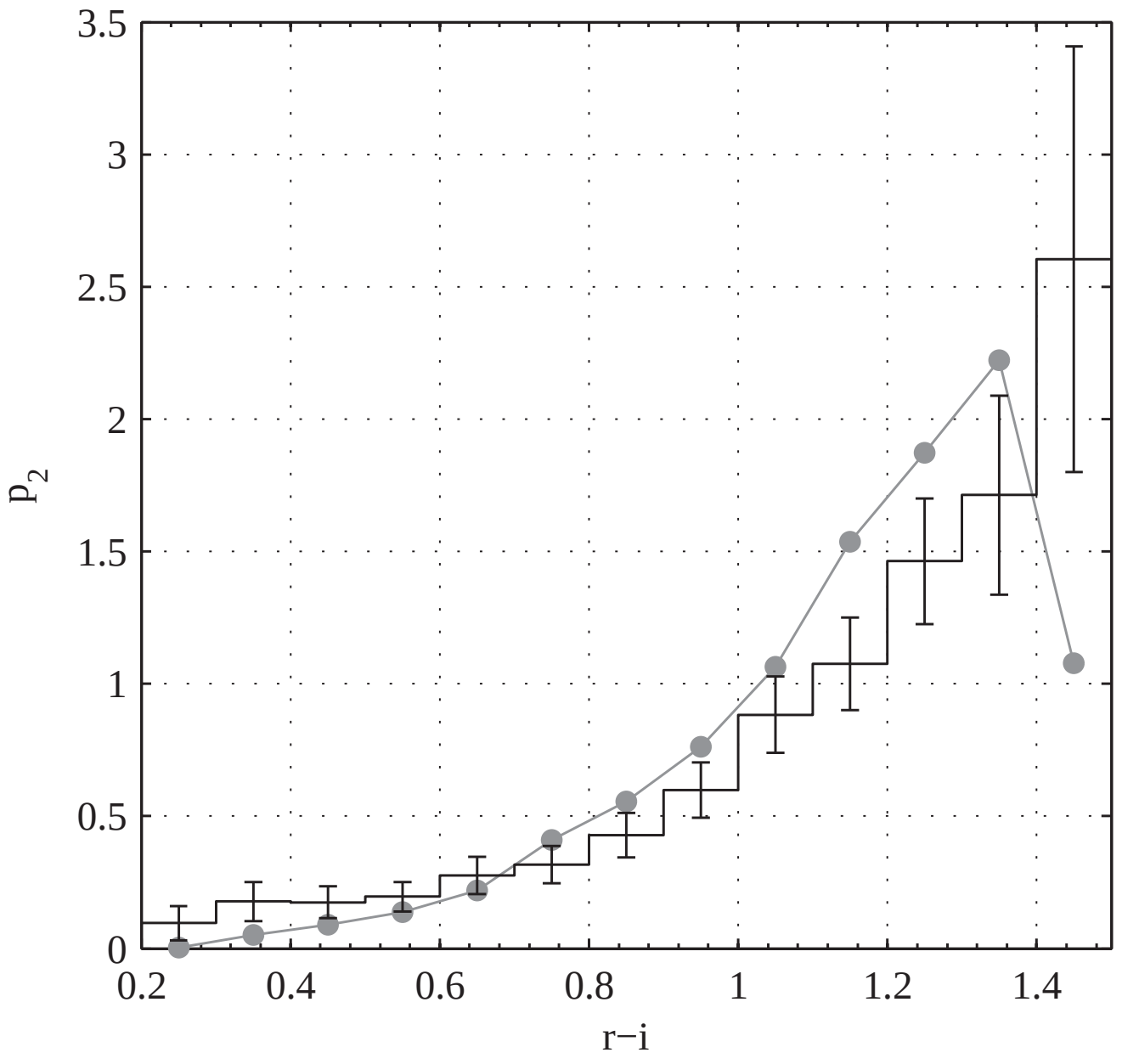

Figure 3.8: Corrected colour distribution (probability density) of *secondary* WB components (histogram). The error bars reflect statistical uncertainties. The (corrected) distribution expected from random pairing of field stars is shown as grey circles.

Finally, we estimate the relative statistical error of the joint probability density by

$$\frac{\delta p}{p} \approx \left(\widetilde{N}_{\rm obs}\right)^{-1/2} , \tag{3.58}$$

which we find to be of the order of 5%.

The most prominent feature in Fig. 3.6 is the maximum density around $ri_{1,2} \approx 1.5$ mag reflecting the peak of the LF at $M_r \approx 11$ mag. Another slight overdensity can be spotted at the blue end around $ri_{1,2} \approx 0.3$ mag. This overdensity can be attributed to the Wielen dip (Wielen et al. 1983) that depletes the counts in the surrounding redder bins.

From the joint probability density $p(ri_1, ri_2)$ we can infer the colour distributions of the WB members. Let $p_1(ri)$ be the probability density of finding a primary WB component in a $ri$-bin and $p_2(ri)$ analogously for a secondary component. Then

$$p_1(ri) = \Delta ri \sum_{ri_2} p(ri, ri_2) \tag{3.59}$$

and

$$p_2(ri) = \Delta ri \sum_{ri_1} p(ri_1, ri) \, . \tag{3.60}$$

The colour distributions $p_1$ and $p_2$ are shown in Figs. 3.7 and 3.8, respectively, as solid lines. While the colour distribution of secondaries rises monotonically towards redder colour (Fig. 3.8), that of the primaries is fairly constant for $ri \lesssim 1$ mag and drops only for redder colour indices (Fig. 3.7). The statistical uncertainties are calculated according to

$$\delta p_1(ri) = \sqrt{\sum_{ri_2} \delta p(ri, ri_2)^2} \tag{3.61}$$

and

$$\delta p_2(ri) = \sqrt{\sum_{ri_1} \delta p(ri_1, ri)^2}\,. \tag{3.62}$$

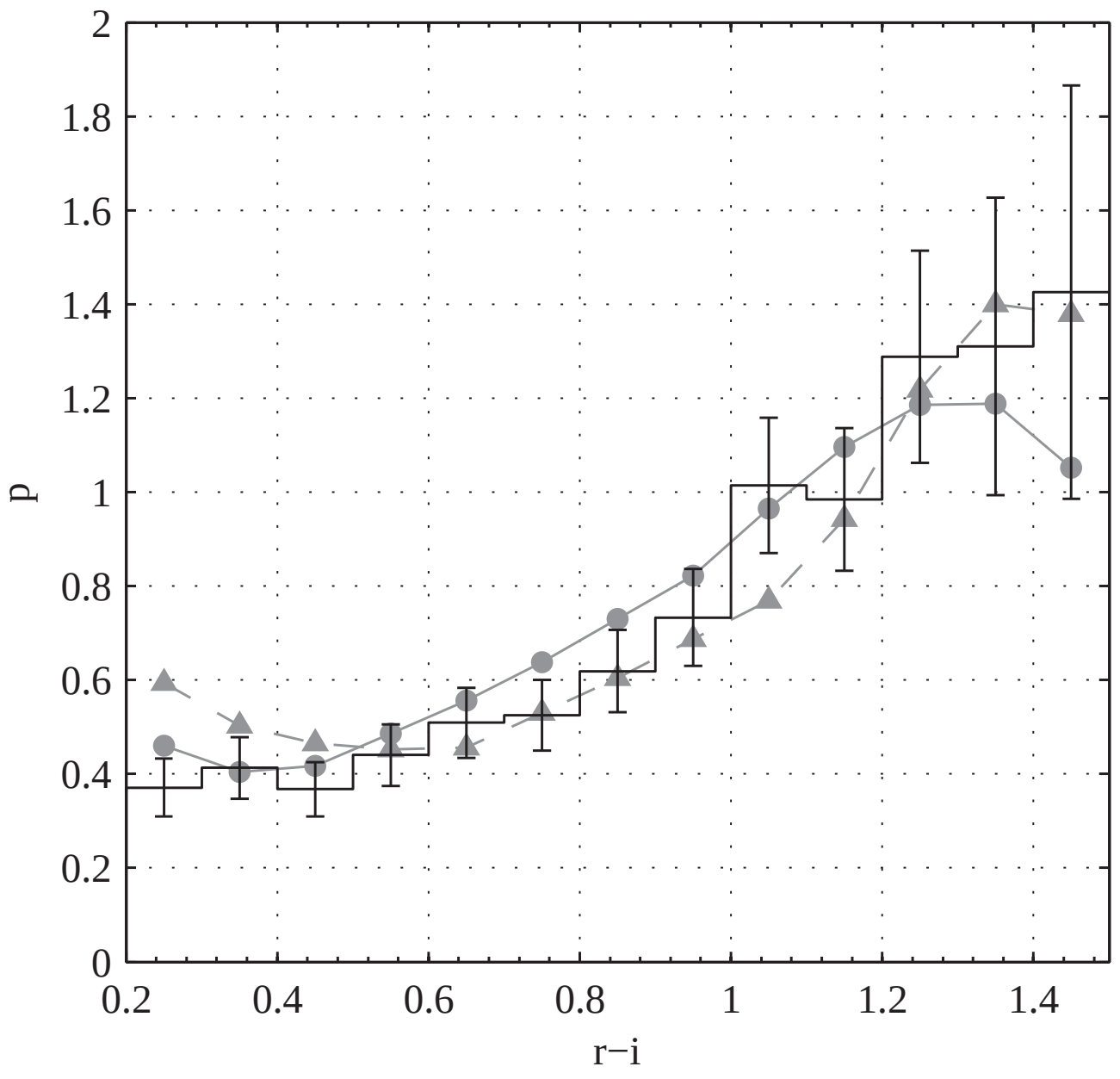

Figure 3.9: Corrected colour distribution (probability density) of *all* WB components (histogram). The error bars reflect statistical uncertainties. The corrected distribution $p_{\rm all}$ of the 456 089 stars in our restricted sample is shown as grey circles. The grey triangles show the Jahreiß and Wielen LF transformed into $r - i$.

In the same Figs. 3.7 and 3.8, we also show as grey dashed lines the colour distributions of primaries and secondaries that are expected from random pairings of the field stars. These expected distributions are generated by a simple Monte Carlo simulation, drawing pairs from the corrected colour distribution $p_{\rm all}$ of all the 456 089 stars in our restricted sample at random. (For consistency reasons we find it more appropriate to use $p_{\rm all}$ instead of the Jahreiß and Wielen LF.)

Looking at Fig. 3.7, we note a deficiency of blue ($ri \lesssim 1$ mag) primary components compared to the expectation from random pairing of field stars, compensated by a relative excess of red ($ri \gtrsim 1.2$ mag) primaries. The errors are, however, quite large, especially at the red end. The colour distribution of the secondaries (Fig. 3.8) agrees well with the prediction from random pairing of field stars. A slight excess at the bluest colours and a small deficiency at redder colours ($ri \gtrsim 1.1$ mag; except for the reddest bin) is, however, present, in contrast to the findings for the primaries.

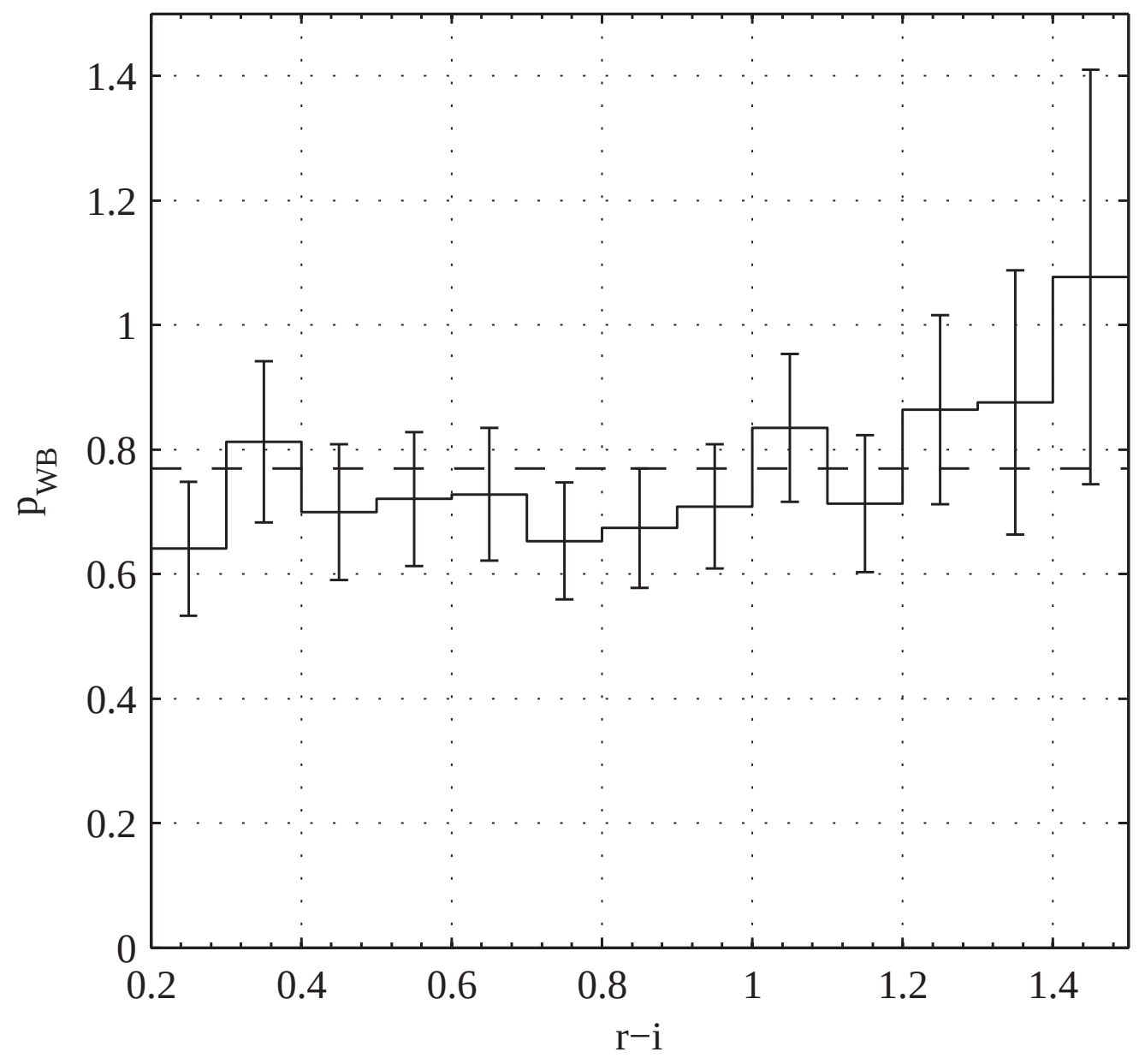

Figure 3.10: Normalized (excess) abundance of WBs relative to field stars, as a function of $r-i$. The error bars reflect statistical uncertainties. The horizontal dashed line indicates the equal probability density.

Next, we consider the probability density $p_{ri}(ri)$ that any WB component falls in a $ri$-bin

$$p_{ri}(ri) \propto p_1(ri) + p_2(ri) - p(ri, ri)\Delta ri\,. \tag{3.63}$$

Here, the last term assures that the probability that primary and secondary components fall in the same colour bin (the diagonal in Fig. 3.6) is not counted twice. The proportionality constant is fixed by the normalisation

$$\Delta ri \sum_{ri} p_{ri}(ri) = 1\,. \tag{3.64}$$

The uncertainty in $p_{ri}$ is

$$\delta p_{ri}(ri) \propto \sqrt{\delta p_1^2(ri) + \delta p_2^2(ri) - \delta p^2(ri, ri)}\,, \tag{3.65}$$

where the last term is subtracted because otherwise it would be included twice: once in $\delta p_1$ and once $\delta p_2$ (see Eqs. 3.61 and 3.62). The proportionality constant is, of course, the same as in Eq. 3.63.

We show $p_{ri}$ in Fig. 3.9 as histogram together with the Jahreiß and Wielen LF transformed into $ri$ as grey triangles and $p_{\rm all}$ as circles. If the components of the WB stars were drawn randomly from the LF, we would expect $p_{ri}$ to resemble the LF and, equivalently, $p_{\rm all}$. We see in Fig. 3.9 that all the three curves are quite similar and most deviations of $p_{ri}$ from the LF or $p_{\rm all}$ at intermediate colours are within the error bars. At the blue end $p_{ri}$ seems to slightly underestimate the LF while it is in line with $p_{\rm all}$, whereas at the red end $p_{ri}$ agrees well with the Jahreiß and Wielen LF but appears to overestimate $p_{\rm all}$. The underestimation of LF at the blue end might be due to the cut at $g - r = 0.5$ mag that excludes also some stars with colour indices up to $ri \approx 0.3$ mag because of the spread of the stellar locus (see Fig. 2 of Paper I).

The reddest stars with $ri \gtrsim 1.4$ mag appear to reside more likely in a WB than it would be expected by a random pairing. But given the large uncertainties at the red end (blown up by the correction function $F_{\rm corr}$), no firm conclusion can be drawn.

The ratio of $p_{ri}$ and $p_{\rm all}$ is proportional to the excess abundance of WBs relative to field stars as a function of colour, given as probability density $p_{\rm WB}(ri)$:

$$p_{\rm WB}(ri) \propto \frac{p_{ri}(ri)}{p_{\rm all}(ri)}\,. \tag{3.66}$$

The proportionality constant is again fixed by the normalisation of the probability density

$$\Delta ri \sum_{ri} p_{\rm WB}(ri) = 1\,. \tag{3.67}$$

Thus, we can infer $p_{\rm WB}$ from Fig. 3.9 simply by dividing $p_{ri}$ through $p_{\rm all}$ and renormalising the area under the resulting curve to unity, as it was already practiced by Sesar et al. (2008). We show $p_{\rm WB}$ in Fig. 3.10. The dashed horizontal line indicates the distribution we would have if $p_{ri} = p_{\rm all}$ for all $ri$. As a direct consequence of the good agreement between $p_{ri}$ and $p_{\rm all}$, the probability $p_{\rm WB}$ that a star chosen by chance is a member of a WB system appears to be independent of colour. This was also found by Sesar et al. (see their Fig. 17). This good agreement between $p_{ri}$ and $p_{\rm all}$ indicates again (as did Fig. 3.9) that the deviations of the primaries and secondaries alone, shown in Figs. 3.7 and 3.8, largely cancel each other. As mentioned, there is, however, a barely significant excess at the red end. Can this excess be attributed to certain combination of primary and secondary colour index? To answer this question we calculate the *conditional* probability density in the next Section.

### 3.4.3 Conditional colour distributions

The conditional probability density $p(ri_B|\,ri_A)$ that a WB component with colour $ri_A$ has a companion with colour $ri_B$, is defined by

$$p(ri_B|\,ri_A) \equiv \frac{p(ri_A \cap ri_B)}{p_{ri}(ri_A)}\,. \tag{3.68}$$

Here, $p(ri_A \cap ri_B)$ is the joint probability density that one component of a WB star (either primary or secondary) has colour index $ri_A$ and the other component has $ri_B$. It can be expressed as the sum of the probability that a WB primary component with colour $ri_A$ has a (secondary) companion with colour $ri_B$ and the probability that a WB secondary with a colour index $ri_A$ has a (primary) companion with colour $ri_B$

$$p(ri_A \cap ri_B) \propto p(ri_A, ri_B) + p(ri_B, ri_A) - p(ri_A, ri_A)\delta_{ri_A, ri_B}\,, \tag{3.69}$$

where the last term with the Kronecker delta assures that the joint probability densities are not counted twice when $ri_A = ri_B$. In terms of Fig. 3.6, the conditional probability density $p(ri_B|\, ri_A)$ can be inferred by adding the column $ri_1 = ri_A$ to the row $ri_2 = ri_A$, while paying attention that no cell is counted twice. Due to the normalisation of the conditional probability density

$$\Delta ri \sum_{ri_B} p(ri_B|\, ri_A) = 1 \tag{3.70}$$

we also have

$$\begin{aligned} p_{ri}(ri_A) &= \Delta ri \sum_{ri_B} p(ri_A \cap ri_B) \\ &\propto \Delta ri \sum_{ri_B} p(ri_A, ri_B) + \Delta ri \sum_{ri_B} p(ri_B, ri_A) \\ &\quad - p(ri_A, ri_A)\Delta ri\,, \end{aligned} \tag{3.71}$$

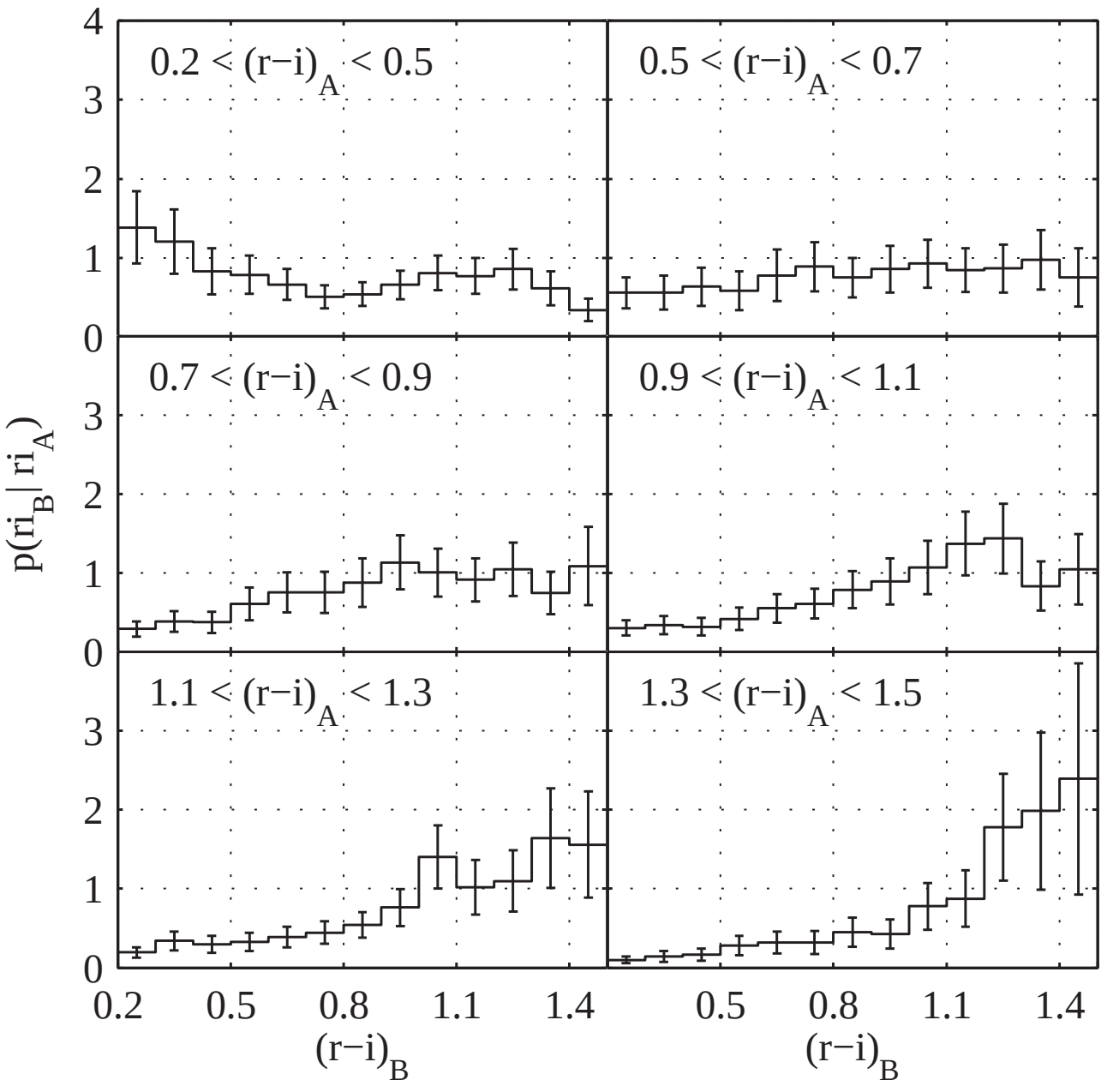

Figure 3.11: Conditional probability density that a WB component within the indicated $ri_A$-colour range has a companion with colour index $ri_B$. The error bars reflect statistical uncertainties.

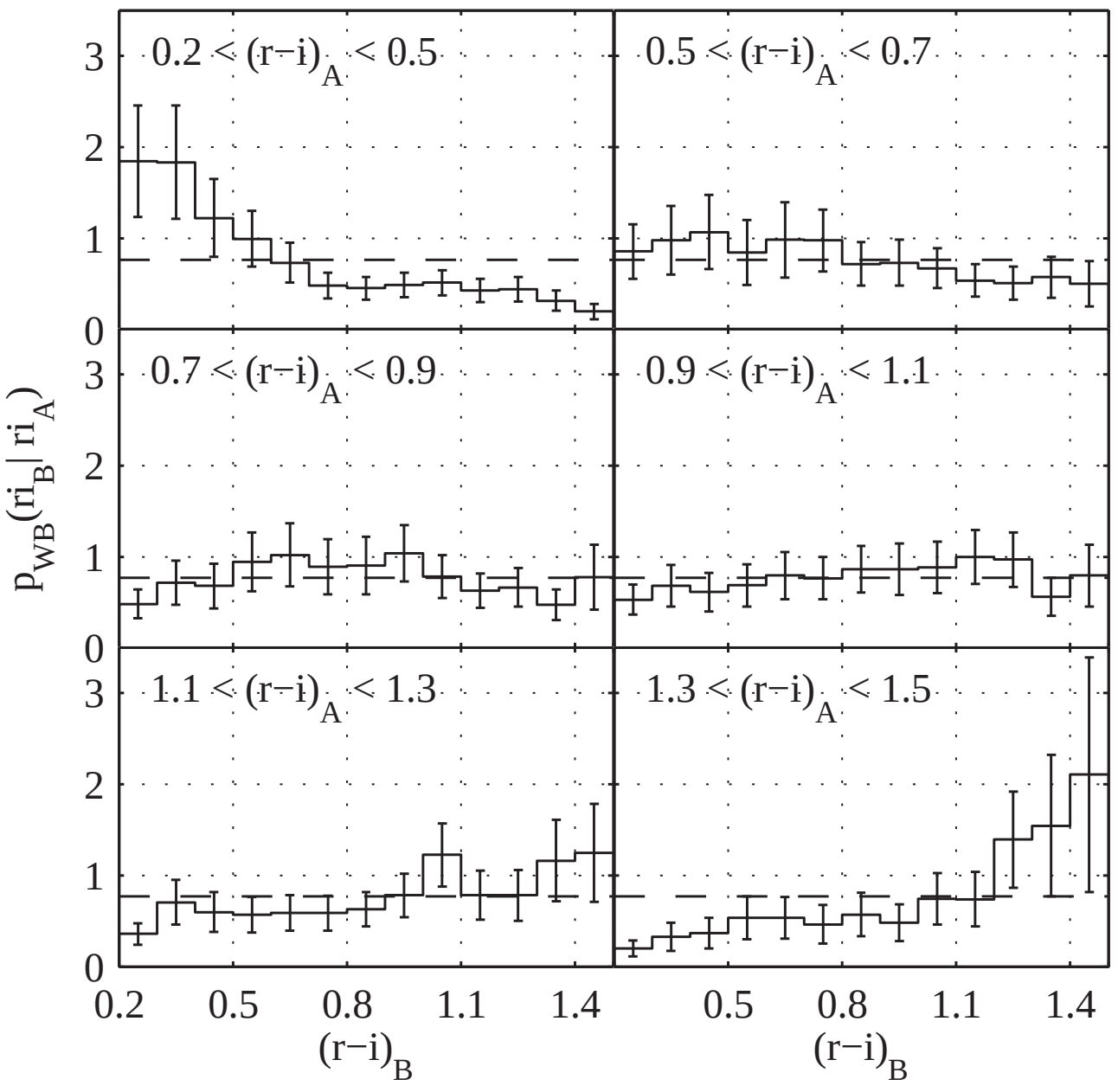

Figure 3.12: Normalised excess abundance (probability density) of WBs with one component in colour index bin $ri_A$ and the other in $ri_B$ relative to field stars in colour index bin $ri_B$. The error bars reflect statistical uncertainties.

which is consistent with Eq. 3.63.

The conditional probability densities $p(ri_B|\, ri_A)$ for various ranges in $ri_A$ are shown in Fig. 3.11. If each component of a WB is chosen at random from the LF, i.e. from $p_{\rm all}$, every curve in Fig. 3.11 should resemble $p_{\rm all}$. At first glance it becomes, however, evident that the conditional colour distribution is not the same for every range in $ri_A$. While the conditional probability density rises towards redder colours like the LF for $ri_A \gtrsim 0.7$ mag, it is almost flat in the range $0.5 < ri_A < 0.7$ mag and shows even a falling tendency towards red colours for the bluest range in $ri_A$ examined ($0.2 < ri_A < 0.5$ mag).

To illustrate the deviation from the $p_{\rm all}$ more clearly, we divide the conditional probability densities through $p_{\rm all}$ and define in analogy to Eq. 3.66

$$p_{\rm WB}(ri_B|\, ri_A) \propto \frac{p(ri_B|\, ri_A)}{p_{\rm all}(ri_B)}\,. \qquad (3.72)$$

This gives the excess abundance (probability density) of WBs with one component in colour index bin $ri_A$ and the other in $ri_B$ relative to field stars in colour index bin $ri_B$.

The resulting curves are shown in Fig. 3.12. We see that at intermediate colour indices ($0.5 < ri_A < 1.3$ mag), $p_{\rm WB}(ri_B|\, ri_A)$ agrees well with $p_{\rm all}$ within the error bars. The bluest ($0.2 < ri_A < 0.5$ mag) and the reddest ($1.3 < ri_A < 1.5$ mag) colours, however, show a (barely) significant deviation. While the bluest WB components seem to have an excess in blue companions relative to $p_{\rm all}$ at the expense of red companions, the opposite is true for the reddest WB components. So, it seems that at the extremes of the colour range studied, companions with similar colour are preferred and that the (barely significant) excess we noted in Fig. 3.10 is largely due to pairs where both components have $ri \gtrsim 1.2$ mag (see, however the discussion in §3.5).

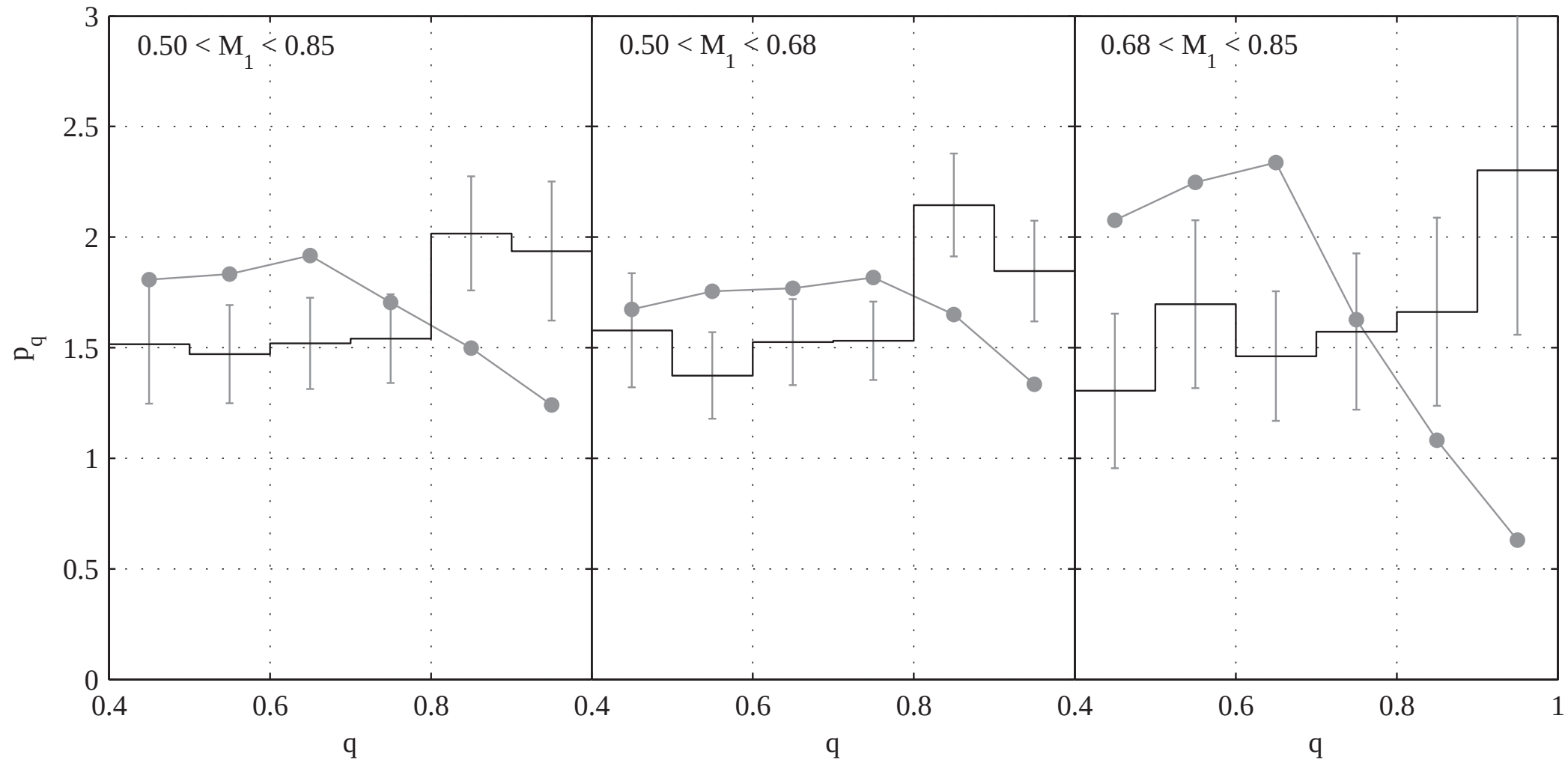

Figure 3.13: Probability densities of WB mass ratios (histogram). The indicated range of the primary mass is in solar masses $\mathcal{M}_\odot$. The error bars reflect statistical uncertainties. The corresponding distributions expected from random pairing of field stars are shown as (grey) circles.

### 3.4.4 Distribution of mass ratios

Using the KTG93 MLR, we assign a mass to each star in the restricted total sample a mass. In analogy to Eq. 3.54, we then infer the (corrected) probability density $p(\mathcal{M}_1, \mathcal{M}_2)$ that a WB system has a primary component with mass $\mathcal{M}_1$ and a secondary with mass $\mathcal{M}_2$. To express the correction function $F_{\rm corr}$, (Eq. 3.30), as a function of mass $\mathcal{M}$, we construct a 'colour-mass relation', $ri(\mathcal{M})$, by combining the PPR (Eq. 3.1) and the KTG93 MLR. We express $ri(\mathcal{M})$ as a sixth order polynomial in $\mathcal{M}$ measured in solar masses $\mathcal{M}_\odot$:

$$\begin{aligned} ri(\mathcal{M}) &\approx 92.2\mathcal{M}^6 - 370.5\mathcal{M}^5 + 586.7\mathcal{M}^4 - 461.8\mathcal{M}^3 \\ &\quad +188.2\mathcal{M}^2 - 39.6\mathcal{M} + 4.8\,, \end{aligned} \tag{3.73}$$

which is valid in the range $0.2 \lesssim \mathcal{M} \lesssim 0.85\mathcal{M}_\odot$, corresponding to $0.2 \leq ri \leq 1.5$ mag.

The probability density $p(\mathcal{M}_1, \mathcal{M}_2)$ can be used to infer the distribution of the mass ratio $q \equiv \mathcal{M}_2/\mathcal{M}_1 \leq 1$ of WB stars via (e.g. Warner 1961; Tout 1991)

$$p_q(q) = \int\limits_{\mathcal{M}_{\min}}^{\mathcal{M}_{\max}} \mathcal{M}_1\, p(\mathcal{M}_1, q\mathcal{M}_1)\, \mathrm{d}\mathcal{M}_1\, . \tag{3.74}$$

The integration limits, i.e. the range of the primary mass $\mathcal{M}_1$, and the range of mass ratios $q$ studied, must be chosen with some caution: the red end of the colour range under consideration here corresponds to a minimum mass of about $0.2\mathcal{M}_\odot$ and the mass of any WB component can not be below that value. Companions to low-mass primaries have therefore a restricted range of possible masses causing the mass-ratio distribution $p_q$ to be biased towards high values (see, e.g., §8.4 in Fischer and Marcy 1992). To overcome this problem we restrict both the range of $\mathcal{M}_1$ and the studied range of $q$. We set somewhat arbitrarily $0.5 < \mathcal{M}_1 < 0.85\mathcal{M}_\odot$, defining the integration limits in Eq. 3.74, and $0.4 < q < 1$. These restrictions make sure that $q\mathcal{M}_1 \geq 0.2\mathcal{M}_\odot$ and avoid a possible bias of the mass-ratio distribution towards unity.

The resulting mass-ratio distribution is shown in Fig. 3.13 (left) as histogram. The distribution is quite flat for $q < 0.8$ and shows a slight enhancement for larger $q$. This is quite different from the expectation from random pairing (plotted as circles) showing a decreasing distribution for $q \gtrsim 0.7$ with increasing $q$, but is in line with the previous Section where we found a deficiency of pairs whose components have a different colour ($\Delta ri \gtrsim 1$ mag, corresponding to $\Delta\mathcal{M} \gtrsim 0.5\mathcal{M}_\odot$), compensated by an excess of pairs with similar colour (i.e. similar mass).

To further explore the mass-ratio distribution and to get an insight into the pairing mechanism, we show its variation with *primary* mass in Fig. 3.13 (middle and right), as suggested by Kouwenhoven et al. (2009). The mass-ratio distribution of the less massive half of primary components ($0.5 < \mathcal{M}_1 < 0.68\mathcal{M}_\odot$) is pretty similar to that shown in Fig. 3.13 (left). The difference to the random distribution is, however, somewhat less pronounced. The more massive half, on the other hand, shows a slightly increasing trend over the hole $q$-range, whereas a strong discrepancy with the distribution expected from random pairing is evident.

### 3.4.5 Distribution of secondary masses

In Fig. 3.14 we present the distribution of secondary masses for pairs with a primary mass larger than $0.5\mathcal{M}_\odot$ as a histogram. It has a broad maximum around $0.45\mathcal{M}_\odot$ and declines rapidly, approximately as $\mathcal{M}^{-4}$, towards larger masses. This decline is considerably steeper than Salpeter's (1955) initial mass function (IMF) with $\mathcal{M}^{-2.35}$, which is found to give a good description for $\mathcal{M} > 0.5\mathcal{M}_\odot$ (e.g. Kroupa 2008). Note that below $\mathcal{M} \approx 0.8\mathcal{M}_\odot$, stars are still on the (initial) main sequence, anyway. The decrease towards low masses, on the other hand, is much shallower.

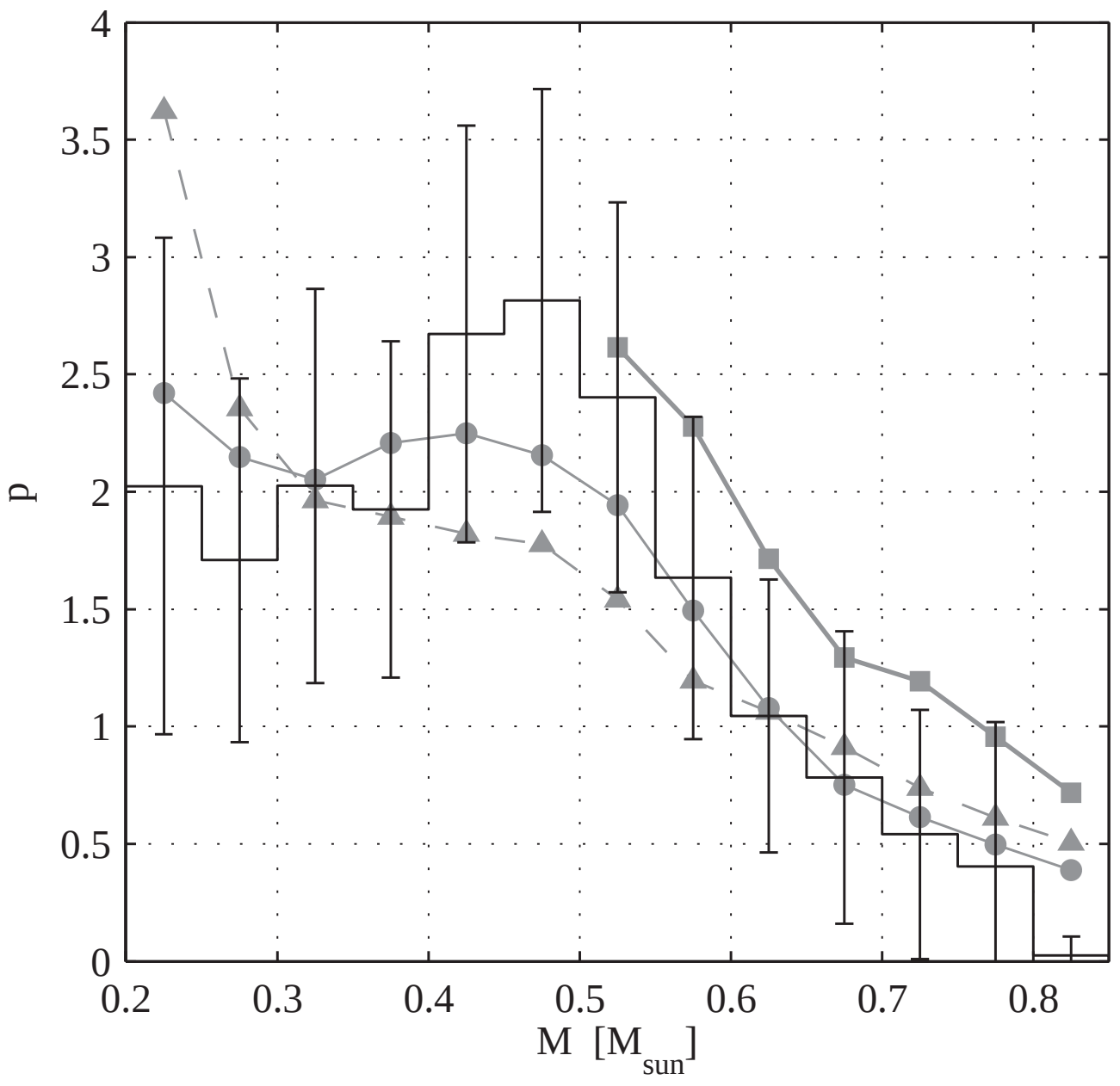

Figure 3.14: Mass distribution of secondaries with a primary component heavier than $0.5\mathcal{M}_\odot$ (histogram). The error bars reflect statistical uncertainties. The (arbitrarily normalised) distribution of primary masses are shown as grey sqares. For comparison are plotted also the mass distribution of field stars (circles) as well as the Jahreiß and Wielen LF transformed into a mass function using the KTG93 MLR (triangles).

For comparison, we also show the (arbitrarily normalised) distribution of primary masses (squares), the mass distribution of the field stars (circles) and the Jahreiß and Wielen LF (triangles) transformed into a mass distribution using the KTG93 MLR. Firstly, we note that the distribution of secondary masses agrees with that of the field stars, suggesting that the secondary masses are drawn from the field population. Secondly, the decline of the distribution of primary masses shows a decline towards high masses of similar steepness as the distribution of secondary masses and that of the field stars, suggesting that primary masses larger than $0.5\mathcal{M}_\odot$ are drawn from the field population as well. Thirdly, unlike the distribution of secondary masses, the mass distribution derived from the Jahreiß and Wielen LF (triangles) has no maximum around $0.45\mathcal{M}_\odot$ but shows an overall decreasing trend towards high masses, where the decline for $\mathcal{M} > 0.5\mathcal{M}_\odot$ is in agreement with Salpeter's IMF. Given the large uncertainties of the secondary-mass distribution and the relatively narrow mass range under consideration ($0.2 < \mathcal{M} < 0.85\mathcal{M}_\odot$), any apparent inconsistency with the canonical stellar IMF (e.g. Kroupa 2008) should probably not be overemphasised.

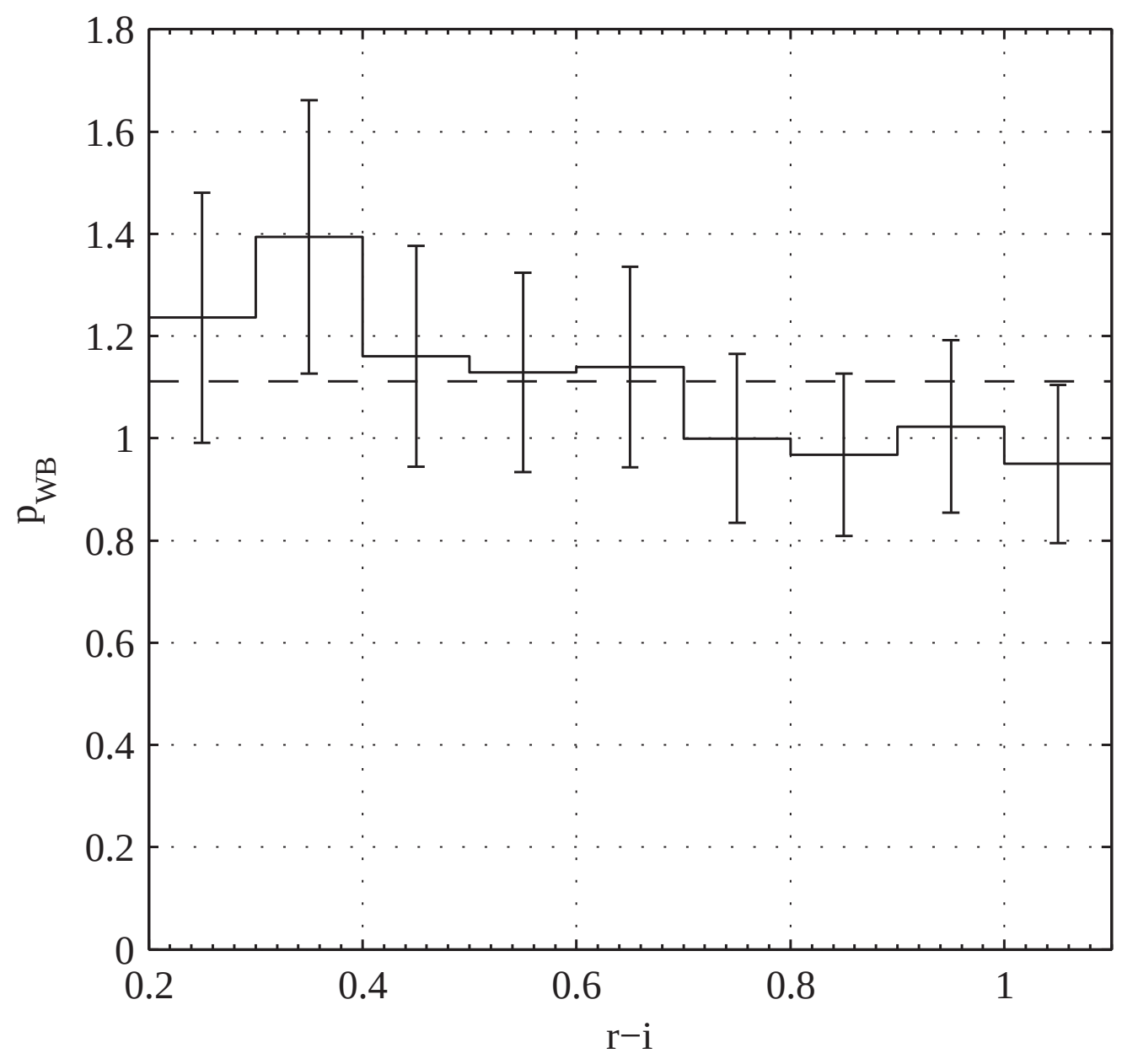

Figure 3.15: Probability density of WBs as a function of $r - i$ in the range $0.2 < r - i < 1.1$ mag relative to the results of Gould et al. (1995). The error bars reflect statistical uncertainties. The horizontal dashed line indicates the equal probability density.

## 3.5 Discussion

### 3.5.1 Colour distributions

The distribution of colours of WB candidates in a volume-complete sample was studied in detail by Sesar et al. (2008) (see their §4.2). As we take quite a different approach to construct the colour distributions, while using similar data from the SDSS and the same PPR, a comparison with Sesar et al. should be a valuable consistency test. The joint colour distribution in Fig. 3.6 can be directly compared to Fig. 15 (bottom) of Sesar et al. We note a good overall agreement. A closer look reveals, however, a difference: Sesar et al. find a *local* maximum around $ri_{1,2} \approx 1.0$ mag, beyond which the probability drops towards redder bins. According to Sesar et al., their sample is complete up to $g - i = 2.8$ mag, which corresponds to $ri \approx 1.4$ mag (using Eq. 10 from Sesar et al.). Thus, the drop beyond the maximum in Sesar et al. should be real. We, on the other hand, find that the probability density continues to rise up to the red end at $ri \approx 1.5$ mag. The LF does not seem to have any significant drop in the range corresponding to $1.0 \lesssim ri \lesssim 1.5$ mag either (see the grey dashed line in Fig. 3.9). However, given the uncertainties of the Galactic model and the PPR as well as the heuristic method we use to transform the LF from $M_V$ into $M_r$ (see §4.2.2 of Paper I), this difference is probably not significant.

Performing a "search for faint, common proper motion companions of *Hipparcos* stars", Lépine and Bongiorno (2007) found a clear deficiency of secondaries relative to the single star field population in the low-luminostiy range $8 < M_V < 14$ mag, which "rules out the idea that the luminosity function of the secondaries is comparable to the field luminosity function". This low-luminosity range corresponds to $0.4 \lesssim ri \lesssim 1.6$ mag. The deficiency of secondaries found in the present study (evident from Fig. 3.8), on the other hand, occurs in a narrower colour range and is much less significant. We therefore conclude that we can not fully reproduce this particular result from Lépine and Bongiorno.

At first sight, the (barely significant) excess of red pairs at $ri \approx 1.5$ mag evident from Fig. 3.10 appears to be contradictory to the findings of Gould et al. (1995). Based on a sample of 13 WB stars discovered in the HST Snapshot Survey, they find that these binaries "have bluer colors than would be expected for random pairs of field stars". The stars observed in the Snapshot Survey typically have absolute magnitudes between 6 and 11 mag, corresponding to $0.2 \lesssim ri \lesssim 1.1$ mag. Hence, the Snapshot sample is not sensitive to the colour range of our reddest bins. We recalculated $p_{\rm WB}$ after having excluded the bins with $ri > 1.1$ mag, while keeping the same weights $w_{ij}$ as before. The resulting probability density is shown here in Fig. 3.15. Again the distribution is consistent, within the uncertainties, with the equal probability density (the dashed horizontal line), but we do note a declining trend towards redder colours. So, in the colour range $0.2 \lesssim ri \lesssim 1.1$ mag blue stars have a slightly larger probability to be a member of a WB, in line with the findings of Gould et al. This also shows that the probability for an arbitrary star to reside in a WB depends somewhat on the colour range under consideration.

The conclusion we draw from Fig. 3.12 that WB components at the blue and the red end have a preference for companions of similar colour stands, however, not on a very firm basis, given the quite large uncertainties, especially for the red pairs. More significant than the excesses of red and blue pairs, respectively, is the apparent absence of pairs with very different colours: at the bluest $ri_A$-range we note a lack of red companions relative to the LF, i.e. to $p_{\rm all}$. Consequently, blue companions are missing in the reddest $ri_A$-range. This suggest that pairs with a large colour difference, $|ri_A - ri_B| \gtrsim 1$ mag, are underrepresented compared to the expectation from random pairing of field stars. This "apparent preference for similar brightness" was already noted by Gould et al. (1995). Their sample was, however, too small to make a secure statement.

### 3.5.2 Mass ratios and secondary masses

Duquennoy and Mayor (1991) found a mass-ratio distribution with a declining trend towards equal mass binaries for every period range examined, which is not necessarily inconsistent with a flat distribution, though. Their study is, however, restricted to solar-type stars (spectral range F7 to G9), whereas our sample includes stars of spectral type between G5 and M5 ($0.2 < \mathcal{M} < 0.85\mathcal{M}_\odot$). As late type stars outnumber those of earlier type, we expect that the mass-ratio distribution we have inferred is dominated by later types. Indeed, Fischer and Marcy (1992) found a flat mass-ratio distribution in the region $q > 0.4$ for M dwarf binaries with separations smaller than $10^4$ AU, more in line with our results. Most recently, and in accord with our study, Dhital et al. (2010) found a mass-ratio distribution rising towards equal masses for binaries with a separation between $10^3$ and $10^5$ AU and a spectral type between K5 and M5. A similar mass-ratio distribution with a peak at $q \geq 0.8$ was found much earlier by Reid and Gizis (1997). However, most of the binaries in their sample had separations smaller than 200 AU, making a direct comparison difficult.

Both Duquennoy and Mayor and Fischer and Marcy found that the distribution of secondary masses is similar to the field mass function, in general agreement with our results. The former authors found a secondary-mass distribution continuously increasing towards small secondary masses, whereas the companion mass distribution for M dwarfs derived by Fischer and Marcy extending from 0.075 to 0.375 $\mathcal{M}_\odot$ is roughly flat. This is consistent with our secondary-mass distribution in Fig. 3.14 which is also roughly flat in the range $\mathcal{M}_2 < 0.4\mathcal{M}_\odot$. But since our distribution is for pairs with primaries of higher mass ($\mathcal{M}_1 > 0.5\mathcal{M}_\odot$), whereas Fischer and Marcy included only primaries with masses between 0.3 and 0.55 $\mathcal{M}_\odot$, a direct comparison is not straightforward. The decline of the secondary-mass distribution towards large masses is in line with the results of Duquennoy and Mayor, but again, because of the different range in primary mass and separation probed, a comparison must be done with care.

### 3.5.3 Comparison with theoretical predictions

Using the largest hydrodynamical simulation to date, Moeckel and Bate (2010) recently studied the formation of single and multiple stars, including WBs, in the expanding halo of a star cluster after gas dispersal. They find a relatively flat mass-ratio distribution for binaries with a primary mass $\mathcal{M}_1 > 0.5\mathcal{M}_\odot$, especially for $q > 0.4$, whereas binaries with $0.1 < \mathcal{M}_1 < 0.5\mathcal{M}_\odot$ show a increasing trend towards high mass ratios over the whole range in $q$. More importantly, the outcome of their simulation suggests that binaries with small separations tend to have large mass ratios, whereas WB stars preferably consist of unequal mass components. Taken at face value, this would clearly contradict our findings. A direct comparison with the simulation by Moeckel and Bate is, however, difficult for the following two reasons:

*i)* their simulation has been stopped after 10 Myr, while the field WB population we study is, on average, much older – of the order of Gyr. So, there is an (at present unavoidable) temporal gap between the simulation and the observations, during which the properties of the field binaries evolve under the influence of dynamical encounters with other stars, giant molecular clouds and (possibly) MACHOs. Regarding the mass-ratio distribution, low-$q$ binaries are indeed more easily disrupted by gravitational encounters than are equal-mass binaries (Kroupa 1995).

*ii)* the results of the simulation are inferred using *one* definite set of initial conditions for *one* collapsing molecular cloud (see Bate 2009, §2.3), whereas the field population must be regarded as the result of many collapsed molecular clouds with a whole range of different properties.

The dependence of wide ($a > 10^3$ AU) binary properties on the initial conditions of a dissolving star cluster was studied by Kouwenhoven et al. (2010). While they find that the "WB fraction decreases strongly with increasing cluster mass", the mass-ratio distribution does not seem to vary much for different initial conditions. Kouwenhoven et al. (2010) find a mass-ratio distribution decreasing towards high mass ratio: high-mass primaries ($\mathcal{M}_1 > 1.5\mathcal{M}_\odot$) peak at $q \lesssim 0.2$; low-mass primaries ($\mathcal{M}_1 < 1.5\mathcal{M}_\odot$) show a more gently decreasing mass-ratio distributions. Their results from the Plummer model in virial equilibrium seem to be also consistent with a flat mass-ratio distribution in the range $q > 0.4$. As mentioned, previous work by (Kroupa 1995) predicted an enhancement of the mass-ratio distribution at large $q$-values, because low-$q$ tend to be disrupted. As we are unable to decide whether the mass-ratio distribution is rising for $q < 0.4$ or not (we can only give lower limits in that range), a clear-cut conclusion whether or not the mass-ratio distribution we have inferred agrees with the most recent simulations by Moeckel and Bate and Kouwenhoven et al. (2010) is impossible. Overall, our results are certainly not inconsistent with these studies.

## 3.6 Summary and conclusions

We have combined our angular correlation analysis of WBs in Paper I with distance information to statistically filter out unwanted optical pairs. To this end we have applied a new 'weighting procedure' based on the binding probability of a pair of stars, as drawn from their angular separation ($2'' \leq \theta \leq 30''$) and distance estimates given by the colour-magnitude relation of J08. Out of a sample of about 450 000 SDSS stars in a NGP-centered field of 675 square degrees, 37 610 stellar pairs in the angular separation range mentioned are assigned a statistical weight. Our sample, and thus the whole study, is restricted to a colour range of $0.2 < r-i < 1.5$ mag, translating into a mass range of $0.2 \lesssim \mathcal{M} \lesssim 0.85\mathcal{M}_\odot$, and to a maximum distance of 2 000 pc.

For a principal lack of WB orbital data, our weighting procedure inherently requires the choice of an average relative velocity assigned to each pair in our sample. Its value is calibrated in such a way that the weighted number of pairs in each angular separation bin reproduces the clustering signal as inferred by the angular 2PCF, and it turns out to be 370 m/s. Fixing the relative velocity, however, limits our study to pairs with projected separations smaller than a maximum projected separation, whose value depends on the total mass of the pair and is found to be between 0.008 and 0.05 pc (1 600 and 10 000 AU, respectively).

The observed colour distributions are corrected for selection effects (Malmquist bias) using the WW technique. The bias-corrected colour distribution of WB stars derived by our weighting procedure is in good agreement with the (corrected) colour distribution of the field stars, in line with the findings of Sesar et al. (2008).

There is a general lack of pairs with very different colours: pairs with a colour difference $\Delta ri \gtrsim 1$ mag, corresponding to a mass difference $\Delta\mathcal{M} \gtrsim 0.5\mathcal{M}_\odot$, seem to be systematically underrepresented compared to the expectation that the components of a WB system are drawn randomly from the field mass function. This preference for pairs of similar colour or mass is also reflected in the distribution of mass ratios $q$, which is inferred from the colour distributions of the WB components using a standard mass-luminosity relation Kroupa et al. (1993). Our mass-ratio distribution for primary masses between 0.5 and 0.85 $\mathcal{M}_\odot$ shows a slight enhancement at $q > 0.8$, while it is nearly uniform in the range $0.4 < q < 0.8$, i.e. there is an overabundance of equal-mass binaries.

Previous observations were broadly consistent with a more uniform $q$ and random pairing, notably the classical studies by Duquennoy and Mayor (1991) and Fischer and Marcy (1992), while the most recent study by Dhital et al. (2010) supports our finding. A comparison between different observational studies is generally difficult because different ranges of separation and mass are studied. In particular, as we have restricted our $q$ range to $q > 0.4$, to avoid bias against low mass ratios, we cannot rule out a rising of the mass-ratio distribution for $q < 0.4$. Such a (inverse) trend is in fact predicted by the most recent star-formation simulations (Moeckel and Bate 2010). However, it is probably not yet appropriate to directly compare the outcome of any star-formation simulation to the observed properties of the field binary population, as stressed by Goodwin (2010).

Using only positional correlations and distance information from photometric parallaxes, our weighting procedure, based on the binding probabilities, produces reliable colour distributions without identifying individual WBs. Circumventing the limitations of proper motion surveys in this way, it is possible to take into account information from a much larger number of WBs and probe their population to greater distances. Furthermore, the procedure allowed us to create a 'ranking list' of WB candidates (sorted by their weights), which can be used for follow-up studies and cross identification (Table 3.1; the full table is available at CDS).

The drawback of the method is clearly the need for a sophisticated modeling to allow for selection effects and the requirement to introduce an average relative velocity in the absence of kinematic information of the pairs. It is not foreseeable that the orbital parameters of WBs can be measured directly in the near future. We therefore believe that the novel procedure presented in this paper is a viable method to infer the statistical properties of WB stars and constitutes an approach complementary to common proper motion studies.

Table 3.1: Top WB candidates[a]

| primary | secondary | $r_1$ [mag] | $r_2$ [mag] | $M_{r_1}$ [mag] | $M_{r_2}$ [mag] | $D_1 \pm \sigma_{D_1}$ [pc] | $D_2 \pm \sigma_{D_2}$ [pc] | $\mathcal{M}_1$ [$\mathcal{M}_\odot$] | $\mathcal{M}_2$ [$\mathcal{M}_\odot$] | $\theta_{ij}$ [arcsec] | $s_{ij}$ [pc] | $w_{ij}$ |
|---|---|---|---|---|---|---|---|---|---|---|---|---|
| J114621.72+305741.0 | J114621.78+305802.7 | 15.1 | 15.4 | 10.0 | 10.2 | 104±16 | 106±16 | 0.41 | 0.39 | 21.7 | 0.011 | 0.00219 |
| J113527.95+363158.1 | J113527.31+363150.5 | 15.2 | 15.3 | 9.4 | 9.5 | 147±21 | 147±21 | 0.48 | 0.47 | 10.8 | 0.008 | 0.00217 |
| J124247.66+313322.0 | J124247.69+313328.1 | 15.4 | 15.5 | 10.0 | 10.2 | 120±18 | 119±19 | 0.41 | 0.40 | 6.1 | 0.004 | 0.00216 |
| J113149.37+351306.7 | J113149.72+351300.7 | 15.8 | 15.8 | 10.9 | 11.0 | 93±14 | 92±13 | 0.30 | 0.28 | 7.4 | 0.003 | 0.00209 |
| J132828.90+292351.9 | J132828.97+292349.9 | 15.0 | 15.2 | 9.1 | 9.2 | 149±22 | 158±26 | 0.50 | 0.50 | 2.2 | 0.002 | 0.00204 |
| J124813.96+302219.9 | J124814.35+302226.5 | 15.5 | 16.3 | 10.6 | 11.3 | 99±14 | 102±15 | 0.34 | 0.25 | 8.3 | 0.004 | 0.00195 |
| J123307.54+361422.2 | J123306.49+361417.0 | 15.1 | 15.5 | 9.9 | 10.6 | 109±15 | 94±13 | 0.43 | 0.33 | 13.7 | 0.006 | 0.00195 |
| J115006.76+272253.7 | J115006.60+272253.4 | 15.8 | 16.0 | 11.2 | 11.2 | 84±13 | 90±16 | 0.27 | 0.26 | 2.1 | 0.001 | 0.00173 |
| J111259.12+335255.4 | J111259.56+335257.5 | 16.1 | 16.5 | 10.7 | 11.1 | 120±18 | 119±17 | 0.33 | 0.27 | 5.9 | 0.003 | 0.00169 |
| J121443.58+320809.6 | J121443.49+320811.8 | 15.9 | 16.0 | 10.0 | 10.1 | 152±25 | 153±23 | 0.41 | 0.40 | 2.5 | 0.002 | 0.00169 |
| J120652.71+392057.5 | J120652.96+392105.3 | 15.2 | 15.3 | 8.4 | 8.5 | 225±33 | 226±34 | 0.57 | 0.56 | 8.3 | 0.009 | 0.00162 |
| J130350.00+321000.7 | J130349.38+320954.8 | 15.2 | 15.5 | 8.6 | 8.9 | 207±30 | 207±33 | 0.55 | 0.52 | 9.8 | 0.010 | 0.00161 |
| J124130.26+411854.7 | J124130.32+411851.8 | 15.7 | 15.9 | 9.6 | 9.7 | 166±26 | 173±30 | 0.46 | 0.45 | 2.9 | 0.002 | 0.00158 |
| J112012.18+252821.5 | J112013.60+252811.9 | 15.3 | 15.9 | 10.5 | 10.8 | 89±13 | 104±15 | 0.35 | 0.31 | 21.6 | 0.009 | 0.00157 |
| J132648.33+343215.0 | J132648.13+343216.1 | 15.2 | 15.8 | 8.8 | 9.2 | 190±27 | 207±35 | 0.53 | 0.50 | 2.8 | 0.003 | 0.00155 |
| J123453.60+344704.3 | J123453.95+344703.8 | 15.6 | 16.2 | 9.8 | 10.2 | 150±22 | 156±32 | 0.44 | 0.39 | 4.4 | 0.003 | 0.00150 |
| J112944.55+355324.1 | J112944.79+355326.6 | 15.8 | 16.2 | 11.2 | 11.5 | 84±13 | 88±20 | 0.26 | 0.24 | 3.8 | 0.002 | 0.00147 |
| J124632.06+364902.4 | J124632.70+364901.5 | 15.3 | 16.8 | 9.1 | 10.8 | 173±25 | 158±24 | 0.50 | 0.31 | 7.8 | 0.006 | 0.00145 |
| J122008.67+280549.0 | J122008.67+280551.2 | 15.1 | 16.0 | 9.9 | 10.9 | 106±16 | 108±32 | 0.42 | 0.30 | 2.2 | 0.001 | 0.00144 |
| J130327.14+363858.6 | J130326.71+363900.9 | 15.9 | 16.5 | 9.5 | 10.3 | 190±28 | 177±27 | 0.47 | 0.38 | 5.7 | 0.005 | 0.00144 |
| J114408.78+370315.4 | J114409.48+370317.6 | 15.4 | 16.3 | 8.6 | 9.5 | 229±34 | 229±34 | 0.55 | 0.47 | 8.6 | 0.010 | 0.00142 |
| J115425.04+312441.6 | J115424.38+312438.8 | 15.2 | 17.0 | 8.9 | 10.7 | 181±28 | 188±29 | 0.53 | 0.33 | 8.8 | 0.008 | 0.00142 |
| J110102.50+241840.4 | J110102.60+241838.4 | 15.7 | 15.7 | 10.1 | 10.3 | 134±22 | 120±28 | 0.40 | 0.38 | 2.4 | 0.001 | 0.00141 |
| J115317.01+412027.6 | J115317.92+412028.8 | 15.9 | 16.4 | 9.2 | 9.7 | 212±30 | 224±32 | 0.49 | 0.45 | 10.3 | 0.011 | 0.00136 |
| J133252.61+301014.3 | J133251.89+301024.0 | 15.6 | 16.9 | 9.2 | 10.4 | 191±27 | 200±29 | 0.50 | 0.36 | 13.5 | 0.012 | 0.00131 |
| J115229.70+265459.1 | J115230.09+265503.9 | 15.4 | 15.7 | 8.2 | 8.6 | 272±42 | 263±40 | 0.58 | 0.55 | 7.1 | 0.009 | 0.00131 |
| J121638.83+345117.2 | J121639.25+345119.8 | 16.3 | 16.8 | 11.0 | 11.4 | 115±18 | 125±20 | 0.29 | 0.25 | 5.8 | 0.003 | 0.00130 |
| J133045.63+245417.9 | J133046.12+245419.1 | 16.1 | 16.9 | 9.8 | 10.4 | 187±27 | 201±29 | 0.44 | 0.36 | 6.8 | 0.006 | 0.00130 |
| J133625.57+375056.9 | J133626.27+375053.2 | 16.7 | 16.9 | 11.3 | 11.4 | 125±18 | 124±18 | 0.26 | 0.24 | 9.0 | 0.005 | 0.00130 |
| J124439.80+404922.7 | J124439.63+404921.8 | 15.9 | 15.6 | 10.1 | 10.3 | 142±27 | 116±18 | 0.40 | 0.37 | 2.2 | 0.001 | 0.00128 |

[a] The full table is available at CDS.

# Chapter 4

# Conclusion

## 4.1 Summary

The present work is devoted to the study of the statistical properties of wide binary stars – stellar pairs with separations larger than 200 AU. Such weakly bound pairs are easily disrupted by encounters with (non-luminous) massive objects, making them an interesting tool to constrain both the nature of dark matter and star-formation theory.

Here, the focus is on late-type main sequence stars having 20% to 85% of the Sun's mass. Nearly 670 000 stars with apparent magnitudes between 15 and 20.5 mag were selected from a homogeneous sample covering about 675 square degrees in the direction of the Northern Galactic Pole. The data were taken from the *Sloan Digital Sky Survey*, which are freely accessible online (`www.sdss.org`).

In the first part of this work (Chap. 2), the *angular two-point correlation function* (2PCF) is used to investigate the clustering properties of the stars in the sample. The 2PCF measures the excess probability with respect to a random distribution of observing two stars close together in the sky. It relies only on the stellar positions. A clear signal due to the presence of wide binary systems emerged at angular separations smaller than $10''$ (Fig. 2.5 on p. 39).

The observed 2PCF was modeled by means of a modified Wasserman and Weinberg (1987) technique. In this way, it was possible to infer simultaneously the power-law index $\lambda$ of the semi-major axis distribution, which was assumed to obey a single power law, and the local wide binary number density $n_{\rm WB}$. The semi-major axis distribution was found to follow Öpik's law ($\lambda = 1$), whereas the best-fit density $n_{\rm WB}$ corresponds to 5 wide binaries per 1 000 $\rm pc^{-3}$ implying that about 10% of all stars in the solar neighbourhood belong to a late-type wide binary system.

Prior studies found evidence for a break in the separation distribution attributed to the disruptive effects of massive encounters. Due to the increasing noise from optical pairs at larger separations, the 2PCF method is not sensitive enough to reliably distinguish between single and a broken power-law distribution. Neither the single nor the broken power-law model could have been rejected by the data with confidence. The question of a break in the separation distribution is one of the most interesting questions in the field of wide binary research because of the implications to the dark matter enigma.

The second part of this work (Chap. 3) was aimed at filtering out the optical pairs from the sample and, hence, increasing the sensitivity of the analysis at larger separations. To this end, distance information from photometric parallaxes were included in a statistical manner using a novel weighting procedure. Based on the binding probability every double star in the sample nearer than 2 kpc was assigned a statistical weight, which is large for real pairs and small for optical pairs. Consequently, the derived colour and mass-ratio distributions are dominated by the real double stars.

The colour distribution, corrected for selection effects, is found to be in good agreement with the colour distribution of single field stars (Fig. 3.9 on p. 80). However, pairs with a large colour difference seem to be systematically underrepresented as compared to a random pairing of field stars (Fig. 3.12 on p. 84). This preference of pairs with similar colour or mass, respectively, is also apparent in the mass-ratio distribution, which shows an enhancement for large mass ratios ($q > 0.8$). In the range $0.4 < q < 0.8$ the mass-ratio distribution is found to be rather flat (Fig. 3.13 on p. 85). The mass-ratio distribution inferred is contrary to the expectation from random pairing of field stars. While the results of the present work are in broad agreement with previous wide binary studies, they are difficult to compare to the most recent star-formation simulations.

The drawback of the statistical methods used in this work is clearly the need for a sophisticated model both to interpret the clustering signal measured by the 2PCF, and to allow for observational bias. In the absence of any information on the relative velocities of the components in a wide binary system, it became necessary to assign an *average* relative velocity to every pair. The average orbital velocity was chosen to reproduce the observed 2PCF. The best-fit value turned out to be 370 m/s artificially introducing a maximum separation around 4 000 AU beyond which the pairs get assigned negligible weights. This limitation again prevents us from drawing any conclusion on the shape of the separation distribution at large separations.

On the other hand, the statistical approach successfully circumvents the restrictions of common proper motion studies, which are limited to relatively nearby stars with high proper motions. Not aiming at identifying individual wide binary systems opens the possibility to include into the analysis information from a much larger number of wide binaries and probe their population to greater distances. Here, information from about 4 000 wide binary systems were included statistically. Yet, the method put forward allows to compile a 'ranking list' of wide binary candidates (table 3.1 on p. 93), which is useful for follow-up studies and cross identifications with other wide binary catalogues. In summary, the novel weighting procedure presented in this work constitutes a viable method to infer the statistical properties of wide binary stars and provides an approach complementary to common proper motion studies.

## 4.2 Outlook

The most obvious next step would be to compare our list of WB candidates (table 3.1) to other double star catalogues available, like for example the *Washington Double Stars Catalog* (WDS, Mason et al. 2001). A first glance at the WDS using the *VizieR Service* from CDS suggests that most of our WB candidates are not included yet. In the region around the NGP studied in this work, the WDS lists 525 pairs with angular separations between 2″ and 30″. But most of them are brighter than the stars in our sample: less than 80 (60) have both their components fainter than 14 (15) mag. A careful cross-identification of these pairs with our WB candidates remains to be done. Possibly more rewarding will be the cross-identification of our candidates with the catalogues compiled by Sesar et al. (2008) and by Dhital et al. (2010), both using SDSS data as well but consulting also proper motion information from the USNO-B catalogue (Munn et al. 2004)[1].

To confirm or disprove the genuineness of our WB candidates follow-up studies are needed. Some of the stars in our sample may have their radial velocity measured, e.g. by the *RAdial Velocity Experiment* (RAVE, Steinmetz 2003). Radial velocity measurements can be used to discriminate against optical pairs, since WB components are expected to have similar radial velocities. Up-coming large-scale survey, like for example ESA's *Gaia* mission (Turon et al. 2005), whose launch is scheduled for 2012, will provide a wealth of data (accurate distances, proper motions and also radial velocities) that can be used to distinguish between optical and real pairs.

Another issue that is relatively straightforward to investigate is related to the clustering properties of the WBs themselves. Studying the distribution of WBs of intermediate brightness at the Galactic poles, Saarinen and Gilmore (1989) showed that the pairs "are not distributed in a Poissonian manner, but that significant and not understood structure is apparent in the stellar distribution on the sky." The differentiation of our sample in terms of direction (§2.5.3) gave no evidence of any significant WB density variation (apart from subsample E). But a more detailed analysis is surely worthwhile.

Most interestingly, Saarinen and Gilmore found only in the NGP a significant WB 'lumpiness'; the WBs in the SGP seemed to be distributed more randomly. It would be interesting to repeat our study also for the SGP, as not expected differences between the stellar distributions at the Galactic poles might be attending. The SDSS mapped only a small fraction of the southern sky and is not suitable for an analogous WB study around the SGP. However, such a study will become feasible with the data from *Southern Sky Survey* conducted with the *SkyMapper* telescope at the Siding Spring Observatory in Australia (Keller et al. 2007).

An important prediction by $n$-body simulations of dissolving clusters is that a large fraction of WBs are in fact hierarchical triple or quadruple systems (e.g. Kouwenhoven et al. 2010). Again, photometric and spectroscopic follow-up studies of our WB candidates are needed to confirm this prediction. Another way to identify unresolved pairs among our WB candidates might be to use the *colour-induced displacement* method (CID, Pourbaix et al. 2004), but its very low efficiency casts doubt whether the CID is practicable here. Also first attempts to find unresolved binaries by a peculiar position in a colour-colour diagram in the spirit of Smolčić et al. (2004) did not yield any promising result. We do, however, not expect that our WB candidates are heavily contaminated by unresolved binary and multiple stars. The photometric parallax method tends to underestimate the distance of unresolved binary stars, especially when the components are of similar brightness. A wrong distance estimate most likely results in a too large difference in the component's distances, which in turn results in a small weight within our weighting procedure.

[1]See also the erratum Munn et al. (2008).

One of the major difficulties in studying the properties of WB stars is to properly take into account all sorts of observational biases and selection effects. With the data from *Gaia* it will become feasible to handle these problems in a model-independent way by constructing large, complete WB samples (Halbwachs et al. 2003). It will be interesting to compare the WB statistics inferred from *Gaia* data to the results presented in this work. Even if *Gaia* will map the sky not in the same depth as the SDSS[2], the unprecedented accuracy of its parallax and proper motion measurements will complete the census of faint wide companions to stars in the solar neighbourhood. In this way, also the question whether the Sun itself is a member of a WB (the 'Nemesis-Hypothesis', Davis et al. 1984; Whitmire and Jackson 1984, see also Appendix B.2) will be settled once and for all[3]. We also expect that with *Gaia* the inability to probe the separations distribution at large separations encountered in this study will be overcome. Hence, the question of a break in the WB semi-major axis distribution can be readdressed and, eventually, it becomes possible to place severe constraints on the nature of dark matter (DM) using WBs.

The DM paradigm will soon have to pass a test where WBs play a central role. Both Hernandez and Lee (2008) and Peñarrubia et al. (2010) predict that WBs with separations larger than about 0.1 pc should be absent or strongly depleted in dwarf spheroidal (dSph) galaxies, because dynamical friction has tightened them or because they have been disrupted by encounters with dark substructures. This prediction is in principle already testable for the local dSph using deep, high-resolution exposures of the *Hubble* ACS camera and can be verified with up-coming deep large-scale surveys such as *Pan-STARRS*, *LSST* and *Gaia*. If abundant WB populations are found in local dSphs – possibly through the 2PCF method – the CDM picture would be seriously challenged. Such a finding could, however, be in favour of MOND-like theories (e.g. Milgrom and Bekenstein 1987) that go without DM.

In his (unpublished) Master thesis at the University of Basel, Marc Horat analysed *Hubble* ACS data of the two globular cluster *47 Tucanae* and *ω Centauri* with the 2PCF method. Unlike close binaries that are 'hard' in the sense of Heggie (§1.2), wide (soft) binaries play no significant role in the evolution of stellar clusters. In fact, it can be shown that the equilibrium number of very soft binaries in clusters is of order unity, independent of the number of stars in the cluster (Binney and Tremaine 2008, §7.5.7). Therefore, no clustering signal due to WBs would be expected in any globular cluster. Most interestingly, Horat found evidence for pairs in excess of a random distribution in both clusters. While the signal for *47 Tuc* is barely significant and questionable, $\omega$ *Cen* shows a clear signal at angular separations smaller than $2''$. In a distance of about 5.5 kpc this signal corresponds to pairs with separations of less than 0.05 pc (10 000 AU). The qualitative difference in the 2PCF inferred from the two clusters is striking and possibly supporting the notion that $\omega$ *Cen* is a former nucleus of a dissolved dwarf galaxy (Hilker and Richtler 2000). A major problem in the study was, however, the extreme sensitivity of the results to the exact density profile of the random sample that, of course, must match the profile of the real cluster. Finding evidence for WB in globular clusters was unexpected and future studies should investigate this preliminary result in more detail.

[2]It seems that *Gaia*'s limiting magnitude will be considerably less than the quoted limit around 20 mag (H. Jerjen, personal communication).

[3]This might be anticipated by NASA's *WISE* mission (Wright et al. 2010).

In a recent theoretical work Jiang and Tremaine (2010) studied the orbital evolution of nearby WBs due to gravitational perturbations from passing stars. They predict a second peak in the separation distribution around 200 pc stemming from formerly bound systems that are slowly drifting apart (the first peak is due to bound pairs). This peak causes a significant correlation in the positions and velocities of the stars and should be detectable in large astrometric surveys like *Gaia*. Another prediction is the emergence of 'tidal tails' in the direction of the binaries' Galactocentric orbit, which can extend several thousands parsec. An alignment of the components of disrupted binaries can possibly be measured using the *alignment correlation function* introduced by Faltenbacher et al. (2009) – an extension of the 2PCF that takes into account the orientation of the object (here a stellar pair) with respect to a given frame of reference (here the Milky Way Galaxy).

Of all known stars hosting a planetary system about 20% are known to have a stellar companion, most of them with separations larger than 200 AU (e.g. Raghavan et al. 2006). This is most likely a observational selection effect: The majority of the known exoplanetary systems have been discovered by the radial velocity technique. But in close (separations smaller than a few AU) binary systems the signal from a supposed extrasolar planet would be obscured through the large orbital speed of the two stars. Therefore, such tight pairs are usually excluded from exoplanet searches. Numerical simulations suggest that close pairs can truncate the protoplanetary disk, preventing the formation of planets beyond a few AU, while binaries wider than about 100 AU are indistinguishable from single stars regarding their planetary systems (Duchêne 2010). However, as pointed out by Mugrauer et al. (2007), wide companions with high orbital eccentricities get close to the planet hosting star at periastron[4]. In addition, WBs were 'softened' through numerous weak gravitational encounters and might have been closer in the past. It can be assumed that these fragile systems have been spared from catastrophic encounters, potentially making them even more hospitable to the formation of larger planetary systems than single stars that have been ejected from their birth cluster by a violent encounter (Boss 2000).

[4]In fact, the periastron distance is the most important parameter in assessing the influence of a companion on planetary formation and orbital evolution (Quintana et al. 2007). Assuming that the eccentricities of the WB are distributed thermally ($f(e) = 2e$), the average periastron distance is equal to a third of the semi-major axis: $\bar{r}_{\rm peri} = a/3$.

A companion star may scatter small bodies (comets, asteroids, etc.) in the direction of the planetary system resulting in repeated impacts onto the planet's surface – possibly facilitating the emergence of life (Burgasser 2007). In the case of the Earth, icy bodies might have brought necessary water ice as well as the chemical precursors to biotic life (e.g. Delsemme 2001). It is believed that a cataclysmic impact has formed the Moon (e.g. Canup and Asphaug 2001), whose tidal forces are stabilising the Earth's axis and thereby reducing unpleasant climate variations (Williams and Pollard 2000). Furthermore, mass extinctions (caused by impacts) had a profound – and sometimes constructive – influence on the evolution of life (Raup 1994). On the other hand, too many disastrous impacts are obviously destructive and the protective role of Jupiter has often been pointed out[5]. As the formation of planets at distances from the host star comparable to that of Jupiter ($\sim$ 5 AU) appears to be inhibited by close stellar companions, the search for terrestrial extrasolar planets among WB components might be especially rewarding. Thus, the WB candidate's list (table 3.1) that we have compiled, presumably contains promising targets for the search of habitable exoplanets and extraterrestrial life.

[5]Recent studies suggest, however, that this idea is probably only true for cometary impacts of long period comets stemming from the Oort cloud. Impact rates of asteroids and Kuiper belt objects may even be enhanced by the presence of Jupiter (Horner and Jones 2008, 2009; Horner et al. 2010).

# Appendix A

# Top wide binary candidates

Figure A.1 shows SDSS images from the top six wide binary candidates centered at the primary component. The edge length of an image is 200″. We see that they all have a reddish colour corresponding to the late spectral type (all have masses below $0.5\mathcal{M}_\odot$; see table 3.1). With an angular separation of only 2.2 arcseconds, the two components of our candidate number 5 (at the centre of Fig. A.1e) are barely separable.

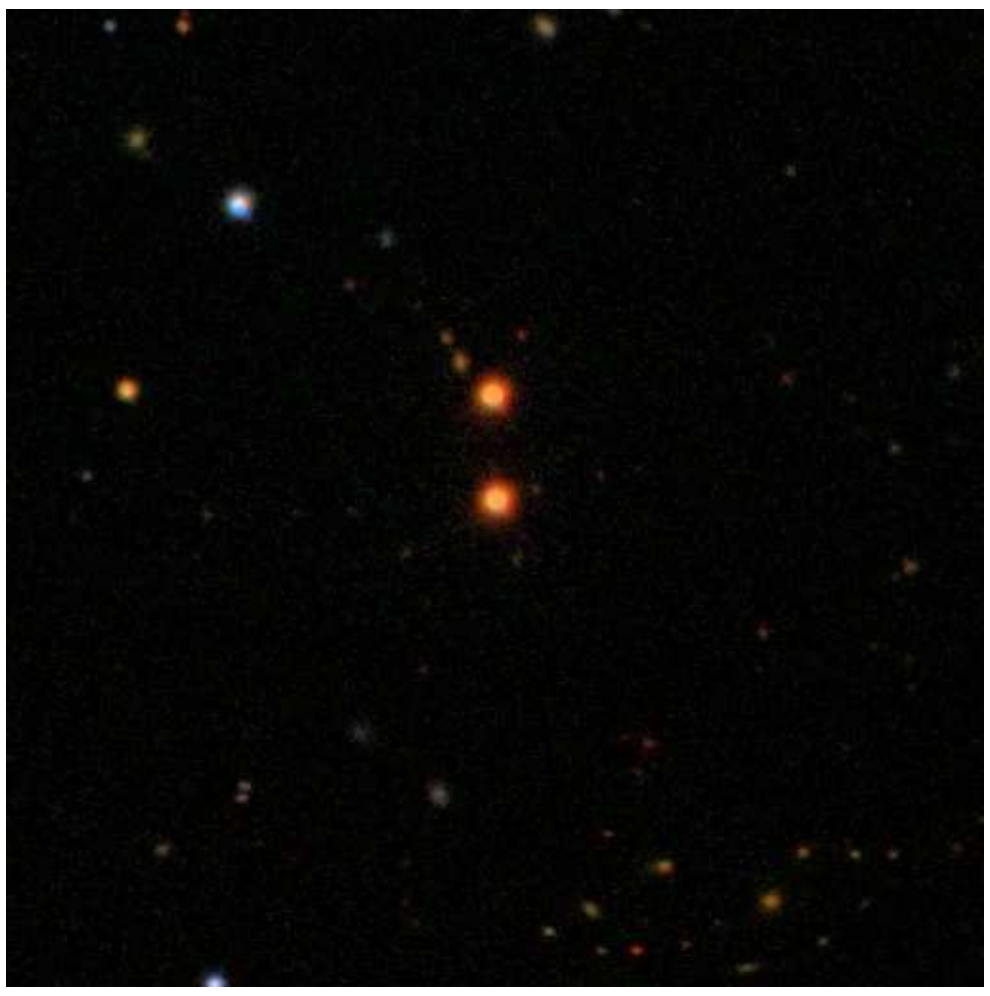

(a)

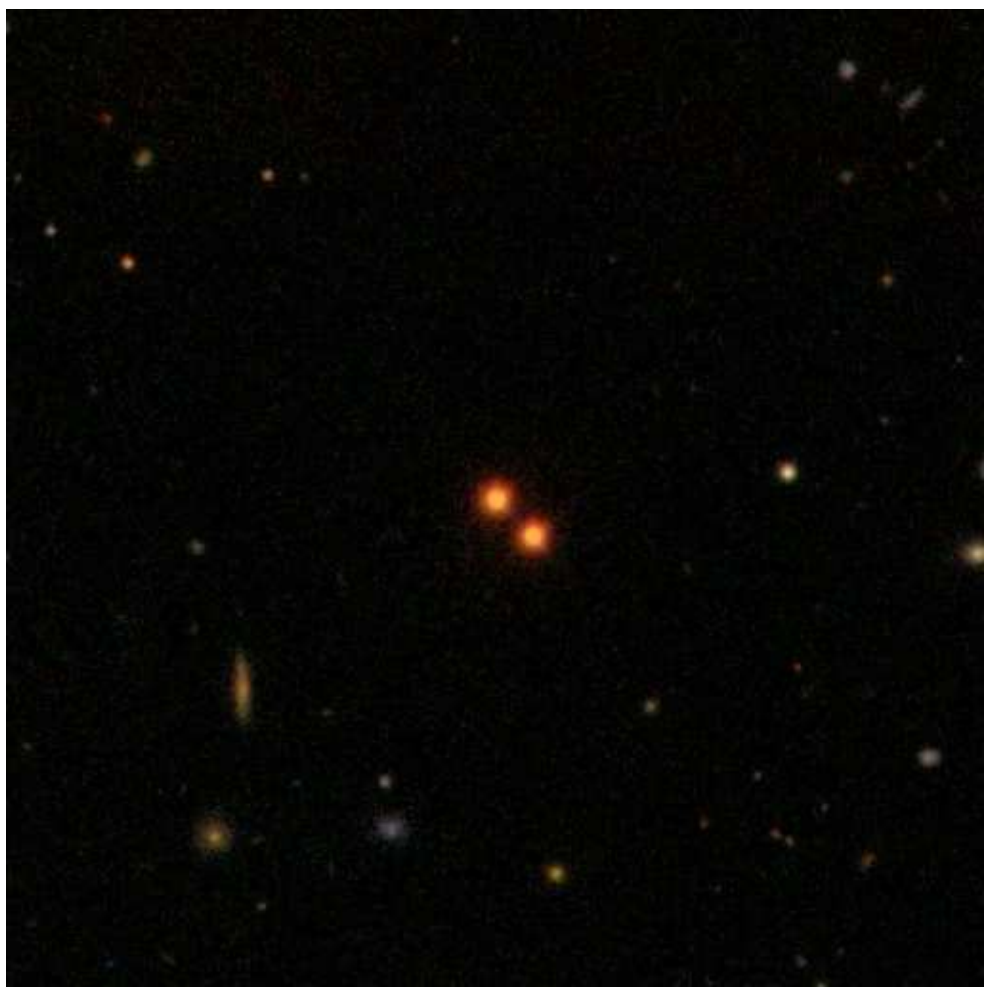

(b)

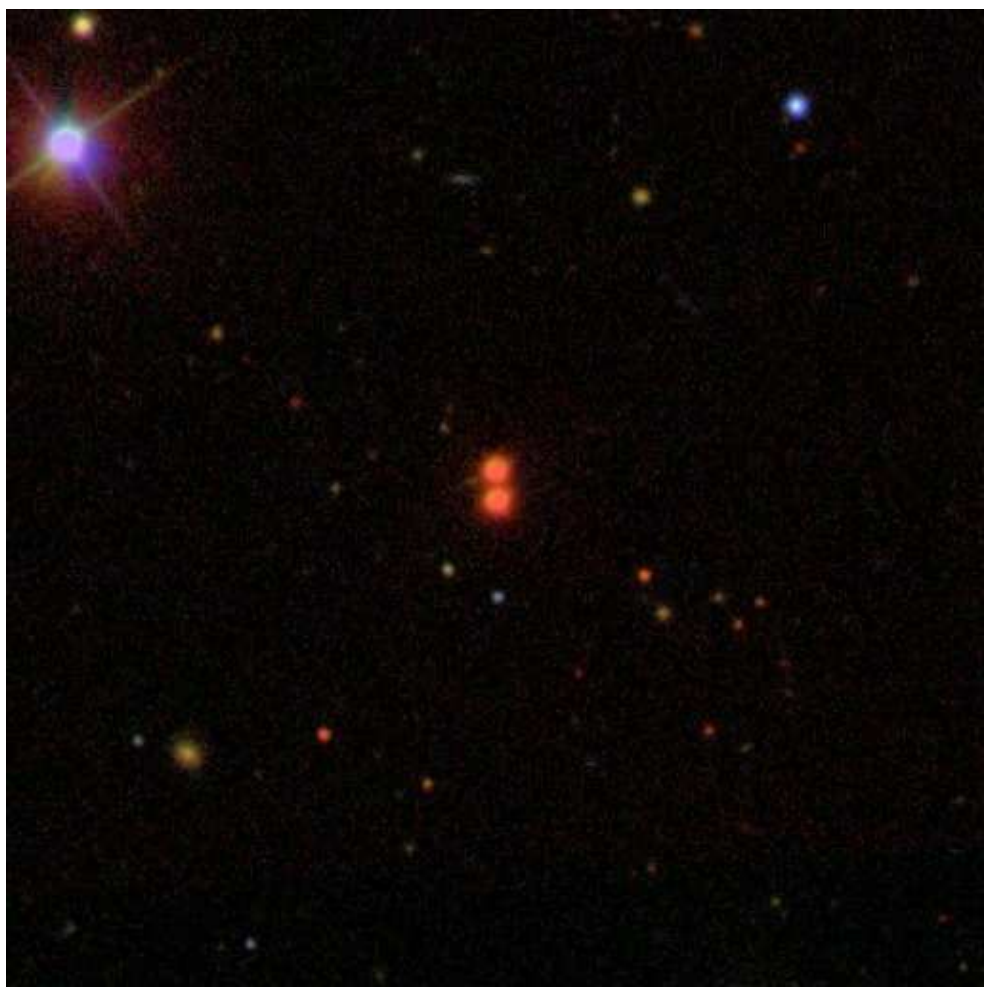

(c)

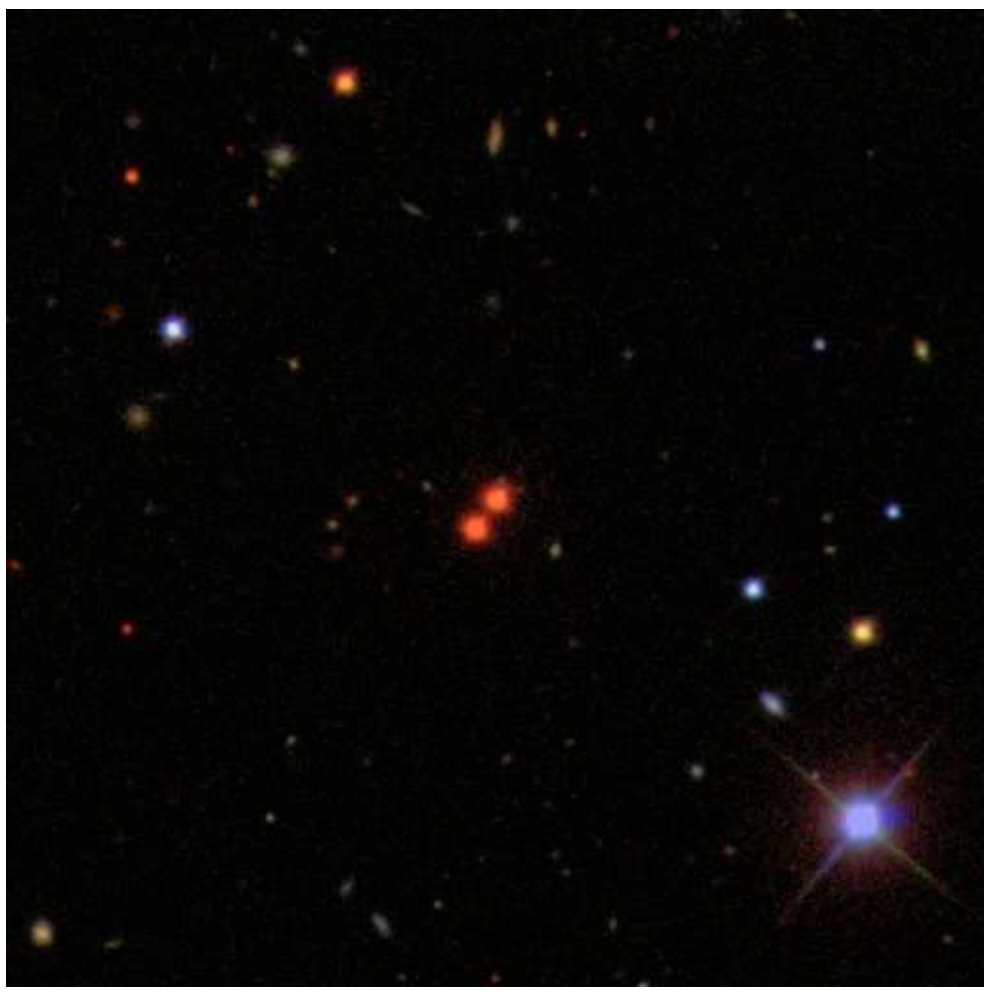

(d)

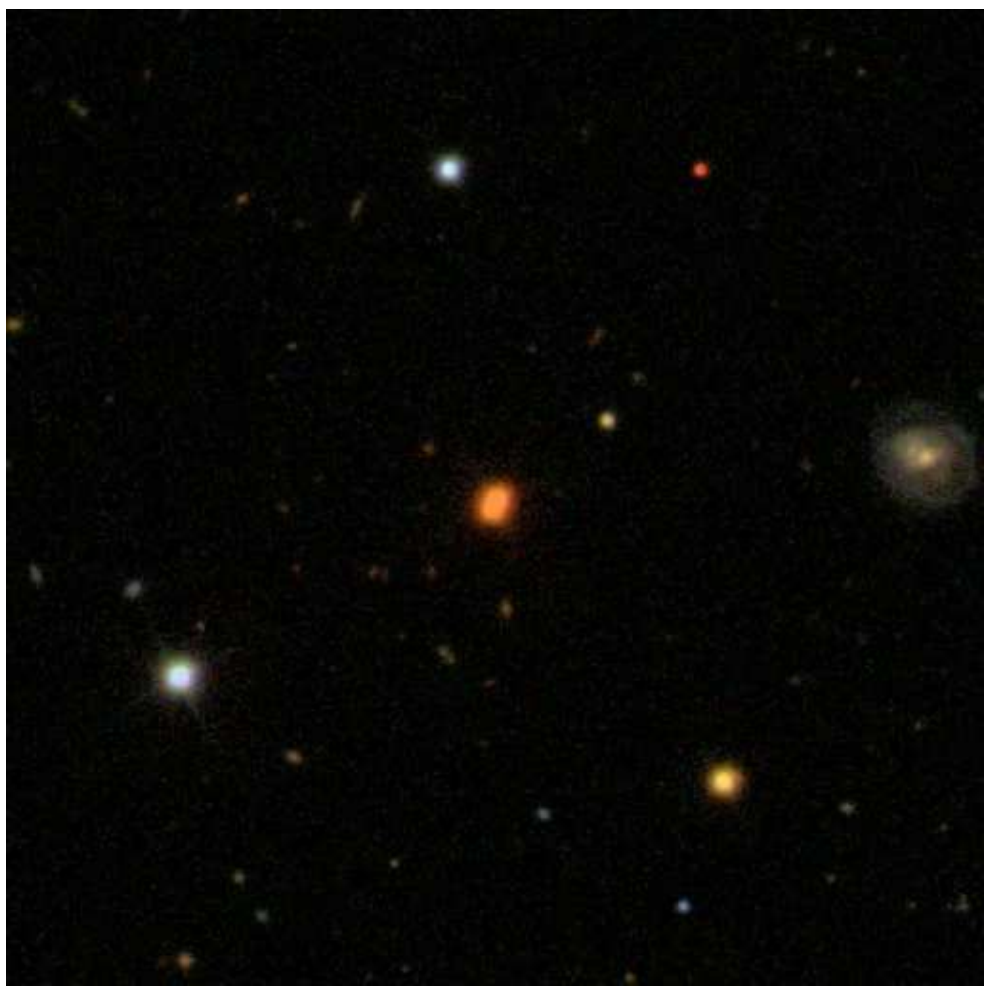

(e)

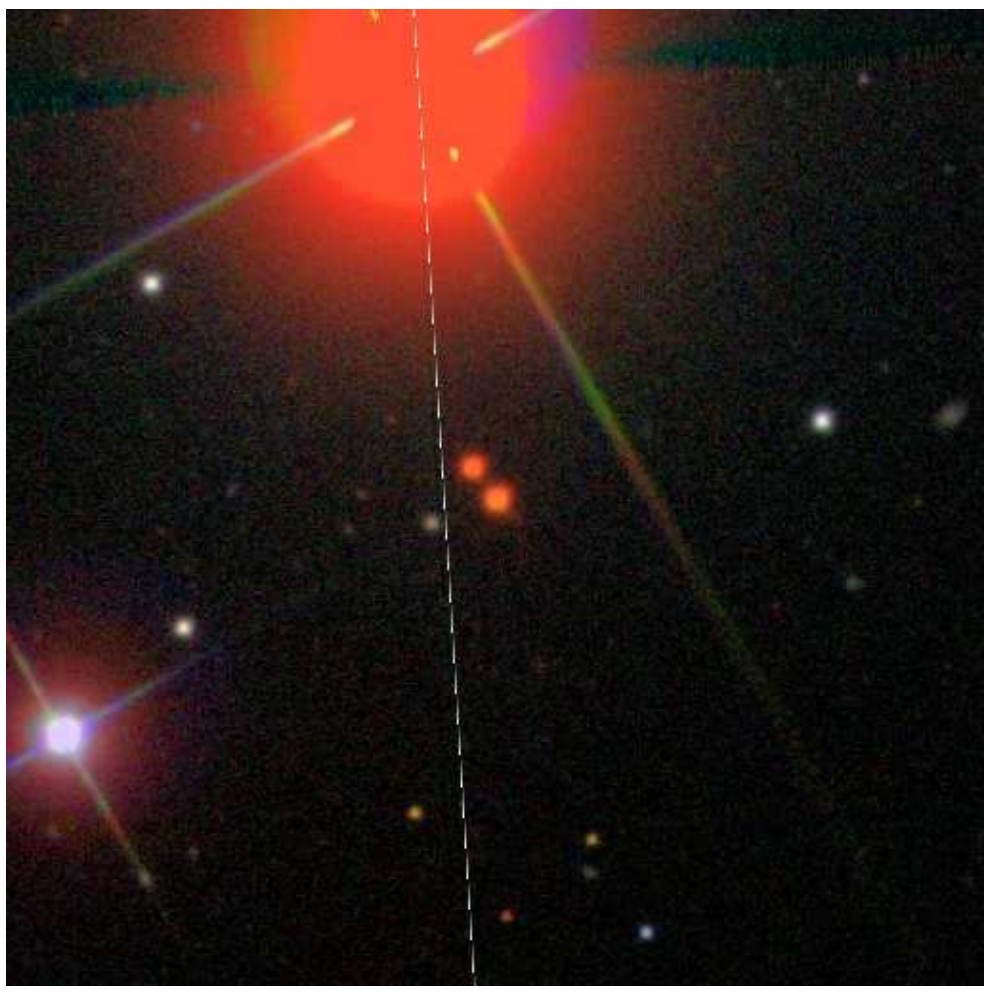

(f)

Figure A.1: SDSS images of the top six WB candidates from table 3.1. The edge length is 200″.

# Appendix B

# Talks

## B.1 The widest binary stars – A statistical approach

This talk was given at the international conference *Binaries – key to comprehension of the Universe* ('Binkey') held in Brno, Czech Republic, 8–12 June 2009. The proceedings will be published as a special issue of the ASP Conference Series.

Dear colleagues,

I will report on our statistical study on wide binary stars and its preliminary results. Let me start with a short motivation why wide binaries are interesting to study.

Wide binaries may shed light on one of the most pressing mysteries of modern astronomy and cosmology: the nature of dark matter. Very wide pairs, with projected separations of more than 0.1 pc, are easily disrupted by encounters of massive objects. So, the today's separation distribution of the widest binary stars contains fossil information of the dynamical history of the Galaxy. However, the available data on very wide pairs are quite sparse and their implications on the density of 'MAssive Compact Halo Objects' (MACHOs) are still disputed. We will hear more about that by Damien Quinn on Friday.

Wide binaries are also interesting in and of themselves, as their existence is difficult to explain in the context of star formation theory. A possible formation mechanism in terms of dissolving young star clusters will be presented by Thijs Kouwenhoven tomorrow. The study of the wide binary stars may provide important clues to the conditions of their formation and of star formation in general.

A very interesting approach to test the predictions of the dark matter hypothesis was put forward by Hernandez and Lee. According to their calculations, wide binaries should now be absent in high-density dark matter halos with low velocity dispersion as inferred for local dwarf spheroidal galaxies. There, wide binaries should have evolve into close binary stars because of the orbital tightening due to the dynamical friction caused by dark matter particles. We have restricted our study to the wide binary population in our Galaxy so far.

The long periods of wide binaries make it virtually impossible to identify them by their orbital motion. We therefore rely on statistical techniques. In the past, two related methods where used to study wide binaries: First, the angular two-point correlation function, which was pioneered by Bahcall and Soneira in the early 80s and relies only on the positions of the stars. Second, common proper motion, which adds proper motion information to the astrometry of the first.

We do not aim at identifying the wide pairs individually. So, the angular two-point correlation function appears to be suitable for our purposes. It is a quite straightforward and well-established clustering measure, and has been widely used to study the distribution of galaxies. No proper motion or radial velocities are needed – it relies entirely on photometric data, which are available and readily accessible from modern large-scale surveys, such as the Sloan Digital Sky Survey (SDSS).

The correlation function has also some drawbacks. The noise due to random associations increases rapidly with angular separation and, therefore, a large sample of stars is needed to get a statistically significant signal. The approach is purely statistical in the sense that the pairs are not individually identified. Additionally, we need a theoretical model to allow for selection effects and to interpret the signal measured by the two-point correlation function.

For our study we chose a rectangluar region around the Northern Galactic Pole covering about 675 square degrees. We selected all point-like sources classified as 'star' from the sixth data release of the SDSS having a $r$-band magnitude between 15 and 20.5 magnitudes.

Here (Fig. 2.2 on p. 25), we see a colour-colour diagramm of the nearly one million point-like sources within our chosen region and magnitude limits. In order to minimise the contamination with non-stellar point-like objects, we exclude all SDSS-quasar candidates as well as all moving objects, such as asteroids. Even after removing all quasar candidates, a considerable scatter around the expected stellar locus remain. This scatter is probably due to further quasars and misidentified galaxies. We therefore decided to reject all objects having $g - r < 0.5$. About 670 000 stars remain in our final sample all of them having a spectral type later than that of the sun.

Finlator et al. estimated that only about 1% of all the stars observed by the SDSS are not on the main sequence. This justifies our basic assumption that all the stars in our sample are main sequence stars. So we state that our final sample contains about 670 000 main sequence stars with spectral type later than G5.

The correlation function measures the excess of pairs with respect to a random distribution. The simplest estimator, neglecting edge effects due to the finite sample size, is the ratio of the observed number of distinct pairs, $F$, and the number of pairs expected from a random distribution, $P$, minus one

$$\hat{w}(\theta) = \frac{F(\theta)}{P(\theta)} - 1\,.$$

This figure (Fig. 2.5 on p. 39) shows the two-point correlation function as inferred from our final sample as solid circles. The corresponding correlation amplitude is shown on the left ordinate. The errors are Poissonian. We can see that there is a strong clustering signal at angular separations smaller than $10''$, down to the resolution limit of the SDSS at $2''$.

In the same figure we also have plotted the cumulative number of pairs in excess of a random distribution as open circles. Herefrom, it is evident that there are pairs in excess of random up to the maximum angular separation we have examined, that is, up to half an arcminute. This appears to be contradictory to an independend study on wide binaries using the SDSS database by Sesar et al. They find that there are essentially no physical bound pairs with angular separations larger than 15 arcseconds. Unfortunately, all our attempts to track down the cause of this apparent discrepancy failed so far.

To model the clustering signal, we rely on a versatile technique, developed in the late 80's by Wasserman and Weinberg. It was designed "for comparing wide binary observations with theoretical semimajor axis distributions". In this framework, the number of observed binaries as a function of the projected separation, $s$, can be expressed as a product of the reduced separation distribution, $Q$, and the effective volume, $V$, times the total number density of wide binaries in the solar neighbourhood $n_{\mathrm{WB}}$

$$\psi(s) = n_{\mathrm{WB}} Q(s) V(s)\,.$$

Without going into details, let me mention that the reduced separation distribution contains the physical properties of the wide binary stars, such as the semi-major axis distribution, the distribution of eccentricities, the orientations of the orbital planes, and projects them onto the sky. The effective volume, on the other hand, allows for selection effects. It accounts for the characteristics of our sample such as the area covered by it, its magnitude limits and the resolution limit. Furthermore, it includes the stellar density distribution and a luminosity function for single stars.

Regarding the reduced semi-major axis distribution, we assume that it obeys a single power-law, which is normalised in the range, where this single power-law model holds. This range provides at the same time the definition of what we consider to be a 'wide binary star'. Following Wasserman and Weinberg we take the lower limit to be 200 AU. A natural choice for the upper limit is provided by the Galactic tidal limit. We find it to be of the order of 1 parsec.

There has been some discussion whether the data require a break in the semi-major axis distribution or not. While Wasserman and Weinberg could not give a conclusive answer, more recent studies by Lpine and Bongiorno and by Sesar et al. indicate that the data indeed is best described by a broken power-law model. So the question that arises is whether the assumption of a single power-law is rejected by the data.

As the original Wasserman-Weinberg technique deals with projected separations, whereas we are dealing with angular spearations alone, some modifications are required. Without going into detail again, we find this expression for the number of wide binaries, $\varphi$, as a function of angular separation

$$\varphi(\theta) = n_{\mathrm{WB}}\Omega \int\limits_{\langle s\rangle_{\min}/\theta}^{\langle s\rangle_{\max}/\theta} \mathrm{d}D D^3 \tilde{\rho}(D) Q(D\theta) \iint\limits_{M_{\min}(D)}^{M_{\max}(D)} \mathrm{d}M_1 \mathrm{d}M_2 \widetilde{\Phi}(M_1)\widetilde{\Phi}(M_2)\,.$$

The model correlation function is then obtained by adding the number of physical pairs, $\varphi$, to the number of pairs expected from a random distribution, $P$

$$w_{\mathrm{mod}}(\theta) = \frac{\varphi(\theta) + P(\theta)}{P(\theta)} - 1\,.$$

The model has two free parameters - the local number density of wide binaries, $n_{\mathrm{WB}}$, and the power-law index of the semi-major axis distribution, $\lambda$. We infer them simultaneously by least squares fitting. The uncertainties of the best-fit values are estimated by Monte Carlo confidence regions. We find a local wide binary density of about 5 wide binaries in 1 000 cubic parsecs and a power-law index of 1, consistent with Öpik's law. The relative errors are roughly $\pm 10\%$, where we have quoted 95% confidence intervals.

Here (Fig. 2.5 on p. 39), the model curves corresponding to the best-fit values are plotted together with the observed correlation function from the SDSS data. We see that at larger angular separations the cumulative number of pairs in excess of random is slightly overestimated. This could be interpreted as a hint that the semi-major axis distribution is indeed broken. Having a steeper decrease from a certain semi-major axis on would result in a flatter model curve. However, the assumption of a single power-law model cannot be rejected with confidence.

So, we conclude that the data is consistent with a single power-law up to the tidal limit, whereas the power-law index agrees with Öpik's law. I need to emphasise that this finding is not necessarily contradictory to previous studies, which found a break in the separation distribution. However, whether the data is also consistent with a specific broken power-law model, like that found by Lépine and Bongiorno for example, remains to be checked. The derived wide binary density tells us that about 10% of all the stars in our sample are a member of a wide binary.

Once we have determined the two free parameters in our model, we can calculate the number of physical pairs that can be observed in our sample as a function of the projected separation. In doing so, we find that about 800 very wide binaries with projected separations larger than 0.1 pc can presumably be observed in our sample, whereas none are expected to be found beyond 0.8 given the range in angular separation we have examined. This does, however, not mean that pairs with projected separations larger than 0.8 pc do not exist at all in the Galaxy. Indeed, the existence of an extremly wide binary with a projected separation of 1.1 pc has been recently confirmed. I guess, we will hear more about this outstanding discovery by Damien Quinn.

Here (Fig. 2.8 on p. 43), we see the number of wide binaries as a function of the projected separation, which are expected to be observed in our sample having angular separations between $2''$ and $30''$. The shape of this distribution is largely determined by selection effects.

In view of their implications on star formation and the nature of dark matter, wide binaries may indeed constitute a key to the comprehension of the Universe.

Thank you!

## B.2 Nemesis: Hat die Sonne einen Begleiter?

Dieser Vortrag wurde im Rahmen einer vom Astronomischen Verein Basel anlässlich des Internationalen Jahr der Astronomie organisierten Vortragsreihe am 11. November 2009 in Basel gehalten. In den 80er Jahren trug die Nemesis-Hypothese massgeblich zum Aufschwung der Forschung an weiten Doppelsternen bei.

Sehr geehrte Damen und Herren, liebe Freunde der Astronomie,

Ich danke Herrn Binggeli für die Einleitung und bedanke mich für Ihr zahlreiches Erscheinen. Vielleicht mögen sich einige von Ihnen daran erinnern als vor etwa 25 Jahren in vielen Zeitungen Meldungen über einen noch unentdeckten Sonnenbegleiter zu lesen waren; einen Stern, der um unsere Sonne kreist, ähnlich wie die Erde. *Nemesis – der Todesstern*; so oder so ähnlich lauteten damals die Schlagzeilen, denn Nemesis, wie der Schwesterstern der Sonne getauft wurde, soll in regelmässigen Abständen Tod und Verderben über die Welt bringen. Welche Indizien für eine solch kühne Hypothese sprechen und was heute aus der Nemesis-Hypothese geworden ist, möchte ich in meinem heutigen Vortrag erläutern.

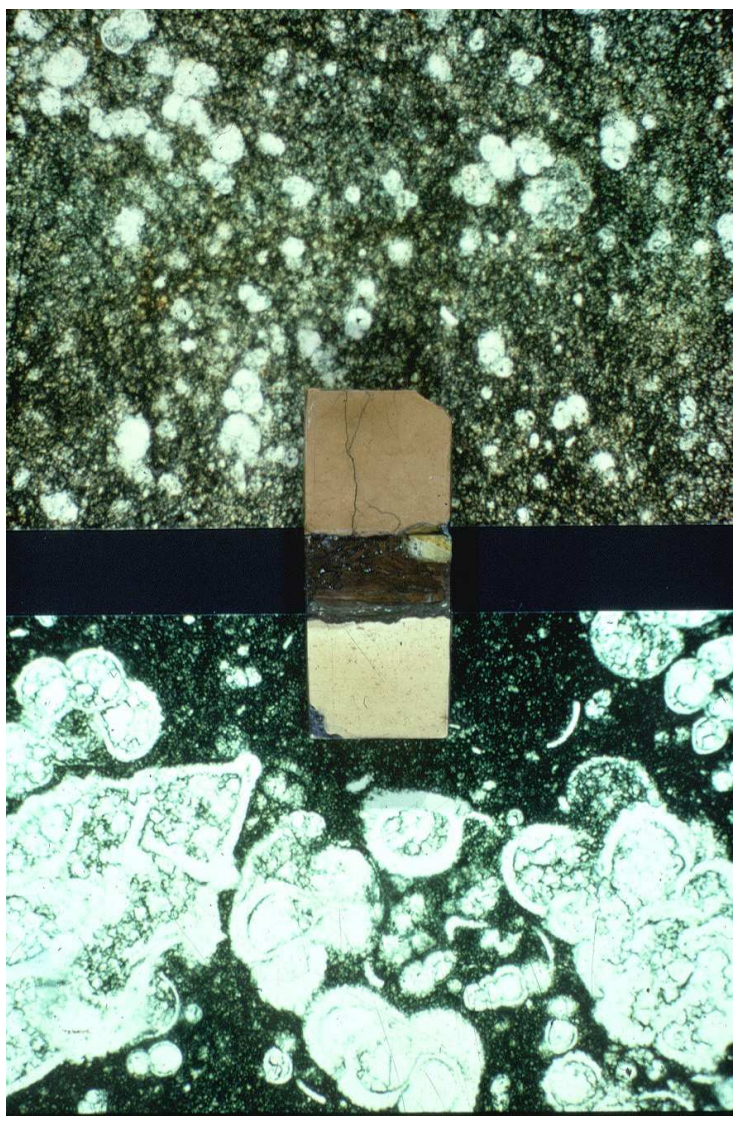

Abb. B.1: Ausschnitt aus dem Gestein an der Kreide-Tertiär Grenze (Mitte). Darum herum sind mikroskopische Aufnahmen (grünlich) des Sediments zu sehen. © A. Montanari.

Die Geschichte des Lebens auf der Erde ist keineswegs geradlinig verlaufen. Immer wieder starben fast schlagartig ein Grossteil aller Tier- und Pflanzenarten aus. Das wohl bekannteste dieser Massenaussterbeereignisse ereignete sich vor etwa 65 Millionen Jahren am Ende der Kreidezeit. Etwa drei Viertel aller irdischen Lebensformen sind damals ausgestorben, unter ihnen auch die Dinosaurier. Massenaussterbeereignisse sind durch fossile Funde gut belegt, doch deren Ursache ist in den meisten Fällen noch völlig unklar.

Es gilt heute jedoch als gesichert, dass die Ursache des Untergangs der Dinosaurier ein verheerenden Meteoriteneinschlag war. Ein Asteroid oder ein Komet von der Grösse des Mount Everest schlug damals mit ungeheurer Wucht auf der Erde auf. Die Folgen eines solchen Einschlags sind katastrophal: Die Gesteinsmassen, welche beim Aufprall in die höchsten Schichten der Atmosphäre geschleudert wurden, verteilten sich in Form von dunklen Staubwolken um den gesamten Erdglobus. Eine Nacht ohne Mond und ohne Sterne brach über die Welt herein. über viele Monate sollte es keinen Morgen geben.

Das wichtigste Indiz für einen Meteoriteneinschlag liefert eine dünne, etwa zwei Zentimeter dicke Tonschicht, die das Ende der Kreidezeit in den Gesteinsschichten markiert. Im Bild hier (Abb. B.1) sehen wir einen Ausschnitt aus dem Gestein an der Kreide-Tertiär Grenze. Das hellere Sediment unten stammt aus der Kreidezeit, während die dunklere

Gesteinsschicht oben bereits aus dem darauf folgenden Zeitabschnitt, dem Tertiär, stammt. Dazwischen liegt jene aufschlussreiche Tonschicht, die die Grenze zwischen den beiden geologischen Zeitabschnitten markiert. In den grünlichen mikroskopischen Aufnahmen können wir zahlreiche Fossilien, von winzigen Lebewesen erkennen. Es handelt sich um so genannte Foraminiferen, kurz Forams. Diese einzelligen, meist gehäusetragenden Meeresbewohner kamen während der Kreidezeit in vielen verschieden Formen und Grössen vor. Den Übergang zur Kreidezeit überlebten aber nur ganz wenige, kleine Arten. Riesige auf dem Land lebende Dinosaurier und winzige einzellige Meeresbewohner starben zur selben Zeit aus. Was war da nur geschehen? Der Schlüssel zur Antwort muss in dieser Tonschicht stecken.

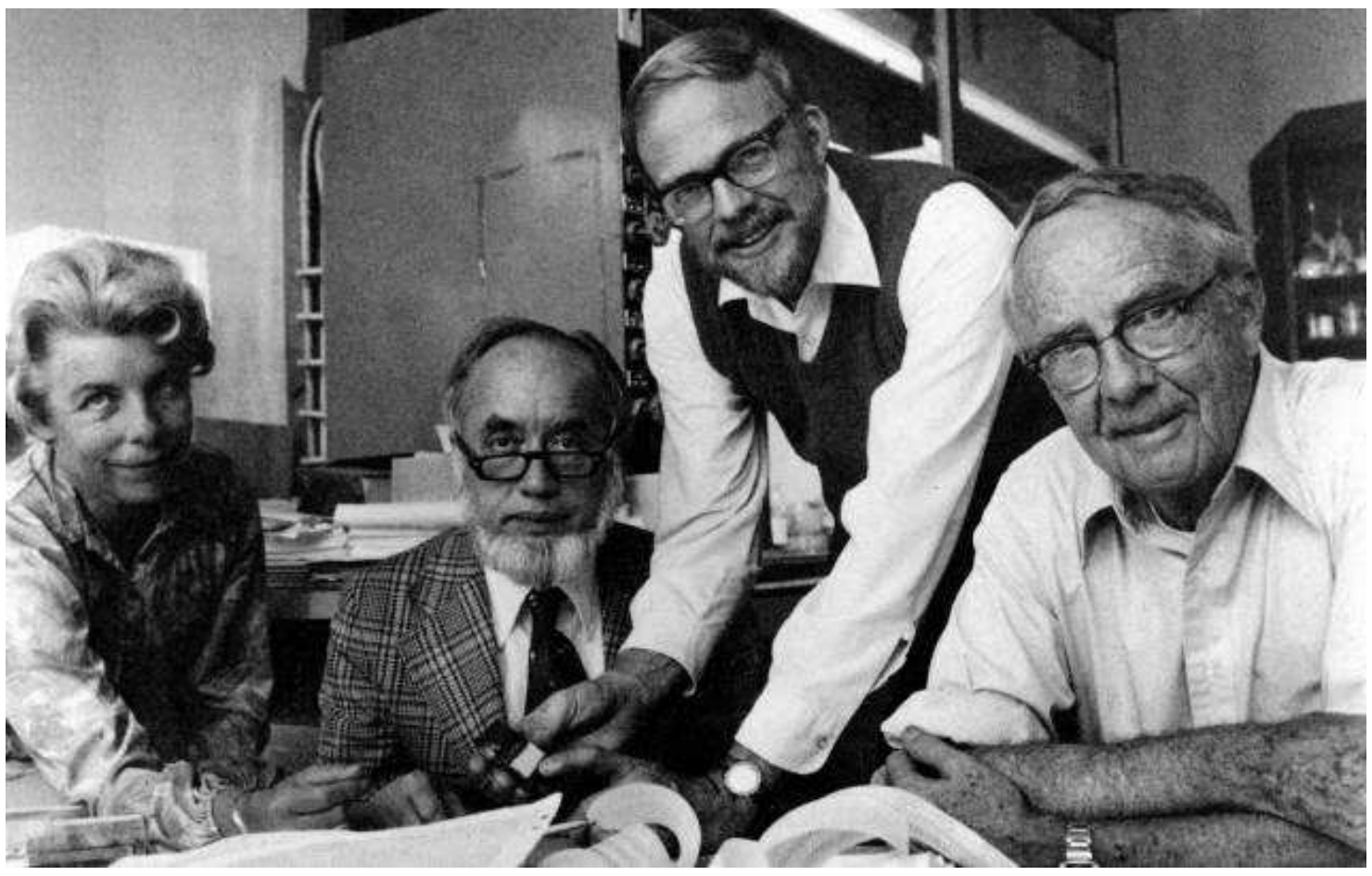

Abb. B.2: Das "Alvarez-Team" um 1980. Von links nach rechts: Helen Michel, Frank Asaro, Walter Alvarez und Luis Alvarez. © Lawrence Berkeley National Laboratory.

Der Physik-Nobelpreisträger Luis Alvarez, hier ganz rechts im Bild (Abb. B.2), stellte Ende der 70er Jahre zusammen mit seinem Sohn Walter, einem Geologen, und den beiden Chemikern Frank Asaro und Helen Michel bei der Analyse dieser Tonschicht fest, dass der Iridiumgehalt in dieser Schicht etwa 600 mal höher als in den umgebenden Sedimenten ist. Dieses Edelmetall kommt in der Erdkruste kaum vor. Es findet sich aber in erhöhter Konzentration in Meteoriten und Kometen. In Folge wurden weltweit Gesteinsproben analysiert und die "Iridium-Anomalie" in der besagten Tonschicht wurde vielfach bestätigt. Es konnte auch gezeigt werden, dass all das Iridium in jener Schicht von ein und derselben Quelle stammt. Die Arbeit des Alvarez-Team hat gezeigt, dass zumindest ein Massenausterbeereignis höchst wahrscheinlich durch einen Meteoriteneinschlag verursacht wurde.

Doch wie jede revolutionäre Idee, stiess die "Alvarez-Hypothese" auf heftigen Widerstand, war man doch der Ansicht, dass die Dinosaurier deshalb ausstarben, weil sie sich nicht mehr gegen die Säugetiere behaupten konnten, welche sich besser an veränderte Umweltbedingungen angepasst haben. Nur langsam, während immer mehr Indizien zusam-mengetragen wurden, konnte sich die neue Theorie des Untergangs der Dinosaurier durchsetzen. Möglicherweise ist die Alvarez-Theorie aber nur die Spitze des Eisbergs.

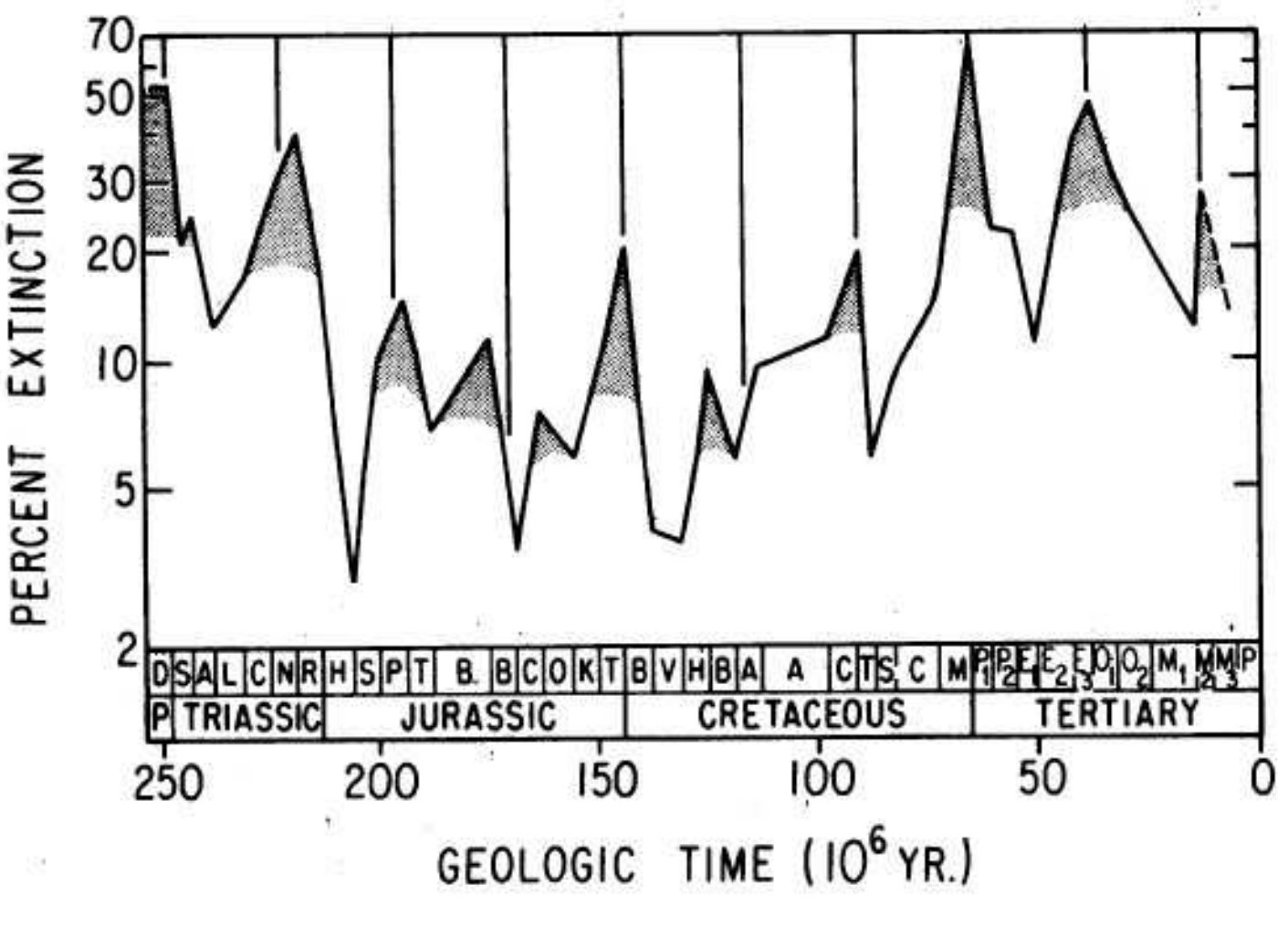

Abb. B.3: Verteilung der Aussterbeereignisse nach Raup and Sepkoski (1984). Die (logarithmische) Skala auf der Ordinate gibt an wie viele Prozent der Lebewesen in welchem Zeitabschnitt ausgestorben sind. Die grau schraffierten Spitzen deuten Massenausterbeereignisse an. Die strikte 26 Millionen Jahre Periodizität ist als vertikale Linien eingezeichnet. © National Academy of Sciences.

Im Februar 1984 veröffentlichten die beiden angesehenen Paläontologen David Raup und John Sepkoski, beide damals an der University of Chicago tätig, einen folgenschweren Artikel. Darin untersuchten sie zum ersten mal mit Hilfe eines Computers eine Liste von über 30'000 fossilen Gattungen von Meerestieren. Die Liste wurde über Jahrzehnte von Sepkoski zusammengetragen und stellt bis heute die vollständigste Datenbasis der marinen Fossilien dar. Raup und Sepkoski entdeckten, dass die Massenausterbeereignisse nicht zufällig in der Erdgeschichte verteilt waren, sondern ein erschreckend regelmässiges Muster an den Tag legten: In regelmässigen Abständen von ungefähr 26 Millionen Jahren sind fast alle Tier- und Pflanzenarten auf der Erde ausgestorben.

Diese Figur (Abb. B.3) ist aus dem Originalartikel von Raup und Sepkoski entnommen. Sie zeigt wie viele Prozent der Lebewesen in welchem Zeitabschnitt ausgestorben sind. Das rechte Ende der horizontalen Achse entspricht der heutigen Zeit, während wir ganz links gerade noch das Ende des Perms vor etwa 250 Millionen Jahren sehen. Dort an der Perm-Trias Grenze fand das grösste Massenausterbeereignis der Erdgeschichte statt. Auch jenes Aussterbeereignis, bei dem die Iridium-Anomalie gefunden wurden, an der Kreide-Tertiär Grenze, wo die Dinosaurier ausstarben, ist deutlich sichtbar. Als vertikale Linien ist die strikte 26 Millionen Jahre Periodizität eingezeichnet.

Es war nicht sehr überraschend, dass es so häufig zu Massenaussterbeereignisse kam, aber die Regelmässigkeit mit der sie offenbar stattfanden war sehr wohl eine grosse Überraschung. Was könnte der Antrieb dieses tödlichen Uhrwerks sein, das im Rhythmus von etwa 26 Millionen Jahren tickt? Aufgrund der buchstäblich astronomischen Zeitspanne, halten Raup und Sepkoski rein biologische oder irdische Ursachen für unwahrscheinlich und bevorzugen extraterrestrische Erklärungen. Nur zwei Monate spaeter, am 19. April 1984, publizierte das renommierte Fachmagazin *Nature* gleich fünf Beiträge, in welchen verschiedene astrophysikalische Szenarien vorgeschlagen wurden, um die beobachtete Regelmässigkeit des Aussterbens in der Erdgeschichte zu erklären. Ich möchte Ihnen, meine Damen und Herren, eine dieser Szenarien etwas näher vorstellen. Jenes, das unter dem Namen "Nemesis-Hypothese" bekannt wurde.

In der Griechischen Mythologie ist Nemesis die Göttin der Vergeltung, welche unerbittlich jegliches Vergehen gegen die Götter rächt. Sie bestraft vor allem die Überheblichen und Anmassenden und wird oft als unbestechliche Rachegöttin dargestellt. In diesem Gemälde von Alfred Rethel (Abb. B.4) verfolgt Nemesis einen Dieb dessen Opfer im Hintergrund verwundet am Boden liegt. Aber es gibt für den Schurken kein Entkommen: Seine Zeit ist bereits abgelaufen, wie die Sanduhr in der linken Hand des Racheengels andeutet. Das Schwert der Vergeltung in Nemesis' rechter Hand wird das Schicksal des Fliehenden besiegeln.

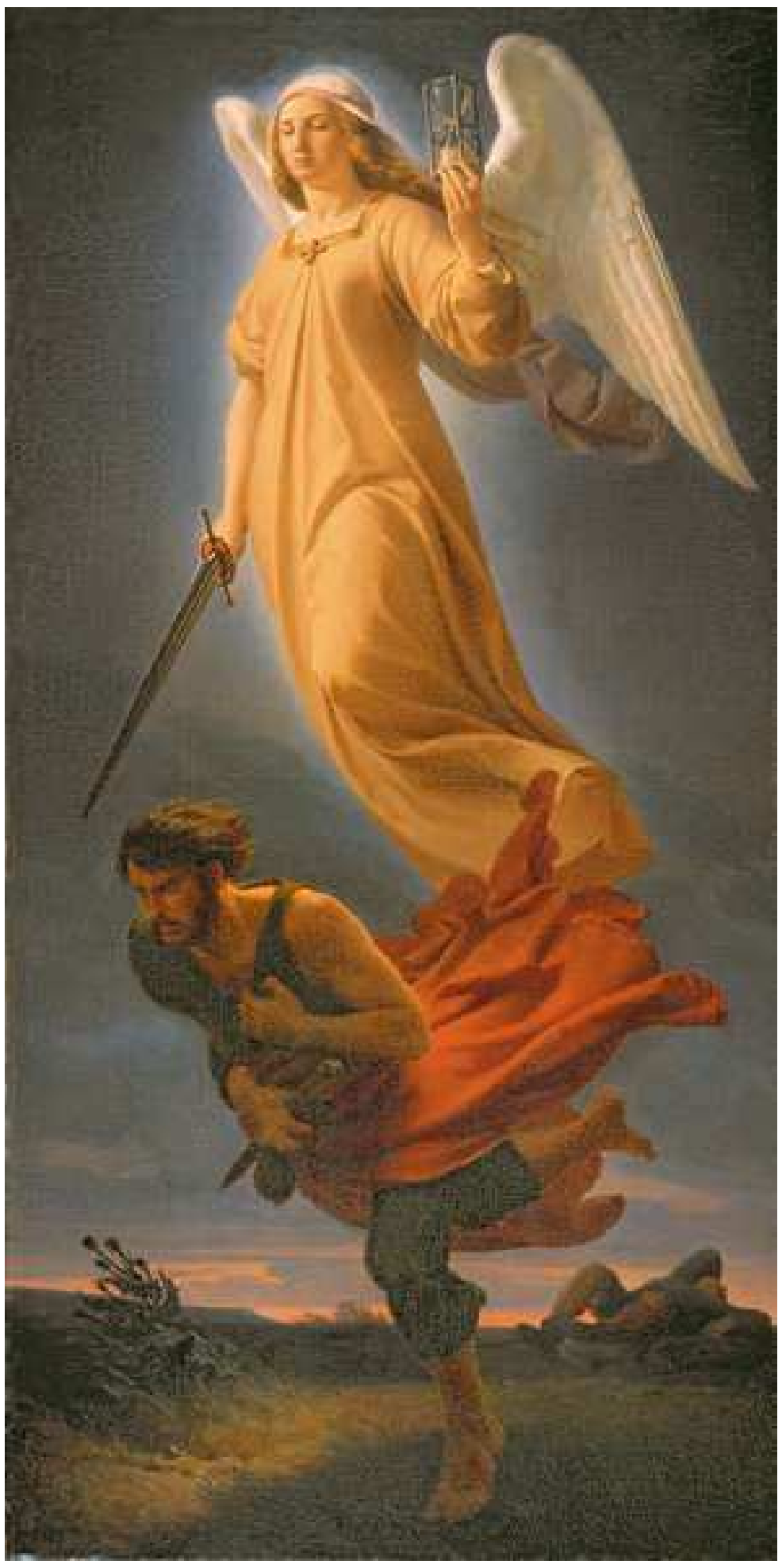

Abb. B.4: *Nemesis*, Alfred Rethel, 1837. Öl auf Leinwand.

Was haben aber nun die Dinosaurier und all die anderen ausgestorbenen Tiere und Pflanzen verbrochen, dass Nemesis an ihnen Vergeltung übte? Der Vater der Nemesis-Hypothese, Richard Muller, Professor für Physik an der University of California in Berkeley, schreibt über die Namensgebung:

> Für mich spiegelt der Name die Tatsache wider, dass die Dinosaurier erfolgreiche Kreaturen waren, die durch ein Ereignis zerstört wurden, das vom Himmel kam.

Haben die Dinosaurier und die anderen Lebewesen es gewagt die Autorität der Götter in Frage zu stellen? Lassen Sie uns lieber einen genaueren Blick auf die Nemesis-Hypothese werfen.

Der Originalartikel (Davis, Hut, and Muller 1984) ist in der besagten *Nature* Ausgabe unter dem Titel "Extinction of species by periodic comet showers" – also etwa "Aussterben von Arten durch regelmässige Kometenregen" – erschienen. Er ist neben Muller noch von zwei weiteren Physikern unterschrieben: Marc Davis, einem theoretischen Astrophysiker der in Princeton und Harvard tätig war und jetzt, wie Muller, Professor für Physik in Berkeley ist und Piet Hut, auch Astrophysiker und Professor am Institute for Advanced Study in Princeton.

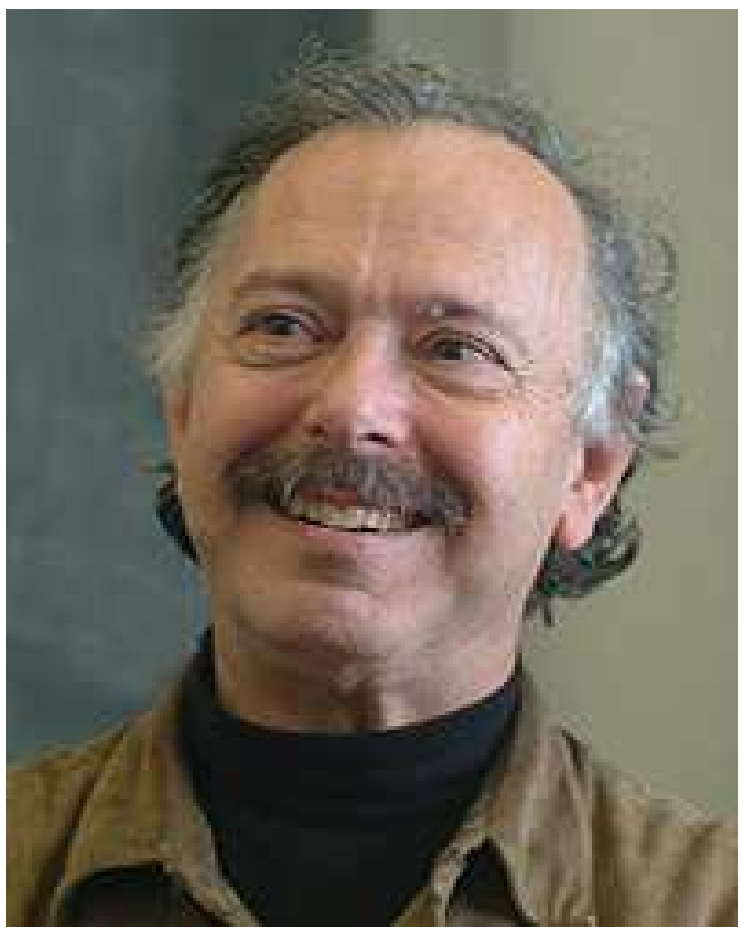

Abb. B.5: Richard A. Muller, Professor für Physik an der University of California in Berkeley. Er ist der Vater der Nemesis-Hypothese. Das Bild stammt von seiner Homepage: `http://muller.lbl.gov/`.

Die Idee ist, dass ein noch unentdeckter Stern auf einer lang gezogenen Bahn mit einer Umlaufzeit von etwa 26 Millionen Jahren um unsere Sonne kreist (Abb. B.6). Demnach kommt dieser Stern alle 26 Millionen Jahre uns so nahe, dass er durch die Oortsche Kometenwolke fliegt. Die Oortsche Wolke ist eine riesige Ansammlung von Myriaden von Kometenkernen, welche unser Sonnensystem schalenförmig in einem Abstand von etwa 50 000 AU umschliesst. (Eine Astronomische Einheit ist der mittlere Abstand zwischen Erde und Sonne – etwa 150 Millionen Kilometer.) Die Oortsche Wolke wurde 1950 vom niederländischen Astronomen Jan Hendrik Oort postuliert, um den Ursprung langperiodischer Kometen zu erklären. Obwohl die Oortsche Wolke nicht direkt beobachtet werden kann, zweifelt heute kaum ein Astronome an ihrer Existenz.

Nemesis wird also alle 26 Millionen Jahre, so die Idee, die Oortsche Wolke durchfliegen und dabei zahlreiche Kometenkerne aus ihrer Bahn werfen. Die meisten betroffenen Kometen werden in die Weiten des Weltalls katapultiert. Ein kleiner Teil hingegen, der immer noch mehrere hundert Millionen Kometen umfassen kann, wird in Erdrichtung abgelenkt, wo sie einen fürchterlichen “Kometensturm” hervorrufen. Normalerweise herrscht im Inneren des Sonnensystems “Windstille”, denn Jupiter und Saturn sorgen dafür, dass Kometen die sich in die Nähe der Sonne verirrt haben, schleunigst wieder herausgeschleudert werden. Die riesigen Gasplaneten unseres Sonnensystem formen also, so zu sagen, das “Auge” des Kometensturms, worin wir uns im Moment befinden. So sind Schweifsterne, wie die Kometen auch genannt werden, relativ seltene Himmelserscheinungen – und das ist auch Gut so.

Zwar mag sich für einen irdischen Beobachten einen wundervollen Anblick ergeben, wenn jede Nacht mehrere neue Kometen sichtbar werden. Doch wird die scheinbare Idylle eines Kometensturmes mit Sicherheit nicht all zu lange andauern. Früher oder später wird einer der Kometen mit der Erde kollidieren mit katastrophalen Folgen für die Erdbewohner. Nach der Nemesis-Hypothese sollte also nicht nur ein Massenaussterbeereignis durch einen riesigen Einschlag verursacht worden sein, sondern alle!

Da seit dem letzten Massensterben etwa 11 Millionen Jahre vergangen sind, glauben wir, dass Nemesis, falls sie existiert, etwa einen halben Umlauf seit ihrer letzten Durchquerung der Oortschen Wolke durchlaufen hat und sich daher nahe dem Punkt grösster Entfernung zur Sonne, ihrem Aphel, befindet. Nach Keplers Gesetz müsste Nemesis eine Entfernung von etwa 23 Billionen Kilometer (2,4 Lichtjahre) haben. Das ist gut die Hälfte der Entfernung zu Proxima Centauri, des sonnennächsten bekannten Sterns.

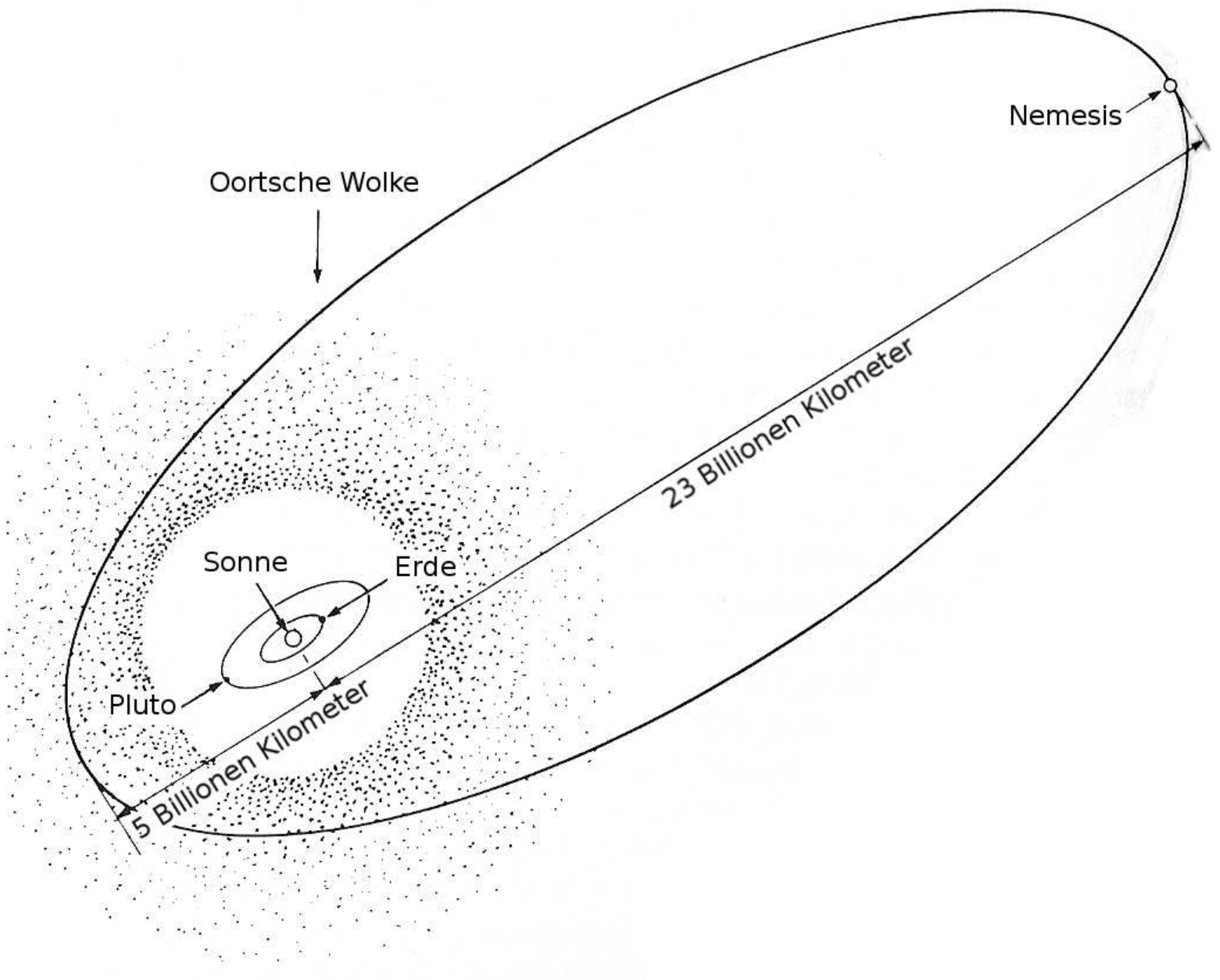

Abb. B.6: Das Sonnensystems mit der Sonnenbegleiterin Nemesis. Der innere Teil der Oortschen Wolke ist frei von Kometenkernen und bildet das "Auge" des Sturmes. Die Grösse des Sonnensystems ist stark übertrieben dargestellt. Abbildung nach Muller (1988), modifiziert durch den Autor.

Als Richard Muller im Streitgespräch mit Luis Alvarez die Idee eines Sonnenbegleiters in den Sinn kam, war die erste Frage, die sich die beiden Physiker stellten, ob denn diese ungewöhnlich weite Umlaufbahn überhaupt stabil sein kann. Je weiter zwei massereiche Körper voneinander entfernt sind, desto schwächer sind sie durch die Gravitationskraft aneinander gebunden. Bei dem riesigen Orbit, den Nemesis haben müsste, um die Sonne mit einer Umlaufszeit von 26 Millionen Jahren zu umkreisen, scheint die Frage nach der Stabilität der Umlaufbahn berechtigt. Zunächst stellt sich die Frage, ob der Orbit gegenüber den Gezeitenkräften der Milchstrasse stabil ist. Jedes ausgedehnte Objekt, das sich in einem (inhomogenen) Gravitationsfeld befindet, erfährt eine Gezeitenkraft. Meine Füsse sind beispielsweise etwas näher bei der Erde als mein Kopf. Daher werden meine Füsse auch ein bisschen stärker an die Erde gezogen als mein Kopf. Unter dem Strich ergibt das eine Kraft, die mich etwas auseinander zieht. Natürlich ist diese Kraft winzig und spielt im Alltäglichen Leben kaum eine Rolle. Anders ist das aber bei den Gezeitenkräften des Mondes, die auf der Erde das bekannte Phänomen von Ebbe und Flut verursachen.

Auch ein Doppelsternsystem, das im Gravitationsfeld der Milchstrasse kreist, erfährt eine Gezeitenkraft. Derjenige Stern, der sich näher zum Galaktischen Zentrum befindet, wird etwas stärker in dessen Richtung gezogen als sein Begleiter, einfach weil es dort mehr Sterne hat als im Aussenbereich der Milchstrasse. In diesem Bild (Abb. B.7) nehmen wir einfach mal an, dass das Galaktische Zentrum irgendwo weit entfernt rechts oben liegt. Die Länge der Pfeile entspricht der Anziehungskraft der Milchstrasse. Wir sehen, dass Nemesis etwas stärker zum Zentrum der Milchstrasse gezogen wird als die Sonne. Auch hier ergibt sich eine Nettokraft, die die beiden Sterne auseinander zu ziehen droht. Nur wenn ihre gegenseitige Anziehungskraft gross genug ist, können sie sich gegen die Gezeitenkräfte der Milchstrasse behaupten und bleiben aneinander gebunden. Trotz des riesigen Durchmessers, ist Nemesis' Orbit stabil gegenüber den Gezeitenkräften der Milchstrasse. Dennoch sind die Gezeitenkräfte keineswegs zu vernachlässigen.

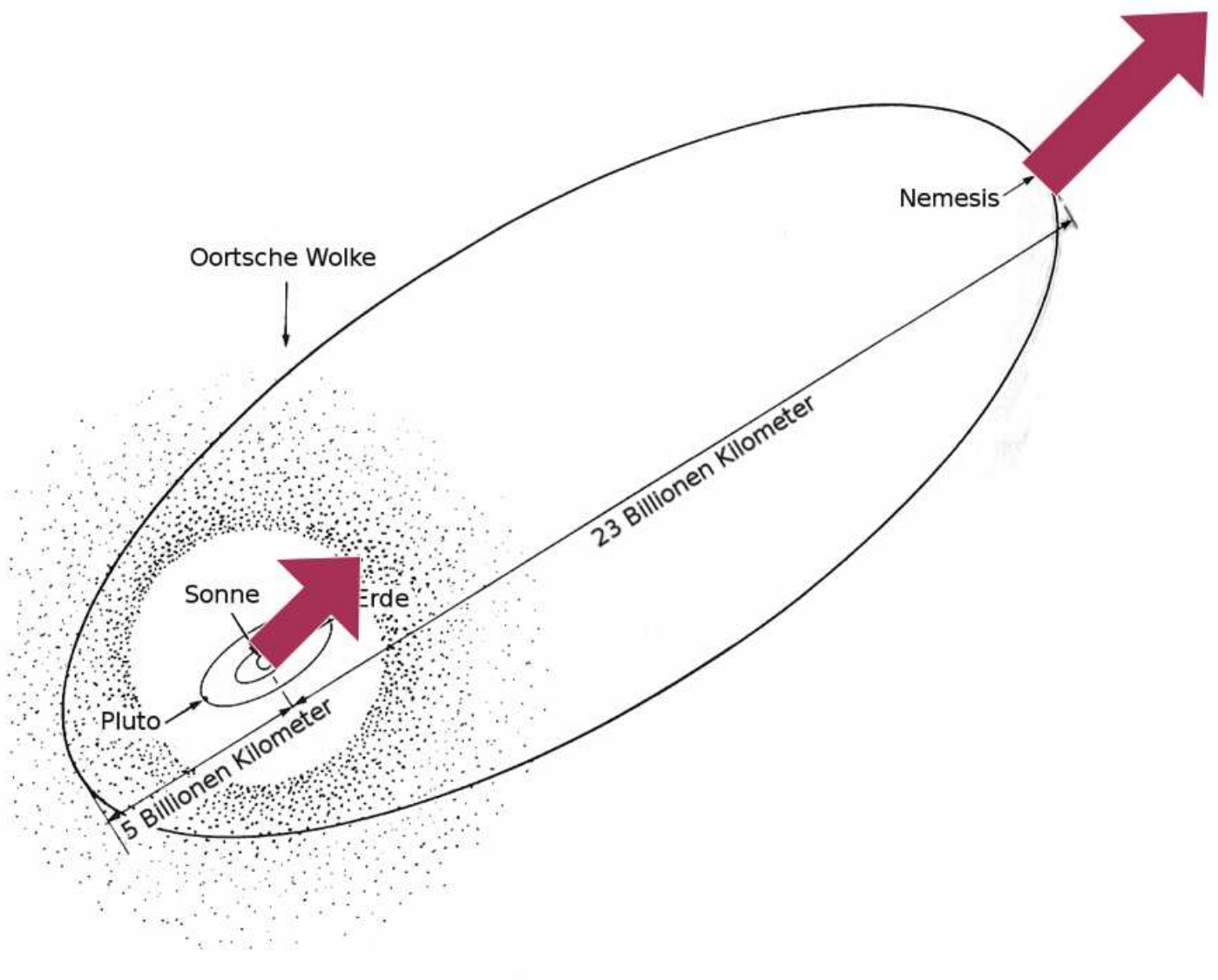

Abb. B.7: Gezeitenkräfte, welche auf das Sonne-Nemesis System wirken. Das gravitative Zentrum der Milchstrasse wurde willkürlich irgendwo weit entfernt rechts oben angenommen. Dann ist Nemesis etwas näher am Zentrum als die Sonne und wird folglich auch ein bisschen stärker angezogen. Dieser Gezeiteneffekt droht weite Doppelsterne auseinanderzureissen. Abbildung nach Muller (1988), modifiziert durch den Autor.

Der Nemesis- Orbit muss nicht nur momentan stabil sein, sondern über mindesten 250 Millionen Jahre, diejenige Zeitspanne, in welcher Raup und Sepkoski die periodischen Aussterbeereignisse beobachteten, mehr oder weniger unverändert bleiben. Wegen des grossen Abstandes ist Nemesis nur schwach an die Sonne gebunden und ihre Bahn ist deshalb anfällig auf Störungen durch vorbeiziehende Sterne. Ein naher Stern kann bei seinem Vorbeiflug Nemesis' Bahn leicht stören, wie hier (Abb. B.8) völlig übertrieben dargestellt. Es ist schwierig sich vorzustellen wie durch eine solch wiederholt gestörte Bahn ein regelmäiges Muster erzeugt werden soll.

Mit Hilfe von detaillierten Berechnungen konnte man zeigen, dass der gemeinsame Einfluss naher Sterne und der Gezeitenkräfte dazu führt, dass Nemesis' Umlaufbahn langsam anwächst und somit immer weniger stabil wird. Schliesslich wird Nemesis sich von der Sonne loslösen und in die Weiten des Weltalls entschwinden. Man schätzt, dass dies in etwa einer Milliarde Jahre geschehen wird. Die Umlaufbahn von Nemesis scheint also auf sehr lange Sicht tatsächlich instabil zu sein. Zudem führt das Wachsen der Umlaufbahn natürlich dazu, dass die Umlaufzeiten immer länger werden.

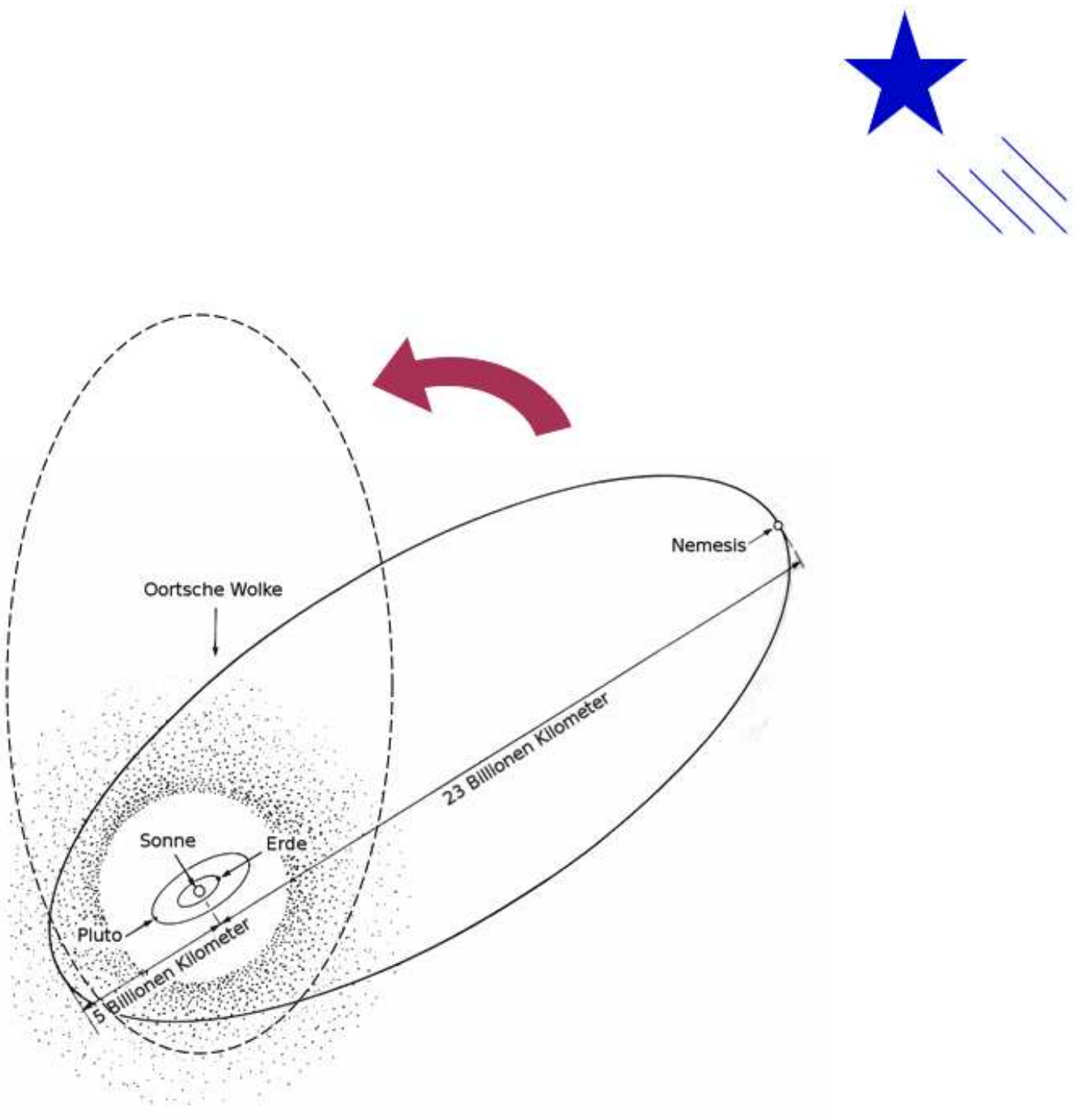

Abb. B.8: Ein vorbeifliegender Stern (blau) stört Nemesis' Bahn (stark übertrieben dargestellt). Abbildung nach Muller (1988), modifiziert durch den Autor.

Es muss betont werden, dass die Nemesis-Theorie niemals eine exakte Regelmäigkeit vorausgesagt hatte. Die Störungen von vorbei fliegenden Sternen verursachen Schwankungen in der Periodizität und, wie schon gesagt, einen leichten Trend hin zu längeren Umlaufzeiten. Dies steht aber nicht unbedingt im Widerspruch zu den fossilen Befunden.

Das genaue Alter von Sedimenten zu bestimmen ist kein leichtes Unterfangen und Unsicherheiten von mehreren Millionen Jahren sind durchaus üblich. In der Zeitspanne von 250 Millionen Jahren erwartet man eine Zunahme der Umlaufszeit von lediglich 2 Millionen Jahren – zu wenig um bei den Unsicherheiten geologischer Altersbestimmungen ins Gewicht zu fallen. Auch die Schwankungen der Umlaufszeit, welche auf etwa 15 Prozent geschätzt wurden, sind durchaus mit den Unsicherheiten der Gesteinsdatierungen verträglich (Hut 1984).

Zusammenfassend kann man die dynamische Geschichte Nemesis' wie folgt erzählen: Vor etwa 4.5 Milliarden Jahren entstand unser Sonnensystem und mit ihm zusammen auch Nemesis. Zu Beginn hatte Nemesis einen viel kleineren Abstand zur Sonne und eine Umlaufszeit von nur etwa 2 Millionen Jahren. Die steten Störungen der benachbarten Sterne und der Einfluss der Gezeitenkräfte hat in Folge den Orbit "aufgeweicht" bis die Umlaufszeit den heutigen Wert von etwa 26 Millionen Jahren erreicht hat. In Zukunft wird sich Nemesis' Bahn weiter vergrössern und in etwa einer Milliarde Jahren wird Nemesis das Sonnensystem für immer verlassen haben.

Die Berechnungen haben also gezeigt, dass ein solcher Orbit über den fraglichen Zeitraum stabil sein kann, also prinzipiell möglich ist und nicht im Widerspruch zu den geologischen Befunden steht. Trotz der Unsicherheiten in der Datenlage ist es keineswegs leicht ein Modell zu finden, das nicht im Widerspruch zu irgend einer Beobachtungstatsache steht. "Beobachtungstatsache" – das führt mich zu einer drängenden Frage, die Sie sich wahrscheinlich schon lange stellen: "Kann es sein, dass wir Nemesis übersehen haben?"

Die Kurzantwort auf diese Frage lautet: "Ja." Zwar wurden praktisch alle nahen Sterne beobachtet und katalogisiert, aber nur bei den aller hellsten wurde auch die Distanz gemessen. Es kann also durchaus sein, dass wir zwar ein Bild von Nemesis haben, aber nicht bemerkt haben, wie nahe dieser Stern unserer Sonne ist.

Man kann grob abschätzen wie hell Nemesis etwa ist. Falls Nemesis existiert, ist sie ein relativ leuchtschwacher Stern, denn wäre sie heller, hätten wir ihre Distanz bereits gemessen. Sie ist also sicherlich viel leuchtschwächer als die Sonne und daher auch viel kleiner und masseärmer. Andererseits darf Nemesis aber auch nicht zu massearm sein, denn sonst würde sie nicht genügend Kometenkerne in der Oortschen Wolke stören, um auf der Erde einen Kometenregen zu verursachen. Wägt man alle Parameter gegeneinander ab, kann man schliessen, dass Nemesis höchstwahrscheinlich ein Brauner Zwerg ist.

Braune Zwerge sind streng genommen schon zu massearm um als Stern gelten zu können. Vielmehr bilden sie das Bindeglied zwischen Sternen und Planeten. Mit einer Masse von weniger als 5% der Sonnenmasse kann ein Brauner Zwerg nicht genügend Druck und Temperatur in seinem Inneren aufbauen um Wasserstoff zu Helium zu verschmelzen. Dieser nukleare Prozess ist charakteristisch für Sterne. Er stabilisiert sie und bringt sie zum leuchten. Dennoch sind Braune Zwerge nicht völlig schwarz, denn ihre Gravitationsenergie reicht aus um sie zum Glimmen zu bringen. Sie sind aber im allgemeinen viele 10 000 male leuchtschwächer als die Sonne und daher sehr schwierig zu beobachten.

Zu der äusserst schwachen Leuchtkraft kommt noch die weitere Schwierigkeit hinzu, dass Nemesis wohl kaum eine Eigenbewegung aufweist. Die meisten nahen Sterne fallen durch ihre relative grosse Eigenbewegung auf. Sie kennen das vom Zug fahren: Wenn Sie aus dem Fenster schauen, sehen Sie wie die nahen Objekte rasch an Ihnen vorbeihuschen, während Sie die in der Ferne liegende Landschaft geniessen können, die nur allmählich vorbeizieht. Aus dem gleichen Grund scheinen weit entfernte Sterne praktisch unbeweglich, weswegen sie auch Fixsterne genannt werden. Nähere Sterne hingegen bewegen sich sehr langsam aber durchaus messbar am Himmelszelt.

Nemesis wäre zwar sehr nahe, aber weil sie an die Sonne gebunden ist, bewegt sie sich mit ihr durch die Milchstrasse. So wie ein in gleicher Richtung und gleicher Geschwindigkeit auf dem Nachbargleis fahrender Zug für Sie scheinbar still steht, so wird auch Nemesis keine Eigenbewegung vorweisen und daher auch nicht besonders auffallen.

Man mag sich vielleicht fragen, ob denn Doppelsternsysteme bekannt sind, die vergleichbare Eigenschaften wie das Sonne-Nemesis-System haben, oder ob wir ein Spezialfall wären. Doppelsternsysteme, also zwei durch die Gravitationskraft aneinander gebundene

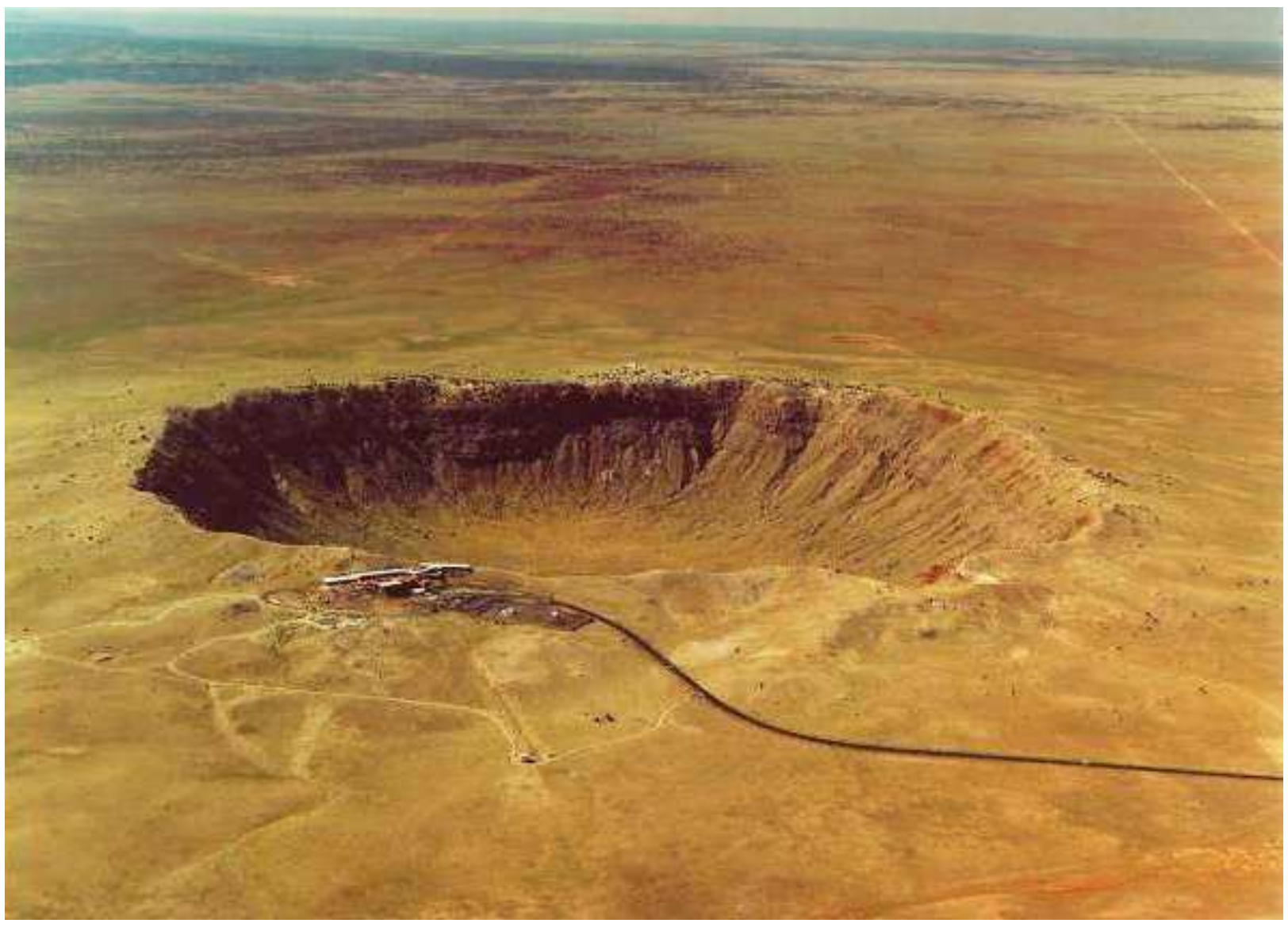

Abb. B.9: Der Barringer Krater in Arizona, USA. Mit 50 000 Jahren ist er viel zu jung um auf das Konto von Nemesis zu gehen. Das Bild stammt von `http://www.donnersteine.de/`. © Stephan Decker.

Sterne, gibt es sehr viele. Die meisten davon sind aber viel näher zusammen als Nemesis von der Sonne entfernt sein müsste. Doppelsterne bei welchen die beiden Sterne eine Abstand von über zwei Lichtjahren voneinander haben, sind sehr selten in unserer Galaxis – kommen aber vor. Solche weiten Doppelsterne beherbergen jedoch, soweit bekannt ist, keine Braunen Zwerge. Aber angesichts der Schwierigkeiten Braune Zwerge zu beobachten, verwundert es aber auch nicht, dass wir noch keine in weiten Doppelsternsystemen entdeckt haben.

Viele Wissenschaftler sind der Ansicht, dass eine neue Theorie nur dann brauchbar ist, wenn es die Möglichkeit gibt sie zu widerlegen. Gibt es also irgendwelche Konsequenzen der Nemesis-Theorie, die wir überprüfen könnten? Wenn wirklich in regelmässigen Abständen von 26 Millionen Jahren ein Kometenregen auf der Erde niederging, dann müssten Spuren davon finden sein.

Der Barringer-Krater in Arizona geht mit Sicherheit nicht auf das Konto von Nemesis. Mit einem Alter von nur etwa 50'000 Jahren ist er viel zu jung. Er ist aber für die Nemesis-Theorie insofern von Belang, als dass er der erste Krater ist, bei welchen man schlüssig zeigen konnte, das er durch einen Meteoriteneinschlag entstanden ist. Die älteren Einschlagskrater auf der Erde sind weit weniger auffällig. Sie wurden durch Wind und Wasser Erosion sowie durch kontinentale Plattenverschiebungen buchstäblich vom Erdboden verschluckt. Einige alte Krater haben aber trotz allem überlebt und etwa hundert wurden bisher von den Forschern entdeckt und datiert. Als die Geologen sich die Verteilung der Alter dieser Krater anschauten, stellten sie fest, dass die grössten Krater sich mit ungefähr der gleichen 26 Millionen Jahren Periodizität häuften als die Massenaussterbeereignisse (Alvarez and Muller 1984). Da die Alter der Krater aber nur sehr ungenau bestimmt sind, ist dieses Ergebnis noch sehr umstritten in der Fachwelt. Es zeigt jedoch, dass man die Nemesis-Hypothese im Prinzip testen und auch falsifizieren kann.

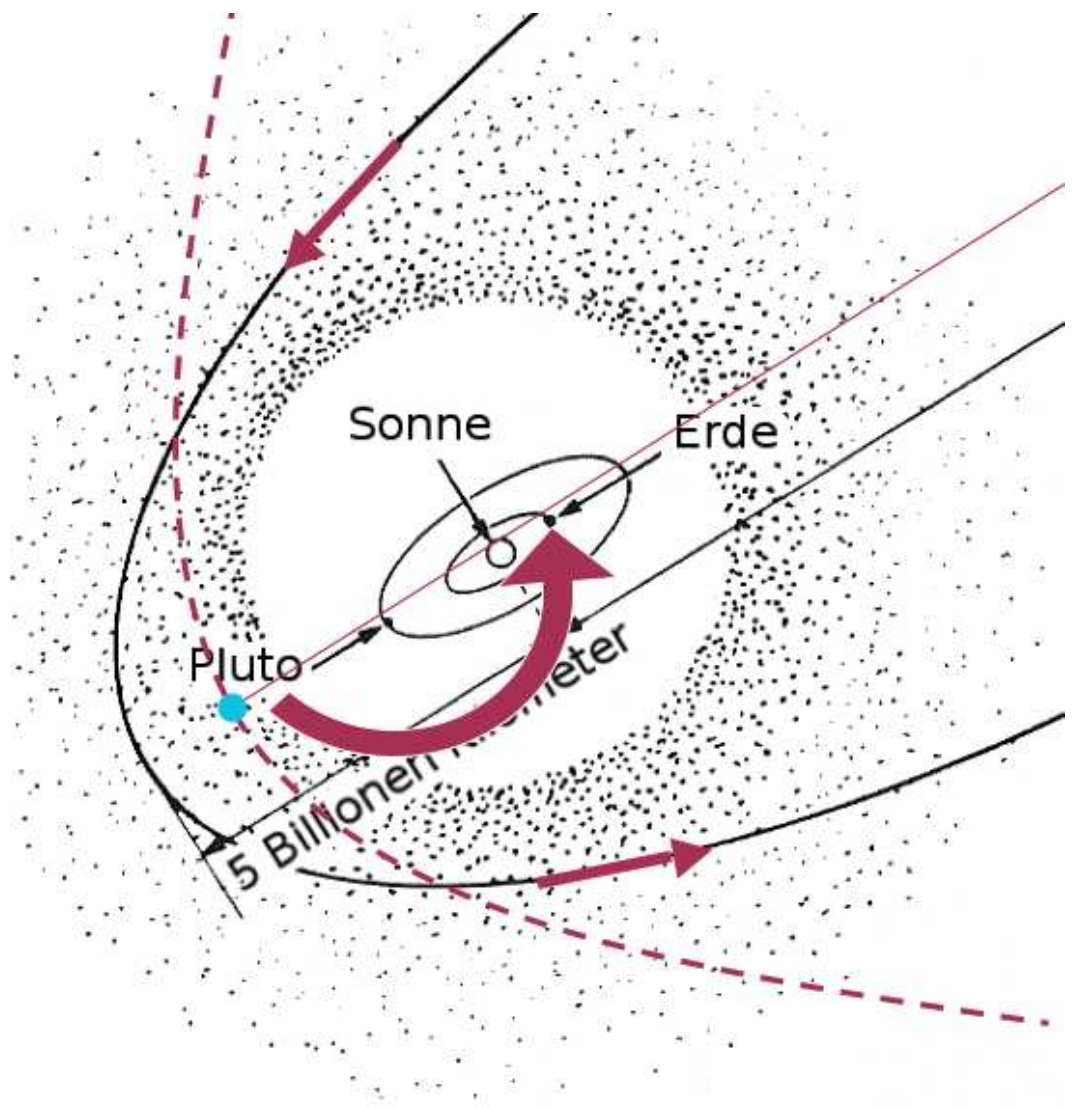

Abb. B.10: Ein Objekt, das durch die Oortsche Wolke fliegt, verleiht den Kometenkernen einen bevorzugten Drehsinn um die Sonne. Die Bahnebene dieses Objektes definiert auch die Ebene, in welcher die Kometen kreisen. Nemesis' Perihel (blauer Punkt in verbesserter Darstellung) ist gleichsam das Aphel der Kometenbahnen. Abbildung nach Muller (1988), modifiziert durch den Autor.

Ein weiterer Hinweis auf die Existenz von Nemesis könnte von den Kometen kommen. Ein massereicher Körper, der durch die Oortsche Wolke fliegt, zieht die dortigen Kometenkerne ein Stück mit sich. Somit erhalten die Kometen eine bevorzugte Richtung, in der sie um die Sonne kreisen (Abb. B.10). Man muss hier zu diesem Schema anmerken, dass ist der Orbit von Nemesis schlecht gezeichnet ist; er sollte vielmehr so aussehen, dass der sonnennächste Punkt, das Perihel (hier blau eingezeichnet), auf der grossen Achse der Ellipse liegt. Dieser Punkt darf man auch in erster Näherung als Ausgangspunkt des Kometensturms betrachten.

Nemesis verursacht also einen "Kometenwirbelsturm", der vorzugsweise in die gleiche Richtung dreht, wie Nemesis um die Sonne kreist. Wir würden also erwarten, dass es einen überschuss an langperiodischen Kometen gibt, die in einer bestimmten Ebene in einem bestimmten Sinn kreisen. Diese Ebene wäre dann natürlich identisch mit der Ebene, in welcher Nemesis ihre Bahnen zieht. Weiterhin würden wir erwarten, dass die sonnenentferntesten Punkte dieser Kometenbahnen, deren Aphelien, sich an einer bestimmten Position am Himmel häufen; und zwar an jener Stelle, wo sich Nemesis' Perihel befinden würde.

Es liegt also auf der Hand, dass man die bekannten Kometenbahnen etwas genauer betrachten sollte. Dies wurde auch gemacht und man hat tatsächlich eine Gruppe von langperiosischen Kometen gefunden, die alle ungefähr in der selben Ebene im gleichen Sinn kreisen und deren Aphelien sich an einer Stelle am Himmel zu häufen scheinen (Delsemme 1986). Mit Hilfe dieser Kometengruppe konnte man berechnen, dass die Störung der Oortsche Wolke, die diese Gruppe erzeugt hatte, vor höchstens 15 Millionen Jahren gewesen sein kann, was nicht im Widerspruch zur Nemesis-Hypothese steht. Ausserdem zeigen die Rechnungen, dass das Objekt, welches durch die Oortsche Wolke flog sehr langsam unterwegs gewesen sein muss. Ansonsten hätten die Kometen, salopp gesprochen, keine Zeit gehabt auf den fahrenden Zug aufzuspringen und von ihm mitgenommen zu werden. Wegen den Unsicherheiten der Beobachtungen kann man nicht genau sagen wie langsam das Objekt gewesen sein muss. Aber wahrscheinlich war es langsamer als die Fluchtgeschwindigkeit, die es benötigt unser Sonnensystem zu verlassen, was nichts anderes heisst, als dass das Objekt um unsere Sonne kreist.

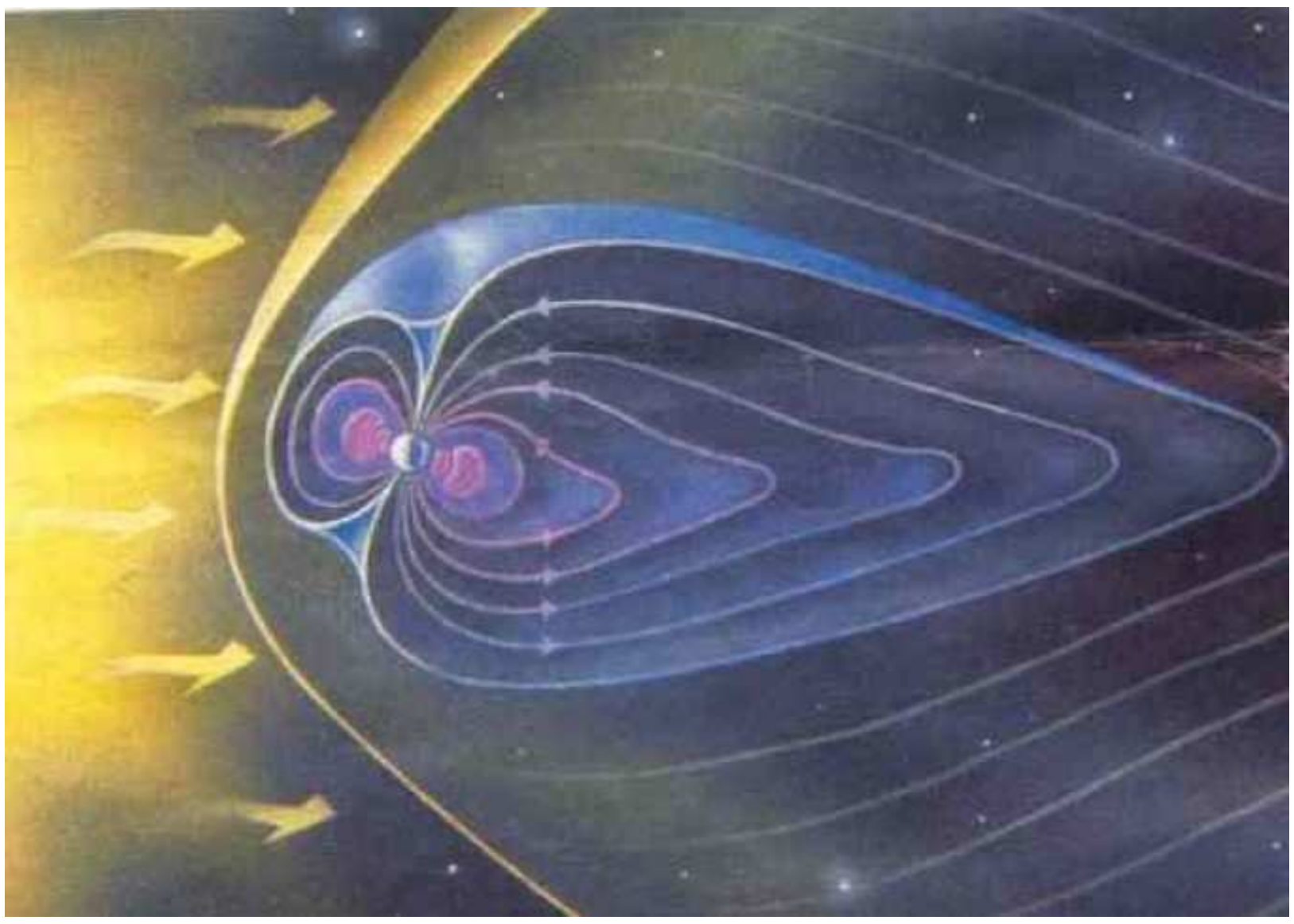

Abb. B.11: Das Erdmagnetfeld wirkt als Schutzschirm gegen die energiereichen Teilchen des Sonnenwindes. Das Bild stammt von `http://surf.to/planeten`.

Es gibt noch weitere Indizien, die für die Nemesis-Hypothese sprechen. So haben manche Wissenschaftler versucht die rätselhafte Bahn von Sedna, einem transneptunischen Objekt, das 2003 entdeckt wurde, mit einem noch nicht entdeckten Sonnenbegleiter zu erklären.

Mein Lieblingsindiz ist wohl gleichzeitig auch das umstrittenste: Erdmagnetfeldumkehrungen. Wie Sie wissen, zeigt die Nadel eines Kompasses in Richtung Norden. Das war aber nicht immer so. Aus den Gesteinsschichten können wir ablesen, dass im wieder in der Erdgeschichte die Kompassnadel nach Süden gezeigt hätte. Im Mittel finden solche Erdmagnetfeldumkehrungen etwa alle 250 000 Jahre statt. Nun gibt es Hinweise, dass sich Erdmagnetfeldumkehrungen im gleichen 26 Millionen Jahren Rhythmus wie die Massenaussterbeereignisse häufen. Es wurde auch ein plausibler Mechanismus vorgeschlagen, der erklärt wie ein Meteoriteneinschlag das Erdmagnetfeld vorübergehend ausschalten kann (Muller and Morris 1986). Schaut man sich die Verformung des Erdmagnetfeldes unter dem Einfluss des Sonnenwindes an (Abb. B.11), möchte man sich nicht ausmalen, was ein Versagen dieses Schutzschildes für Folgen für das Leben auf der Erde hätte.

Man darf aber nicht vergessen, dass geologische Befunde oft mit grossen Unsicherheiten behaftet sind und man sie deshalb auf vielfältige Art und Weise interpretieren kann. Deshalb sind auch die meisten Wissenschaftler sehr skeptisch gegenüber der Idee eines Sonnenbegleiters. Ein gesundes Mass an Skepsis ist hier auch sicherlich angebracht.

Der ultimative Test der Nemesis-Theorie ist natürlich die Entdeckung des Sonnenbegleiters selbst. Der Schlüssel zur Entdeckung von Nemesis ist sicherlich ihre grosse Nähe zu uns und folglich ihre grosse Parallaxe. Als Parallaxe bezeichnet man die scheinbare änderung der Position eines Objekts wenn der Beobachter sich bewegt. Wenn Sie Ihren ausgestreckten Zeigefinger vor Ihr Gesicht halten und ihn abwechslungsweise einmal mit dem linken und einmal mit dem rechten Auge betrachten, können Sie beobachten wie Ihr Finger vor dem weiter entfernten Hintergrund hin und her springt. Genauso springt ein naher Stern gegenüber den weit entfernten hin und her, wenn wir ihn einmal im Winter und einmal im Sommer beobachten, weil die Erde in dieser Zeit von einer Seite der Sonne zur gegenüberliegenden gewandert ist.

Die Parallaxe ist umso grösser je näher der Stern ist. In der Astronomie wird die Parallaxe normalerweise in Bogensekunden angegeben. Ein Bogensekunde ist der 3 600ste Teil eines Winkelgrades. In einer Entfernung von 2.4 Lichtjahren würde Nemesis eine Parallaxe von etwa 1.3 Bogensekunden aufweisen. Das ist in etwa der Winkel unter dem man einen Golfball auf der Spitze des Empire State Building in New York sieht – und zwar von hier aus. Das hört sich nach wenig an, ist aber heutzutage problemlos messbar.

Neben der schwachen Leuchtkraft und der kleinen Eigenbewegung kommt noch die Schwierigkeit hinzu, dass wir nicht genau wissen wo wir am Himmel nach Nemesis suchen sollen. Wo sollen wir also anfangen zu suchen? Die vorher erwähnten Computerrechnung haben gezeigt, dass Nemesis' Bahn stabiler ist, wenn sie in der Galaktischen Ebene liegt. Der Grund hierfür sind hauptsächlich die Gezeitenkräfte der Milchstrasse. Der Nemesis-Orbit darf also bezüglich der Galaktischen Ebene nicht zu stark geneigt sein. Eine Inklination von mehr als 30 Grad scheint mit den geologischen Befunden nur schwer vereinbar zu sein (Torbett and Smoluchowski 1984). Wir erwarten also, dass Nemesis irgendwo vor dem Hintergrund der Milchstrasse zu finden ist, was die Suche nicht gerade einfacher macht.

Auch mithilfe der Bahnen der vorher erwähnten Gruppe langperiodischer Kometen kann man die Position von Nemesis einschränken. Diese Kometenbahnen weisen eine mittlere Neigung von etwa 28 Grad gegenüber der Milchstassenebene auf. Nehmen wir an, dass diese Kometengruppe von Nemesis geschickt wurden, können wir schliessen, dass Nemesis in der selben Ebene um die Sonne kreist. Eine Neigung von 28 Grad liegt, was die Stabilität der Bahn angeht, gerade noch drin. Weiterhin weist uns die Konzentration der Kometen-Aphelien am Himmel die Stelle von Nemesis' Perhihel. Da wir wissen, dass Nemesis vor etwa 11 Millionen Jahren zuletzt der Sonne am nächsten war, können wir also ihre heutige Position am Himmel grob abschätzen.

So kann man schliessen, dass Nemesis, falls sie existiert, wahrscheinlich im Sternbild des Drachen zu finden ist, das ganz in der Nähe des kleinen Bären, auch kleiner Wagen genannt, liegt. Das systematische Suchprogramm, dass von Muller und anderen Mitte der 80er Jahre begonnen wurde, wurde wegen technischen Schwierigkeiten bald darauf wieder eingestellt. Glücklicherweise sind für die nahe Zukunft mehrere Himmelsdurchmusterungen geplant, wie zum Beispiel "Pan-STARRS", dass nächstes Jahr auf Hawaii in Betrieb gehen soll und "LSST" das voraussichtlich ab 2014 von Chile aus den Himmel durchforsten wird. Falls Nemesis existiert, wird sie durch einen dieser Himmelsdurchmusterungen gefunden werden. Wird sie dabei nicht gefunden, darf ihre Existenz mit ruhigem Gewissen ausgeschlossen werden.

Falls wirklich ein Begleiter der Sonne gefunden wird, wäre das eine Sensation. Die Bedeutung einer solchen Entdeckung kann nicht genug betont werden. Nicht nur dass wir die Geschichte der Entwicklung des Sonnensystem neu schreiben müssten, auch die Evolution des Lebens auf der Erde würde plötzlich in einem völlig neuen Licht stehen. Es zeigt sich, dass die Zeit nach einem Massenausterben geprägt ist von einem munteren Spriessen neuen Lebens. Arten, die die Katastrophe überlebten, füllen in Windeseile die frei gewordenen ökologischen Nischen. Neue Arten entstehen – fast so rasch wie sie zuvor verschwanden. Manche Evolutionsbiologen halten es durchaus für möglich, dass die Evolution stagniert wäre, wenn nicht wiederholt ein Grossteil des Lebens vernichtet worden wäre.

Es sieht so aus, als ob zu Darwins Prinzip, "survival of the fittest", dem Überleben des am besten Angepassten, noch ein zweites wichtiges evolutionäres Prinzip hinzukommt: "survival of the first", das Überleben des Ersten. Neue Arten können nicht gedeihen wenn bereits alle Nischen besetzt sind. Wären die Dinosaurier nicht ausgestorben, hätten sich die Säugetiere wahrscheinlich niemals so erfolgreich ausbreiten können und auch wir Menschen hätten uns wahrscheinlich nicht entwickelt. Die Erkenntnis, dass die Entwicklung des Lebens regelmässig "von aussen" entscheidend beeinflusst wurde, ist von grösster Bedeutung. In einem Artikel aus der New York Times liest man im Jahre 1984 vor dem Hintergrund des Kalten Krieges:

> Möglicherweise hat uns Darwin dazu verleitet zu denken, dass unser überleben vom Konkurrenzkampf mit anderen Lebewesen abhängt. Die Existenz von Nemesis, dem Todesstern, könnte genau das sein, was unsere Spezies braucht um zu erkennen, dass die wahre Bedrohung unserer Existenz nicht andere Menschen sind.

Auch bei der Entstehung des Lebens selbst könnte Nemesis eine Rolle gespielt haben. So wird von manchen Wissenschaftlern vermutet, dass das gesamte Wasser und die Atmosphäre unseres Planeten sowie die für die Entstehung von Leben nötigen chemische Ausgangsstoffe von Kometen stammt (Delsemme 2001).

Auch wenn Nemesis nicht gefunden wird, hat uns diese Geschichte doch gezeigt, wie eng die Evolution des Lebens auf der Erde mit unserer kosmischen Nachbarschaft verknüpft sein könnte. Sie zeigt auch, dass die grossen Rätsel der Wissenschaft nicht innerhalb einer einzigen Fachdisziplin gelöst werden können. Die Nemesis-Theorie überspannt viele Fachbereiche: Von der Astronomie über die Geologie bis hin zur Paläontologie. Ich glaube, dass genau einen solch interdisziplinärer Ansatz nötig ist, um das Geheimnis unseres Ursprungs zu lüften.

Dieses Jahr ist sowohl Galileo Galilei als auch Charles Darwin gewidmet. Beide Wissenschaftler haben unser Weltbild revolutioniert. Beide zeichneten sich durch zwei wichtige Eigenschaften aus (ausser dem Bart): Sie waren neugierig und hatten eine grosse Geduld. Zwei Eigenschaften, die auch auf Sie zutreffen, meine Damen und Herren. Sie waren neugierig genug um hier herzukommen und Sie hatten die Geduld mir bis zum Schluss zuzuhören.

Dafür danke ich Ihnen!

## B.3 Cycles in fossil diversity and extragalactic cosmic rays

This talk was given in the Journal Club at the Physics Department of the University of Basel on March 12, 2009.

Gentlemen,

I welcome you to this year's spring semester journal club. As you know, 2009 is dedicated to two men, both of them has fundamentally changed the way we look at the world. Galileo Galilei founded modern astronomy; Charles Darwin founded the theory of evolution. On this occasion, I would like to start this year's journal club, with a topic that – in some sense – connects the fields of this two bearded men.

While looking for such a topic I came across two articles. First a Nature article by Robert Rohde and Richard Muller published almost exactly four years ago (Rohde and Muller 2005). In that article the authors reveal an unexpected and highly significant 62-Myr cycle in fossil diversity. They could, however, not find a plausible explanation for this cycle.

In the second paper, published 2007 in the Astrophysical Journal, Mikhail Medvedev and Adrian Melott propose a mechanism that explains the periodicity in biodiversity in terms of periodic cosmic ray (CR) enhancements (Medvedev and Melott 2007). Let me first briefly describe Rohde & Mullers study before I focus on Medvedev & Melotts astrophysical explanation of the periodicity in life's diversity.

The study by Rohde and Muller is based on Sepkoski's Compendium, which was published posthumously in 2002 (Sepkoski 2002)[1]. It is the most extensive compilation available, listing more than 36 000 marine genera together with the corresponding first and last stratigraphic appearances. Sepkoski's database has a very simple structure, which I illustrated here with the genus *Aviculopecten*:

D (Fame-u) – P (Dora)

These bivalve molluscs are first observed in the upper substage of the Famenian stage of the Devonian period, whereas its last appearance is in the Dorashamian stage of the Permian period. Here (Fig. B.12), we see a fossil of a member of Aviculopecten found in the lower Carboniferous.

Devonian, Carboniferous, Permian, and so on, are names given by geologists and paleontologists to distinguish between different layers of Earth. However, without assigning an age to those layers no analysis of the fossil diversity is possible. This becomes especially clear when we look at the definition of "diversity" used by Rohde & Muller: "Diversity is the number of distinct genera alive at any given time."

[1]Online accessible at `http://strata.geology.wisc.edu/jack/`

'Time'. We need to assign a geologic time scale to the stratigraphic layers. Such a time scale is provided by the International Commission on Stratigraphy (ICS). Rohde & Muller adopted the latest available at that time, the ICS2004 time scale. It gives dates for layers of Earth as far back as the beginning of the Cambrian period, which according to the ICS2004 time scale was 542 Myr ago. The dating is mainly based on the Potassium-argon technique; a radiometric dating method often used in geochronology and archeology. With this calibration Aviculopecten's first appearance in the upper Famian was between 364.3 and 359.2 Myr ago, whereas its last breath was somewhere between 253.8 and 251.0 Myr ago. Thus, Aviculopecten has lasted for more than 100 Myr. We see that Aviculopecten's first appearance, its origination, is given within 5 Myr, whereas its extinction is resolved within 3 Myr. Some genera in Sepkoski's database are given at a much lower resolution, at period level or even only at epoch level. Rohde & Muller exclude all such poorly resolved genera from the analysis. Furthermore, they exclude all genera known only from a single stratum. This refinement process discards fully half of the data set; about 18 000 well-resolved genera remain.

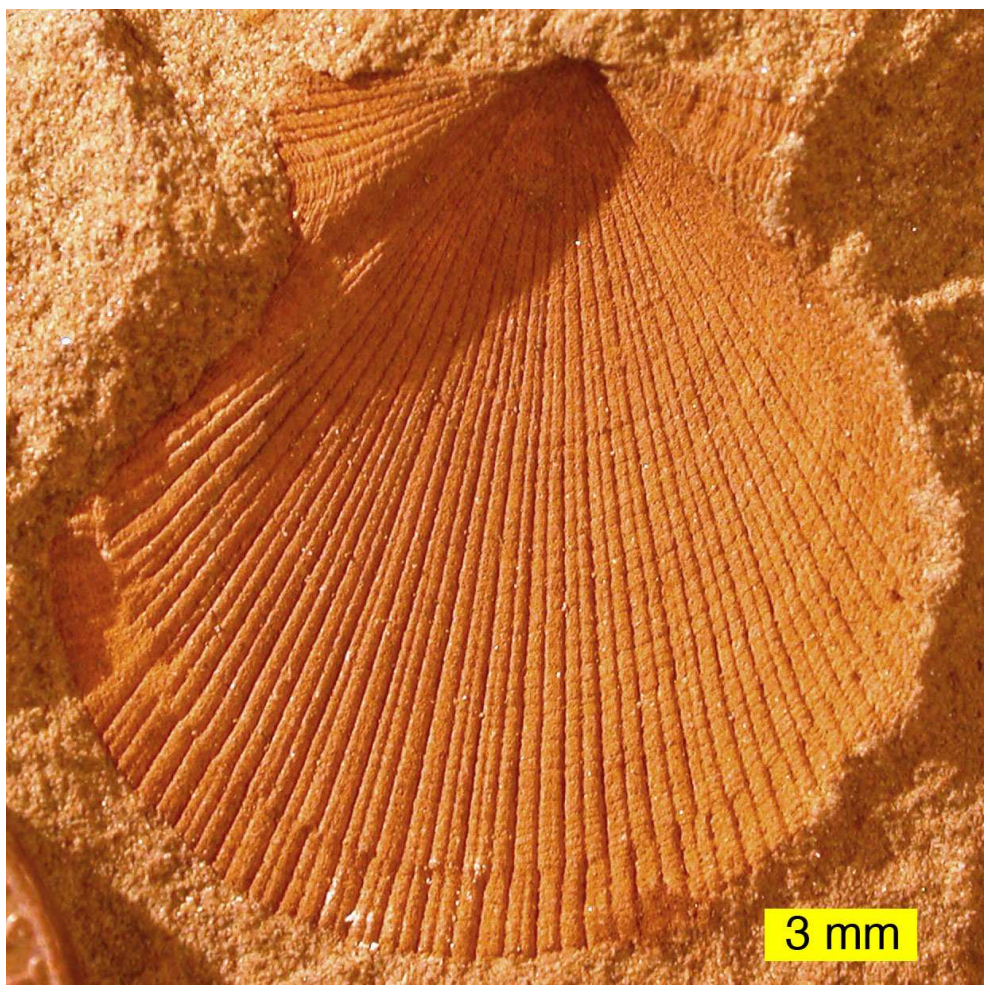

Figure B.12: *Aviculopecten subcardiformis* from the Logan Formation, Mississippian, Ohio. Photograph taken by Mark A. Wilson, Department of Geology, The College of Wooster.

The story about the question whether life's diversity drops and rises periodically, goes back to the 70s. Rohde & Muller revive this question and address it by means of Fourier analysis. Before doing a Fourier analysis, the authors make sure to filter out the lowest-frequency components of the signal. For this purpose, they calculate the best-fitting cubic polynomial and subtracted it from the data. This 'detrending' method imposes a lower limit to the Fourier spectrum, which corresponds in this case here to periods greater than 200 Myr. The low-frequency components would otherwise dominate the spectrum and obscure everything else.

But let us now finally look at Rohde & Muller's figure 1, which shows the fossil diversity against time (Fig. B.13). The green plot 'a' is for all genera in Sepkoski's Compendium. The black plot 'b' represents the data from the well-resolved subsample. The smooth blue curve is the best-fitting cubic polynomial, which is subtracted from the black plot to detrend the data. The residual plot 'c' left by this subtraction shows no large-scale features. This is the data that is submitted to Fourier analysis and its Fourier spectrum is shown in the inset 'e' as black line. A strong peak at a period of 62 Myr dominates the spectrum. There is a second spectral peak, with a period of 140 Myr.

Rohde & Muller determined the statistical significance of the cycles by estimating the background in two different ways, denoted by 'R' (red line) and 'W' (blue line), respectively. I do not want to go into the details of the background calculations here. However, on the basis of these backgrounds, the probability of observing a peak at least as strong as the 62-Myr and 140-Myr spectral peaks were computed and are shown in Rohde & Muller's table 1 (table B.1 here). Both the probability of finding the indicated peak at the specified frequency and more generally the probability of finding a similar peak at any frequency is considered. The 62-Myr peak is highly significant: the probability of occurring anywhere in the spectrum by chance is less than 1%. By contrast the 140-Myr cycle can plausibly result from purely random processes.

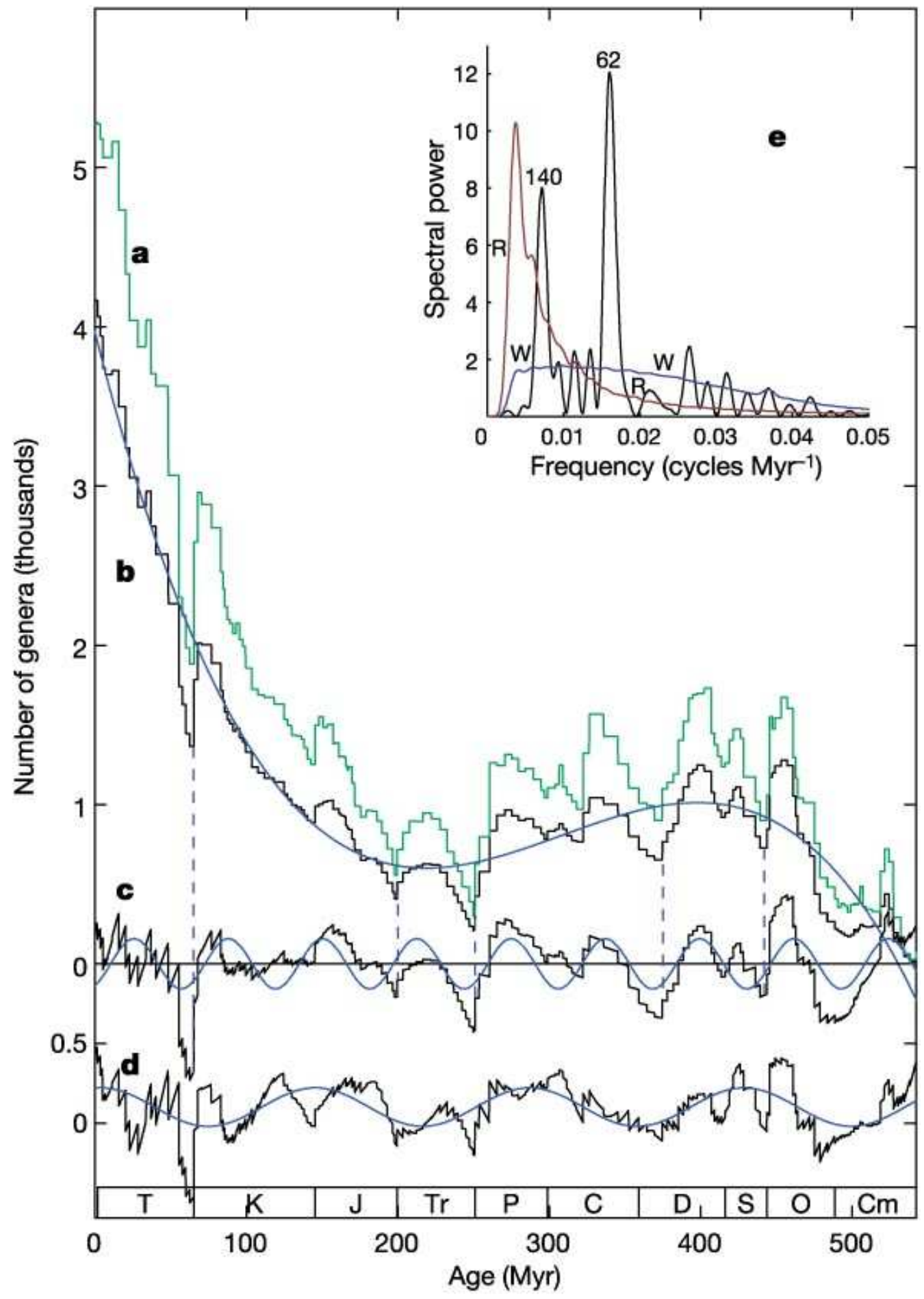

Figure B.13: Genus diversity reproduced from Rohde and Muller (2005). The green plot ('a') shows all genera from Sepkoski's Compendium, converted to the 2004 Geologic Time Scale. The black plot ('b') shows the same data, with single occurrence and poorly dated genera removed. The blue line is a cubic polynomial fitted to the data and used for 'detrending'. 'c' shows the detrended data from 'b' with a 62-Myr sine wave superimposed. 'd' shows the detrended data ('c') after subtraction of the 62-Myr cycle with a 140-Myr cycle overplotted. Dashed vertical lines indicated the five major extinctions. The insert ('e') shows the Fourier spectrum of 'c'. Estimates of the backgroud noise signal are shown as blue ('W') and red ('R') lines. At the bottom are shown conventional symbols for major stratigraphic periods. © Nature Publishing Group.

Rohde & Muller also show that the 62-Myr cycle is robust to various analysis techniques and emerges clearly also if all genera are included in the analysis, provided that the ISC2004 time scale was used. They emphasise that "observing a 62-Myr cycle in fossil diversity is a necessary and unavoidable consequence of combining the Sepkoski compendium and the ICS2004 geologic time scale". In fact, when an older time scale was used, the 62-Myr peak would be of only questionable significance.

Rohde & Muller discuss several processes that might cause a periodicity in the fossil data. However, none of them appears to be suitable to explain the strong 62-Myr cycle. We also do not know whether this cycle is a variation in true diversity or only in observed diversity. But either case requires explanation.

Table B.1: Likelihood of similar cycles (reproduced from Rohde and Muller 2005)

| Probability of peaks | At this frequency | | Anywhere in spectrum | |
|---|---|---|---|---|
| | R | W | R | W |
| 62 Myr | $< 5 \times 10^{-5}$ | $3.6 \times 10^{-4}$ | $< 0.0013$ | 0.010 |
| 140 Myr | 0.12 | 0.0056 | 0.71 | 0.13 |

As already mentioned, Medvedev & Melott have proposed an explanation of the 62-Myr cycle in terms of EGCRs. Let us briefly discuss how CRs might affect biodiversity on Earth.

CRs are high-energy charged particles that travel at nearly the speed of light and strike the Earth from all directions. When penetrating into the Earth's atmosphere, they produce avalanches of energetic secondary particles that are dangerous or lethal to some organisms (Fig. B.14). Especially muons can reach the Earth's surface and damage the DNA even in deep-sea and deep-earth animals.

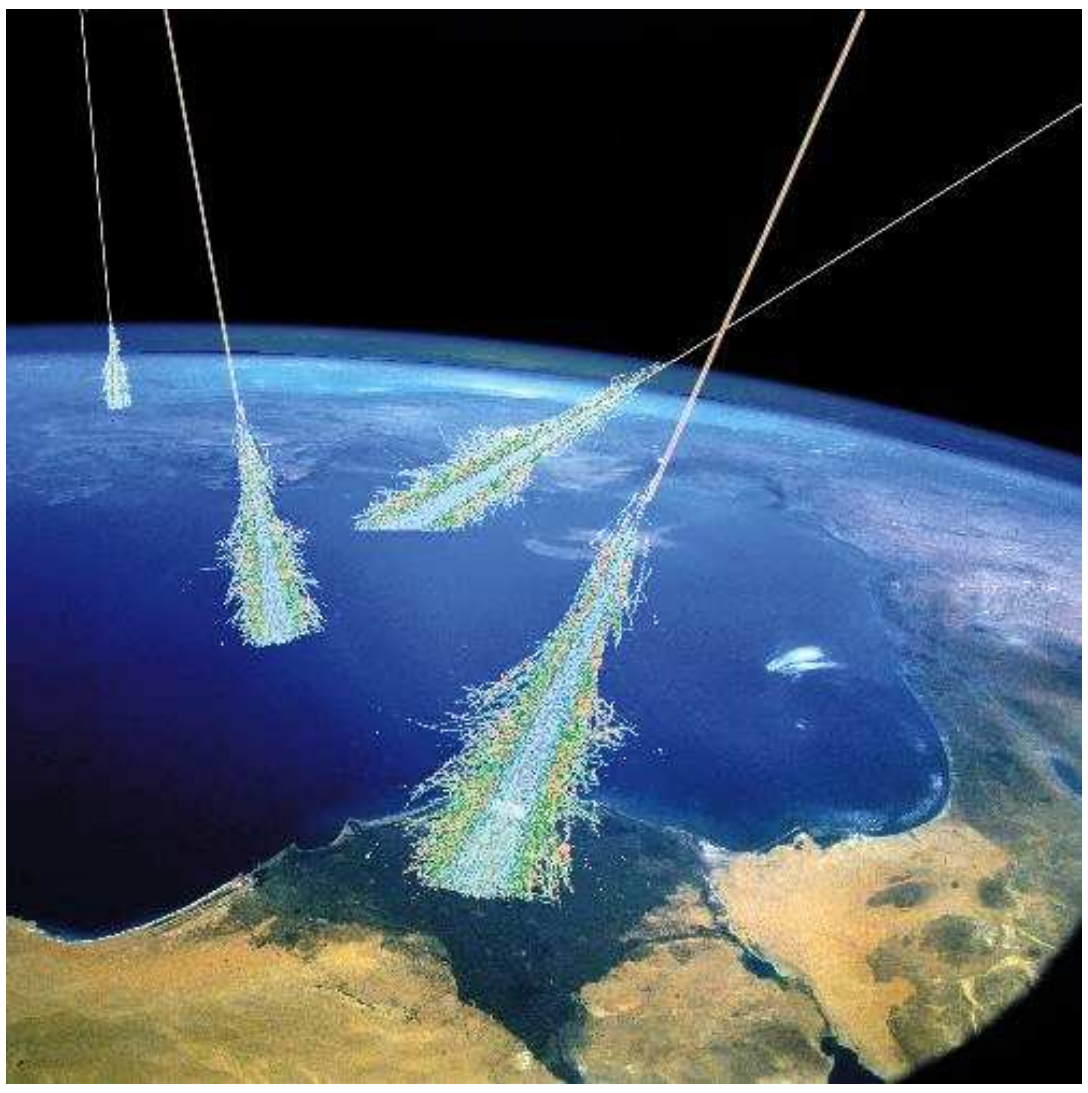

Figure B.14: When energetic cosmic rays strike the Earth's atmosphere avalanches of energetic secondary particles are produced that can be dangerous or lethal to some organisms. © Simon Swordy (University of Chicago), NASA.

Another mechanism is climate change. There is good evidence that the ions produced by CRs in the atmosphere support cloud formation. There is also an ongoing experiment at CERN, called "CLOUD", designed to study the seeding of clouds by CRs. More clouds will increase the planetary albedo, what in turn is believed to cool down the global climate. That a cooler climate is associated with a lower biodiversity is rather well documented: The highest diversity is found in the tropics, for example.

CR ionisation also triggers lightening discharges. This in turn affects the atmospheric chemistry due to the ozone and nitrogen oxides produced by lightening. Both are known to have harmful effects on animals, whereas the rainout of nitrogen oxide as nitric acid can be particularly destructive for life. On the other side, nitrogen oxides are also known to damage the protecting ozone layer. An increased solar UVB radiation can kill the phytoplankton, which is the base of most of the marine food chain.

So, a strong increase in CR flux may affect biodiversity in several ways. Medvedev & Melott do not advocate any particular mechanism and they emphasise that detailed research is required to quantify the effects of the mentioned mechanisms. Let us now turn to Medvedev & Melott's explanation of the 62-Myr cycle in fossil diversity.

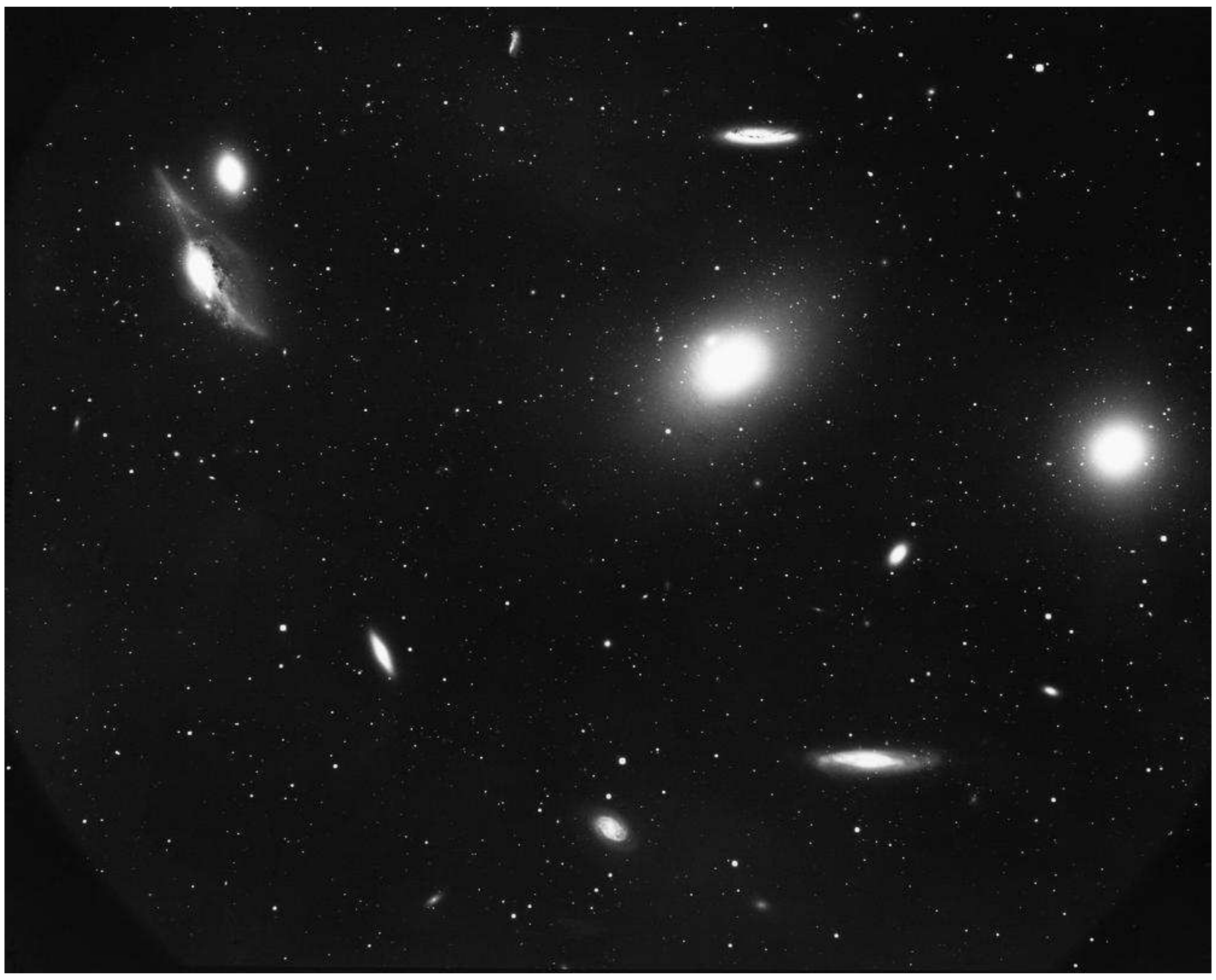

Figure B.15: The Virgo Galaxy cluster is the biggest mass concentration in our galaxy's neighbourhood and lies in the direction of the NGP. The bright elliptical galaxy near the centre is M86. Photograph taken with the 4-meter Mayall Telescope of Kitt peak National Observatory in 1974. © NOAO/AURA/NSF.

As the solar system orbits once in 250 Myr around the Galactic centre, it oscillates up and down, vertically to the Galactic disk. It has long been known that this vertical oscillation has a period of about 63 Myr and amplitude of circa 70 pc ($\sim$ 230 ly). This coincidence of the period with the 62-Myr cycle in fossil diversity is very interesting. Medvedev & Melott have noticed that times of major fossil diversity drop coincide with the times when the solar system was located north most in the Milky Way. The authors estimate that this coincidence has a chance of 1 in 10 000 000 to occur at random.

As the title of their paper already suggests, they consider extragalactic cosmic rays (EGCR) as a possible cause of the 62-Myr cycle in fossil diversity. CRs are most dangerous to the Earth biota at energies around a TeV because they and their secondaries have the largest flux in the lower atmosphere. Lower energy CRs are attenuated by the Earth's magnetosphere, whereas the flux of the higher energy particles rapidly decreases with energy. Near the Galactic mid-plane life on Earth is protected from those EGCR by the Galactic magnetic field. This shielding effect diminishes when the Sun is far away from the Galactic mid-plane. As mentioned before, an increased CR flux can cause all kinds of problems with the climate as well as directly damage the DNA of animals.

But why should there be more dangerous cosmic rays on the north side than in the south? What is particular about the northern Galactic hemisphere?

In the direction of the NGP lies the biggest mass concentration in our galaxy's neighbourhood: the Virgo Galaxy cluster. The image (Fig. B.15) shows a smaller section of the

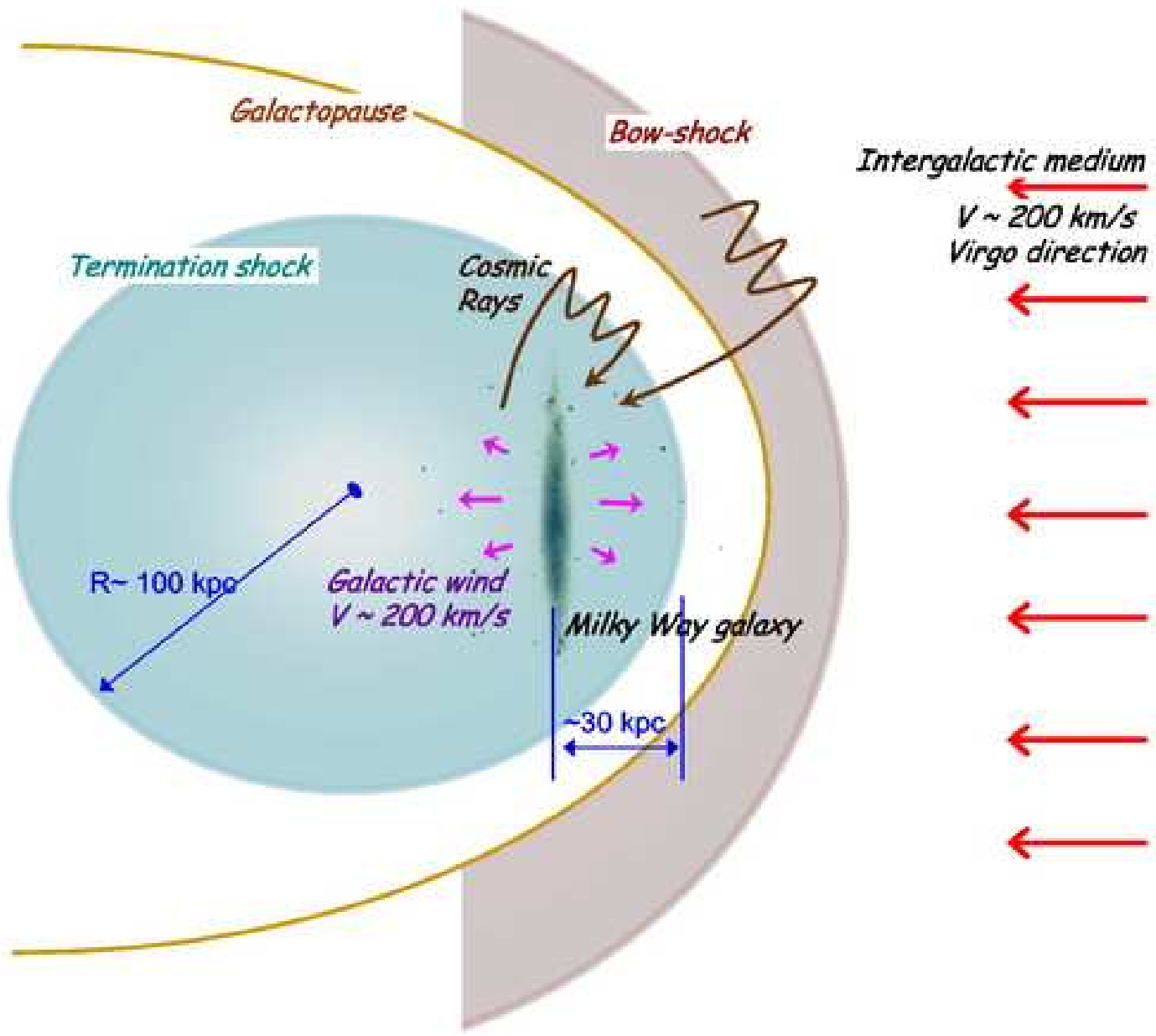

Figure B.16: Cartoon of the "galactosphere" reproduced from Medvedev and Melott (2007). The termination and bow shocks are sources of EGCR. Due to the asymmetry caused by the Virgo Cluster, which is nearly at the NGP, the cosmic-ray flux on the north side of the Milky Way Galaxy is larger than at the south side. © The American Astronomical Society.

Virgo cluster. The two bright elliptical galaxies are M86 and M84. The enormous mass of the Virgo cluster dominates our intergalactic neighbourhood and the Milky Way is falling with 200 km/s towards the Virgo cluster.

Let us look at Medvedev & Melott's Figure 1 (Fig. B.16 here) to illustrate how this motion of our Galaxy affects the global geometry of the "galactosphere". Like in a gigantic spaceship our Galaxy is plowing with 200 km/s through the intergalactic medium (IGM). This supersonic motion breaks the north-south symmetry of the galactosphere and pushes the center of the Galactic wind halo, indicated by the blue halo in the figure, well south of the Galactic disk. Like a ship that generates a bow wave when cruising the sea, it is believed that the Milky Way is headed by a huge bow-shock front. Medvedev & Melott expect this shock front to produce CRs energetic enough to make their way to the Galactic disk against the plasma wind.

This Galactic wind is produced by young massive stars and their deaths in form of supernova explosions, the so-called starburst wind. It consists of charged particles, mainly protons, having as well a supersonic velocity of about 200 km/s and it has blown away the intergalactic gas around our Galaxy. In the region where the pressure of the Galactic wind equals that of the intergalactic gas, the particles are slowed down abruptly. This produces a shock front, the so-called termination shock, much like the shock wave of a supersonic jet. According to the estimations of Medvedev & Melott, the northern part of the termination shock, which is closer to the Galaxy, constitutes a natural source of CRs with energies around the critical value of a TeV.

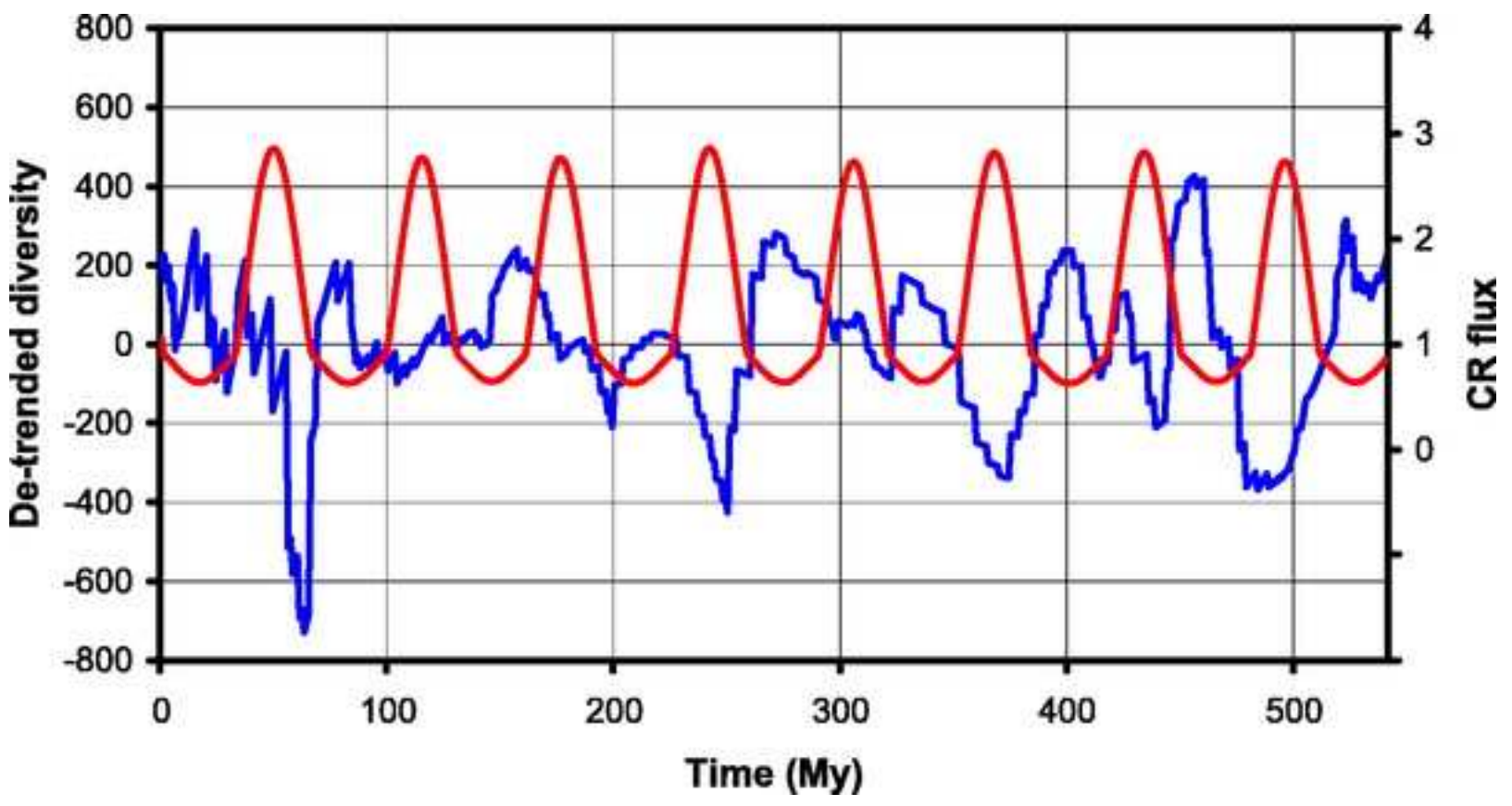

Figure B.17: Detrended diversity variation from Rohde and Muller (blue line) and EGCR flux at the Earth calculated from the Medvedev and Melott model (red line). Each maximum in CR flux coincides with a minimum of the diversity cycle within few Myr. Figure reproduced from Medvedev and Melott (2007). 

The termination and bow-shock are presumed to be the dominant sources of EGCRs reaching the Earth. Given the proximity of the Milky Way's north side to the two shock fronts, it becomes clear why the solar system would experience a maximum flux of CR whenever it is at its farthest northward excursion. At our present location – near the Galactic mid-plane – we are well shielded from EGCR due to the Galactic magnetic field. Therefore, it will be very difficult to observe the anisotropy in EGCR flux predicted by Medvedev & Melott, in particular because the observed flux is dominated by Galactic sources, such as supernova explosions. But let me come back later on to the observational predictions of Medvedev & Melott's model.

The next figure (Fig. B.17) shows again the detrended diversity from Rohde & Muller in blue and the computed EGCR flux on Earth versus time. For each CR maximum there is always a diversity minimum within few Myr. A number of statistical tests have been performed in order to address the significance of the correlation between fossil data and modeled CR flux. The tests show that the observed correlation is most probably not due to chance.

Now, the Sun does not only oscillate vertically, up and down in the disk, but also in radial direction, back and forth to the Galactic centre. This radial motion causes variations in the vertical amplitude, because the density of the Galactic disk increases towards the Galactic centre. A variation in amplitude causes, according to Medvedev & Melott's model, also a variation of the EGCR flux maxima. The radial oscillations of the solar system were included in the calculation, causing the modulation of the CR maxima that you can see in the figure.

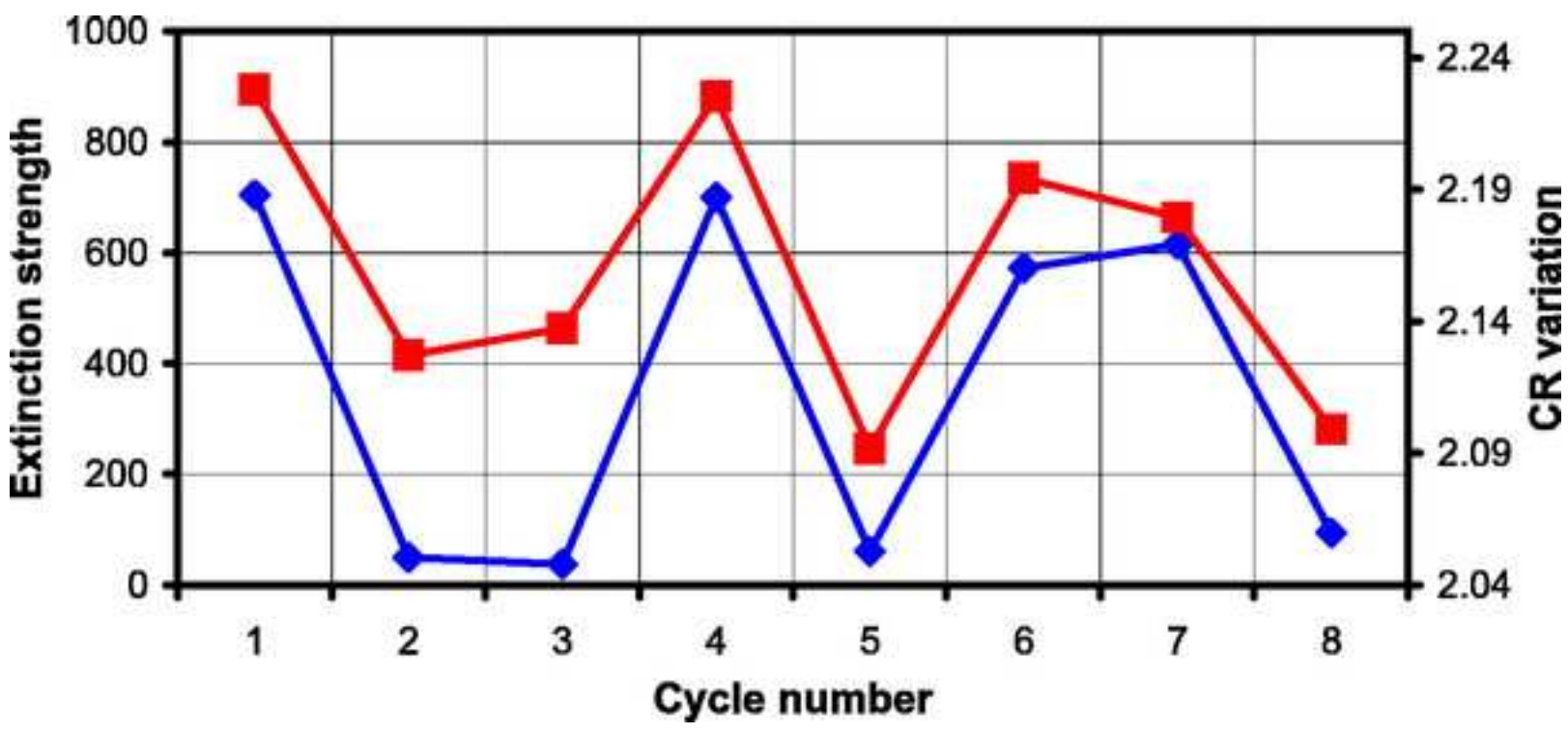

Figure B.18: Extinction strength (blue line) and the amplitude of the EGCR flux (red line). The correlation is about 93% and has a chance of less than 1 in 1 000 to occur at random. Figure reproduced from Medvedev and Melott (2007). © The American Astronomical Society.

It is now interesting to ask, whether the amplitude of the CR flux variation in each cycle is correlated to the magnitude of the corresponding drop of diversity. To answer this question, the authors define the 'extinction strength' as the difference between the minimum in diversity nearest to the corresponding CR maximum and its preceding diversity maximum. The magnitude of the CR flux variation is calculated analogously: Maximum flux minus preceding minimum flux.

The results are shown in this figure (Fig. B.18). It shows an impressively strong correlation between the amplitude of CR flux variation (red line) and the corresponding diversity drop (blue line) in each cycle. Medvedev & Melott write that "This provides a very solid and independent confirmation of the model, which provides a natural mechanism for observed cycles in fossil diversity."

The anisotropy in CR flux predicted by Medvedev & Melott's model could possibly be observed. The authors discuss various scenarios how to verify or falsify their hypothesis. The most promising among them is to look for an excess of pion decay products, i.e. gamma rays and neutrinos, from the north Galactic hemisphere. Pions are produced when CR protons interact with the interstellar medium in the Galaxy. The authors predict the emission of $\sim 2$ GeV photons with a flux of $\sim 1\%$ of the cosmic gamma-ray background at this energy. Detection of this excess emission seems a feasible task for *GLAST* (which is now called 'Fermi Gamma-ray Space Telescope'). Similarly, they predict the excess of $\sim 900$ MeV neutrinos which may be a good target for the *IceCube Neutrino Observatory* which is currently under construction at the South Pole.

Even if the work of Medvedev & Melott is merely a first step towards an explanation of the cycles in fossil diversity, I think it is a beautiful example of how the evolution of life on Earth may be related to astrophysical processes.

Thank you!

# Curriculum Vitae

Geboren wurde ich, Marco Longhitano, Bürger von Reinach (BL) und Bronte (CT, Italien), als Sohn des Giuseppe Longhitano und der Lucie, geborene Klenner, am 1. Juni 1981 in Basel (BS). Hier trat ich 1992 – nach dem Besuch der Primarschule Vogelsang – in das Gymnasium Kirschgarten (ehem. Realgymnasium) ein und legte dort im Sommer 2000 die Maturitätsprüfung Typus B ab. Im Wintersemester 2000/01 immatrikulierte ich mich an der Philosophisch-Naturwissenschaftlichen Fakultät der Universität Basel, wo ich im Hauptfach Theoretische Physik und in den Nebenfächern Experimentalphysik, Mathematik, Informatik und Astronomie belegte. Während des Schuljahres 2003/04 unterrichtete ich Mathematik und Informatik an der Oberstufe des Gymnasiums Kirschgarten. Im Laufe des Wintersemesters 2005/06 arbeitete ich unter Anleitung von Herrn PD Dr. Kai Hencken eine Diplomarbeit im Bereich der Kernphysik aus. Dazu hielt ich mich während drei Monaten am CERN bei Genf auf. Im Sommer 2006 schloss ich das Studium mit Erlangung des Diploms für Theoretische Physik ab. Anschliessend begann ich im August 2006 im Rahmen eines Projekts des Schweizerischen Nationalfonds unter der Anleitung von Herrn Prof. Dr. Bruno Binggeli an meiner Dissertation über die statistischen Eigenschaften weiter Doppelsterne zu arbeiten.

Meine Lehrerinnen und Lehrer an der Universität Basel waren folgende Professoren und Dozenten:

*Astronomie:* Bruno Binggeli, Roland Buser, Ortwin Gerhard, Eva K. Grebel, Gustav A. Tammann. *Theoretische Physik:* Andreas Aste, Wolfgang Belzig, Christoph Bruder, Kai Hencken, Thomas Heim, Daniel Loss, Thomas Rauscher, Friedrich-Karl Thielemann, Dirk Trautmann. *Experimentalphysik:* Hans-Joachim Güntherodt, Hans-Josef Hug, Jürg Jourdan, Bernd Krusche, Christian Schönenberger, Ingo Sick, Ludwig Tauscher†. *Mathematik:* Catherine Bandle, Norbert A'Campo, Marcus Grote, Lorenz Halbeisen, Hans-Christoph Im Hof, Wolfgang Reichel. *Informatik:* Helmar Burkhart, Martin Guggisberg, Olaf Schenk.

# Erklärung

Ich erkläre, dass ich die Dissertation *Wide Binary Stars in the Galactic Field* nur mit der darin angegebenen Hilfe verfasst und bei keiner anderen Universität und keiner anderen Fakultät der Universität Basel eingereicht habe.

Basel, den 3. September 2010

Marco Longhitano